\documentclass[a4paper,11pt,leqno,openbib,oldfontcommands]{memoir} 
\usepackage{subcaption}
\usepackage{afterpage}
\usepackage{caption}
\usepackage{nicefrac}
\usepackage{xcolor}
\usepackage{multirow}
\usepackage{etoolbox}
\usepackage{nonumonpart}

\usepackage{ifpdf}
\ifpdf
\fi
\ifdraftdoc 
	\usepackage{draftwatermark}				
	\SetWatermarkScale{0.3}
	\SetWatermarkText{\bf Draft: \today}
\fi
\newsubfloat{figure}
\newsubfloat{table}
\settrimmedsize{297mm}{210mm}{*}
\settypeblocksize{634pt}{448.13pt}{*} 
\setulmargins{4cm}{*}{*} 
\setlrmargins{*}{*}{1.5} 
\setmarginnotes{17pt}{51pt}{\onelineskip} 
\setheadfoot{\onelineskip}{2\onelineskip} 
\setheaderspaces{*}{2\onelineskip}{*} 
\checkandfixthelayout
\usepackage{fouriernc}
\usepackage[T1]{fontenc}
\OnehalfSpacing 
\setsecnumdepth{subsection} 
\maxsecnumdepth{subsubsection}
\usepackage{calc,soul,fourier}
\makeatletter 
\newlength\dlf@normtxtw 
\newsavebox{\feline@chapter} 
\newcommand\feline@chapter@marker[1][4cm]{%
	\sbox\feline@chapter{%
		\resizebox{!}{#1}{\fboxsep=1pt%
			\colorbox{gray}{\color{white}\thechapter}%
		}}%
		\rotatebox{90}{%
			\resizebox{%
				\heightof{\usebox{\feline@chapter}}+\depthof{\usebox{\feline@chapter}}}%
			{!}{\scshape\so\@chapapp}}\quad%
		\raisebox{\depthof{\usebox{\feline@chapter}}}{\usebox{\feline@chapter}}%
} 
\newcommand\feline@chm[1][4cm]{%
	\sbox\feline@chapter{\feline@chapter@marker[#1]}%
	\makebox[0pt][c]{
		\makebox[1cm][r]{\usebox\feline@chapter}%
	}}
\makechapterstyle{daleifmodif}{

	\renewcommand\printchapternum{\null\hfill\feline@chm[2.5cm]\par}

} 
\makeatother 
\chapterstyle{daleifmodif}
\makepagestyle{myvf} 
\makeoddfoot{myvf}{}{\thepage}{} 
\makeevenfoot{myvf}{}{\thepage}{} 
\makeheadrule{myvf}{\textwidth}{\normalrulethickness} 
\makeevenhead{myvf}{\small\textsc{\leftmark}}{}{} 
\makeoddhead{myvf}{}{}{\small\textsc{\rightmark}}
\newcommand{\clearemptydoublepage}{\newpage{\thispagestyle{empty}\cleardoublepage}}
\makeindex
\usepackage{import}

\usepackage{amsfonts} 					
\usepackage[centertags]{amsmath}			
\usepackage{stmaryrd}					
\usepackage{amssymb}					
\usepackage{amsthm}					
\usepackage{newlfont}					
\usepackage{layouts}					
\usepackage{graphicx}					
\usepackage{longtable,rotating}			
\usepackage[utf8]{inputenc}			
\usepackage{colortbl}					
\usepackage{wasysym}					
\usepackage{mathrsfs}					
\usepackage{float}						
\usepackage{verbatim}					
\usepackage{latexsym}					
\usepackage[square,numbers,
		     sort&compress]{natbib}		
\usepackage{url}						
\usepackage[spanish,english]{babel}		
\usepackage{color}                    				
\usepackage[colorlinks=true,
		     allcolors=black]{hyperref}              
\usepackage{memhfixc}					
\usepackage{enumerate}					
\usepackage{microtype}					
\usepackage{rotfloat}					
\usepackage{alltt}						
\usepackage{pgfplots}
\graphicspath{{figs/pgfs/}{figs/}}
\makeatletter
\def\input@path{{figs/pgfs/}}
\makeatother

\pdfpageattr{/Group << /S /Transparency /I true /CS /DeviceRGB>>}

\usepackage{lettrine}
\newcommand{\initial}[1]{%
	\lettrine[lines=3,lhang=0.33,nindent=0em]{
		\color{gray}
     		{\textsc{#1}}}{}}
\theoremstyle{plain}

\theoremstyle{plain}

\theoremstyle{plain}
\theoremstyle{definition}

\theoremstyle{plain}

\theoremstyle{plain}

\theoremstyle{plain}

\begin{document}
\frontmatter
\pagenumbering{roman}
\begin{titlingpage}
\begin{SingleSpace}
\calccentering{\unitlength} 
\begin{adjustwidth*}{\unitlength}{-\unitlength}
\vspace*{13mm}
\begin{center}
\rule[0.5ex]{\linewidth}{2pt}\vspace*{-\baselineskip}\vspace*{3.2pt}
\rule[0.5ex]{\linewidth}{1pt}\\[\baselineskip]
{\HUGE A First Principles Approach to the 100,000-year Problem}\\[4mm]
{\Large \textit{Examining the fundamental dynamics of Earth's glaciers\\over the
past 800,000 years}}\\
\rule[0.5ex]{\linewidth}{1pt}\vspace*{-\baselineskip}\vspace{3.2pt}
\rule[0.5ex]{\linewidth}{2pt}\\
\vspace{6.5mm}
{\large By}\\
\vspace{6.5mm}
{\large\textsc{Liam Wheen}}\\
\vspace{11mm}
\includegraphics[scale=0.6]{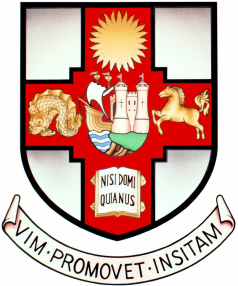}\\
\vspace{6mm}
{\large Department of Engineering Mathematics\\
\textsc{University of Bristol}}\\
\vspace{11mm}
\begin{minipage}{10cm}
A dissertation submitted to the University of Bristol in accordance with the requirements of the degree of \textsc{Doctor of Philosophy} in the Faculty of Engineering.
\end{minipage}\\
\vspace{9mm}
\large\textsc{February 2025}\\
\vspace{12mm}
\end{center}
\end{adjustwidth*}
\end{SingleSpace}
\end{titlingpage}

\clearemptydoublepage
\begingroup
\setlength{\beforechapskip}{-2.3cm} 
\chapter*{Abstract}
\endgroup
\begin{SingleSpace}
  \initial{T}he past 800,000 years of Earth's climate have been characterised by
  repeated transitions between glacial and interglacial states. These
  glacial-interglacial cycles have occurred with a dominant period of
  approximately 100,000 years. The 100,000-year problem refers to the apparent
  correlation between these cycles and the periodic variation in Earth's orbital
  eccentricity, which impacts the amount of solar radiation received by the
  planet. Although the period and phase of these two cycles are in agreement,
  it is thought that eccentricity does not vary the incoming solar radiation
  enough to directly drive a significant climate response. This thesis examines
  the fundamental dynamics governing Earth's climate over the past 800,000
  years, focussing on the 100,000-year problem.

  Two primary theories seek to explain the glacial-interglacial cycles. The
  astronomical theory suggests that orbital variations drive the cycles, with
  amplification from Earth system feedbacks. In contrast, the geochemical theory
  suggests that internal Earth system dynamics play the dominant role, with
  orbital forcing acting primarily to synchronise the oscillations. We explore
  these theories with a number of conceptual models that hold varying alignment
  with each theory. The Budyko energy balance model is originally a steady state
  system that illustrates how ice-albedo feedback can create climate
  bi-stability. Attempts to enhance the realism of this model proved
  unsuccessful in reproducing the 100,000 year period seen in the data,
  suggesting that the formulation of the model itself is the limiting factor. We
  then move to models explicitly designed to replicate ice volume dynamics.
  These employ switching mechanisms or highly non-linear dynamics to achieve the
  observed behaviour in the data. However, we show that linearised versions of
  these models perform comparably, suggesting that the data does not necessitate
  non-linear dynamics. We then present two simple linear models, one feedforward
  and one feedback, aligning with the astronomical and geochemical theories
  respectively.
  
  Our feedforward model is able to reproduce the ice volume data well, providing
  a novel explanation for the absence of eccentricity's 400,000-year period in
  the ice volume record. Through a physical interpretation of the
  phenomenological model, we suggest the omission of this period could arise
  from the interaction of two Earth system components, oceanic heat storage and
  tropospheric energy, which respond to eccentricity with different phase lags.
  Using conservative parameter estimates, we show that bulk ocean temperature
  variation over the past 800,000 years can be sufficiently explained by
  eccentricity forcing alone, challenging the geochemical theory's core
  assumption that eccentricity is too weak to drive a significant climate
  response. Our findings also demonstrate that the widespread use of $Q_{65}$
  as a forcing function may unintentionally bias models toward geochemical
  explanations by under representing eccentricity's influence. The feedback
  model is found to improve upon the fit achieved by the
  feedforward model. However, this improvement is largely confined to the
  interglacial period labelled as Marine Isotope Stage 11. This suggests that,
  rather than indicating a general requirement for feedback mechanisms
  throughout the glacial cycles, the anomalous behaviour during this
  interglacial may be explained by specific Earth-based events, such as enhanced
  volcanic activity.

  We find that the past 800,000 years of glacial-interglacial cycles can be
  largely reproduced by a linear astronomical forcing model, challenging the
  necessity of more complex Earth system feedbacks. We therefore cannot rule out
  the astronomical theory from the ice volume data alone, and emphasise the
  importance of minimising model complexity when interpreting limited
  palaeoclimate data.

\end{SingleSpace}
\clearpage

\clearemptydoublepage
\chapter*{Dedication and acknowledgements}
\begin{SingleSpace}
  \initial{T}here are many people without whom this thesis would be either unfinished or
unreadable. I would first like to thank my supervisor, Oscar Benjamin, who
provided an immeasurable amount of technical guidance, along with supportive
advice during our meetings at The Robin Hood. I would also like to thank the
other academics that helped shape this project and provided insightful
perspectives from their own areas of expertise: Cameron Hall, Jerry Wright, and
Thomas Gernon.

Academic support is only half of the requirements for a bearable PhD experience,
and I would like to express my deepest thanks to those around me that provided
the other half. My mum and dad have been a constant source of support and
encouragement, keeping me optimistic about the end goal and routinely asking
when I would be graduating from around second year. My brother also deserves a
mention as the family member who was most willing and able to listen to the
uninteresting details of how the PhD was going, and also for sorting present
ideas for mum, dad, and himself for the entire duration. The person who has
possibly been the biggest contributor to a balanced and happy life during the
PhD is my girlfriend, Ellie. She has celebrated all the little breakthroughs and
commiserated all the many setbacks, never once complaining about the midday
starts and late nights that I often adopted. Love you. 

Finally, I would like to thank all of the friends from Maggs house and beyond
who made each day a delight, with quiz lunches, after-work pints, and 3 hour
debates about nothing. Kit Simmonds, Paul Fuchter, Ben Warmington, Dan Marris,
John Moloney, Toby Kay, Mark Blyth, Will Leeney, Ricky Hunter, James Cass, Dan
Stocks and Pip Sears, Will Simpkins and Bridget Odette, to name a few.
Honourable mentions go to Sophie Landon for teaching me that I don't like Go,
Stasiu Biber for teaching me everything about the university, and Elena Fillola
for teaching me that temporary tattoos are still cool. Thank you all for making
the last four years a pleasure.

\end{SingleSpace}
\clearpage

\clearemptydoublepage
%
%
%
%
%
%
%
\chapter*{Author's declaration}
\begin{SingleSpace}
\begin{quote}
\initial{I} declare that the work in this dissertation was carried out in accordance with the requirements of  the University's Regulations and Code of Practice for Research Degree Programmes and that it  has not been submitted for any other academic award. Except where indicated by specific  reference in the text, the work is the candidate's own work. Work done in collaboration with, or with the assistance of, others, is indicated as such. Any views expressed in the dissertation are those of the author.

\vspace{1.5cm}
\noindent
\hspace{-0.75cm}\textsc{SIGNED: .................................................... DATE: ..........................................}
\end{quote}
\end{SingleSpace}
\clearpage
\clearemptydoublepage
\renewcommand{\contentsname}{Table of Contents}
\maxtocdepth{subsection}
\vspace*{-3.1cm}
\tableofcontents*
\addtocontents{toc}{\par\nobreak \mbox{}\hfill{\bf Page}\par\nobreak}
\clearemptydoublepage
\listoftables
\addtocontents{lot}{\par\nobreak\textbf{{\scshape Table} \hfill Page}\par\nobreak}
\clearemptydoublepage
\vspace*{-2.8cm}
\listoffigures
\addtocontents{lof}{\par\nobreak\textbf{{\scshape Figure} \hfill Page}\par\nobreak}
\clearemptydoublepage
%
%
\mainmatter
\chapter{Introduction}
\label{chap:intro}
\initial{T}he Pleistocene epoch, spanning the past 2.6 million years, is
characterised by periodic transitions between glacial and interglacial states
\cite{benthic_data}. These cycles are well-documented through proxy data, yet
their underlying mechanisms remain unresolved \cite{review_paper_2_theories}.
Around one million years ago, the dominant period of these cycles shifted from
41 thousand years (kyr) to 100\,kyr \cite{mpt_review}, a phenomenon known as the
mid-Pleistocene transition, which presents a key challenge in palaeoclimatology
and is often referred to as the 100,000-year problem \cite{imbrie_inertia}. This
problem arises from the apparent correlation between the glacial cycles and
Earth’s orbital eccentricity, despite the latter’s relatively weak influence on
incoming solar radiation \cite{high_prec_reduces_ecc_impact}.

Earth's orbital configuration varies periodically, with prominent periods of 41
and 100\,kyr, affecting the magnitude and distribution of solar energy on Earth
\cite{astronomical_theory}. Two primary theories seek to explain the
relationship between these orbital parameters and the glacial-interglacial
cycles. The astronomical theory suggests that orbital parameters drive the
cycles, with amplification from Earth system feedbacks
\cite{orbital_freqs_without_tuning}. In contrast, the geochemical theory
suggests that internal Earth system dynamics play a dominant role, with orbital
forcing acting primarily to synchronise the oscillations
\cite{saltzman_intrinsic_first}.

This thesis will explore these theories through the use of previous conceptual
models, and two of our own, that allow for a clear and systematic analysis of
the dynamics. Understanding the fundamental mechanisms that are at play is
crucial for our broader comprehension of the climate system and its response to
external forcing.

\section{Background}
\begin{figure}
    \centering
    \input{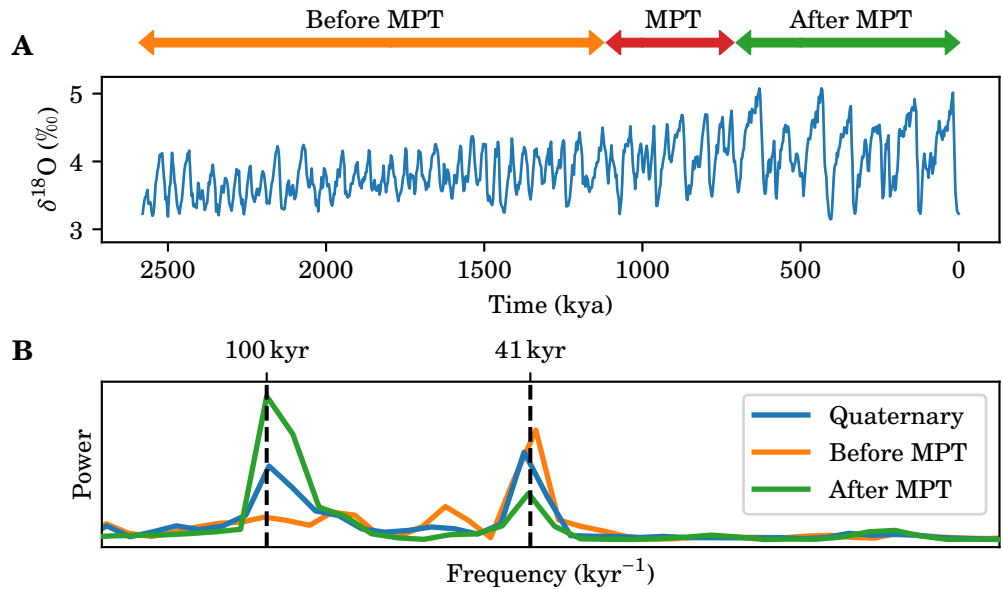}
    \caption[Benthic $\delta^{18}$O]{Time series (\textbf{A}) and power spectra
      (\textbf{B}) for the benthic $\delta^{18}$O stack from Lisiecki and Raymo
      \cite{benthic_data}. This data covers from present to 2600 thousand years
      ago (kya). The benthic $\delta^{18}$O ratio is commonly used as a proxy
      for global ice volume. The Mid-Pleistocene Transition (MPT) spans
      approximately 550\,kyr and marks a distinct change in dominant period,
      from 41\,kyr to 100\,kyr, as shown in the power spectra.}
  \label{fig:benth_and_power_specs}
\end{figure}

Earth's ice volume history has been reconstructed by many researchers using a
range of models and proxy data. A combined average of one of these proxies, the
benthic $\delta^{18}$O stack from Lisiecki and Raymo \cite{benthic_data}, is
shown in Figure \ref{fig:benth_and_power_specs}\textbf{A}. This data is commonly
used as a proxy for global ice volume \cite{shack_bwt_eq} and is explained fully
in Chapter \ref{chap:data}. The data spans the past 2.6 million years, known as
the Quaternary period. This period encompasses the Pleistocene epoch, which
started 2.6 million years ago, but ended 11.7 thousand years ago (kya) at the
start of the current interglacial period \cite{quaternary_definition}. The past
11.7\,kyr is known as the Holocene epoch and marks a period of relatively stable
climatic conditions and the development of human civilisation
\cite{holocene_definition}.

The Mid-Pleistocene Transition (MPT) is a key event in Earth's climatic history
that occurred around 1 million years ago, delimited by the red arrows in Figure
\ref{fig:benth_and_power_specs}\textbf{A}. This transition marks a change in the
dominant period of the glacial-interglacial cycles from around 41\,kyr to around
100\,kyr. This is more clearly shown in the power spectra in Figure
\ref{fig:benth_and_power_specs}\textbf{B}, where the power associated with the
two periods drastically shifts after the MPT.

The underlying cause of the MPT remains unclear. One hypothesis is that it
resulted from glacial erosion gradually removing a sediment layer beneath the
glaciers~\cite{mpt_cause_regolith}. This regolith would have facilitated more
rapid ice growth and decay, leading to the 41\,kyr cycles seen in the data.
However, once this layer was sufficiently eroded, the ice volume dynamics
transitioned to the 100\,kyr cycles seen today. This hypothesis sometimes
includes a decrease in atmospheric CO$_2$ as a contributing factor to the
MPT~\cite{mpt_cause}.

Another hypothesis suggests that the MPT was driven by changes in Earth's
orbital parameters. Eccentricity is currently at a minimum of its 400\,kyr cycle
and was at a similar minimum 800\,kyr ago, as shown in Figure
\ref{fig:orbital_and_benthic_time_series_power_specs}. Imbrie showed that the
MPT can be reproduced with a model using only the orbital parameters as input,
proposing that this shift was due to a non-linear interaction between the
orbital parameters and the Earth system \cite{imbrie2011}. The MPT has been
reproduced in a number of orbitally forced models, such as the one from Paillard
and Parrenin, however this required a slowly varying parameter to be fit, once
again including the decrease in atmospheric CO$_2$ as a contributing factor
\cite{pp04}. Both of these models are discussed in detail in Section
\ref{sec:conceptual_models}

Since the Mid-Pleistocene Transition (MPT) is thought to represent a fundamental
regime shift in the Earth system, we consider the post-MPT climate system to be
dynamically distinct from its pre-MPT counterpart. For this thesis, we focus
primarily on ice volume dynamics following the MPT, as these post-MPT dynamics
are more directly relevant to the current climate state. By examining this
period, we aim to gain deeper insights into the mechanisms driving the
glacial-interglacial cycles that characterise the modern climate system.

The 100,000-year problem arises from the observation that the dominant period of
the glacial-interglacial cycles in the post-MPT period aligns with the period
and phase of Earth's orbital eccentricity. However, the small magnitude of
eccentricity's influence on the total energy reaching Earth has led to 
ongoing debate over its role in driving these cycles. This has given rise to two
main theories. The first proposes that the weak orbital forcing is amplified by
internal Earth system processes, and is often referred to as \textit{The
Astronomical Theory of Climate Change} \cite{astronomical_theory}. The opposing
theory posits that the glacial-interglacial cycles are primarily driven by
feedbacks within the Earth system, with the orbital parameters merely setting
the phase of the oscillations. This is known as \textit{The Geochemical Theory
of Climate Change} \cite{review_paper_2_theories}.

Earth's orbital configuration can be defined by three parameters; eccentricity,
obliquity, and precession, which modulate the magnitude and distribution of
incoming solar radiation, known as insolation. Eccentricity is the shape of
Earth's orbit around the Sun and varies with a period of around 100\,kyr as well
as a second period of around 400\,kyr. Obliquity is the tilt of Earth's axis and
varies with a period of around 41\,kyr. Precession is the orientation of Earth's
axis relative to the sun and varies with a period of around 23\,kyr. These will
be discussed in more detail in Section \ref{sec:orbital_parameters}.

In 1842, Adh{\'e}mar first suggested that precession would result in the
hemisphere pointed towards the sun experiencing warmer conditions than the other
hemisphere~\cite{original_orbital_data}. They proposed that when one hemisphere
is maximally angled away from the sun, its conditions are cool enough to allow
ice to accumulate, leading to glaciation. To support this theory, they produced
orbital calculations that showed the precession parameter modulates the
intensity of the seasons, with the effects on each hemisphere being opposite.

Croll extended this work by including the effects of all three orbital
parameters and introducing the idea that Earth system feedbacks may also
contribute to the glacial-interglacial cycles~\cite{croll_before_milanko}. Given
that these theories revolve around the uneven distribution of insolation between
hemispheres, it is reasonable that both concluded the ice sheets would vary
asynchronously across the hemispheres. However, with the advent of more reliable
data, Milankovitch was able to show that the glacial periods occur
simultaneously around the globe, with ice volume variations being more
pronounced in the northern hemisphere. This led them to suggest that a reduced
intensity in northern summer insolation plays an important role in the onset of
glacial periods~\cite{milankovitch}.

\section{Literature Review}
\label{sec:literature_review}
Since Milankovitch, theories of Earth's ice volume dynamics have bifurcated.
We will now explore key literature that has shaped our understanding of Earth's
ice volume dynamics, focusing on how these dynamics align with the two primary
schools of thought: the astronomical theory and the geochemical theory. The next
two sections will outline the history of these theories and how they have
developed through to the start of the 21st century, at which point the two
theories start to overlap and become more intertwined. This leads us onto the
final section where we introduce six models from mostly recent studies that we
will be analysing in this thesis. These models are chosen to reflect the two
theories to differing extents and will inform the development of our own two
models.

\subsection{Astronomical Theory}
\label{sec:astronomical_theory}
As mentioned, two of the pioneers of the astronomical theory were Croll and
Milankovitch. The changes to Croll's theory that Milankovitch proposed are now
considered more accurate and so we will outline this theory in more detail.

Milankovitch used Adh{\'e}mar's orbital calculations to model insolation over
time and estimated the temperature changes that this may produce
\cite{milankovitch}. They suggest that it is the intensity of summer insolation
in the high northern latitudes that triggers glacial periods. As is discussed
in Section \ref{sec:precession}, the precession parameter modulates the duration
and intensity of the seasons, with the effects on each hemisphere being
opposite. At one extreme of precession, the northern hemisphere experiences a
long and mild summer. K{\"o}ppen and Wegener proposed that this mild summer does
not sufficiently melt the ice accumulated during the winter, leading to a net
increase in snow and ice volume each year during this period
\cite{milankovitch_support}. Milankovitch and others suggest that the increase
in Earth's surface albedo as a result of this growing snow and ice cover further
reduces the amount of solar radiation absorbed by the Earth, leading to a
positive feedback loop that amplifies the initial cooling. This albedo feedback
mechanism is examined more closely by Budyko, discussed in Section \ref{sec:budyko}.

In 1969, Budyko \cite{budyko}, and contemporaneously Sellers
\cite{orbital_not_enough}, employed a model based approach to investigate the
role of ice albedo in amplifying the effect of Earth's orbital parameters on the
glaciers. Budyko found that the current extent of Earth's glaciers are stable,
but if enough forcing were to bring them to a critical point, the ice-albedo
feedback could cause runaway ice growth. Sellers found that the Earth's surface
temperature would be too high for ice to accumulate if it were not for the
ice-albedo feedback. These models both support the astronomical theory, however
they found the climatic response to orbital forcing to be too weak to explain
the dynamics seen in the data. However, these models have been criticised for
over simplifying important physical processes such as temperature diffusion
\cite{astronomical_theory}. 

In 1976, Hays, Imbrie, and Shackleton modelled the insolation-climate
system as linear and explored whether Earth's ice volume dynamics could be
explained by the astronomical theory~\cite{orbital_freqs_without_tuning}. They
found evidence that the presence of precession and obliquity frequencies in the
ice volume data can be explained by assuming the climate system responds
linearly to orbital forcing. However, they suggest that the correlation between
eccentricity and ice volume likely requires an assumption of non-linearity. In
contrast to this explanation for the eccentricity frequency in the ice volume
data, some researchers suggest it could instead emerge from multiples of the
obliquity periods \cite{obliquity_making_100kyr,obliquity_subharmonics}, or
combinations of obliquity and precession \cite{no_ecc_just_bet_and_rho}. In
Chapter \ref{chap:feedforward_model}, we demonstrate that even eccentricity
could be described as having a broadly linear effect on ice volume dynamics.

Shortly after, in 1978, Berger produced a more accurate model of Earth's orbital
parameters over the past million years and used this to develop a versatile
insolation model~\cite{early_insol_calculation}. This model is able to produce
insolation solutions for different months and latitudes, as well as forecasting
into the future. Berger was able to demonstrate that summer solstice insolation at
65$^\circ$ north is not unique in its predictive power of the
glacial-interglacial cycles, with a minimum in daily insolation during the autumnal
equinox coinciding with the last glacial maximum, and the following peak in
insolation coinciding with the post-glacial warming that led to our current
climate. This raises the question of whether the northern latitude summer
solstice insolation is the best predictor of the glacial-interglacial cycles, or
if it is instead a less straightforward function of the orbital parameters. We
explore this further in Section \ref{sec:insolation}.

In 1992, Imbrie et al. used Berger's insolation model to revisit the topic of
linearity between the orbital parameters and ice volume \cite{imbrie_inertia}.
In their comprehensive study of the available proxy data, they found evidence to
support the linear assumption for obliquity and precession from Hays et al., but
not for eccentricity. They also identify that there is a small phase lag
($\approx 1$\,kyr) between the two orbital parameters and the relevant
frequencies in the ice volume data. This is due to physical processes taking
place after insolation reaches Earth's surface. They propose that northern
summer insolation is still the main driver, however this is directly acting more
on the arctic ocean than the glaciers. The arctic ocean is then warmed, melting
sea ice from winter and feeding into the southern hemisphere via the North
Atlantic Deep Water current. This is a key result as it extends the astronomical
theory, which originally proposed ice albedo feedback as the main Earth system
mechanism contributing to the glacial-interglacial cycles. By adding components
like ocean circulation, sea ice dynamics, and CO$_2$, they opened up a new
avenue for research into the Earth system model, bridging the gap between the
astronomical and geochemical theories.

In 1999, Berger and Loutre used a model to investigate the role of Earth's
orbital parameters in different CO$_2$
regimes~\cite{linear_co2_supports_orbital}. They were able to produce key
features of the ice volume data, such as the growing ice volume variation
post-MPT, and the transition from a dominant period of 41\,kyr to 100\,kyr. This
was achieved by using a linearly decreasing CO$_2$ concentration, based on
average measured values, over the past 3000\,kyr. They were also able to show
that orbital forcing alone can generate the glacial-interglacial cycles, however
this is not possible if CO$_2$ concentration was held constant, suggesting again
that internal dynamics in the Earth system play an important role in the
glacial-interglacial cycles.

The astronomical theory has two commonly cited limitations. Firstly,
eccentricity is the only parameter that oscillates with the 100\,kyr period seen
in the ice volume data. However, as we show in Section \ref{sec:eccentricity}, it
can only vary the magnitude of total annual insolation by 0.2\%. This has led
many to suggest that an Earth based amplification mechanism is necessary in
order to explain the effect of
eccentricity~\cite{imbrie_inertia,saltzman,nonlinear_amplifier}. Another
limitation of the astronomical approach relates to the second prominent
frequency of eccentricity. As shown in
Figure~\ref{fig:orbital_and_benthic_time_series_power_specs}, the eccentricity
signal also contains a period of 400\,kyr. This is not clearly discernible in
the ice volume data, leading some studies to suggest that the 100\,kyr period
instead results from the interplay of obliquity and precession
\cite{obliquity_subharmonics,intrinsic_no_eccentricity,no_ecc_just_bet_and_rho}.

\subsection{Geochemical Theory}
The geochemical theory of climate change emphasises that internal Earth system
dynamics drive the glacial-interglacial cycles. The theory acknowledges that the
orbital parameters set the phase of the cycles, but suggests that the Earth
system is capable of producing the cycles independently of external forcing.
Hypotheses aligned with the geochemical theory can vary significantly, as it
encompasses a wide range of internal Earth system mechanisms proposed to cause
the planet's intrinsic oscillations. Some examples of these mechanisms are: the
interactions between the ice sheet and
bedrock~\cite{intrinsic_precession_freq_diff}, the atmospheric response to sea
ice \cite{sea_ice_glacial_cycles}, and dust impacting Earth's surface
albedo~\cite{intrinsic_dust}. However, most geochemical hypotheses will include
the carbon cycle to some degree. This is because it can significantly impact a
wide range of Earth system processes and is closely linked to the Earth's
surface temperature \cite{carbon_cycle_effect}.

As early as 1824, Fourier proposed that the Earth's surface temperature is
regulated by the atmosphere in some way~\cite{fourier1824_greenhouse}, though it
was Babbage in 1847 who more formally proposed what we now call the greenhouse
effect \cite{greenhouse_origin}. Babbage had not yet identified gases such
as CO$_2$ and CH$_4$ (methane) as contributing to the greenhouse effect,
focussing only on water vapour as a mechanism. Dumas established the first
notion of a carbon cycle in 1842, primarily focussed on the role of living
organisms in the exchange of carbon between the ground and the atmosphere
\cite{carbon_cycle}. The role of the carbon cycle in the Earth's climate system
was demonstrated by Arrhenius in 1896, who concluded that a doubling of
atmospheric CO$_2$ would lead to a $\sim6^\circ$C increase in global
temperature via the greenhouse effect~\cite{first_carbon_calculation}.

By 1925, a more complete picture of the carbon cycle was established by
Lotka~\cite{full_carbon_cycle}. They show that the carbon cycle is made up of
biological, physical, and chemical processes that exchange carbon between the
atmosphere, ocean, and land. This was further developed by Revelle and Suess in
1957, who showed that the ocean is a able to absorb a large amount of
atmospheric CO$_2$ \cite{ocean_absorb_carbon}. They also demonstrated that the
proportion of CO$_2$ that the ocean can store is dependent on its temperature,
with a warmer ocean being less effective at absorbing CO$_2$. This is a key
result as it provides a feedback mechanism whereby atmospheric CO$_2$ can affect
the Earth's surface temperature, and vice versa. A number of geochemical models
have been developed to explore the effect of ocean carbon storage and release on
the Earth's climate \cite{saltzman_intrinsic_first,saltzman_intrinsic,pp04}.

In 1982, Saltzman looked at the effect of stochastic climate variations
on the ice volume dynamics \cite{saltzman_stochastic}. The source of these
variations is not certain, but they suggest it could be volcanic activity
changing the chemical composition of the atmosphere. They provide two key models
of the climate. One assumes the ice volume to oscillate about a stable
equilibrium with either fast or slow damping, the cause of the oscillations in
this case is the orbital and stochastic forcing. The other model assumes the ice
volume to oscillate around an unstable equilibrium, unable to settle. This model
includes negative feedbacks that increase their effect as the ice volume
deviates from the equilibrium, resulting in a limit cycle. Saltzman finds that
the presence of stochastic forcing in this model can lead to both irregular and
quasi-periodic variations as seen in the ice volume data. This is an interesting
branch of the geochemical theory as it does not directly rely on an oscillatory
mechanism to produce the glacial-interglacial cycles, rather the cycles emerge
from an unstable system being constantly perturbed as well as tempered by the
Earth system. This has been further developed by Mitsui and Aihara 2013 who
analysed a number of conceptual models and found most of them to exhibit a
strange nonchaotic attractor \cite{chaos_in_glacial_models}. This means the
models produce complex quasi-periodic dynamics that are not highly sensitive to
the initial conditions. They propose this balance between chaotic and orderly
dynamics is representative of what we see in the data.

In 2000, Shackleton analysed the isotope ratio ($\delta^{18}$O) from ocean
sediment cores, similar to that shown in Figure \ref{fig:benth_and_power_specs},
as well as the same isotope ratio from trapped air bubbles in an Antarctica ice
core \cite{shack_ice_lags_co2}. They were able to separate the contributions of
ice volume and deep ocean temperature to the ocean sediment core, which is
commonly used as a proxy for ice volume. They found that deep ocean temperature
has a significant contribution to the signal, and that ice volume lags behind
the signals from CO$_2$, Antarctic air temperature, and deep ocean temperature,
which are all in phase with eccentricity. They suggests that the 100\,kyr period
in the ice volume data is not directly driven by eccentricity, but rather by the
global carbon cycle, which is in turn influenced by eccentricity.

In 2005, Bintanja reproduced the $\delta^{18}$O data shown in Figure
\ref{fig:benth_and_power_specs} using a coupled model of ice volume and deep
ocean temperature, with the aim of isolating the contribution of each. In
agreement with Shackleton, they found that contributions from ice volume can
vary between 10 and 60\% of the total signal, meaning that deep ocean temperature
plays a significant role in the $\delta^{18}$O data. Although this could be seen
to discredit the $\delta^{18}$O data as a reliable proxy for ice volume, Figure
\ref{fig:ice_part_of_O18} shows that the ice volume component of the
$\delta^{18}$O data still closely resembles the original data after a linear
rescaling, explaining 96\% of the variance. The same is true for the deep ocean
temperature component, which explains 90\% of the variance. This suggests that
the two components covary with the original data, reducing the degree to which
ocean temperature can be seen as a confounding factor in the $\delta^{18}$O
data. Nevertheless, we will use this isolated ice volume component in our
analysis of key conceptual models in Section \ref{sec:conceptual_models}.

There are both theoretical and practical limitations to the geochemical school
of thought. The geochemical theory does not offer a clear explanation of how the
100\,kyr period in the ice volume data is phase locked to eccentricity, this is
due, in part, to wide range of mechanisms proposed to cause the intrinsic
oscillations. A key practical limitation of the theory is that geochemical
processes are generally more complex than the amplification mechanisms proposed
within the astronomical theory. This often means that for a geochemical model to
produce Earth-intrinsic oscillations, it will either require a large number of
variables and parameters, or a large number of simplifying assumptions.

\subsection{Ice Volume Models}
\label{sec:intro_ice_volume_models}
We will now introduce some significant ice volume models and place them in the
context of the aforementioned theories. In Chapters \ref{chap:models}, we examine
them more closely, reproducing their results and comparing their dynamic
behaviour.

This thesis primarily focusses on conceptual models as they are the simplest
form of climate model. These take a heavily simplified approach to modelling the
Earth system, usually relying on fewer than 4 state variables. The benefit of
this simplicity is that we can clearly present and understand the dynamics of
the model in an analytical form. On the opposite end of the complexity spectrum,
there are Earth System Models (ESMs) that aim to capture the full range of
interactions between the atmosphere, ocean, and cryosphere. These often comprise
two sub-models; a general circulation model, which uses the Navier–Stokes
equations to model the atmosphere and ocean, such as \textit{CCSR/NIES AGCM5.4}
\cite{abe_ouchi_gcm}, and an ice sheet model, which models ice sheet dynamics in
three dimensions, preserving mass and momentum, such as
\textit{IcIES}~\cite{abe_ouchi_ice_model}.

These complex models have been used to attain a more realistic geographical
representation of the Earth system \cite{abe_ouchi_full_model}, though can be
computationally expensive and difficult to interpret. A compromise between these
two extremes is the use of intermediate complexity models, such as the
\textit{LLN-2D} model, from Gall{\'e}e et al. \cite{intermed_complex_model},
which was used by Berger for their study outlined in Section
\ref{sec:astronomical_theory}. An intermediate complexity model allows for a
somewhat realistic representation of the Earth system, with spatial
representation of atmospheric CO$_2$ and insolation, whilst still being more
computationally efficient and interpretable than ESMs.

Despite the benefits of more complex modelling approaches, the prevailing class
for palaeoclimate studies is the conceptual model, due to the computational and
interpretative simplicity. We will now introduce some key conceptual models that
have been developed to simulate ice volume dynamics. They are chosen to
represent both the astronomical and geochemical theories, as well as a range of
complexity, within the conceptual model class.

In 1969, Budyko modelled the Earth's annually averaged surface temperature as a
function of latitude and time \cite{budyko}. This is referred to as an energy
balance model as it comprises a simple representation of the Earth's incoming
and outgoing energy, through insolation and outgoing radiation respectively.
There is also a heat transport term that redistributes energy from the warmer
equator to the poles. This form of model has been studied many times since, with
some proposing augmentations to improve its physical representation
\cite{budyko_with_diffusion,budyko_with_diffusion_2,insol_latlon}. Budyko only
analysed their model in an equilibrium context, setting the rate of change of
the surface temperature to zero. This was used to demonstrate that a critical
point exists in the model, where positive feedbacks from ice-albedo can lead to
a runaway ice growth or ablation. An important augmentation was made by Widiasih
in 2013, who included a dynamic ice line with a much slower rate of change than
the surface temperature \cite{widiasih2013}. This improved the model's
representation of the Earth system and allowed it to be dynamically analysed. We
discuss this model in more detail in Section \ref{sec:budyko}.

A common measure used to represent orbital forcing is the daily averaged
insolation on the summer solstice at 65$^\circ$ north, known as $Q_{65}$. This
stems from Milankovitch proposing the northern summer intensity to be the main
driver of the glacial-interglacial cycles.

In 1980, Imbrie and Imbrie modelled the change in ice volume as proportional to
$Q_{65}$, supporting the astronomical theory~\cite{imbrie}. Realising that the
rate of ice growth is slower than its recession, they included a conditional
time constant that switches to capture the two rates.

In 2004, Paillard and Parrenin proposed a more complex system that covered a
longer timespan and was able to reproduce the MPT with the inclusion of a tuned
time-varying parameter~\cite{pp04}. This model also took $Q_{65}$ as an input
and included a switching mechanism that depended on the direction of ice volume
change. The model uses a variable for deep-water stratification, which depends
on ice volume induced changes in salinity. This feeds into a variable for
atmospheric CO$_2$ which, in turn, provides feedback to the ice volume variable.
This model is therefore capable of producing unforced oscillations and so aligns
with the geochemical theory.

The 2011 model from Imbrie, Imbrie-Moore, and Lisiecki focussed on the interplay
between the orbital parameters and how this affects ice volume, again switching
regimes depending on the direction in which the ice is
changing~\cite{imbrie2011}. Though this model is intended to support the
astronomical theory, it can also produce unforced oscillations, meaning intrinsic
dynamics could be captured by the model. An important result of this paper is
that it was able to recreate the MPT without using a time dependent parameter.
This model produces the closest fit of the three models discussed so far, but is
very sensitive to parameter perturbations.

In the same year, Crucifix developed a Van der Pol style model that produces
relaxation oscillations with a 100\,kyr period, which is entrained by the
$Q_{65}$ signal~\cite{crucifix_original}. This model aligns with the geochemical
theory. This model is included in our comparison as it uniquely reproduces the
ice volume data well despite using the fewest parameters of any of the models.
However, this model differs from the others by the fact that it does not propose
a physical interpretation of the equations as it is instead used to demonstrate
that the ice volume dynamics are easily produced by a carefully tuned model that
is capable of producing the 100\,kyr period internally. Crucifix shows that the
strong performance of the model is highly sensitive to parameter perturbations,
calling into question the predictive power of this class of model.

Aside from Budyko, all of the models discussed so far in this section include the
assumption of a switching mechanism that depends on whether the ice volume is
growing or ablating. This is reasonable, given the sawtooth nature of the ice
volume data shown in the last third of Figure \ref{fig:benth_and_power_specs}A,
however there is no consensus on what this mechanism is. The inclusion of a
switching mechanism introduces non-linearity and makes the model sensitive
to the choice of switching condition.

In 2018, Verbitsky, Crucifix, and Volobuev developed a model comprising state
variables for glacial area, basal temperature, and climate
temperature~\cite{verbitsky2018}. The model is once again forced by $Q_{65}$
and emphasises the importance of internal climatic feedbacks in producing the
glacial-interglacial cycles, supporting the geochemical theory. Although this
model is capable of reproducing the ice volume data well, it does not rely on a
switching mechanism to capture the different rates of ice growth and decay. Our
own models presented in this thesis also omit a switching mechanism. This shared
feature allows for a more direct comparison between the three models in Section
\ref{sec:comparison_verbitsky}.

\begin{figure}
  \centering
  \def\svgwidth{0.6\textwidth}
  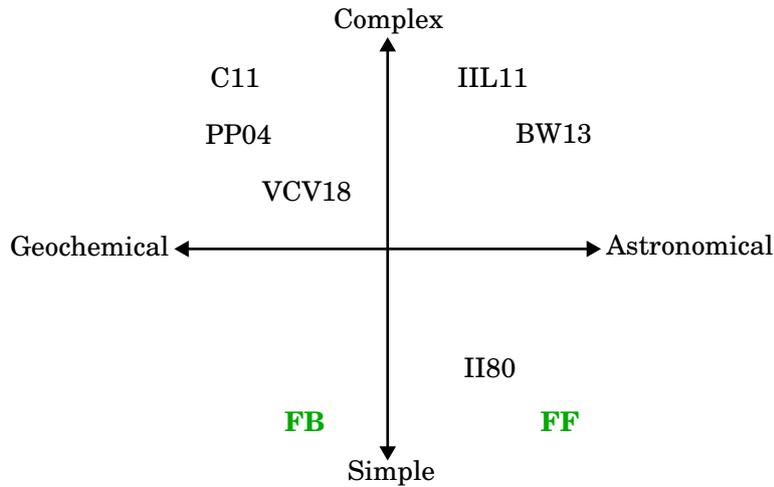
  \caption[Conceptual Model Landscape]{A simplified representation of the
    conceptual model landscape for those discussed in this thesis. The
    horizontal axis represents the extent to which the model aligns with either
    the astronomical theory or the geochemical theory. The vertical axis
    represents the complexity of the model. The feedback (FB) and feedforward
    (FF) models that we have developed are shown in green. The models are
    \textbf{C11}: Crucifix (2011)~\cite{crucifix_original}, \textbf{IIL11}:
    Imbrie, Imbrie-Moore, and Lisiecki (2011)~\cite{imbrie2011}, \textbf{PP04}:
    Paillard and Parrenin (2004)~\cite{pp04}, \textbf{BW13}: Budyko
    (1969)~\cite{budyko} augmented by Widiasih (2013)~\cite{widiasih2013},
    \textbf{VCV18} Verbitsky, Crucifix, and Volobuev
    (2018)~\cite{verbitsky2018}, and \textbf{II80}: Imbrie and Imbrie
    (1980)~\cite{imbrie}.}
\label{fig:conceptual_model_landscape}
\end{figure}

All of the models discussed so far are shown in Figure
\ref{fig:conceptual_model_landscape}, with the models we have developed
shown in green. All of the models have been placed on a conceptual model
landscape, where the horizontal axis represents the extent to which the model
aligns with either the astronomical or geochemical theory. The vertical axis
represents the relative complexity of the model, where complexity in this case is an
approximate measure of: the number of state variables, the non-linearity, and
the sensitivity to parameter perturbations. We emphasise that all of these
models aim to represent the Earth system in a highly simplified form, and so
complexity is relative to this class of model.

\section{Outline}
\label{sec:outline}
Here we will give a brief overview of the structure of this thesis, outlining
the purpose of each chapter and how it contributes to the overall goal of
understanding Earth's ice volume dynamics.

In Chapter \ref{chap:data}, we introduce the data used in this thesis. This
starts with our choice of ice volume proxy. A popular choice in this field is
the benthic $\delta^{18}$O stack from Lisiecki and Raymo \cite{benthic_data}, as
it produces a consistent, high-resolution record over the past 800\,kyr,
stacking isotopic ratio records from 57 deep-sea sediment cores from around the
globe. A limitation of benthic $\delta^{18}$O measure is that it is sensitive
to both the global ice volume as well as the local water temperature in which
the sediment was deposited. To combat this, we use the data produced by
Bintanja, who developed a model to isolate the ice volume contribution to
Lisiecki and Raymo's benthic stack.

We then introduce the orbital parameters and examine their impact on the
insolation profile. This is followed by a derivation of the equations for both
daily and annually averaged insolation. We discuss how these solutions will be
used as input to the models presented in Chapter \ref{chap:models}.

In Chapter \ref{chap:models}, we first introduce the Budyko model and its
augmented form from Widiasih. This comprises a one-dimensional representation of
Earth's surface temperature across the northern hemisphere as a function of time
and latitude, as well as an ice line that depends on the surface temperature in
its vicinity. We discuss what can be learned from this model in the context of
the astronomical theory, in particular how the ice-albedo feedback mechanism can
amplify the orbital forcing. We then introduce a number of augmentations with
the aim of improving the model's representation of the Earth system. We show
that the model is capable of producing ice volume oscillations but not to the
extent seen in the data. This leads us onto a number of zero-dimensional models
that represent ice volume as a single value. These cover a range of complexity
and align with both the astronomical and geochemical theories. We reproduce the
models and examine their underlying dynamics, comparing them to the data.

In Chapter \ref{chap:feedforward_model}, we introduce a new model that is
capable of reproducing the ice volume data with a similar level of accuracy as
the models discussed in Chapter \ref{chap:models}, but with the simplest
possible form. We focus on simplicity in the following chapters due to the
limited number of glacial-interglacial cycles that are available for analysis.
With only eight complete cycles in the post-MPT regime, we wish to avoid either
overfitting to the data, or including mechanisms that are not necessary to
reproduce the data. To guarantee our model is of the simplest form whilst
adequately reproducing the data, we perform a systematic pruning of the model
components and evaluate the performance in each case. This model is intended to
represent an Earth system that is incapable of producing the
glacial-interglacial cycles without orbital forcing, supporting the astronomical
theory. We analyse its dynamics and show that it is able to reproduce the ice
volume data well despite omitting a switching mechanism, and assuming a linear
relationship between the orbital parameters and ice volume. We also show that
the absent 400\,kyr period of eccentricity in the ice volume data can be
explained due to the offset time response of two terrestrial mechanisms to
eccentricity. We follow this up with a physical interpretation of the model in
which we propose surface air temperature and bulk ocean temperature are the two
contributing state variables in the system. Using a simple model of bulk ocean
temperature, we estimate the range of warming rates that can result from orbital
forcing. Even with conservative parameter estimates, we find that eccentricity
can cause a large enough difference in warming rates to explain the full range
in the ocean temperature data, without the need for amplification. This
addresses a key criticism of the astronomical theory, that eccentricity is too
weak to drive the ice volume dynamics without amplification.

Chapter \ref{chap:feedback_model} introduces a second new model that is capable
of producing feedback, representing an Earth system that is capable of producing
relaxation oscillations, with orbital forcing playing a secondary role. This is
intended to support the geochemical theory. We show that this model is able to
reproduce the ice volume data slightly better, whilst maintaining the same
simplicity as the first model. Although this may suggest that the geochemical
theory is better supported by the data, in Chapter \ref{chap:analysis}, we show
that the difference in ice volume solutions are mostly contained within a
particular period known as Marine Isotope Stage 11 (MIS 11). This suggests that
the second model is better capturing this period as it can combine both the
relaxation oscillations and the orbital forcing, meaning we cannot rule out the
astronomical theory from the data alone. A second result of producing these
models is that we show the ice volume data can be mostly explained by assuming a
linear relationship between orbital forcing and the Earth system, regardless of
the internal feedbacks that may occur on Earth. This calls into question how
significant any switching mechanism in the Earth system is, and therefore
whether it should be so frequently included in conceptual models. To further
support this, we examine Verbitsky's non-linear model from Chapter
\ref{chap:models} as it also omits a switching mechanism. We show that its
linearised form is able to reproduce the data with a similar level of accuracy
to the original model, as well as our highly simplified models.

We conclude the thesis with Chapter \ref{chap:discussion}, where we summarise the
key results from this thesis. We evaluate the strengths and limitations of our
models and discuss the implications they have of the two opposing theories of
ice volume dynamics. We also suggest areas for future research that could build
on the work presented here.

\clearemptydoublepage
\chapter{Data}
\label{chap:data}

\initial{T}his chapter introduces the data sources used in this thesis. We begin
by discussing the possible sources of proxy data used to estimate global ice
volume over the past 800\,kyr. Although there are a number of methods by which
to estimate ice volume, they are not all equally reliable. We focus on benthic
$\delta^{18}$O data from ocean sediment cores, which is considered one of the
best proxies for ice volume \cite{benth_is_best_proxy}. The main limitation of
this source is that it produces a signal that is a combination of global ice
volume and local bottom water conditions. We therefore use the work done by
Bintanja to isolate the ice volume component of this data using a model that
includes global sea level and mean surface air temperature
\cite{benthic_model_contributions}. This model is then verified with independent
surface temperature data from the Dome C ice core in Antarctica \cite{dome_C}.
We finally perform a linear transformation on the ice volume component of the
$\delta^{18}$O data to convert it into global ice volume in km$^3$.

We then introduce each of the orbital parameters that are known to influence
Earth's climate and justify the use of Laskar's 2004 orbital solutions over
alternative sources \cite{laskar2004}. We discuss the effects of each of these
parameters on the magnitude and distribution of insolation reaching Earth. We
show that eccentricity has a small effect on the annually averaged insolation
reaching Earth, obliquity has a significant effect on the latitudinal
distribution of insolation across Earth's surface, and precession has a
significant effect on the length and severity of the seasons in each hemisphere.

Lastly, we derive equations for the daily and yearly averaged insolation
reaching Earth's surface as a function of the orbital parameters and latitude,
drawing from the work of McGehee and Lehman \cite{insol_latlon}. This allows us
to examine the impact of each orbital parameter on the insolation profile. We
are also to produce specific signals such as $Q_{65}$, which will be used for
many of the models discussed in Chapter \ref{chap:models}.

\section{Ice Volume Proxy Data}
\label{sec:ice_volume_data}

We wish to model the global ice volume over the past 800\,kyr. The only way to
understand the ice volume over this time period is through proxy data. There are
several sources of data that can be used to estimate ice volume, such as glacial
geomorphology, speleothems, loess deposits, ice cores, and ocean sediment cores.

Glacial geomorphology is the study of geological evidence of past glaciations,
such as moraines and glacial striations. This can be used to estimate the
extent of ice sheets in the past. However, it can be difficult to determine
the exact timing of these events as they result from cumulative effects
\cite{glacial_geomorphology}.

Speleothems are mineral deposits that form in caves, such as stalactites and
stalagmites. In addition to containing isotopes that can be used for climate
reconstruction, the speleothem layers contain uranium, which decays into thorium
at a known rate, allowing for a precise chronology to be established
\cite{speleothem}. However, the data is localised to individual caves and the
resolution and span of the isotopic record is governed by the timescale of the
speleothem growth, which can result in inconsistency when data from multiple
locations are combined.

Loess deposits are layers of fine silt and clay particles transported and
deposited by wind, typically found in regions with dry conditions and strong
winds. Analysing the characteristics, such as grain size and isotopic
composition, of these deposits can provide information on the past climate and
provide insights into how ice volume was influenced \cite{loess_ice_proxy}.
However, the data this produces is an indirect proxy for ice volume and it can
be difficult to establish a precise chronology for the deposits as they do not
contain time markers like speleothems.

\begin{figure}[h]
  \centering
  \input{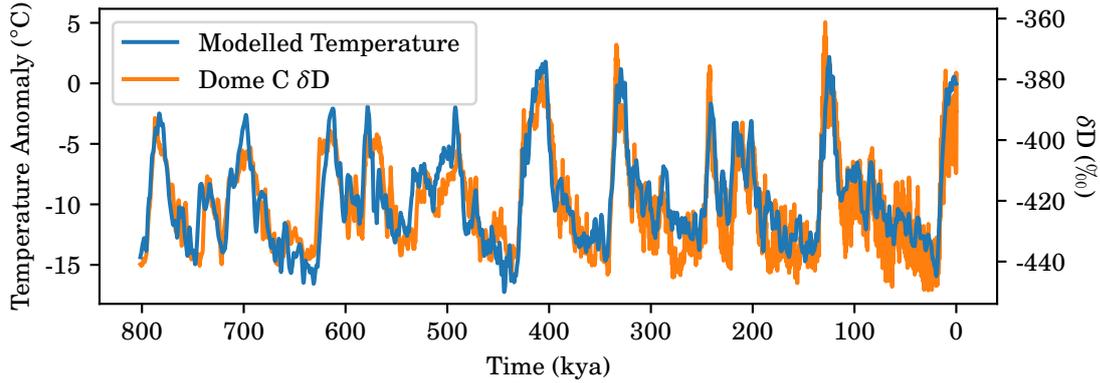}
  \caption[Bintanja's Modelled Temperature vs Dome-C Deuterium]{Comparing the
    modelled surface temperature anomaly from Bintanja's model
    \cite{benthic_model_contributions} with the $\delta$ Deuterium from Dome-C in
  Antarctica, a proxy for surface temperature \cite{dome_C}.}
  \label{fig:model_temp_vs_dome_c}
\end{figure}

Ice cores are drilled out of ice sheets, providing a direct record of the
palaeoclimate. Similar to speleothems and loess deposits, the ice contains
isotopes, such as $^{18}$O and deuterium, that can be used to infer past
temperatures. An example of this can be seen in Figure
\ref{fig:model_temp_vs_dome_c}. In addition, the ice contains trapped air
bubbles from the time of formation, which can help reconstruct the composition
of the atmosphere at that time \cite{petit1999}. A glacial model can be used to
convert the depth of the core into age, providing a chronology for the data
\cite{dome_c_chronology}. One limitation of this approach is that, since the
cores must be drilled from deep ice in order to cover hundreds of thousands of
years, the data sources are limited to the polar regions, which restricts the
global coverage of the data.

This has led to benthic $^{18}$O data from ocean sediment cores being suggested
as one of the best proxies for ice volume \cite{benth_is_best_proxy}. The main
advantages of this data source are: it contains a global ice volume signal
within the isotopic composition, the resolution is relatively high whilst still
spanning a long time period ($\approx$3000\,kyr), and the data is available from
many locations around the world, allowing for individual records to be combined
into a global stack.

Benthic Foraminifera are single-celled organisms that live on the seabed and
can be used for palaeothermometry. From drilling deep cores out of the seabed, we
can access foraminifera from millions of years ago, giving us insight about the
state of the world over that time. Their calcite shells are sensitive to the
physiochemical characteristics of their environment. Although variations in the
temperature and chemical make-up of the ambient water impact the composition
of foraminiferal shells, by sampling from a wide range of locations and
combining the data, we can expect the effect to diminish. Lisiecki and Raymo
collected core sample data from 57 distinct locations around the world and
combined them using an automated graphic correlation algorithm
\cite{benthic_data}. This is a commonly used source for $\delta ^{18}$O data in
papers that study the Quaternary period
\cite{insol_latlon,benthic_model_contributions,crucifix_original,imbrie2011}

The ratio of $^{18}$O to $^{16}$O isotopes, known as $\delta ^{18}$O, in the
shell provides information about the local bottom water temperature (BWT) in
which the foraminifera lived, as well as the global ice volume. The lighter
$^{16}$O isotopes evaporate more easily due to a process known as Rayleigh
distillation \cite{original_d18O_to_temp}. This causes a larger proportion of
them to end up in the precipitation formed ice-sheets. As a result, an increased
$\delta ^{18}$O value indicates increased global ice volume. However, it is also
known that a decrease in ambient BWT, which would occur with increased ice
volume, results in an increased $\delta ^{18}$O value \cite{delta_O_to_temp}.

There is an ongoing effort to isolate the effects of global ice volume from the
BWT on this isotopic record. The foremost method for this is to measure the
ratio of Mg to Ca in the same Foraminifera. 

Rosenthal presented a 3\,myr dataset, using a linear model to relate Mg/Ca to
BWT \cite{linear_mgca_temp}. However, Yu critiqued this work, saying that the
linear model's calibration used only 2 data points from only one drilling site
\cite{crit_linear_mgca_temp}. These points are modern BWT and Mg/Ca from the top
of the drilled core, and an estimate for the BWT at the last glacial maximum
along with the corresponding Mg/Ca in the core. Yu applied this linear model to
the same species of benthic foraminifera from other cores and demonstrated an
average estimation error of $\pm2.7^\circ$C. This error is significantly greater
than Rosenthal's quoted value of $\pm1.1^\circ$C and casts doubt on the validity
of their model.

Martin compared 30 core tops and corresponding modern temperatures to produce an
exponential fit of Mg/Ca to BWT \cite{exponent_mgca_temp}. This includes cores
from the Bahamas Transect, which provides a BWT range up to 18$^\circ$C.
Although this exponential fit appears to work well over the large BWT range,
there is still poor agreement between the lower BWT values and the model. A
potential cause for this is dissolution of the calcite. This reduces the Mg/Ca
ratio within the foraminifera shells, with the degree of this effect increasing
with depth \cite{dissolution_with_depth}.

Martin looked at the coarse fraction variability of sand in the same core
sample as the benthic foraminifera. This measures the fraction of grains in a
sample greater than 250\,$\mu$m and has been shown to decrease with increased
dissolution \cite{sand_variability_dissolution}. The sand data shows an increase
in dissolution during glacial periods. Given that dissolution reduces the Mg/Ca
ratio, which we have so far considered a proxy for BWT, we could expect the
estimated BWT record to show a larger temperature range than is accurate.

Since it appears unreliable to extract global ice volume purely from core sample
analysis, we instead look towards modelling. Bintanja uses the same 57 site
benthic stack produced by Lisiecki and Raymo to create an internally consistent
model of mean surface air temperature, linked with ice volume and global sea
level \cite{benthic_model_contributions}. This then produces time series for the
ice volume and BWT contributions to the $\delta ^{18}$O data.

Although we cannot know the validity of these two components, the mean surface
temperature time series that is produced from the model can be compared to
other, independent surface temperature metrics. Bintanja compares their modelled
surface temperature to $\delta$\,Deuterium data from the Dome C ice core in
Antarctica, which ranges back to 740\,kya \cite{dome_C}. 

The $\delta$\,Deuterium data describes the ratio of deuterium ($^2$H) to the
more common hydrogen isotope, protium ($^1$H). There is an approximately linear
relationship between this ratio in snow and the mean temperature at which the snow
was formed, allowing $\delta$\,Deuterium to operate as a proxy for surface air
temperature \cite{deuterium_air_temp}. Hoffman conducted a series of experiments
with the ECHAM atmospheric general circulation model with water isotopes for
different climate stages \cite{model_temp_with_det_excess}. These were used to
constrain the Dome C deuterium record to an expected temperature range over the
past 800\,kyr of ${\sim}15^\circ$C, with temperature anomaly from present day
being linearly mapped to $[-10.3^\circ$C,$+4.5^\circ$C].

Bintanja's modelled surface temperature agrees well with this estimated
temperature range and shows good qualitative agreement with the Dome C deuterium
data as shown in Figure \ref{fig:model_temp_vs_dome_c}. As a second
verification, the modelled sea-level time series is compared with that of
Siddall's, which was attained through combining fossil coral reef terraces with
salinity and $\delta^{18}$O data from the Red Sea basin
\cite{red_sea_level_data}. This independent dataset also agrees closely with the
modelled output from Bintanja, suggesting the isolation of $\delta^{18}$O
contributions is reasonable.

\begin{figure}[h]
  \centering
  \input{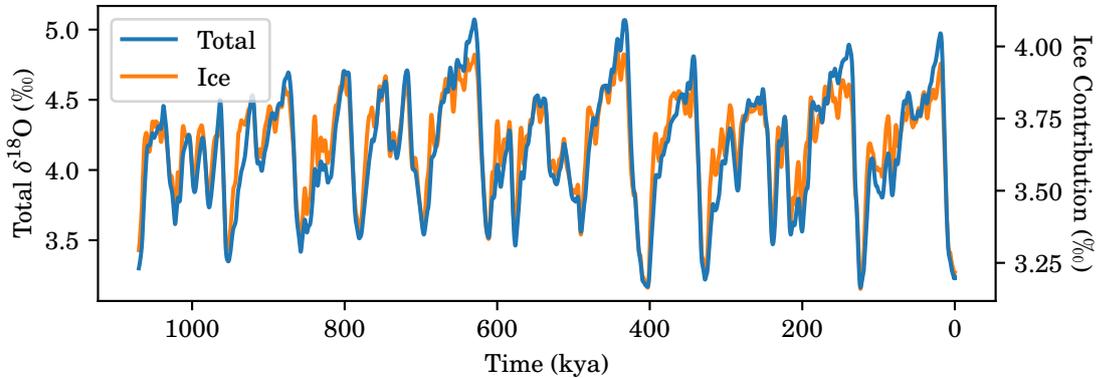}
  \caption[Modelled Ice Contribution to $\delta^{18}$O]{Comparing the
    contribution of ice volume, as modelled by Bintanja
    \cite{benthic_model_contributions}, to the $\delta^{18}$O data from Lisiecki
  and Raymo \cite{benthic_data}.}
  \label{fig:ice_part_of_O18}
\end{figure}

Looking now at the isolated ice volume contribution to the $\delta^{18}$O data,
as shown in Figure \ref{fig:ice_part_of_O18}, we can see the two time series are
visually similar. Although BWT is also contributing to the total $\delta^{18}$O
value, this is approximately proportional to the ice volume contribution. Since
we are linearly mapping the ice volume component of $\delta^{18}$O to ice volume
in km$^3$, we are not interested in the absolute value of the signal, only the
relative behaviour.

\begin{figure}[h]
  \centering
  \input{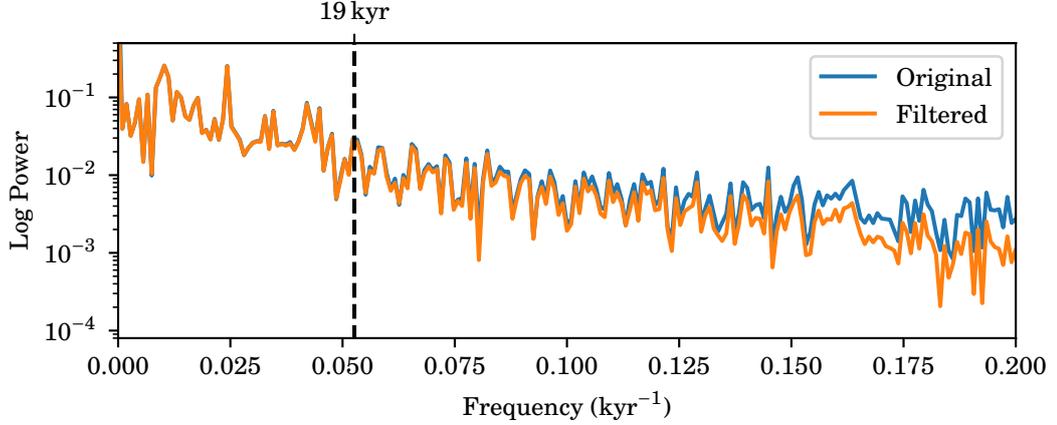}
  \caption[Original and Filtered $\delta^{18}$O Power Spectra]{Power spectra for
    the original $\delta^{18}$O data from Lisiecki and Raymo
    \cite{benthic_data}, and the filtered version that is used as input to
    Bintanja's model \cite{benthic_model_contributions}. Frequencies above that
    of precession (19\,kyr) begin to deviate, however this detail is not
    relevant for the models discussed in Chapter~\ref{chap:models}.}
    \label{fig:fft_O18_vs_model}
\end{figure}

It is worth noting that the input to Bintanja's model is a filtered version of
the $\delta^{18}$O time series from Lisiecki and Raymo. Although the filtering
used is not discussed in their paper, we can compare the power spectra for the
original and filtered signals, as shown in Figure \ref{fig:fft_O18_vs_model}.
Here we have marked the highest frequency that can be explained by the
Milankovitch cycles, corresponding to 19\,kyr, and we can see the two signals
are almost identical for frequencies lower than this. Frequencies above 19\,kyr
can be attributed to Earth based events, such as volcanic eruptions, and noise
in the data, which are out of the scope of the models discussed in Chapter
\ref{chap:models}.

The ice volume data from Bintanja's model is expressed as a proportion of
the total benthic $\delta^{18}$O value, measured in parts-per-thousand
($\permil$). In order to evaluate our model by comparing to other
proxy data, we convert the ice proportion of $\delta^{18}$O into ice
volume. For this, we assume a linear relationship between ice volume and the
contribution to the $\delta^{18}$O data. We then use estimates for the physical
values needed to linearly transform the ice contribution into ice volume.

The current estimates of ice volume by location are approximately
$2.7{\times}10^7$\,km$^3$ in Antarctica~\cite{antarctica_ice_vol},
$2.99{\times}10^6$\,km$^3$ in Greenland~\cite{greenland_ice_vol}, and
$1.58{\times}10^5$\,km$^3$ in all other regions~\cite{other_ice_vol}.
This gives an approximate total of $3.0{\times}10^7$\,km$^3$. Lambeck
estimates that the ice volume during the last glacial maximum was approximately
$5.2{\times}10^7$\,km$^3$ greater than at present~\cite{lgm_ice_vol}. This
occurred around 21\,kya, giving us an anchor point to scale
the range of $\delta^{18}$O to ice volume. We can now convert the ice volume
component of the benthic $\delta^{18}$O data $I_\mathrm{Benth}$ into global ice
volume $I_\mathrm{Data}$ using
\begin{equation}
  I_\mathrm{Data} = mI_\mathrm{Benth} + c,
  \label{eq:ice_data_rescale}
\end{equation}
where
\begin{align}
    m &=
    \frac{I_\mathrm{Data}(0)-I_\mathrm{Data}(-21)}{I_\mathrm{Benth}(0)-I_\mathrm{Benth}(-21)}
    \approx \frac{3.0\times 10^7 - 8.2\times 10^7}{0.0-1.0} = 5.2\times
    10^7\,\mathrm{km}^3/\permil,\\ c &= I_\mathrm{Data}(0) -
    mI_\mathrm{Benth}(0) \approx 3.0\times 10^7 - 5.2\times 10^7\times 0.0 =
    3.0\times 10^7\,\mathrm{km}^3.
\end{align}

\section{Orbital Parameters}
\label{sec:orbital_parameters}
\begin{figure}
  \centering
  \includegraphics[width=0.8\linewidth]{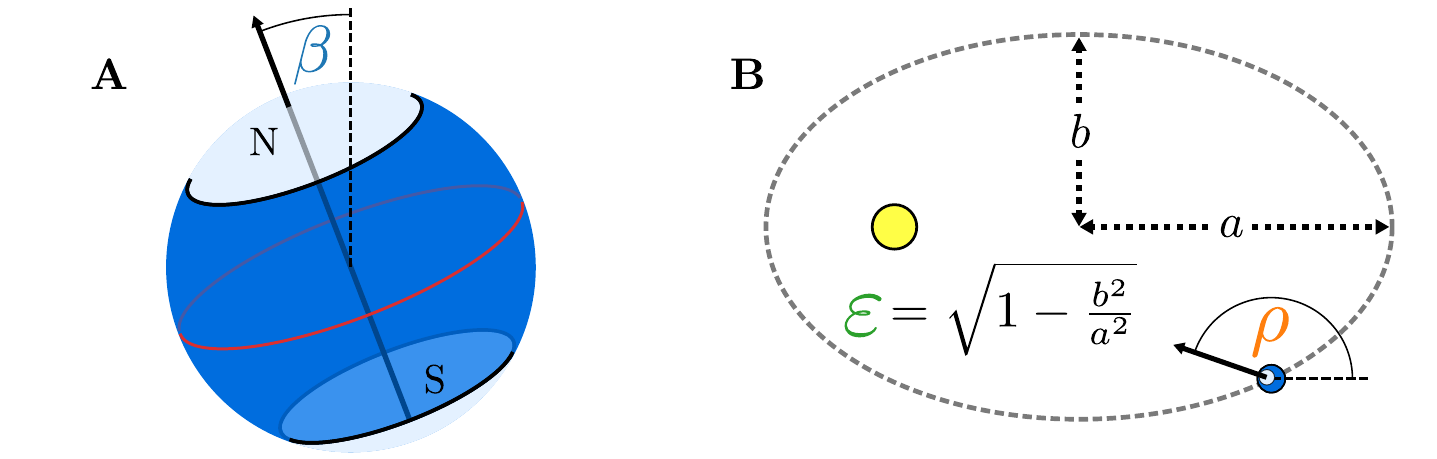}
  \caption[Orbital Parameter Definitions]{Diagram of Earth from two
    perspectives, defining the three orbital parameters, obliquity $\beta$,
    eccentricity $\varepsilon$, and precession $\rho$. Obliquity describes the
    tilt of Earth's rotational axis from vertical in the ecliptic frame.
    Eccentricity is a function of the semi-major and semi-minor axes of Earth's
    orbit, $a$ and $b$ respectively. Precession describes the rotation of
    Earth's rotational axis around the vertical in the ecliptic frame.}
  \label{fig:orbital_params}
\end{figure}
The parameters describing Earth's orbit configuration are outlined in Figure
\ref{fig:orbital_params}. Obliquity and precession describe the orientation of
Earth in the ecliptic frame, whilst eccentricity describes the elongation of
Earth's orbit.

\begin{figure}[p]
    \begingroup
    \thispagestyle{empty}
    \vspace*{-2cm}
    \centering
    \input{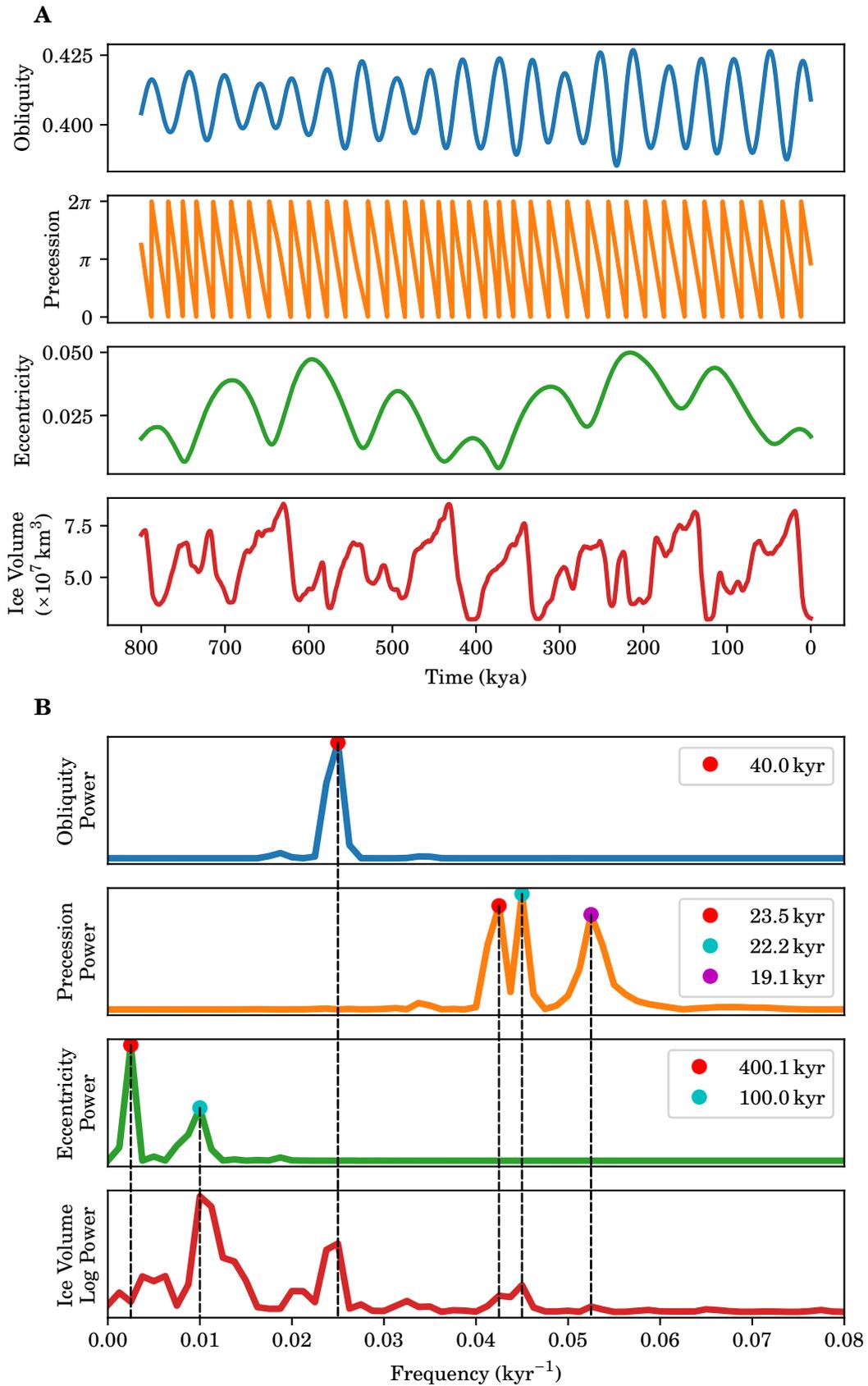}
    \caption[Orbital Signals]{Time series (\textbf{A}) and power spectra
      (\textbf{B}) for the three orbital parameters and ice volume data. The ice
      volume power is logarithmically scaled to highlight the smaller peaks that
      align with precession. The dashed lines in \textbf{B} show all orbital
      frequencies aligning with frequencies in the ice volume data except for
    the 400.1\,kyr peak in eccentricity.}
    \label{fig:orbital_and_benthic_time_series_power_specs}
    \endgroup
\end{figure}

These orbital parameters vary cyclically with an approximately constant
frequency, as shown in Figure
\ref{fig:orbital_and_benthic_time_series_power_specs}. Also included in this
Figure is the ice volume from Bintanja that we have converted to km$^3$ as
discussed in Section \ref{sec:ice_volume_data}. The power spectra of these time
series show all frequencies present in obliquity and precession matching with
frequencies in the ice volume data. There is also a peak in the ice volume power
spectrum around the ${\sim}100$\,kyr period, relating to eccentricity, however
there is a notable lack of the ${\sim}400$\,kyr period. We will discuss a
potential mechanism to explain this in Section \ref{sec:physical_model}.

It is worth noting that the benthic $\delta^{18}$O stack introduced in Section
\ref{sec:ice_volume_data} was converted from core depth to age using an orbitally
forced model, which is introduced in Section \ref{sec:imbrie_imbrie_1980}. Since this
$\delta^{18}$O data is used to produce the ice volume data shown in Figure
\ref{fig:orbital_and_benthic_time_series_power_specs}, using this data on its
own to justify a dependence on orbital parameters would be circular. However,
the orbital parameters having an impact on ice volume is well established
\cite{astronomical_theory,time_series_fix_orbital_tuning,orbital_freqs_without_tuning,orbital_tuning_support}.
Additionally, other core depth to age models exist, such as the EDC3 chronology
produced by Parrenin et al. \cite{dome_c_chronology}. This uses a combination of
age markers (such as volcanic ash layers) and a glaciological model to produce a
depth to age relationship independent of orbital tuning for the past 400\,kyr.
This depth to age conversion is applied to Dome C deuterium data in Section
\ref{sec:ice_volume_data} and still contains the same orbital frequencies as the
Bintanja ice volume data, suggesting that the benthic $\delta^{18}$O stack was
not compromised by orbital tuning.

The source used for the orbital parameters comes from Laskar, whose high precision
celestial models are able to reproduce the orbital parameters over the past 50
million years~\cite{laskar2004}. Although inaccuracy in the ephemeris and
unpredictable perturbations reduce the model's reliability over time, these
effects will be minimal over most recent 800\,kyr.

For all models in this thesis, we use the orbital solutions produced by Laskar in
2004~\cite{laskar2004}. To estimate the reliability of these solutions over the
past 1\,myr, we can compare them to the previous solutions produced by Laskar in
1993~\cite{laskar1993}, as well as the solutions produced by Berger in
1999~\cite{berger1999}. Laskar's 1993 solution for eccentricity deviates from
the more recent equivalent by an average of 0.4\% over the past 1\,myr, whilst
Berger's eccentricity solution deviates by an average of 1.2\%. The obliquity
and precession deviations are significantly less in both cases. The differences
between these solutions suggests an uncertainty on the order of 1\%. This is an
acceptable level for our purposes due to the less accurate nature of the proxy
data we rely on to estimate global ice volume. To justify this, we tried fitting
our model with all three orbital solutions and found that the difference was
negligible when comparing to the ice volume data.
\clearpage
\clearpage
\subsection{Eccentricity}
\label{sec:eccentricity}
Eccentricity $\varepsilon$ is a measure of the deviation of an orbit from a
perfect circle. As is shown in Figure~\ref{fig:orbital_params}, it can be
defined as a function of the semi-major and semi-minor axes, $a$ and $b$
respectively. In a simple two-body system, eccentricity would remain constant.
However, Earth's orbit experiences perturbations due to the gravitational
influence of other planets in our solar system, most notably Jupiter and Saturn.

Eccentricity stands apart from obliquity and precession as the only orbital
parameter that affects the magnitude of annual insolation reaching Earth. In an
elliptical orbit, the sun lies at one of the foci, with Earth's distance from it
varying throughout the orbit. The power of insolation reaching Earth is
inversely proportional to the square of this distance. However, it is not
immediately clear how a more elliptical orbit results in more insolation over
a year than the approximately circular orbit.

To quantify this, we will establish the relationship between eccentricity and
the annually averaged insolation reaching Earth. Let $K$ be the solar output in
Watts, which reaches Earth at distance $r(t)$. At this distance, the power of
the sun is distributed over an area equal to $4\pi r(t)^2$, giving the energy
flux at Earth as
\begin{equation}
  I_\mathrm{E}(t) = \frac{K}{4\pi r(t)^2}.
  \label{eq:energy_flux}
\end{equation}
From Kepler's second law, we have that
\begin{equation}
  r(t)^2 = \frac{2\pi ab(t)}{P}\frac{\mathrm{d}t}{\mathrm{d}\theta},
  \label{eq:r_squared}
\end{equation}
where $P$ is the period of Earth's orbit, $\theta(t)$ is the Earth's angular
position in its orbit, and $a$ and $b(t)$ are the semi-major and semi-minor axes
respectively. From Laskar, we know that the semi-major axis of Earth’s orbit
is essentially constant, whilst $b$ varies with eccentricity~\cite{laskar2004}.

Substituting \eqref{eq:r_squared} into \eqref{eq:energy_flux} gives
\begin{equation}
  I_\mathrm{E}(t) = \frac{PK}{8\pi^2 ab(t)}\frac{\mathrm{d}t}{\mathrm{d}\theta}.
\end{equation}

Rearranging this gives
\begin{equation}
  \frac{1}{P}I_\mathrm{E}(t)\,\mathrm{d}t = \frac{K}{8\pi^2 ab(t)}\,\mathrm{d}\theta.
\end{equation}

Since the semi-minor axis $b$ varies on the timescale of 100\,kyr, we will
treat it as constant and integrate over a year to get
\begin{equation}
  \frac{1}{P}\int_0^P I_\mathrm{E}(t)\,\mathrm{d}t = \int_0^{2\pi}\frac{K}{8\pi^2
  ab}\,\mathrm{d}\theta,
\end{equation}
and so 
\begin{equation}
  \overline{I_\mathrm{E}}(t) = \frac{K}{4\pi ab(t)},
\end{equation}
where $\overline{I_\mathrm{E}}$ is the annually averaged irradiance, in
W/m$^2$, reaching Earth.

The area of flux intercepted by Earth at any one time is equivalent to a disc
with Earth's radius $R_\mathrm{E}$, giving the average power as $\pi
R_\mathrm{E}^2\overline{I_\mathrm{E}}$. However, the power is distributed over
Earth's entire surface, which has area $4\pi R_\mathrm{E}^2$. The annually
averaged insolation over Earth's surface is therefore
\begin{equation}
  \overline{Q_\mathrm{E}}(t) = \frac{K}{16\pi ab(t)},
  \label{eq:annual_insolation_ab}
\end{equation}
with units W/m$^2$.

Kepler's first law tells us that $b(t) = a\sqrt{1-\varepsilon(t)^2}$, which we
can substitute into \eqref{eq:annual_insolation_ab} to get
\begin{equation}
  \overline{Q_\mathrm{E}}(\varepsilon(t)) = \frac{K}{16\pi
  a^2\sqrt{1-\varepsilon(t)^2}}.
  \label{eq:annual_insol}
\end{equation}
We now have an expression for the annually averaged insolation over Earth's
surface, purely as a function of eccentricity. This also shows how the annual
insolation is independent of both obliquity and precession.

The Earth's orbital eccentricity varies between approximately 0 and 0.06. If we
substitute these limits, along with $K=3.8287\times10^{26}$\,W and
$a=1.4960\times10^{11}$\,m \cite{solar_constant}, into $\overline{Q_\mathrm{E}}$
we see the annually averaged insolation ranges from 340.353 to 340.967\,W/m$^2$.
This corresponds to a maximum change of 0.18\% due to eccentricity.

This seemingly insignificant variation has led some researchers to propose the
need for an Earth-based amplifier in order for eccentricity to have any
significant impact on ice volume \cite{imbrie_inertia,saltzman,nonlinear_amplifier}.
Although we do not rule out the existence of these amplifiers, in Section
\ref{sec:ocean_model}, we estimate the potential ocean temperature change that
could result from the direct effect of eccentricity.

\subsection{Obliquity}
\label{sec:obliquity}
Obliquity $\beta$ refers to the tilt of Earth's axis of rotation from vertical
in the ecliptic frame. It is caused by gravitational forces from other celestial
bodies exerting a torque on Earth's equatorial bulge. As shown in Section
\ref{sec:insolation}, obliquity does not affect the total insolation reaching
Earth. This is because the torque applied from external bodies rotates Earth
about its centre of mass, but does not change Earth's position relative to the
sun.

Despite this, the angle by which Earth is tilted significantly affects the
distribution of insolation across the planet's surface. McGehee and Lehman
express the annually averaged insolation for a given latitude $\varphi$
\cite{insol_latlon} as
\begin{equation}
  \overline{Q_\varphi}(\varepsilon,\beta,\varphi) = \frac{K}{8\pi^3a^2\sqrt{1-\varepsilon^2}}
  \int_0^{2\pi}\sqrt{1 - \left(\cos\beta\sin\varphi -
  \sin\beta\cos\gamma\cos\varphi\right)^2}\,\mathrm{d}\gamma,
\end{equation}
where $\gamma$ is the longitude and $K$ and $a$ are as before. The minimum and
maximum values of obliquity over the past 800\,kyr are 0.385 and 0.427 radians.
If we use the north pole as an example and input these to
$\overline{Q_\varphi}$, the integral evaluates to 2.36 and 2.60 respectively.
This means that obliquity can cause annual insolation at the poles to vary by up
to 10\%.

If we assume that the polar glaciers are more sensitive to their local air
temperature than the equatorial temperature, we can expect obliquity to play a
significant role in ice volume dynamics. Evidence of this is shown in
Figure~\ref{fig:orbital_and_benthic_time_series_power_specs}, where the
approximately 40\,kyr period of obliquity is clearly visible in the ice volume
data.
\subsection{Precession}
\label{sec:precession}
Precession $\rho$ is the orientation of Earth's axis of rotation projected into
the ecliptic plane and there are multiple established starting angles from which
to measure it. In this thesis, we measure the anti-clockwise angle from the
farthest point on Earth's orbit, known as the aphelion. This is shown in the
exaggerated orbit in Figure~\ref{fig:orbital_params}, which loosely resembles
the current precession angle.

Since Earth's distance from the sun varies throughout the year. If one
hemisphere is tilted towards the sun during the closest part of the orbit, known
as the perihelion, it will experience greater summer insolation than the other
hemisphere. This is because the opposing hemisphere's summer would coincide with
aphelion, where the insolation is less. Hence, the precession of Earth's
axis can affect the severity of the seasons in each hemisphere.

From Kepler's second law, we know that Earth's orbital velocity is greatest at
perihelion and least at aphelion. This means that one hemisphere can experience
more intense summer insolation if it occurs during perihelion, but that
summer will also be shorter. To formalise how the seasons relate to precession,
Table \ref{tab:precession_effects} outlines its effect on each hemisphere for
the two extreme values of precession. Intermediate values would produce effects
that lie between those shown in the table.

Using the precession angle itself in models is not practical due to its
discontinuity around $\rho = 2\pi$, this can be fixed by using a continuous
representation of the angle. By taking the cosine of the precession angle as we
have defined it, its maximum occurs at $\rho=0$, which coincides with the most
intense summer insolation in the northern hemisphere. This aligns with
Milankovitch, who proposed that northern summer insolation correlates with
glacial retreat~\cite{milankovitch}. This is justified by the greater proportion
of land in the higher latitudes of the northern hemisphere, resulting in a
greater range of glacial movement compared to the south.

Precession's impact on insolation depends both on the eccentricity, and the
obliquity of Earth's orbit. If Earth's axis of rotation is pointing
directly out of the ecliptic plane, meaning obliquity is zero,
the precession angle is undefined. Likewise for eccentricity, if Earth had a
circular orbit, meaning eccentricity is zero, then the precession angle would
have no impact on the insolation. As obliquity and eccentricity increase, the
impact of precession on insolation also increases.

As a result, we propose that the cosine of precession plays a role in the
forcing of the glacial cycles. Since the eccentricity of Earth's orbit is always
small, the effect of precession on insolation is also small, though not
negligible. This is supported by the precession frequencies appearing alongside
those of eccentricity and obliquity in the power spectrum for the ice volume data
shown in Figure~\ref{fig:orbital_and_benthic_time_series_power_specs}.
\begin{table}
  \centering
\caption{Precession's effect on each hemisphere's seasons for its two extreme
values.}
\label{tab:precession_effects}
\begin{tabular}{cc|c|c}
 &&$\rho=0$ &
  $\rho=\pi$ \\\hline\hline
  \multicolumn{1}{c|}{\multirow{2}{*}[-6pt]{North}} &
  \multicolumn{1}{c|}{Summer} &
  \begin{tabular}[c]{@{}c@{}}Intense,\\ Short\end{tabular} &
  \begin{tabular}[c]{@{}c@{}}Mild,\\ Long\end{tabular} \\ \cline{2-4}
\multicolumn{1}{c|}{} &
  \multicolumn{1}{c|}{Winter} &
  \begin{tabular}[c]{@{}c@{}}Intense,\\ Long\end{tabular} &
  \begin{tabular}[c]{@{}c@{}}Mild,\\ Short\end{tabular} \\ \hline
  \multicolumn{1}{c|}{\multirow{2}{*}[-6pt]{South}} &
  \multicolumn{1}{c|}{Summer} &
  \begin{tabular}[c]{@{}c@{}}Mild,\\ Long\end{tabular} &
  \begin{tabular}[c]{@{}c@{}}Intense,\\ Short\end{tabular} \\ \cline{2-4}
\multicolumn{1}{c|}{} &
  \multicolumn{1}{c|}{Winter} &
  \begin{tabular}[c]{@{}c@{}}Mild,\\ Short\end{tabular} &
  \begin{tabular}[c]{@{}c@{}}Intense,\\ Long\end{tabular}
\end{tabular}
\end{table}

\section{Insolation}
\label{sec:insolation}
Insolation is solar irradiance integrated over time, however, due to the daily
fluctuations in insolation, we will average insolation either a day or a year.
As a result, the units for both irradiance and average insolation are Wm$^{-2}$.
We will first look at average daily insolation to see how this varies throughout
a year, then a yearly average will be used to investigate long-term changes to
insolation across different latitudes. We can then compare this with the
approximations used in the Budyko model to see how they differ. All simulations
that depend on the Milankovitch cycles take this data from Laskar
\cite{laskar2004}.

To model the average insolation over a chosen period, we first express the
irradiance arriving at a given point on Earth as a function of the orbital
parameters, latitude, and longitude. For this we will use an Earth centred
Earth-fixed frame. This derivation follows the approach taken by McGehee and
Lehman \cite{insol_latlon}.

Firstly, we express the total irradiance arriving at Earth's atmosphere. This
is calculated using the inverse square law, distributing the total solar
output $K$ over the surface area of an imagined sphere with a radius
equal to Earth's distance from the sun. This gives
\begin{equation}
  \frac{K}{4\pi r^2}\,\,\mathrm{Wm^{-2}}.
\end{equation}

We express a point on Earth using latitude and longitude ($\varphi,\gamma$),
which can be expressed with the unit vector
\begin{equation}
  \mathbf{u} = \begin{bmatrix}\cos\varphi\cos\gamma\\\cos\varphi\sin\gamma\\\sin\varphi\end{bmatrix}.
\end{equation}
The irradiance at ($\varphi,\gamma$) is proportional to the cosine of the angle
between $\mathbf{u}$ and the vector that points from Earth to the sun,
$\mathbf{n}$. This vector uses $\theta$, Earth's angle around the sun in
ecliptic polar coordinates, expressed in the Earth centred Earth-fixed frame as
\begin{equation}
  \mathbf{n} = \left(U_{\rho}U_{\beta}\right)^{-1}\begin{bmatrix}-\cos\theta\\-\sin\theta\\0\end{bmatrix} = 
  \begin{bmatrix}-\cos{\beta} \cos(\rho - \theta)\\
  \sin(\rho - \theta)\\-\sin{\beta} \cos(\rho -\theta)\end{bmatrix},
\end{equation}
where
\begin{equation}
  U_{\beta} = \begin{bmatrix}\cos\beta & 0 & \sin\beta\\0 & 1 & 0\\-\sin\beta & 0 & \cos\beta\end{bmatrix},\quad
  U_{\rho} = \begin{bmatrix}\cos\rho & -\sin\rho & 0\\\sin\rho & \cos\rho & 0\\0
  & 0 & 1\end{bmatrix}
\end{equation}
are the rotation matrices for obliquity and precession respectively.

Multiplying the scalar product of these two unit vectors by the irradiance at
the atmosphere gives
\begin{equation}
  \begin{split}
    I(r,\theta,\beta,\rho,\varphi,\gamma) &= \frac{K}{4\pi
    r^2}\,\mathbf{u}^\top\mathbf{n}\\
  &=\frac{K}{4\pi r^2}
  \begin{bmatrix}\cos\varphi\cos\gamma\\\cos\varphi\sin\gamma\\\sin\varphi\end{bmatrix}^\top
  \begin{bmatrix}-\cos{\beta} \cos(\rho - \theta)\\
  \sin(\rho - \theta)\\-\sin{\beta} \cos(\rho -\theta)\end{bmatrix}\\
  &= \frac{K}{4\pi r^2}[(\sin{\gamma}\sin{(\rho - \theta)}-\cos{\beta} \cos{\gamma}
    \cos{(\rho-\theta)})\cos{\varphi}
  -\sin{\beta}\sin{\varphi} \cos{(\rho -\theta)}].
\end{split}
\label{eq:irradiance_at_given_point}
\end{equation}

Note this expression is only valid for positive values, since the irradiance
value for the dark side of Earth will be negative, when it should be 0. To
calculate the average daily insolation, we can treat the orbital parameters as
constant and integrate with respect to longitude $\gamma$ over the range
$[0,2\pi]$. However, since this expression is only valid for positive values,
we must either augment our expression for irradiance $I$ to be the piecewise
function Max$(0,I)$, or define the limits such that we only integrate over the
range of longitudes that receive sunlight. Through simulation, the former
approach was found to be more computationally costly as the piecewise needed
evaluating at every step of the integration, whereas the piecewise limits only
need to be calculated once. 

To find these limits in terms of the orbital parameters, we consider a plane
passing through the Earth centred origin, separating the light and dark
hemispheres of Earth, rotating so that its normal vector always points towards
the sun. The plane can therefore be expressed as
\begin{equation}
  ax+by+cz=0,
\end{equation}
where $a$, $b$, and $c$ are the components of $\mathbf{n}$.

\begin{figure}
  \centering
  \includegraphics[width=0.5\linewidth]{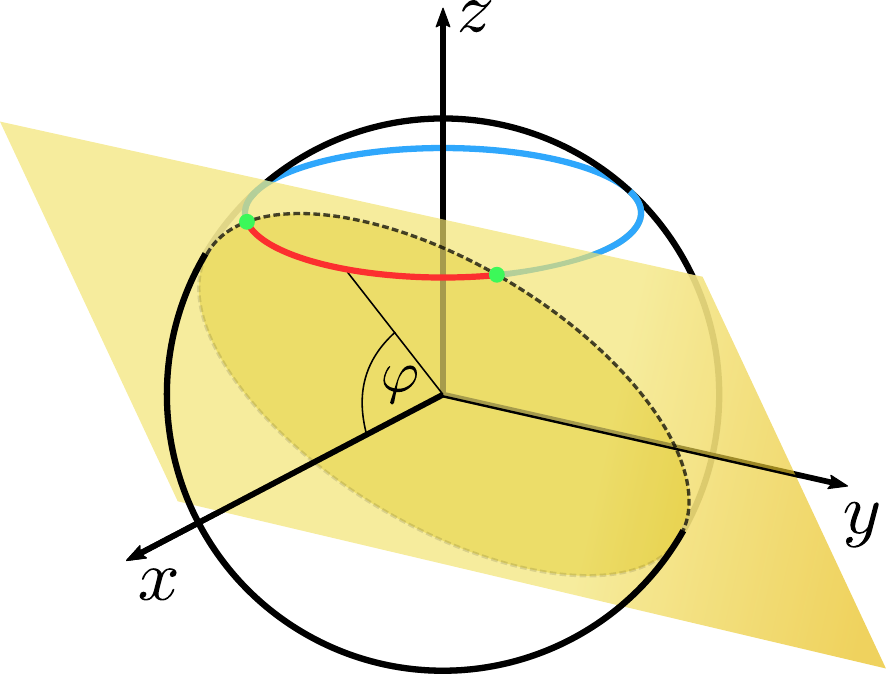}
  \caption[Insolation Calculation Longitude Range]{Diagram showing the day-night
    plane (yellow), currently intersecting the $y$-axis, which rotates about the
    origin to face the sun. The points where the circle of constant latitude
    $\varphi$ intersects the plane (green) indicate the range of longitudes for
    which sunlight will reach Earth at latitude $\varphi$ (red), whilst
    longitudes on the opposite side of the plane (blue) are in darkness so
  receive no insolation.}
  \label{fig:circ_plane_intercept}
\end{figure}

Next we consider a circle of constant latitude $\varphi$, lying parallel to the
Earth centred Earth-fixed $x$-$y$ plane, such that
\begin{equation}
  x^2+y^2 = \cos^2\varphi.
\end{equation}
This derivation is using a unit sphere for convenience since longitude is
invariant with size. The intersections of this circle with the plane separating
day and night indicate the limits over which integration should be performed for
the given latitude and orbital parameters. An example of this can be seen in
Figure \ref{fig:circ_plane_intercept}.

Using the final condition that 
\begin{equation}
  z = \sin\varphi,
\end{equation}
we find the two intersection points to be
\begin{equation}
  \boldsymbol{\mu}_{0,1}=\frac{1}{a^2 + b^2}\begin{bmatrix}- a c \sin{\varphi}\mp b
    \sqrt{\left(a^2+b^2\right) \cos^2{\varphi}- c^2 \sin^2{\varphi}}\\
  - b c \sin{\varphi}\pm a \sqrt{\left(a^2+b^2\right) \cos^2{\varphi}
  - c^2 \sin^2{\varphi}}\\
(a^2 + b^2)\sin{\varphi}\end{bmatrix}.
\label{eq:integration_limits}
\end{equation}

Looking at \eqref{eq:integration_limits}, we see there are two real solutions
when the discriminant, 
\begin{equation}
  \left(a^2+b^2\right) \cos^2{\varphi} - c^2 \sin^2{\varphi} = \Delta,
\end{equation}
is positive.

If $\Delta \le 0$, then there is either no intersection between the circle and
the plane, or only one. In these cases, the entire circle of latitude $\varphi$
will lie on the dark, or light, side of the plane.

To determine which side it is, we return to the scalar product
$\mathbf{u}^\top\mathbf{n}$ where $\mathbf{u}$ will point to some part of the
circle, for simplicity we choose $(\varphi,0)$. If no intersection has
occurred and $\mathbf{u}^\top\mathbf{n}$ is positive, then this point and the
entirety of the circle lies within sunlight, whilst a negative product 
corresponds to darkness. This is written as
\begin{equation}
  \mathbf{u}^\top\mathbf{n}\,\bigr|_{\gamma=0} = -\cos\left(\beta - \varphi
  \right) \cos\left(\rho - \theta \right).
\end{equation}

In the rare case that this is exactly equal to 0, a different point on the
circle is chosen.

The intersection points, $\boldsymbol{\mu}_0$ and $\boldsymbol{\mu}_1$, are
currently expressed in Cartesian coordinates. For the corresponding longitudes
we use the general relation $\gamma = \arctan\left(\frac{y}{x}\right)$. We can now
express the longitudinal limits in piecewise form as
\begin{equation}\gamma_0 = \begin{cases}
    \arctan\left(\frac{\mu_{0y}}{\mu_{0x}}\right), & \Delta>0\\
    0, & \Delta\le 0
  \end{cases},
  \quad\quad\quad
  \gamma_1 = \begin{cases}
    \arctan\left(\frac{\mu_{1y}}{\mu_{1x}}\right), & \Delta>0\\
    2\pi, &\Delta\le 0 \land\mathbf{u}^\top\mathbf{n}\,\bigr|_{\gamma=0}> 0\\
    0, &\Delta\le 0 \land\mathbf{u}^\top\mathbf{n}\,\bigr|_{\gamma=0}< 0
  \end{cases},
\end{equation}
where the limits become either $[0,2\pi]$ or $[0,0]$ if the given latitude lies
entirely in day or night respectively.

The expression for daily average insolation is therefore
\begin{equation}\begin{split}
  Q_{\mathrm{day}} &= \frac{1}{2\pi} \frac{K}{4\pi
  r^2}\int_{\gamma_0}^{\gamma_1} (\sin{\gamma}\sin{(\rho - \theta)}-\cos{\beta} \cos{\gamma}
    \cos{(\rho-\theta)})\cos{\varphi} -\sin{\beta}\sin{\varphi} \cos{(\rho
    -\theta)}\,\mathrm{d}\gamma\\ 
  &=\frac{K}{8\pi^2r^2}[(\gamma_{0} - \gamma_{1}) \sin{\beta}
  \sin{\varphi} \cos{\left(\rho - \theta \right)} +
  (\sin{\gamma_{0}} \cos{\beta}
  \cos{\left(\rho - \theta \right)} - \sin{\gamma_{1}} \cos{\beta} \cos{\left(\rho - \theta \right)}\\
&\quad\quad\quad\quad\quad\quad + \sin{\left(\rho - \theta \right)} 
\cos{\gamma_{0}} - \sin{\left(\rho - \theta
\right)} \cos{\gamma_{1}} ) \cos{\varphi}].
\end{split}
\end{equation}
\begin{table}
  \centering
  \caption{Earth's orbital parameter values in the year 2000 along with their
  approximate ranges.}
\label{tab:current_orbital_params}
\begin{tabular}{c|c|c|c}
Parameter                    & Value  & Range               & Unit   \\ \hline\hline
Eccentricity ($\varepsilon$) & 0.0167 & 0.0002 - 0.0613     & None   \\ \hline
Obliquity ($\beta$)          & 0.4090 & 0.3814 - 0.4255     & Radians \\ \hline
Precession ($\rho$)          & 2.9101 & $-\pi$ -  $\pi$       & Radians
\end{tabular}
\end{table}

\begin{figure}
  \hspace{-40pt}
  \input{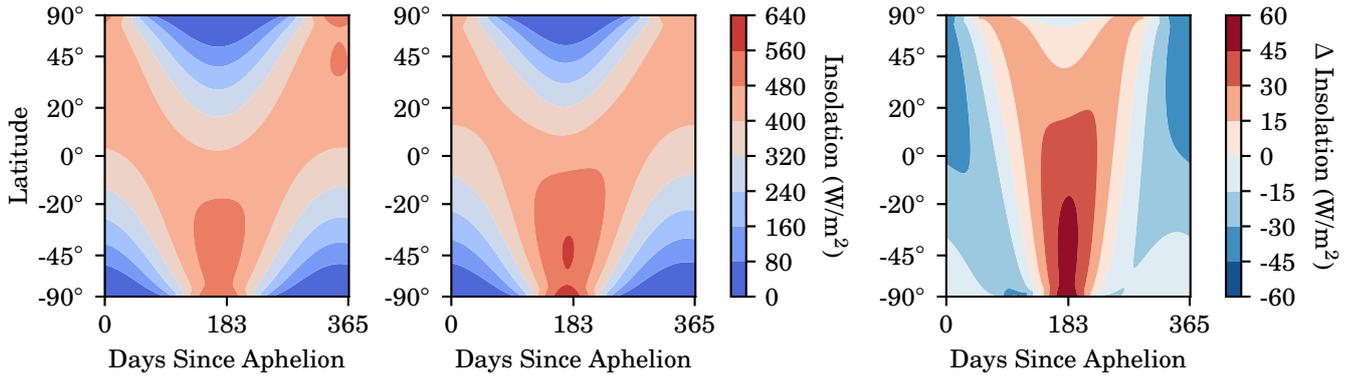}
  \caption[Daily Average Insolation Contours]{Contours showing the daily average
    insolation arriving at the atmosphere for different latitudes over a year
    period. The left plot is based on current orbital parameters, from Table
    \ref{tab:current_orbital_params}. In the second plot, eccentricity is
    increased to 0.06, approximately the maximum for Earth's orbit. The third
    plot shows how these two cases differ, where a positive value corresponds to
    more insolation in the second plot. The $y$-axis uses the domain $y=\sin
    \varphi$ as is used in the Budyko model, however the corresponding latitudes
  are shown for reference.}
  \label{fig:daily_ave_insol_all_lats}
\end{figure}

Using this expression, Figure \ref{fig:daily_ave_insol_all_lats} shows
how insolation varies over a year period, at each latitude, using current orbital
parameters. The second plot in this figure shows the same year, but with
eccentricity increased to 0.06, approximately the maximum eccentricity
of Earth's orbit. The third plot shows the difference between the two cases,
where a positive value corresponds to more insolation in the second plot. This
third plot shows that the maximum eccentricity can result in a 10\% increase in
polar summer insolation, which we will show can have a significant impact on ice
volume.

The summer solstice in the northern hemisphere currently occurs 13 days before
the aphelion of Earth's orbit, meaning Earth is almost the farthest it can be
from the sun. Conversely, summer in the southern hemisphere begins 170 days after the
aphelion, when Earth is almost the nearest it can be to the sun. Since the
irradiance that reaches Earth is proportional to $1/r^2$, where $r$ is the
distance from Earth to the sun, we see more insolation in the southern
hemisphere over its summer solstice than in the northern hemisphere's summer
solstice. The difference plot in Figure \ref{fig:daily_ave_insol_all_lats}
confirms this, showing that greater eccentricity leads to a greater difference
between maximum insolation values in the hemispheres.

Kepler's second law gives the angular speed of Earth around the sun as
\begin{equation}
  \frac{\mathrm{d}\theta}{\mathrm{d}t} = \frac{2\pi a b}{Pr^2},
\end{equation}
where $P$ is the year period, and $a$ and $b$ are the semi-major and semi-minor
axes respectively. On the scale of a year, these can be considered constant,
meaning Earth's angular speed at a given time is proportional to $1/r^2$.

As eccentricity increases, the distance between the perihelion and the sun
decreases and so the angular speed during this part of the orbit increases. This
decrease in distance causes an increase in the intensity of the southern
hemisphere summer, shown by the darkest red region in the difference plot of
Figure \ref{fig:daily_ave_insol_all_lats}. Due to the increased speed, the
southern summer lasts for a shorter period, leading to a decrease in the
insolation intensity either side of the peak. The opposite is true for the
northern hemisphere. Since the aphelion is now farther from the sun, the summer,
which occurs around aphelion, is less intense, but lasts longer. This is shown
by the widening of the red region in the top half of the difference plot, since
summer is extends further towards the middle of the year. Although eccentricity
effects the insolation in each hemisphere differently, we will see later that
they still receive the same total insolation over a year period.

We now express the average yearly insolation as a function of the orbital
parameters. Recall, the expression for irradiance at a given point is
\begin{equation}\begin{aligned}
  I(r,\theta,\beta,\rho,\varphi,\gamma) &= \frac{K}{4\pi r^2}\mathbf{u}^\top\mathbf{n}\\
    &= \frac{K}{4\pi
    r^2}\begin{bmatrix}\cos\varphi\cos\gamma\\\cos\varphi\sin\gamma\\\sin\varphi\end{bmatrix}^\top
    U_{\beta}^{-1}U_{\rho}^{-1}\begin{bmatrix}-\cos\theta\\-\sin\theta\\0\end{bmatrix},
  \end{aligned}
\end{equation}
which is only valid for positive values of $I$.

It will be useful to define 
\begin{equation}
  \begin{bmatrix}\cos\varphi\cos\gamma\\\cos\varphi\sin\gamma\\\sin\varphi\end{bmatrix}^\top
  U_{\beta}^{-1} = \begin{bmatrix}\sin{\beta} \sin{\varphi } + \cos{\beta}
  \cos{\gamma} \cos{\varphi}\\\sin{\gamma} \cos{\varphi }\\\cos{\beta }\sin{\varphi} - \sin{\beta} \cos{\gamma}
\cos{\varphi} \end{bmatrix}^\top =
  \begin{bmatrix}\cos\hat{\varphi}\cos\hat{\gamma}\\\cos\hat{\varphi}\sin\hat{\gamma}\\\sin\hat{\varphi}\end{bmatrix}^\top,
  \label{eq:u_hat_definition}
\end{equation}
where $\hat{\varphi}$ and $\hat{\gamma}$ can be visualised as the latitude and
longitude measured not from the north pole, but from the $z$-axis in the Earth
centred ecliptic frame. The irradiance at a given point can therefore be written
as
\begin{equation}
  I = \frac{K}{4\pi r^2}\begin{bmatrix}\cos\hat{\varphi}\cos\hat{\gamma}\\
  \cos\hat{\varphi}\sin\hat{\gamma}\\\sin\hat{\varphi}\end{bmatrix}^\top 
  U_{\rho}^{-1}\begin{bmatrix} -\cos{\theta}\\-\sin{\theta}\\0\end{bmatrix} 
  = -\frac{K}{4\pi r^2} \cos\hat{\varphi}\cos{\left(\hat{\gamma} + \rho -
  \theta \right)}.
\end{equation}
Treating the orbital parameters as fixed, the average insolation over a year
period $P$ is given by
\begin{equation}
  Q = \frac{1}{P}\int_0^P\frac{-K}{4\pi r^2}
  \cos\hat{\varphi}\cos{\left(\hat{\gamma} + \rho - \theta
  \right)}\,\mathrm{d}t,
  \label{eq:yearly_insol_integral_dt}
\end{equation}
however, since $r$ and $\theta$ have a non-linear dependence on time, we will 
use $\theta$ as the variable of integration. Returning to Kepler's second law,
we know that
\begin{equation}
  \frac{\mathrm{d}t}{\mathrm{d}\theta} = \frac{Pr^2}{2\pi ab},
\end{equation}
where $a$ and $b$ are the semi-major and semi-minor axes of Earth's orbit
respectively.

Substituting $\theta$ for $t$ in (\ref{eq:yearly_insol_integral_dt}) now
gives
\begin{equation}
\begin{split}
    Q &=\frac{1}{P}\int_0^{2\pi}\frac{-K}{4\pi r^2}
    \cos\hat{\varphi}\cos{\left(\hat{\gamma} + \rho - \theta \right)}
    \frac{Pr^2}{2\pi a b}\,\mathrm{d}\theta\nonumber\\ 
    &=\frac{K\cos\hat{\varphi}}{8\pi^2
    a b}\int_0^{2\pi}-\cos{\left(\hat{\gamma} + \rho - \theta \right)}
    \,\mathrm{d}\theta.\label{eq:yearly_insol}
  \end{split}
\end{equation}
Since the expression for $I$ is only valid for positive values, the integral in
(\ref{eq:yearly_insol}) is equivalent to the positive area under $\cos\theta$ 
from 0 to 2$\pi$, which gives 2. The average yearly insolation at a given point is therefore
\begin{equation}
  Q = \frac{K\cos\hat{\varphi}}{4\pi^2 a b}.
\end{equation}
Note that this is independent of precession, meaning that over a year, its
effect on insolation is cancelled out at all latitudes.

In order to express average yearly insolation over a latitude,
we must average $Q$ over longitude, $\gamma$. It will be of more use to
express this in the Earth-fixed frame, using the conventional latitude values.
For this we refer to (\ref{eq:u_hat_definition}), which defines
\begin{equation}
  \sin{\hat{\varphi}} = \cos{\beta}\sin{\varphi} -
  \sin{\beta}\cos{\gamma}\cos{\varphi},
\end{equation}
hence
\begin{equation}
  Q = \frac{K\sqrt{1 -\left( \cos{\beta}\sin{\varphi} -
  \sin{\beta}\cos{\gamma}\cos{\varphi}\right)^2}}{4\pi^2 a b}.
\end{equation}

We now average this over longitude to get
\begin{equation}
  \frac{K}{4\pi^2 a b}\frac{1}{2\pi} \int_0^{2\pi}\sqrt{1 -\left( \cos{\beta}\sin{\varphi} -
  \sin{\beta}\cos{\gamma}\cos{\varphi}\right)^2}\,\mathrm{d}\gamma.
\end{equation}

In order to express this purely as a function of the orbital parameters and
latitude, we use the relation $b = a\sqrt{1-\varepsilon^2}$, where $\varepsilon$
is orbital eccentricity. Laskar found that $a$ remains essentially constant,
regardless of the orbital parameters \cite{laskar2004}. We therefore treat
average yearly insolation for a given latitude as depending only on
$\varepsilon$ and $\beta$, giving
\begin{equation}
  Q_{\mathrm{year}}(\varepsilon,\beta,\varphi) = \frac{K}{8\pi^3 a^2
  \sqrt{1-\varepsilon^2}}\int_0^{2\pi}\sqrt{1 -\left( \cos{\beta}\sin{\varphi} -
  \sin{\beta}\cos{\gamma}\cos{\varphi}\right)^2}\,\mathrm{d}\gamma.
  \label{eq:q_year}
\end{equation}

We now examine the insolation forcing used in the Budyko model, which will be
introduced in the following chapter. Due to computational constraints, the
Budyko model uses an approximation for \eqref{eq:q_year}, which we will examine
here. The Budyko model focusses on just a single hemisphere, assuming
hemispheric symmetry. In addition, the Budyko model converts the spatial domain
from latitude $\varphi$ to the domain $y=\sin\varphi$. Expression our derived
insolation function with this new domain gives
\begin{equation}
  Q_{\mathrm{year}}(\varepsilon,\beta,y) = \frac{K}{8\pi^3 a^2
  \sqrt{1-\varepsilon^2}}\int_0^{2\pi}\sqrt{1 -\left( y\cos{\beta} -
  \sqrt{1-y^2}\sin{\beta}\cos{\gamma}\right)^2}\,\mathrm{d}\gamma.
  \label{eq:q_year_budyko}
\end{equation}

As we will show in the following chapter, the yearly insolation in the Budyko
model is approximated by $Q(\varepsilon)s(\beta,y)$, where
\begin{align}
  Q(\varepsilon) &= \frac{Q_0}{\sqrt{1-\varepsilon^2}},\\
  s(\beta,y) &= 1 + \frac{1}{2}c(\beta)(3y^2-1),\\
  c(\beta) &= \frac{5}{16}\left(3\sin^2{\beta} - 2\right).
\end{align}
The $Q(\varepsilon)$ term represents the average yearly insolation, which
depends on eccentricity, whilst $s(\beta,y)$ distributes this according to
latitude and obliquity. The integral of $s(\beta,y)$ in the range $y\in[0,1]$ is
1, ensuring that the global average insolation will have a magnitude equal to
$Q(\varepsilon)$.

\begin{figure}
  \centering
  \input{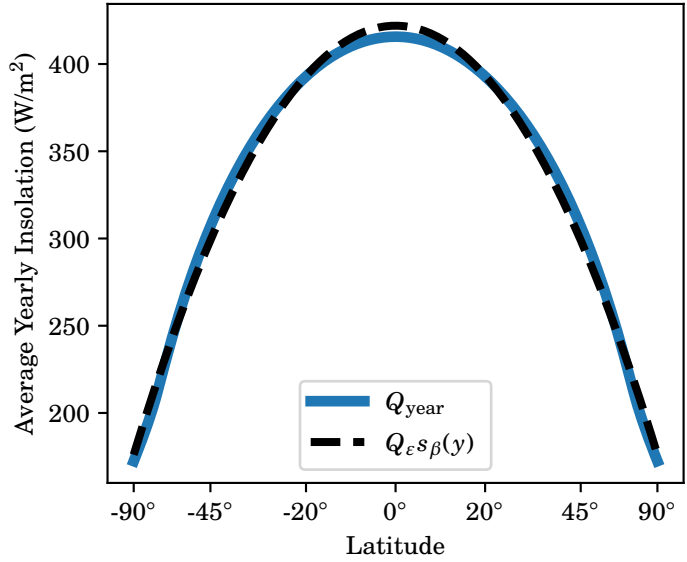}
  \caption[Average Yearly Insolation by Latitude vs Approximation]{The average
    yearly insolation $Q_{\mathrm{year}}$ given by \eqref{eq:q_year_budyko} for
    current orbital parameters, shown in Table \ref{tab:current_orbital_params}.
    The dashed line shows the approximation for yearly insolation used for the
    Budyko model in Section \ref{sec:budyko}. The horizontal axis uses the
    domain $y=\sin\varphi$, however the corresponding latitudes are shown for
  reference.}
  \label{fig:yearly_ave_insol_present}
\end{figure}

The solar constant is given by $\frac{K}{4\pi a^2}$ and is defined as the total
solar irradiance $K$ per unit area at Earth's atmosphere, 1 astronomical unit
(AU) from the sun, where the semi-major axis of Earth's orbit $a$ is 1\,AU. The
insolation constant used in the Budyko approximation $Q_0$, is equal to one
quarter of the solar constant. This is because the area of Earth that intercepts
the sun's rays is equivalent to a disk with the radius of the Earth, but over
a year, this is distributed across the entire area of Earth's surface, which is
4 times the area of the disk. We can therefore relate the magnitude of the
Budyko insolation function to our derived expression by
\begin{equation}
  Q(\varepsilon) = \frac{Q_0}{\sqrt{1-\varepsilon^2}}= \frac{K}{16\pi a^2 \sqrt{1-\varepsilon^2}}.
\end{equation}
By substituting this into (\ref{eq:q_year_budyko}), we see that
$s(\beta,y)$ is approximating the function
\begin{equation}
  \frac{2}{\pi^2}\int_0^{2\pi}\sqrt{1 -\left( y\cos{\beta} -
  \sqrt{1-y^2}\sin{\beta}\cos{\gamma}\right)^2}\,\mathrm{d}\gamma,
\end{equation}
which also equates to 1 when integrated over the range $y\in[0,1]$.
This function for insolation, along with the Budyko approximation, can be see in Figure
\ref{fig:yearly_ave_insol_present}. 
\begin{figure}
  \hspace{-40pt}
  \input{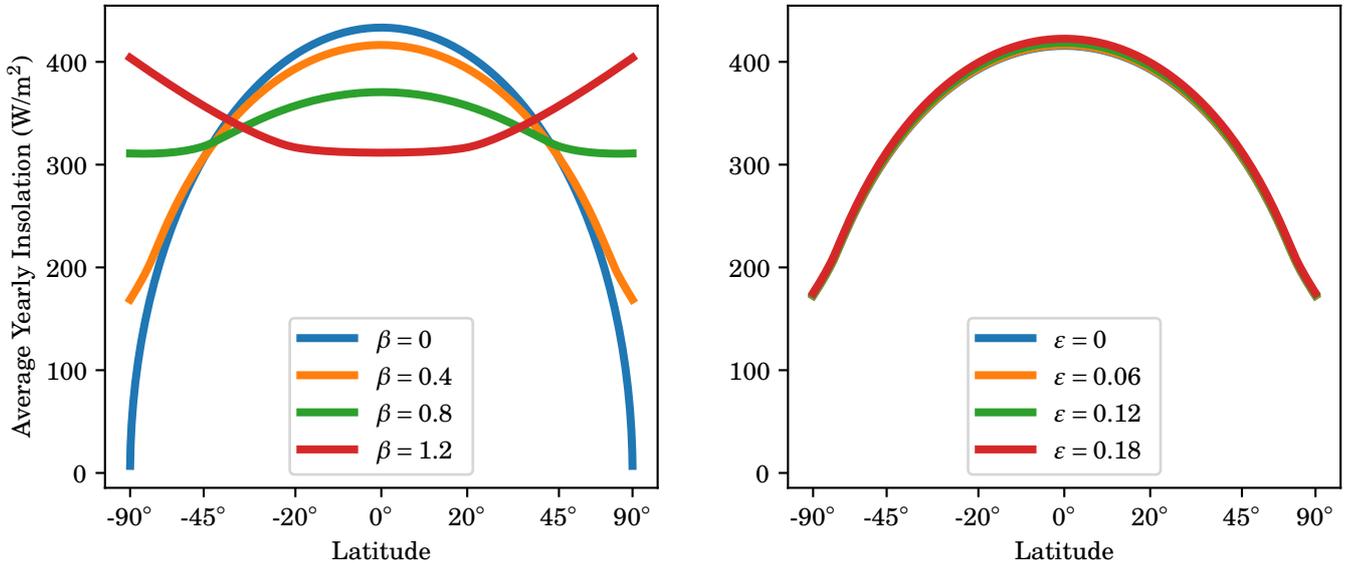}
  \caption[Effect of Obliquity and Eccentricity on Yearly Insolation]{Plots
    showing the effect of obliquity (left) and eccentricity (right) on the
    average yearly insolation across latitudes. Unless stated, parameter values
    are $\beta = 0.409$ and $\varepsilon = 0.0167$. The parameters are given
    values beyond their actual ranges to demonstrate their effect. Their actual
  ranges are give in Table \ref{tab:current_orbital_params}.}
  \label{fig:beta_eps_effect}
\end{figure}

By breaking the insolation expression into two terms, $Q(\varepsilon)$ and
$s(\beta,y)$, the effect of obliquity and eccentricity on yearly insolation can
be isolated. Their differing impacts on insolation are shown in Figure
\ref{fig:beta_eps_effect}, where each plot varies one of the two orbital
parameters. Since $\varepsilon$ only appears in the magnitude component of
global average insolation $Q(\varepsilon)$, we see that eccentricity will only
scale the insolation curve. Conversely, $\beta$ is only present in the function
$s(\beta,y)$, which distributes the insolation according to latitude and always
integrates to 1 over the range $y\in[0,1]$. We therefore see the shape of the
curve varying with $\beta$, but not the area underneath it.

\begin{figure}
  \centering
  \input{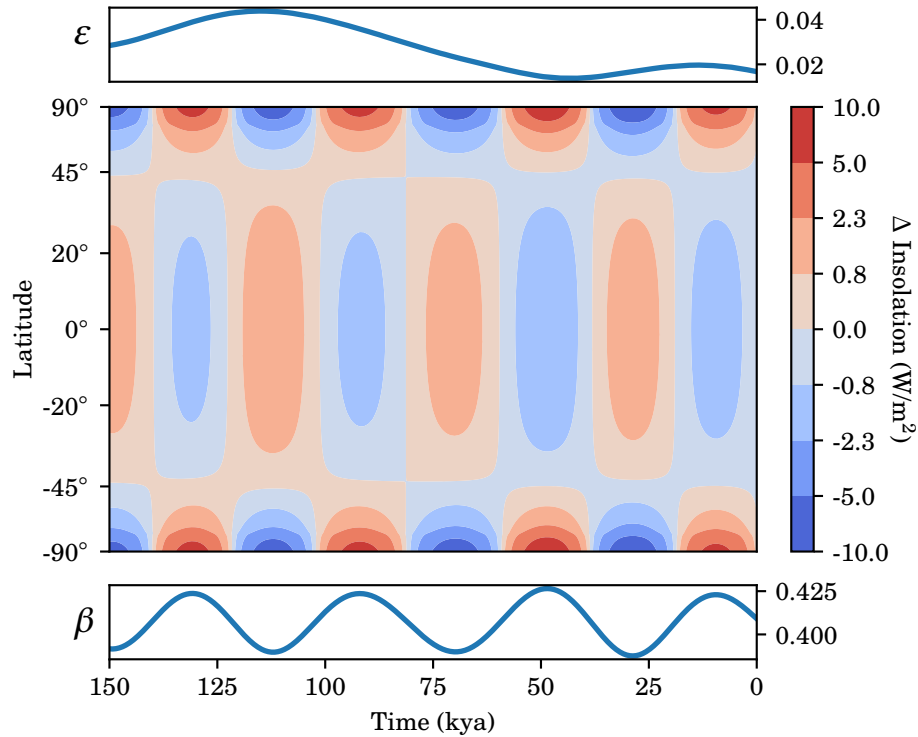}
  \caption[Effect of Obliquity and Eccentricity on Insolation Profile]{Contour
    showing the yearly average insolation anomaly over the 150k year period at
    each latitude. Above this is eccentricity $\varepsilon$, whilst obliquity
    $\beta$ is shown below. These are the two orbital parameters that govern the
  yearly average insolation.}
  \label{fig:yearly_ave_insol_all_lats}
\end{figure}

We can see the actual impact of these orbital parameters in Figure
\ref{fig:yearly_ave_insol_all_lats}, where the change in average yearly
insolation over the past 150\,kyr is shown alongside the two orbital parameters.
When obliquity is greater, Earth's equator aligns less with the ecliptic plane,
absorbing less insolation over the year. However the poles point closer towards
the sun for approximately one half of the year, whilst still receiving no
insolation during the other half. This means greater obliquity relates to
an increase in yearly insolation around the poles and a decrease around the
equator. More subtly, as eccentricity drops to lower values, at around 80\,kya,
we see the overall global insolation lowering slightly, with the predominant
colour turning from red to blue. 

From the past three plots we have shown, it is clear that our derived function
for yearly averaged insolation, and Budyko's approximation of it, are even about
the equator, meaning yearly averaged insolation at some latitude $\varphi$ is
always equal to the yearly averaged insolation at $-\varphi$. However, when
comparing a northern latitude with its southern counterpart in Figure
\ref{fig:daily_ave_insol_all_lats}, where daily insolation is shown throughout
the year, it would appear that this is not the case. The following derivation
aims to dispel this apparent contradiction. As discussed, Kepler's second law
tells us that the angular speed of Earth around the sun at any given time is
proportional to $1/r^2$, meaning it will take longer for Earth to traverse some
orbital range $\psi$ around aphelion of its orbit, where $r$ is at its
maximum value, compared to perihelion.

We also know that the irradiance reaching Earth is proportional to $1/r^2$.
Hence, for $\psi$ centred around aphelion for example, the increased duration
and the reduced irradiance cancel each other out. Applying this to an arbitrary
latitude, we will see how this results in a symmetric insolation curve when
averaged over a whole orbit.

To express insolation over the orbital range $\psi$, we integrate over
$\theta$. We again treat Earth as non-rotating, instead integrating over
longitude, $\gamma$. It is useful to separate our expression for irradiance at
a given point into two components; a magnitude term and a direction term. We
write this as
\begin{equation}
  I(r,\theta,\beta,\rho,\varphi,\gamma) =
  \frac{K}{4\pi r^2}S(\theta,\beta,\rho,\varphi,\gamma),
\end{equation}
where 
\begin{equation}
  S(\theta,\beta,\rho,\varphi,\gamma) = (\sin{\gamma}\sin{(\rho -
  \theta)}-\cos{\beta} \cos{\gamma} \cos{(\rho-\theta)})\cos{\varphi}
  -\sin{\beta}\sin{\varphi} \cos{(\rho -\theta)},
\end{equation}
which has a magnitude less than or equal to 1.

\begin{figure}
  \centering
  \includegraphics[width=0.7\linewidth]{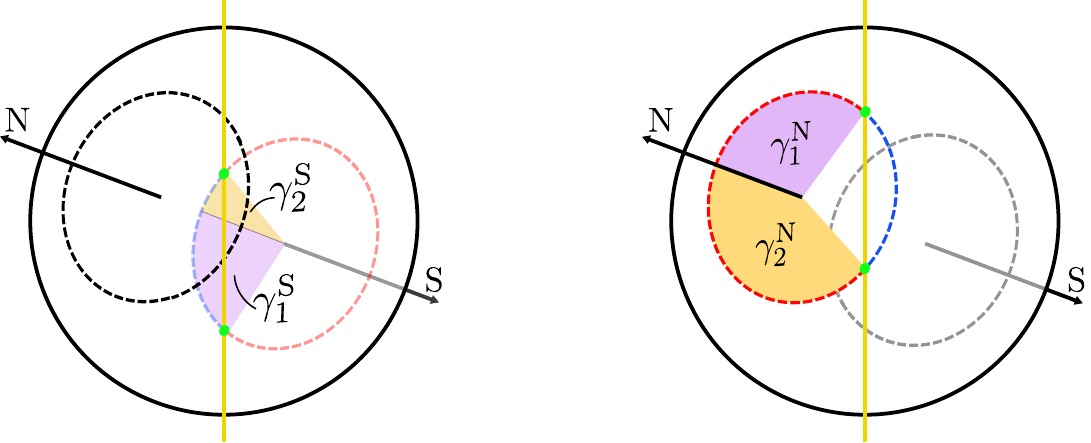}
  \caption[Longitudinal Angle Limits for Integration]{Diagrams of Earth at its
    perihelion (left) and aphelion (right) from a top-down perspective of the
    ecliptic plane with the north pole pointing out of the page. The two dashed
    circles in each diagram lie at latitudes $\varphi$ and $-\varphi$ in the
    northern and southern hemispheres respectively. The yellow line separates
    the day and night on Earth, with the sun illuminating the red sections of the
    circles. The longitudinal boundaries for these sections (green) are
    measured from the Earth centred Earth-fixed $x$-axis.}
  \label{fig:gamma_limits}
\end{figure}

We first measure the insolation over $\psi$ when centred around aphelion, which
occurs at angle $\theta_{\mathrm{A}}$. This is measured at latitude $\varphi$,
which we will choose to be in the northern hemisphere. This gives
\begin{equation}
  Q_{\mathrm{A}} = \int_{t(\theta_{\mathrm{A}}-\frac{\psi}{2})}^{t(\theta_{\mathrm{A}}+\frac{\psi}{2})}
  \int_{\gamma_1^{\mathrm{N}}}^{\gamma_2^{\mathrm{N}}}
  \frac{K}{4\pi r^2} S(\theta,\beta,\rho,\varphi,\gamma)\, \mathrm{d}\gamma\,\mathrm{d}t,
\end{equation}
where $t(\theta)$ is the time that corresponds to orbital angle $\theta$. Note
that the longitudinal limits also depend on $\theta, \beta, \rho$, and
$\varphi$, which vary so as to only span the positive range of $S$. This is shown by the red
regions in Figure \ref{fig:gamma_limits}.

Using Kepler's second law to change the variable of integration to $\theta$, we
obtain
\begin{align}
    Q_{\mathrm{A}} &= \int_{\theta_{\mathrm{A}}-\frac{\psi}{2}}^{\theta_{\mathrm{A}}+\frac{\psi}{2}}
    \int_{\gamma_1^{\mathrm{N}}}^{\gamma_2^{\mathrm{N}}}
  \frac{K}{4\pi r^2}S(\theta,\beta,\rho,\varphi,\gamma)\,\mathrm{d}\gamma\,\frac{\mathrm{d}t}{\mathrm{d}\theta}\,\mathrm{d}\theta\\
  &=\int_{\theta_{\mathrm{A}}-\frac{\psi}{2}}^{\theta_{\mathrm{A}}+\frac{\psi}{2}}\int_{\gamma_1^{\mathrm{N}}}^{\gamma_2^{\mathrm{N}}}
  \frac{K}{4\pi r^2}\frac{Pr^2}{2\pi a b}S(\theta,\beta,\rho,\varphi,\gamma)\, \mathrm{d}\gamma\,\mathrm{d}\theta\\
  &=\frac{KP}{8\pi^2 a
  b}\int_{\theta_{\mathrm{A}}-\frac{\psi}{2}}^{\theta_{\mathrm{A}}+\frac{\psi}{2}}
  \int_{\gamma_1^{\mathrm{N}}}^{\gamma_2^{\mathrm{N}}}
  S(\theta,\beta,\rho,\varphi,\gamma)\, \mathrm{d}\gamma \,\mathrm{d}\theta\\
    \begin{split}&=\frac{KP}{8\pi^2 a b}\biggl[\Bigl[\gamma \sin{\beta} \sin{\varphi}
  \sin(\rho - \theta) \label{eq:north_summer_solstice}\\
 &\quad\quad\quad\quad + (\sin{\gamma} \sin{(\rho - \theta)} \cos{\beta}
-\cos{\gamma}\cos(\rho -
\theta))\cos{\varphi}\Bigr]_{\gamma_1^{\mathrm{N}}}^{\gamma_2^{\mathrm{N}}}
\biggr]_{\theta_{\mathrm{A}}-\frac{\psi}{2}}^{\theta_{\mathrm{A}}+\frac{\psi}{2}}
      \end{split}
\end{align}

If we now compare this to the insolation over $\psi$ when centred around
perihelion, occurring at angle $\theta_{\mathrm{A}}+\pi$, we will see that the
opposite hemisphere, at latitude $-\varphi$, receives the same insolation.
Figure \ref{fig:gamma_limits} shows that when Earth is at two opposite points in
its orbit, the range of longitudes receiving sunlight at latitude $\varphi$ in
one position matches the range at latitude $-\varphi$ in the opposite position,
independent of precession. Since longitude is measured from the Earth centred
Earth-fixed $x$-axis, the southern limits are offset from the northern limits by
$\pi$, hence we can define the southern limits as
$\gamma_1^{\mathrm{S}}=\gamma_1^{\mathrm{N}}+\pi$ and
$\gamma_2^{\mathrm{S}}=\gamma_2^{\mathrm{N}}+\pi$.

Following the same steps as before but for the perihelion
side of the orbit, the insolation at latitude $-\varphi$ is
\begin{align}
  Q_{\mathrm{P}} &=\frac{KP}{8\pi^2 a b}\int_{\theta_{\mathrm{A}}+\pi-\frac{\psi}{2}}^{
  \theta_{\mathrm{A}}+\pi+\frac{\psi}{2}}\int_{\gamma_1^{\mathrm{N}}+\pi}^{\gamma_2^{\mathrm{N}}+\pi}S(\theta,
  \beta,\rho,-\varphi,\gamma)\,\mathrm{d}\gamma \,\mathrm{d}\theta\\
  &= \frac{KP}{8\pi^2 a b}\int_{\theta_{\mathrm{A}}-\frac{\psi}{2}}^{\theta_{\mathrm{A}}+\frac{\psi}{2}}
  \int_{\gamma_1^{\mathrm{N}}}^{\gamma_2^{\mathrm{N}}}
  S((\theta+\pi),\beta,\rho,-\varphi,(\gamma+\pi))\,
  \mathrm{d}\gamma \,\mathrm{d}\theta\\
    \begin{split}&= \frac{KP}{8\pi^2 a b}\biggl[\Bigl[\gamma\sin{\beta} \sin{\varphi}
  \sin(\rho - \theta) \\
  &\quad\quad\quad\quad+ (\sin{\gamma} \sin{(\rho - \theta)} \cos{\beta} -
\cos{\gamma} \cos(\rho -
\theta))\cos{\varphi}\Bigr]_{\gamma_1^{\mathrm{N}}}^{\gamma_2^{\mathrm{N}}}
\biggr]_{\theta_{\mathrm{A}}-\frac{\psi}{2}}^{\theta_{\mathrm{A}}+\frac{\psi}{2}},
        \end{split}
\end{align}
which is equal to (\ref{eq:north_summer_solstice}).

This derivation serves as proof that the insolation at a given latitude is
symmetric about the equator when averaged over a year, a fact that is utilised
by the Budyko model in the following chapter.

\subsection{$Q_{65}$}
\label{sec:Q65}
We have so far introduced an exact expression for both daily and yearly averaged
insolation, along with the approximated yearly average that is used in the
Budyko model. These functions all give a two dimensional function that varies
with time and latitude. Most of the models that we discuss in Chapter
\ref{chap:models} produce a one dimensional solution for ice volume,
representing just the global total. As a result, they all use a one dimensional
insolation function as input. The vast majority of conceptual models of this
form use a measure referred to as $Q_{65}$, which is the insolation at
65$^\circ$ north averaged over the summer solstice each year. The frequency of
this signal is strongly tied to the precession cycle, with its amplitude being
modulated by both eccentricity and obliquity. There has been a consensus within
the literature that the $Q_{65}$ signal is a good predictor for glacial
transitions \cite{milankovitch,imbrie}. Here we will investigate the nature of
this signal and its ability to predict glacial cycles.

From the previous section, we know that the irradiance at a given latitude
$\varphi$ and longitude $\gamma$ can be expressed as 
\begin{equation}
\frac{K}{4\pi r^2}[(\sin{\gamma}\sin{(\rho - \theta)}-\cos{\beta} \cos{\gamma}
    \cos{(\rho-\theta)})\cos{\varphi}
  -\sin{\beta}\sin{\varphi} \cos{(\rho -\theta)}],
\end{equation}
where $K$ is the sun's luminosity and $r$ is Earth's distance from the sun.
Recall this expression is only valid for positive values.
Using Kepler's first law, we can express $r$ as
\begin{equation}
  r = \frac{a(1-\varepsilon^2)}{1-\varepsilon\cos\theta},
\end{equation}
where $a$ is the semi-major axis of Earth's orbit, which can be considered
constant \cite{laskar2004}.

In order to calculate $Q_{65}$, we must average irradiance over the summer
solstice at 65$^\circ$ north. The northern summer solstice occurs when Earth's
north pole aligns with the sun in the ecliptic plane. At this point,
Earth's orbital position $\theta$ is equal to $\rho+\pi$. Substituting this
into the expression for $r$ and irradiance gives
\begin{equation}
  \frac{K(1+\varepsilon\cos\rho)^2}{4\pi a^2(1-\varepsilon^2)^2}
  (\cos{\beta}\cos{\gamma}\cos{\varphi} + \sin{\beta}\sin{\varphi}).
\end{equation}
To average over the summer solstice, we integrate longitude over the range
$[0,2\pi]$. The edge of the arctic circle lies at 66.56$^\circ$. Within this
circle, the sun does not set during the summer solstice. We therefore treat the
insolation function at 65$^\circ$ as remaining non-negative for all longitudes
during the summer solstice, avoiding the added complexity that arises from the
sun setting. This gives
\begin{equation}
\begin{split}
  Q_{65} &= \frac{K(1+\varepsilon\cos\rho)^2}{4\pi a^2(1-\varepsilon^2)^2}\int_0^{2\pi}
  \cos{\beta}\cos{\gamma}\cos{(65^\circ)} + \sin{\beta}\sin{(65^\circ)}\,
\mathrm{d}\gamma\\
    &= \frac{K(1+\varepsilon\cos\rho)^2}{4\pi
    a^2(1-\varepsilon^2)^2}2\pi \sin{(65^\circ)} \sin{\beta}\\
    &= M \frac{(1+\varepsilon\cos\rho)^2}{(1-\varepsilon^2)^2}\sin{\beta},
\end{split}
\end{equation}
where the latitude substitution, $\varphi=65^\circ$, has been made, and the
constant terms absorbed into $M$.

\begin{figure}
  \centering
  \input{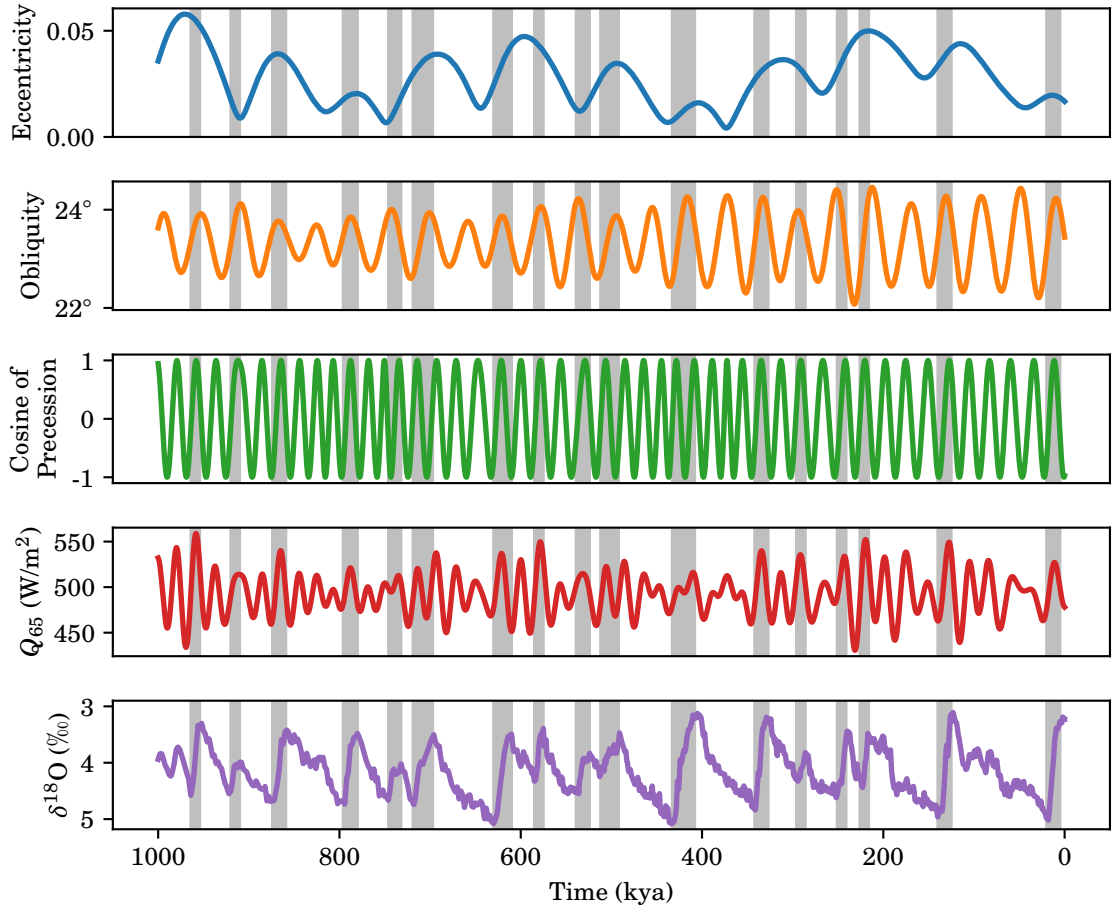}
  \caption[Interglacial Correlation with Orbital Forcing and Insolation]{Orbital
    parameters, $Q_{65}$, and benthic $\delta^{18}\mathrm{O}$ data over the past
    1\,myr. The amplitude of $Q_{65}$ can be seen to modulate according to
    eccentricity, whilst the frequency and phase of the oscillations align with
    those of obliquity and the cosine of precession. The grey regions highlight
    where the ice volume (represented by the benthic $\delta^{18}\mathrm{O}$
    data) transitions to an interglacial period. We can see these aligning with
  peaks in the $Q_{65}$ signal each time.}
  \label{fig:Q65_benthic}
\end{figure}

The time series plots in Figure \ref{fig:Q65_benthic}, show eccentricity,
obliquity, the cosine of precession, $Q_{65}$, and the benthic $\delta^{18}$O
data for the past million years. The grey bars highlight periods of significant
temperature rise in the benthic data. These short rises can be seen to correlate
with peaks in the other signals, aside from eccentricity, which aligns more
loosely with the entire interglacial period, instead of the short transition.
Although this appears to serve as evidence for $Q_{65}$ being a good predictor
of glacial transitions, we must consider the cross-dependencies of these
signals. $Q_{65}$ is a function of the three orbital parameters, which are
themselves well correlated to the ice volume data. It is therefore hard to
conclude that $Q_{65}$ is a better predictor than some other function that
combines these three orbital parameters.

Although each of the ice volume peaks align with a peak in $Q_{65}$, there are
also many peaks in $Q_{65}$ that do not align with a transition in the benthic
data. There is a noticeable correlation between the relative magnitude of the
peaks and their likelihood of aligning with a peak in the ice volume data.
However, these larger peaks come about from constructive interference between
the obliquity and precession signals, suggesting that a linear combination of
the two could produce a signal with similar predictive power. In addition, the
alignment of the 100\,kyr period in eccentricity with that of the ice volume
data suggests that it could play an important role in more generally predicting
the glacial dynamics.

We will revisit this notion with our own models in Chapters
\ref{chap:feedforward_model} and \ref{chap:feedback_model}, using a simple
linear combination of the orbital parameters in place of the $Q_{65}$ insolation
function. This approach reduces the chance for our model to make any unnecessary
assumptions about how the orbital parameters impact the ice volume. Where a
model with $Q_{65}$ assumes that the insolation reaching a specific latitude
during a specific time of year is the principle driver of ice volume, our models
will instead learn the relationship between the orbital parameters and ice
volume directly from the data.

\section{Conclusion}
In this chapter, we have introduced the data and derived the mathematical tools
that form the basis of the models presented in this thesis. We discussed the
available proxy data for global ice volume and justified the use of benthic
$\delta^{18}$O records from deep-sea sediment cores as the most reliable
continuous proxy over the past 800\,kyr. A key limitation of this data is its
sensitivity to both global ice volume and local bottom water temperature. To
address this, we adopted Bintanja's model-based decomposition, which isolates
the ice volume component from the benthic signal. This decomposed ice volume
data serves as the target against which all models in this thesis are evaluated.

We then introduced the three orbital parameters; eccentricity, obliquity, and
precession, and examined their individual effects on the insolation reaching
Earth's surface. We showed that eccentricity has a small but non-negligible
effect on the total insolation received by Earth, obliquity governs the
latitudinal distribution of insolation, and precession determines the seasonal
contrast within each hemisphere. From these, we derived expressions for both
daily and annually averaged insolation as functions of latitude and the orbital
parameters. We also derived the commonly used $Q_{65}$ signal and examined its
relationship with the benthic data. While $Q_{65}$ correlates well with glacial
transitions, we argued that this predictive power may largely stem from the
underlying orbital parameters themselves, rather than from any special
significance of the insolation at 65$^\circ$ north during the summer solstice.
This motivates our approach in Chapters \ref{chap:feedforward_model} and
\ref{chap:feedback_model}, where we use the orbital parameters directly as
forcing inputs, allowing the models to learn their relationship with ice volume
from the data without imposing assumptions inherent in $Q_{65}$.

\clearemptydoublepage
\chapter{Conceptual Models}
\label{chap:models}
\initial{B}efore we introduce our simple ice volume models in the following
chapters, we first provide some context for previous conceptual models within
this field. Conceptual models are characterised by their abstraction of the
dynamical system, simplifying a highly complex processes into the core
mechanisms that drive it. Although there are more detailed model types, such as
General Circulation Models (GCMs), these are computationally expensive and are
infeasible to run over the timescales we are interested in
\cite{abe_ouchi_full_model}.

We begin with the energy balance model from Budyko \cite{budyko}, which focuses
on the feedback that arises from the relationship between ice sheet albedo and
global temperature. We will base our analysis on the augmented version of this
model from Widiasih \cite{widiasih2013}, who brought the model into a dynamic
form, allowing for the ice line to move in response to the temperature profile.
Although this model is beneficial to understand the equilibria and stability of
the ice line, it is not able to adequately model the fluctuations that the ice
volume data shows. In order to explore whether this is possible with the model,
we introduce the following augmentations: sub-year simulation resolution, a
southern hemisphere, a more accurate heat diffusion term, a more accurate albedo
function, and a second ice line to capture land-based ice. These are intended to
improve the model's realism, to reduce the chance of an assumption causing the
model to diverge from the data.

Despite these augmentations, we are unable to fully capture the ice volume
fluctuations seen in the ice volume data. We then move on to reproduce and
analyse a number of conceptual models from the literature. These models differ
from the Budyko model by producing a single value for global ice volume as
opposed to a latitudinal distribution. Three of these models align with the
geochemical theory of climate change, whilst the other two are more closely
aligned with the astronomical theory. We show that these models are able to
reproduce the data well, though each have strengths and limitations. From this
analysis, we are able to inform the development of our own conceptual models,
that will be introduced in the following chapters.

\section{Budyko Model}
\label{sec:budyko}
The Budyko model is an energy balance model (EBM) that explores the positive
feedback effect of ice albedo, whereby the increased reflectivity of polar
ice reduces the insolation that Earth can absorb around these regions. According
to the model, if Earth's ice sheets were to extend sufficiently far down toward
the equator, then a tipping point would be reached and a runaway glaciation
would occur, covering the entire Earth in ice. This model also contains a stable
equilibrium point for the ice sheet, which is approximately consistent with the
current ice line latitude. The model is based on the energy balance of Earth,
with the temperature at a given latitude being determined by the balance of
insolation, re-radiation, and heat transport around the Earth.

\begin{figure}
  \centering
  \includegraphics[width=0.7\linewidth]{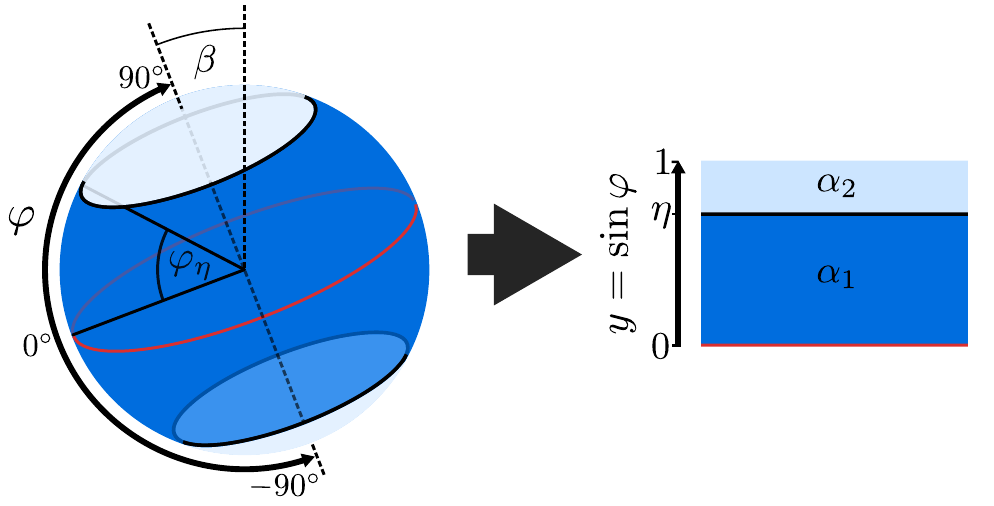}
  \caption[Budyko Model Domain]{The change of domain used for the Budyko model,
    latitude $\varphi$ becomes $y = \sin\varphi$ with only the northern
    hemisphere being considered. The ice line lies at latitude $\varphi_\eta$
    and separates the water from the polar ice. This becomes $\eta$ in the new
    domain with range $[0,1]$. The albedo of the ocean is represented by
  $\alpha_1$, whilst the polar ice has albedo $\alpha_2$.}
  \label{fig:budyko_domain}
\end{figure}

A number of assumptions are made in order for this EBM to work. Temperature is
modelled as varying only across latitude, reducing the domain of the system from
a spherical surface, to a latitudinally distributed temperature profile. In
order to simplify integrals that span the latitudes, the domain is defined
using $y = \sin\varphi$, where $\varphi$ is latitude. To simplify the system
further, the Earth is treated as an entirely water-based planet. This allows for
the heterogeneous reflectivity of Earth to be represented by just its average
albedo, whilst the ice covered region uses the average ice sheet albedo. The
Earth is also modelled as symmetric across the equator, allowing for only one
half of the domain to be considered, i.e. $y \in [0,1]$. This simplified domain
is shown in Figure \ref{fig:budyko_domain}.

The original Budyko model form 1968 consisted of a single equation that
described the change in heat energy at a given latitude over time. We will
introduce this model in a more modern form, drawing from the work of McGehee and
Lehman, who express insolation as a function of the orbital parameters
\cite{insol_latlon}, and Widiasih, who added a dynamic ice line
\cite{widiasih2013}. The change in temperature in this model depends on three
components. They describe the flow of energy into, out of, and around Earth.

These three components are insolation, reradiation, and heat transport and are
expressed in the model as
\begin{equation}
  R\frac{\partial T(y,t)}{\partial t} = \underbrace{Q_\varepsilon s_\beta(y)
  (1-\alpha_\eta(y))}_{\text{Insolation Absorbed}}
  - \underbrace{(A+BT(y,t))}_{\text{Reradiation}}
  + \underbrace{C(\overline{T}(t)-T(y,t))}_{\text{Transport}}
  \label{eq:budyko_temp_init}
\end{equation}
where $R$ is the heat capacity of Earth per unit area, $T(y,t)$ is the
temperature at latitude $y$ and time $t$. The right hand side of the equation is
made up of the three components, which we will now introduce.

The first component of absorbed insolation is the average yearly insolation
reaching Earth, $Q_\varepsilon$, which varies according to the eccentricity of
Earth's orbit, $\varepsilon$, such that $Q_\varepsilon =
Q_0/\sqrt{(1-\varepsilon^2)}$.

As discussed in the previous chapter, $Q_0$ is calculated using the solar
constant. This is attained through satellite measurements and is the solar
irradiance per unit area arriving at Earth's atmosphere, one astronomical unit
from the sun. As explained in Section \ref{sec:insolation}, the average annual
insolation for a given point on Earth is a quarter of the solar constant, as the
intercepted insolation is distributed over four times the area. In the
literature, $Q_0$ has often been given as 343\,Wm$^{-2}$
\cite{insol_latlon,widiasih2013,c_beta} which comes from Tung
\cite{tung2007topics}. However, in this work, Tung uses a solar constant of
1372\,Wm$^{-2}$, which is now considered above the actual value. More recently,
Kopp estimated the current solar constant to be approximately 1361.5\,Wm$^{-2}$
\cite{solar_constant}, which gives $Q_\varepsilon=340.375$\,Wm$^{-2}$, and so
$Q_0 = Q_\varepsilon \sqrt{(1-\varepsilon^2)} = 340.327$\,Wm$^{-2}$ given the
current value for $\varepsilon$, as shown in Table
\ref{tab:current_orbital_params}. Although this is a relatively small
difference, Budyko found that the model is very sensitive to the insolation
constant used, with an increase of just 1\% leading to a temperature increase of
1.5$^\circ$C \cite{budyko}.

This average insolation value then scales a latitudinal distribution function,
as introduced in Section \ref{sec:insolation}. In that section, we first derived
the formula for the annual insolation at a given latitude, which is analytically
intractable. We then introduced an approximation for this function which
produces a curve that is within 2\% of the actual function, and is made up of
the even Legendre polynomials,
\begin{equation}
  p_0(y) = 1,\quad\quad\quad\quad p_2(y) = \frac{1}{2}(3y^2-1),
\end{equation}
to give
\begin{equation}
  s_\beta(y) = 1 + \frac{1}{2}c_\beta(3y^2-1),
\end{equation}
where $c_\beta = \frac{5}{16}(3\sin^2\beta - 2)$ and $\beta$ is obliquity
\cite{c_beta}. This approximation can be seen alongside the actual function in
Figure \ref{fig:yearly_ave_insol_present}. The final component of the insolation
term is $(1-\alpha_\eta(y))$, which describes the degree to which heat is
absorbed at a given latitude. This depends on the albedo of the surface, which
this model assumes to be either water or ice. This piecewise albedo function is
given by
\begin{equation}
  \alpha_\eta(y) = \begin{cases} \alpha_1, & y<\eta\\
                                 \alpha_2, & y>\eta\\
                                 \frac{1}{2}(\alpha_1+\alpha_2), & y=\eta
                   \end{cases},
\end{equation}
where $\eta$ is the position of the ice line, whilst $\alpha_1=0.32$ and
$\alpha_2=0.62$ are established using satellite data to represent albedo for an
ice-free or ice covered latitude \cite{tung2007topics}. This results in less
heat being absorbed over the area where ice is present due to the increased
reflectivity.

The next component of the equation is the outgoing infrared radiation.
Although the energy flow occurring at the boundary between Earth and space is
complex, North performed a linear regression on satellite data to estimate the
net flow of long-wave radiation leaving Earth to the first order
\cite{linear_reradiation}. This gives
\begin{equation}
  A+BT(y,t),
\end{equation}
where, for average cloud cover, $A=202.1$\,Wm$^{-2}$ and
$B=1.9$\,Wm$^{-2}\,^\circ$C$^{-1}$ with a correlation coefficient of 0.90.

The final component describes the latitudinal transport of heat, from the hot
equator to the cold north pole. The heat transport at a given latitude is
determined by the difference in temperature at that latitude and the global
average temperature, which we express as
\begin{equation}
  C(\overline{T}(t)-T(y,t)),
\end{equation}
where
\begin{equation}
  \overline{T}(t) = \int_0^1 T(y,t) \,\mathrm{d}y.
\end{equation}
By calculating the equilibrium temperature, $C$ can be chosen such that the
equilibrium matches current climatic conditions. Tung's work on this yielded the
relation $C=1.6B=3.04$\,Wm$^{-2}\,^\circ$C$^{-1}$ \cite{tung2007topics}.
Once again, these heat flow components are combined to give the full temperature
equation
\begin{equation}
  R\frac{\partial T(y,t)}{\partial t} = \underbrace{Q_\varepsilon s_\beta(y)
  (1-\alpha_\eta(y))}_{\text{Insolation Absorbed}}
  - \underbrace{(A+BT(y,t))}_{\text{Reradiation}}
  + \underbrace{C(\overline{T}(t)-T(y,t))}_{\text{Transport}}.
  \label{eq:budyko_temp}
\end{equation}
This is known as an integro-differential equation, differing from a conventional
partial differential equation in that there is no derivative of the spatial
variable $y$. The only term that relates the temperature at a given position to
the rest of the system is $\overline{T}(t)=\int_0^1 T(y,t) \,\mathrm{d}y$, which
is the average global temperature.

To understand the time constant $R$, which governs the temperature's rate of
change, we inspect the units of the equation's terms. The insolation term has
units Wm$^{-2}$ from the $Q_\varepsilon$ component with the other two components
being dimensionless. The heat loss term describes radiation leaving Earth per
unit area, so again has units Wm$^{-2}$. Since $T$ is in $^\circ$C, the units
for $B$ are Wm$^{-2}$\,$^\circ$C$^{-1}$. $C$ is defined as $1.6B$ so also has
units Wm$^{-2}$\,$^\circ$C$^{-1}$, meaning the heat transport term is in
Wm$^{-2}$ as well. Since $\frac{\partial T(y,t)}{\partial t}$ will be in
$^\circ$C\,s$^{-1}$, the units for $R$ are J\,m$^{-2}$$^\circ$C$^{-1}$, which
equates to the heat capacity of Earth per unit area.

The value that $R$ takes will govern the rate at which the temperature profile
equilibrates. The heat capacity of water is approximately
4$\times10^3$J\,kg$^{-1}$$^\circ$C$^{-1}$ =
4$\times10^6$J\,m$^{-3}$$^\circ$C$^{-1}$. With the assumption that Earth is
entirely water, which must be heated to a depth of 100\,m, we have that
\begin{equation}
  R = 4\times10^8\mathrm{J\,m}^{-2}\,^\circ\mathrm{C}^{-1}.
\end{equation}
This is a very approximate value, however $R$ does not appear in equilibrium
solutions, relating only to the dynamic behaviour of the model.

As we are interested in the dynamics of the models in this chapter, we now
introduce the ice line equation developed by Widiasih, which allowed for the ice
line to move in response to the temperature profile. The equation for this given
by
\begin{equation}
  S\frac{\mathrm{d}\eta(t)}{\mathrm{d}t} = T(\eta,t) - T_{\mathrm{ice}}
  \label{eq:budyko_ice_line}
\end{equation}
where $T_{\mathrm{ice}}$ is an estimate for the critical temperature at which an
ice sheet can form and $S$, which has units C$^\circ$s, scales the ice line's
rate of movement. Widiasih estimates this to be on the order of
$10^{13}$\,C$^\circ$s. Comparing this to the equivalent $R=4\times10^8$ for the
temperature dynamics, we can categorise this as a slow-fast system. This means
that the fast variable, temperature, can be considered equilibrated when
modelling the dynamics of the slow ice line.

Although the value for $S$ is a rough approximation, as with $R$, it is only
required for the dynamic analysis. Before looking at the dynamic behaviour of
the Budyko model, we first look at the equilibria of the model. This will allow
us to understand the stability of Earth's climate, and the role that the ice
line plays in this. To calculate these, we set \eqref{eq:budyko_temp} and
\eqref{eq:budyko_ice_line} equal to 0 and solve for the state variables.

It will be useful to define
\begin{equation}
  I_\eta = \int_0^1 s_\beta(y)(1-\alpha_\eta(y))\,\mathrm{d}y,
\end{equation}
which requires that $\eta$ is known.

First setting \eqref{eq:budyko_temp} equal to 0 and integrating over the range
$[0,1]$, we attain
\begin{equation}
  Q_\varepsilon I_\eta - (A+B\overline{T^*_\eta}) = 0,
  \label{eq:T_bar}
\end{equation}
where $\overline{T^*_\eta}$ is the integral over $[0,1]$, of the equilibrium
temperature $T^*_\eta$.

Substituting this into \eqref{eq:budyko_temp} gives
\begin{equation}
  Q_\varepsilon s_\beta(y)(1-\alpha_\eta(y)) - (A+BT^*_\eta(y)) - C(T^*_\eta(y) -
  \overline{T^*_\eta}) = 0,
\end{equation}
from which we get
\begin{equation}
T^*_\eta(y) = \frac{1}{B+C}\Big[Q_\varepsilon s_\beta(y)(1-\alpha_\eta(y)) -
A+C\overline{T^*_\eta}\Big].
\label{eq:T_star}
\end{equation}

For the ice line we return to $I_\eta$, which we can express as
\begin{equation}\begin{split}
    I_\eta &= \int_0^1 s_\beta(y) (1-\alpha_\eta(y))\,\mathrm{d}y\\
           &= 1 - \int_0^1 s_\beta(y) \alpha_\eta(y)\,\mathrm{d}y\\
           &= 1 - \alpha_2\int_0^1 s_\beta(y)\,\mathrm{d}y +
           (\alpha_2-\alpha_1)\int_0^\eta s_\beta(y)\,\mathrm{d}y,\\
  \end{split}
\end{equation}
using the definition $\int_0^1 s_\beta(y)=1$ and then breaking up the
discontinuous albedo function into two integrals.

Now substituting $s_\beta(y) = 1 + \frac{1}{2}c_\beta(3y^2-1)$, we get
\begin{equation}\begin{split}
    I_\eta &= 1 - \alpha_2 + (\alpha_2-\alpha_1)\left[\left(1-\frac{1}{2}
      c_\beta\right)y + \frac{1}{2}c_\beta y^3\right]^\eta_0\\
           &= 1 - \alpha_2 + (\alpha_2-\alpha_1)\left(1-\frac{1}{2}c_\beta\right)\eta +
           \frac{1}{2}c_\beta(\alpha_2-\alpha_1)\eta^3.
   \end{split}
\end{equation}

Returning to \eqref{eq:T_star}, for $y=\eta$ we have
\begin{equation}
  T^*_\eta(\eta) = \frac{1}{B+C}\bigg[Q_\varepsilon s_\beta(\eta)
  (1-\alpha_0)- A+C \frac{Q_\varepsilon I_\eta - A}{B} \bigg],
  \label{eq:simple_T_star}
\end{equation}
where $\overline{T^*_\eta}$ is substituted using \eqref{eq:T_bar}.

Finally, substituting for $I_\eta$, we attain
\begin{equation}
  \begin{split}
  T^*_\eta(\eta) &= \frac{1}{B+C}\Bigg[Q_\varepsilon\left(1-\frac{1}{2}c_\beta\right)
    (1-\alpha_0) - A\left(1+\frac{C}{B}\right) + \frac{CQ_\varepsilon}{B}(1-\alpha_2) +
    \frac{CQ_\varepsilon}{B}(\alpha_2 - \alpha_1)\left(1-\frac{1}{2}c_\beta\right)\eta\\
    &\quad\quad\quad\quad\quad\quad
    +\frac{3}{2}Q_\varepsilon c_\beta(1-\alpha_0)\eta^2 + \frac{CQ_\varepsilon
  c_\beta}{2B}(\alpha_2-\alpha_1)\eta^3\Bigg].
\end{split}
\end{equation}

To attain $\frac{\mathrm{d}\eta(t)}{\mathrm{d}t} = 0$, we set $T^*_\eta(\eta) =
T_{\mathrm{ice}}$. This produces a cubic equation for the equilibrium ice line
in the form
\begin{equation}
  f(\eta) = a_3\eta^{*3} + a_2\eta^{*2} + a_1\eta^* + a_0 = 0,
  \label{eq:ice_line_equilibrium}
\end{equation}
where
\begin{align}
    &a_3 = \frac{CQ_\varepsilon c_\beta}{2B}(\alpha_2-\alpha_1)\\
    &a_2 = \frac{3}{2}Q_\varepsilon c_\beta(1-\alpha_0)\\
    &a_1 = \frac{CQ_\varepsilon}{B}(\alpha_2-\alpha_1)\left(1-\frac{1}{2}c_\beta\right)\\
    &a_0 = Q_\varepsilon\left(1-\frac{1}{2}c_\beta\right)(1-\alpha_0) -
    A\left(1+\frac{C}{B}\right) + \frac{CQ_\varepsilon}{B}(1-\alpha_2) -
    (B+C)T_{\mathrm{ice}}.
\end{align}

\begin{figure}
  \centering
  \input{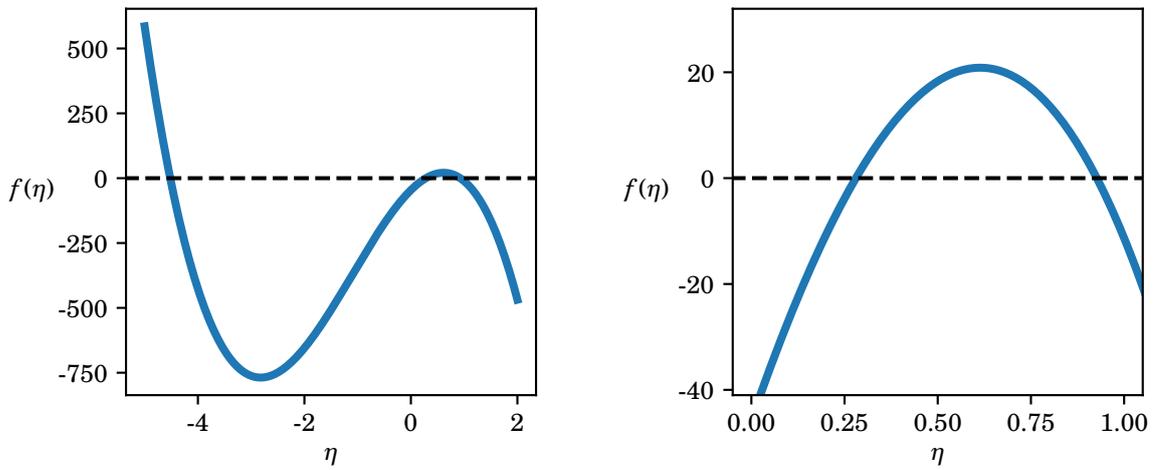}
  \caption[Default Budyko Equilibria]{Two scales of the same plot, using
    (\ref{eq:ice_line_equilibrium}) with the default parameter values shown in
    Table \ref{tab:budyko_params}. The valid range for $\eta$ is [0,1]. The
    right plot shows that with current parameter values, this yields equilibrium
  points 0.2803, and 0.9268 within the valid range.}
  \label{fig:ice_line_equilibrium}
\end{figure}

The equilibrium solutions for the ice line are shown in Figure
\ref{fig:ice_line_equilibrium}, with the two valid roots shown in the plot on
the right. The parameter values used for all Budyko simulations, unless stated,
are all based on physical measurements and are given in Table
\ref{tab:budyko_params}, whilst the orbital values used are given in Table
\ref{tab:current_orbital_params}. The two valid equilibria are 0.2803, and
0.9268, which correspond to latitudes of 16$^\circ$ and 68$^\circ$ respectively.
The equilibrium closer to the equator is unstable, whilst the polar equilibrium
is stable, suggesting that Earth's current ice line latitude, which annually
varies between approximately $60^\circ$ and $80^\circ$, is stable. This ice
line range can be seen in Figure \ref{fig:sea_ice_march_sept}, which shows the
northern sea ice extrema during the year, averaged over the past 133 years.

To understand the role of the unstable equilibrium, we consider what would
happen if the ice line was to grow significantly, extending beyond 0.2803
($16^\circ$) towards the equator. This would require extreme circumstances such
as catastrophic volcanic activity or a large asteroid impact, which could lead to
aerosols and debris in the atmosphere, blocking sunlight and cooling the Earth.
According to the Budyko-Widiasih model, extending beyond $16^\circ$ would lead
to a runaway effect, ending with the entire Earth covered in ice, a state often
dubbed The Snowball Earth. In order to transition out of Snowball Earth, the
average annual insolation would have to increase from 340.374\,Wm$^{-2}$ to
376.5\,Wm$^{-2}$. To put this into perspective, the sun has steadily increased
from 70\% of its current luminosity, 4,700\,myr ago \cite{solar_change}.
Therefore, over the duration of the entire Pleistocene, lasting 2.588\,myr,
solar luminosity has increased by $\frac{2.588}{4700}\cdot30\% = 0.0165\%$ of
the current value, compared to the $10.6\%$ increase needed to provide Earth
with 376.5\,Wm$^{-2}$.

Since solar irradiance is near constant over a 100\,kyr glacial cycle, the only
other factor that could change the irradiance reaching Earth is eccentricity.
The eccentricity required to provide Earth with an average yearly insolation of
376.5\,Wm$^{-2}$, is $\sqrt{1-\frac{Q_0}{376.5}}=0.428$, where $Q_0$ is the
irradiance reaching Earth in an entirely circular orbit. This value is
approximately 7 times larger than the maximum estimated eccentricity of Earth's
orbit.
\begin{figure}
  \centering
  \input{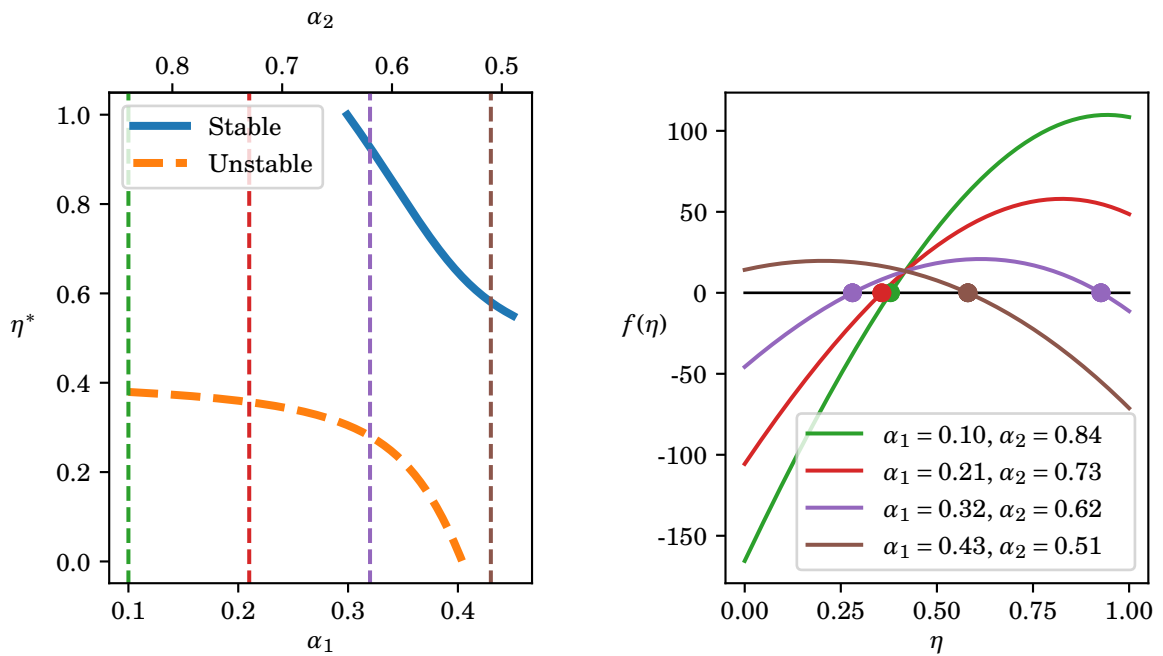}
  \caption[Budyko Equilibria Stability]{Equilibria of the Budyko model with
    varying albedo values. The left plot shows the bifurcation diagram,
    beginning with severely different albedos for the ocean and ice, then
    converging to the estimated values given in Table \ref{tab:budyko_params},
    and then further past this point. The right plot shows four vertical slices
    of the bifurcation diagram, using \eqref{eq:ice_line_equilibrium}, with the
    purple line corresponding to the values used in the model.}
  \label{fig:alpha_bifurcation}
\end{figure}

The equilibria of the Budyko model contribute to our understanding of the
stability of Earth's climate. The stable equilibrium at 0.927 ($70^\circ$) is
consistent with the real current ice line latitude, and suggests that Earth in
its current climatic state is stable. However, it is important to consider that
Earth's climate is not static, and the parameters of the model may change. If we
look at the albedo values used in the model, we see that the ocean has an albedo
of 0.32, whilst the ice has an albedo of 0.62. These values are estimates, and
already assume a homogeneous water planet with uniform ice cover at the pole. In
Figure \ref{fig:alpha_bifurcation}, we vary the two albedo values at a constant
rate in opposing directions. We see that the stable equilibrium at 0.927 (marked
by the purple dashed line) can quickly drop to 0.6 ($37^\circ$) by changing the
albedo values by just 0.1. This new equilibrium would see the ice line reaching
to southern Spain. We must therefore be cautious about the conclusions we draw
from a model that uses a number of estimates and simplifications for parameters
that play a crucial role in its depiction of Earth's climate.
\begin{figure}
\centering
\begin{subfigure}{.4\textwidth}
  \centering
  \includegraphics[width=\linewidth]{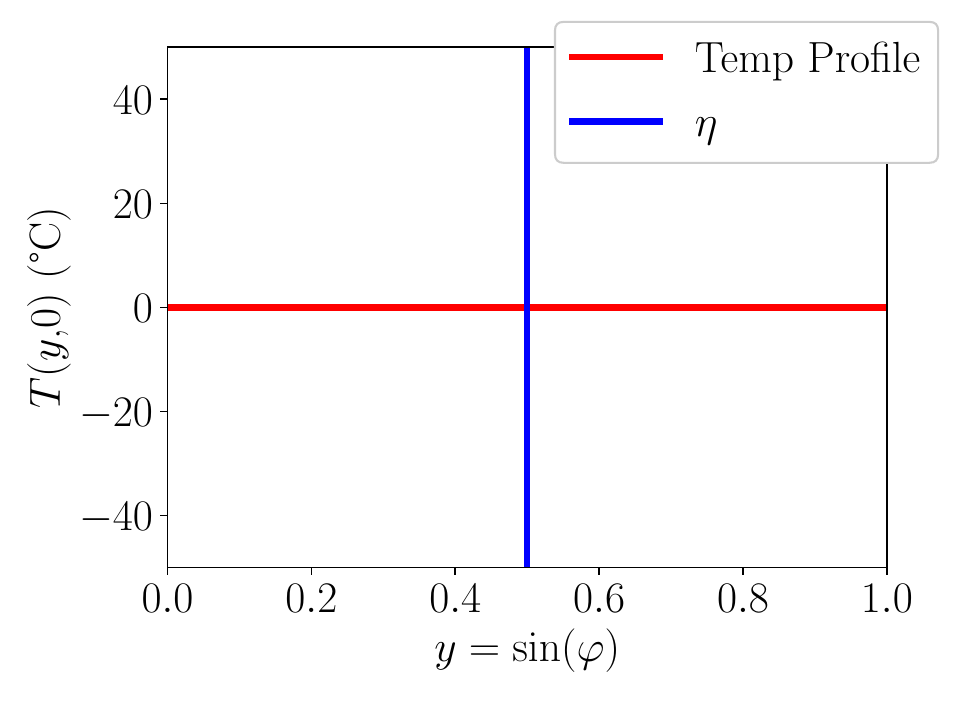}
  \vspace{-23pt}
  \caption{0 Years}
\end{subfigure}%
\vspace{10pt}
\begin{subfigure}{.4\textwidth}
  \centering
  \includegraphics[width=\linewidth]{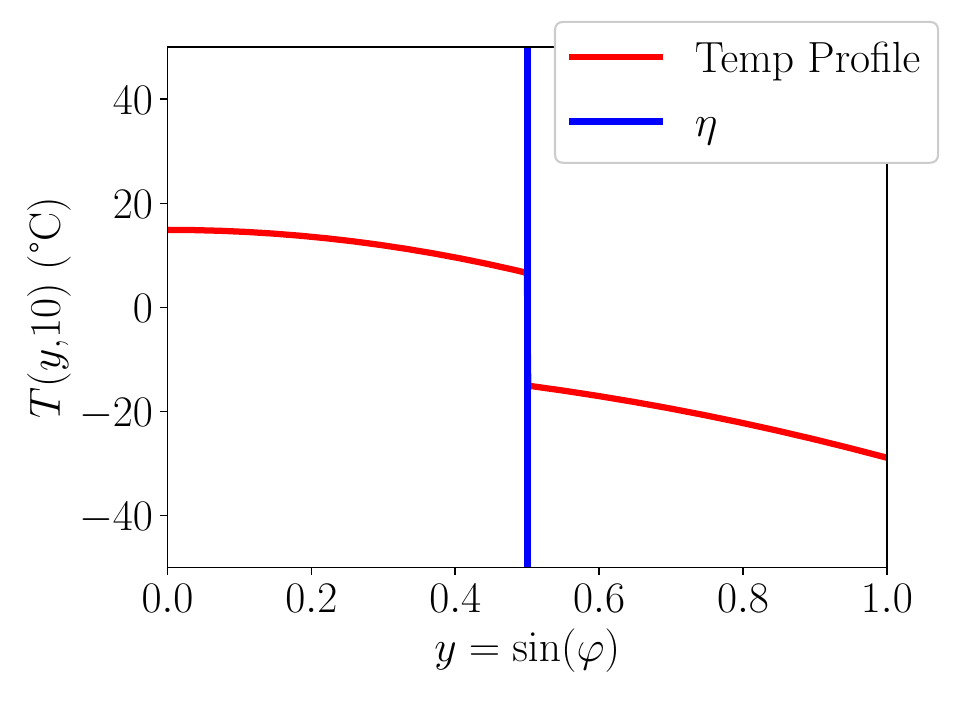}
  \vspace{-23pt}
  \caption{10 Years}
\end{subfigure}
\vspace{10pt}
\begin{subfigure}{.4\textwidth}
  \centering
  \includegraphics[width=\linewidth]{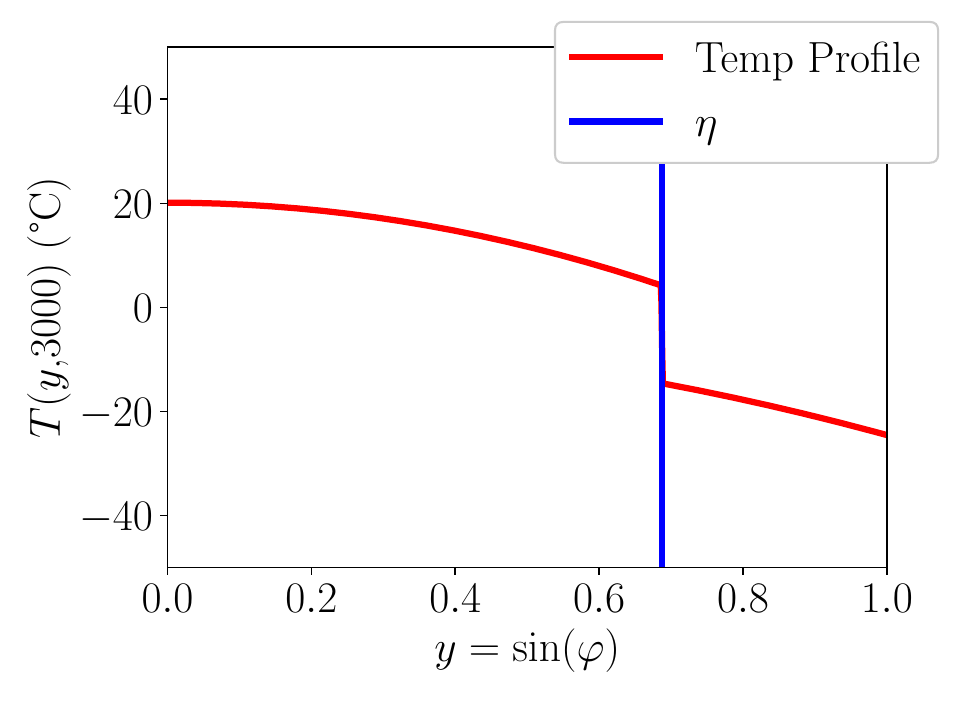}
  \vspace{-23pt}
  \caption{3,000 Years}
\end{subfigure}%
\begin{subfigure}{.4\textwidth}
  \centering
  \includegraphics[width=\linewidth]{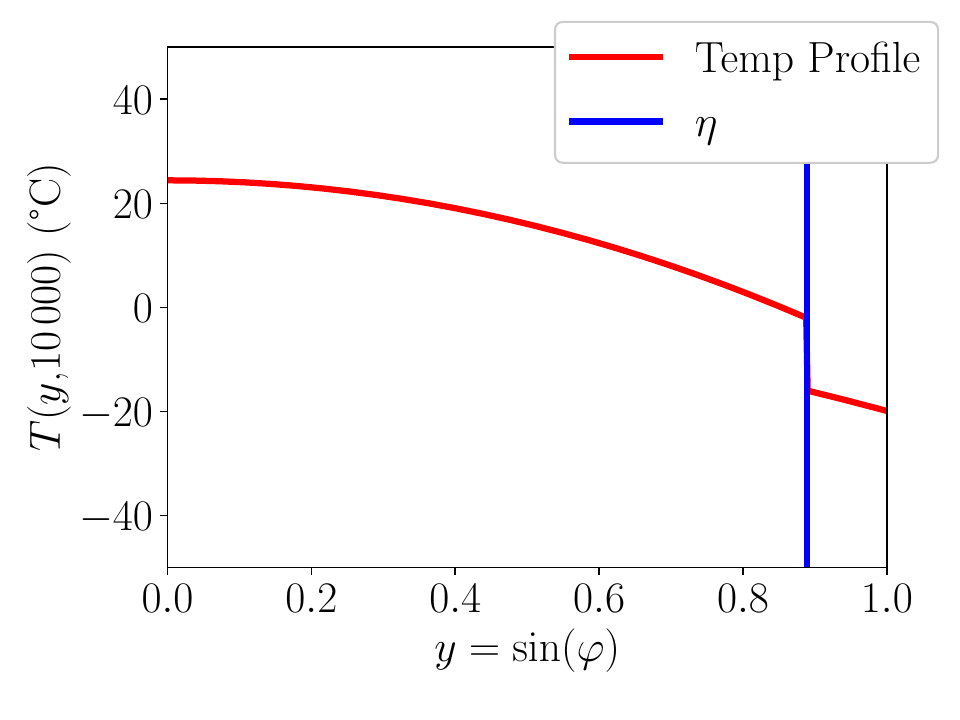}
  \vspace{-23pt}
  \caption{10,000 Years}
\end{subfigure}
\caption[Budyko Simulation Snapshots]{Snapshots of a Budyko simulation starting
  with 0$^\circ$C at all latitudes and the ice line at $\eta=0.5$. The model
  parameters used are shown in Table \ref{tab:budyko_params} whilst the initial
  orbital parameters are in Table \ref{tab:current_orbital_params}. We first see the
  temperature profile equilibrate within 10 years. The ice line then recedes
  towards the stable equilibrium point at $y=0.9268$, getting within 3\% of this
value after 10,000 years.}
\label{fig:budyko_demo}
\end{figure}
\begin{table}[h]
  \centering
  \caption{Parameter values used for the Budyko model.}
  \label{tab:budyko_params}
\begin{tabular}{c|c|c|cl}
Parameter          & Value         & Unit        & Source &  \\ \hline\hline
$Q_0$              & 340.327       & Wm$^{-2}$   & Kopp \cite{solar_constant} \\\hline
$A$                & 202.1         & Wm$^{-2}$   & North \cite{linear_reradiation}  \\\hline
$B$                & 1.9           & Wm$^{-2}$\,$^\circ$C$^{-1}$ &  North \cite{linear_reradiation} \\\hline
$C$                & 3.04          & Wm$^{-2}$\,$^\circ$C$^{-1}$ &  Tung \cite{tung2007topics}\\\hline
$T_{\mathrm{ice}}$ & $-10$           & $^\circ$C   &  Tung \cite{tung2007topics}\\\hline
$\alpha_1$         & 0.32          & None        &  Tung \cite{tung2007topics}\\\hline
$\alpha_2$         & 0.62          & None        &  Tung \cite{tung2007topics}\\\hline
$R$                & $4\times10^8$ & Jm$^{-2}$\,$^\circ$C$^{-1}$ &  McGehee \cite{c_beta}\\\hline
$S$                & $10^{13}$       & C$^\circ$s &  Widiasih \cite{widiasih2013}
\end{tabular}
\end{table}

\begin{figure}
  \centering
  \input{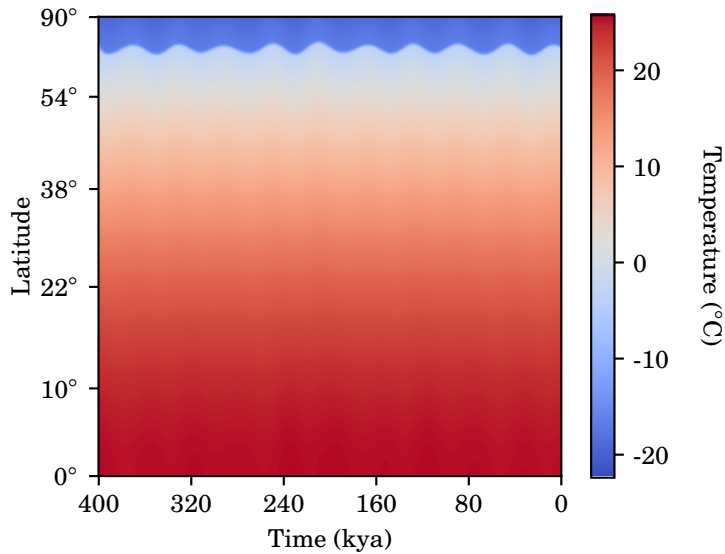}
  \caption[Budyko-Widiasih 400\,kyr]{Temperature profile solution from the
    original Budyko-Widiasih model over the past 400\,kyr. The ice line,
    although not shown, separates the blue and red regions. The dominant period
    of oscillation is 41\,kyr, which is consistent with the obliquity cycle.}
\label{fig:budyko_orig_400k}
\end{figure}

We now look at the dynamic behaviour of the Budyko-Widiasih model. Figure
\ref{fig:budyko_demo} shows snapshots of a simulation starting with a uniform
temperature profile and the ice line at $\eta=0.5$. We see that the temperature
profile equilibrates within 10 years, whilst the ice line takes 10\,000 years to
get within 3\% of the stable equilibrium point at $y=0.9268$. This demonstrates
the slow-fast nature of the system, with the ice line moving much slower than the
temperature profile. This characteristic is common amongst long term ice volume
models. The dynamic response captures what we might expect to physically occur
if such a perturbation were to happen to Earth's climate. However, we wish to
capture the larger ice volume variations that we see over the past 800\,kyr in
the data. Figure \ref{fig:budyko_orig_400k} shows the northern hemisphere
temperature profile over the past 400\,kyr, with the ice line separating the
blue and red regions. This ice line oscillates with a period of 41\,kyr,
aligning with the dominant period of obliquity. This result makes sense as we
are varying the orbital parameters in the insolation term given by
\begin{equation}
  Q_\varepsilon s_\beta(y) = \frac{Q_0}{\sqrt{1-\varepsilon^2}}\left(1+\frac{1}{2}c_\beta(3y^2-1)\right),
  \label{eq:insolation_term_reminder}
\end{equation}
where
\begin{equation}
  c_\beta = \frac{5}{16}(3\sin^2\beta - 2).
\end{equation}
Note that precession is not present in this term. This is because we are using
the yearly averaged insolation and, as shown in Section \ref{sec:insolation},
precession only affects the distribution of insolation within the year. As
discussed in Section \ref{sec:obliquity}, obliquity can vary yearly insolation
by up to 10\% around the poles. Given that the insolation function in the Budyko
model closely approximates the actual function, we can expect obliquity to have
a significant impact on the ice line position. On the other hand, in Section
\ref{sec:eccentricity}, we saw that eccentricity can only vary the magnitude of
insolation reaching Earth by 0.18\%. As a result, it is not surprising that we
do not see eccentricity's 100\,kyr period in the ice line position.

There is an incongruence between this result and what we see in the ice volume
data. Although eccentricity varies insolation by a fraction of the amount that
obliquity does, we see that the dominant period of variations in the ice volume
data is 100\,kyr, aligning with the period and phase of eccentricity. This
suggests that the Earth system is somehow increasing the effect of eccentricity
on ice volume. From \eqref{eq:simple_T_star}, we see that the equilibrium
temperature at the ice line is linearly scaled by the insolation reaching Earth,
given by \eqref{eq:insolation_term_reminder}. Since we are dealing with a
slow-fast system, we can treat the temperature as being equilibrated when
considering the ice line dynamics. This means that the ice line position over
this timescale is approximately just a linear function of the obliquity signal.
Technically, it is a linear function of the squared sine of obliquity. However,
for the range of angles that obliquity takes, this is also approximately linear.
If our current model is essentially representing the ice line as a linear
function of obliquity, it will not be able to reproduce the periods associated
with eccentricity and precession. We can therefore conclude that, if we wish to
reproduce the dynamics seen in the ice volume data, we must augment the model to
try and capture the effects of eccentricity and precession.

\section{Budyko Augmentations}
We now explore some potential augmentations to the Budyko-Widiasih model to see
how they impact the ice line dynamics, with the intention of better aligning
with the ice volume data. We will introduce five main augmentations to the
model, each building on the last unless stated otherwise. Although not shown, we
also tested each augmentation independently and with other valid combinations to
ensure no one augmentation was inhibiting another. The augmentations are shown
to not fully capture the ice volume data, failing to replicate the 100\,kyr
period of eccentricity. However, a number of improvements to the physical
realism of the model are made, and we are able to reproduce the impact of
precession in the ice line dynamics. We therefore present the augmentations here
as they may be of interest to future researchers looking to improve the Budyko
model.

\subsection{Sub-Year Resolution}
\begin{figure}
  \hspace{-45pt}
  \input{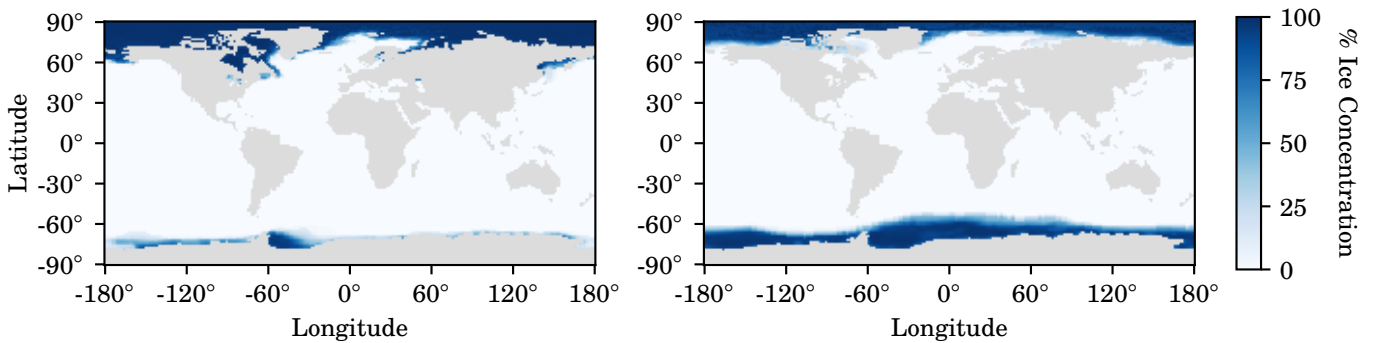}
  \caption[Sea Ice Concentration Map]{Average sea ice concentration for March
    (left) and September (right). This data spans from 1870 to 2003 and is taken
    from the Met Office Hadley Centre \cite{sea_ice_data}.}
  \label{fig:sea_ice_march_sept}
\end{figure}

The Budyko model uses an approximation of the yearly averaged insolation as
input. This is computationally efficient, as time steps are on the scale of
years as opposed to days. However, this means that the model is unable to
capture the dynamics of the ice line throughout a year. This is particularly
important for the polar regions, where sea ice moves significantly over the
course of a year, as shown in Figure \ref{fig:sea_ice_march_sept}. By including
seasonal variations in insolation, we can also capture the impact of mild and
intense summers in the northern hemisphere. Milankovitch suggests this to be the
main driver of the glacial dynamics \cite{milankovitch}. An added benefit of
changing to a sub-year resolution is the introduction of precession to the
insolation function. As we see in \eqref{eq:insolation_term_reminder},
precession averages out of the yearly insolation, so up until now could not
have its frequency appear in the modelled ice line dynamics.

As we are augmenting the Budyko-Widiasih model to simulate Earth's temperature
profile on a daily timescale, we can no longer use the average yearly insolation
approximation given by \eqref{eq:insolation_term_reminder}. Instead, we will use
the analytical solution for daily average insolation $Q_\mathrm{day}$, derived
in Section \ref{sec:insolation}. This results in the system by defined as
\begin{equation}
  \begin{split}
    R\frac{\partial T(y,t)}{\partial t} &= Q_\mathrm{day}(y,t)(1-\alpha_\eta(y))
    - (A + BT(y,t)) + C(\overline{T}(t) - T(y,t)),\\
    S\frac{\mathrm{d}\eta(t)}{\mathrm{d}t} &= T(\eta,t) - T_{\mathrm{ice}},
  \end{split}
  \label{eq:budyko_sub_year}
\end{equation}
where $Q_\mathrm{day}$ is given by
\begin{equation}
  \begin{split}
    Q_{\mathrm{day}}&=\frac{K}{8\pi^2r^2}[(\gamma_{0} - \gamma_{1}) \sin{\beta}
  \sin{\varphi} \cos{\left(\rho - \theta \right)} +
  (\sin{\gamma_{0}} \cos{\beta}
  \cos{\left(\rho - \theta \right)} - \sin{\gamma_{1}} \cos{\beta} \cos{\left(\rho - \theta \right)}\\
&\quad\quad\quad\quad\quad\quad+ \sin{\left(\rho - \theta \right)} 
\cos{\gamma_{0}} - \sin{\left(\rho - \theta
\right)} \cos{\gamma_{1}} ) \cos{\varphi}].
  \end{split}
\end{equation}
The full derivation of $Q_{\mathrm{day}}$ is given in Section
\ref{sec:insolation}, though importantly, it requires calculating the
longitudinal limits of the insolation function $\gamma_0$ and $\gamma_1$ at each
point in time as the Earth orbits the sun. This significantly increases the
computation time of the model, but not prohibitively so, and it allows us to
capture the dynamics of the ice line throughout the year.

In order to allow for seasonal dynamics to occur during a year, some parameters
of the Budyko system must change. The two time constants $R$ and $S$ govern the
rate of temperature and ice line change respectively. So far, we have estimated
these values based on imprecise calculations. However, since we have data for
the yearly variation of both the temperature profile, and the approximate
northern ice line, we can now fit these. Also changed is the free parameter $C$,
which governs the rate of latitudinal heat transport.

\begin{figure}
  \hspace{-45pt}
  \input{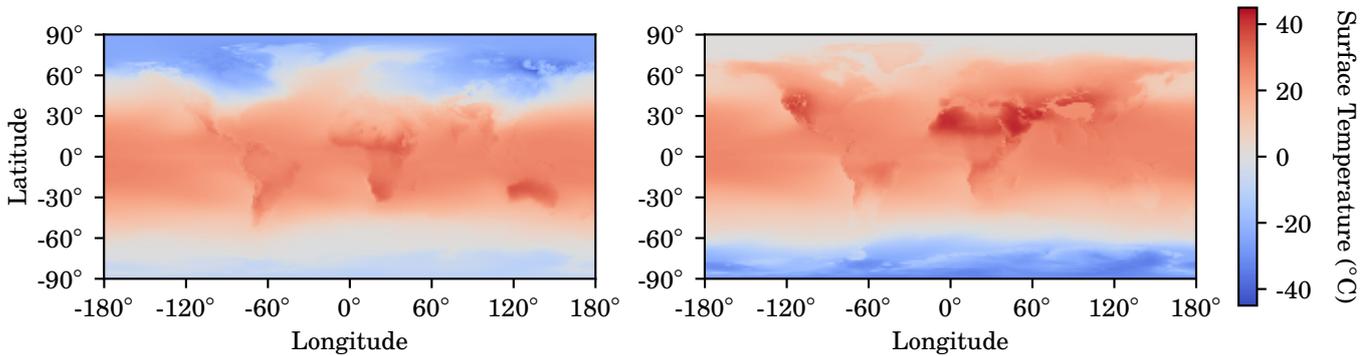}
  \caption[Average Surface Temperature Map]{Average surface temperature for
    January (left) and July (right). This data spans from 2000 to 2010 and is
    taken from The C3S Climate Data Store \cite{surface_temp_data}.}
  \label{fig:surf_temp_jan_jul}
\end{figure}

We first look at the global temperature variation between seasons, as shown in
Figure \ref{fig:surf_temp_jan_jul}. Here we see a map of monthly averaged
surface temperature for the months of January and July, the most extreme months
for temperature. The summer solstice is the day on which the most insolation
reaches one of the hemispheres. In the southern hemisphere, this occurs during
December, and in the northern hemisphere, July. However, due to seasonal lag,
the warmest period follows approximately one month after the summer solstice.
Seasonal lag is predominantly caused by the high specific heat capacity of
water, taking significantly longer than the land to heat up and cool down. This
regulates the Earth's temperature, dampening the effect of maximum and minimum
insolation.

\begin{figure}
  \centering
  \input{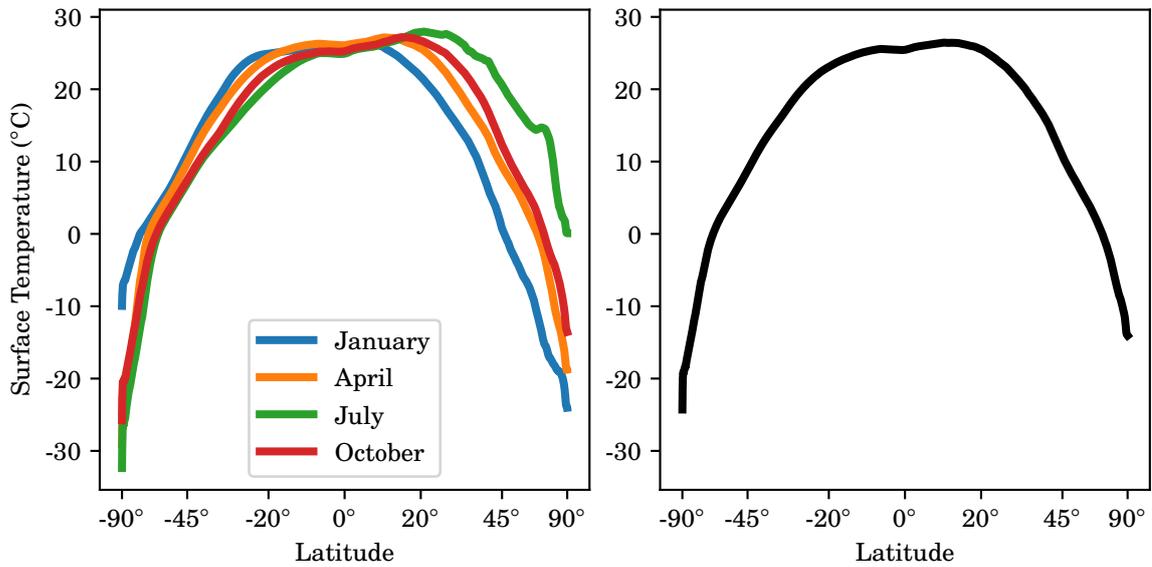}
  \caption[Monthly Average Surface Temperature Latitudinal Distribution]{The
    average monthly surface temperature as a function of latitude for every
    third month (left) and annual average surface temperature (right). The data
    spans from 2000 to 2010 and is taken from The C3S Climate Data Store
  \cite{surface_temp_data}.}
  \label{fig:lats_ave_surf_temp}
\end{figure}

The average monthly temperature as a function of latitude is shown in Figure
\ref{fig:lats_ave_surf_temp}. From this figure, we can expect that over the
year, the equator should remain approximately constant at 25$^\circ$C, whilst
the surface temperature at the north pole ranges between $-25^\circ$C and
$0^\circ$C, with an average of $-15^\circ$C.

The plots in Figure \ref{fig:lats_ave_surf_temp} show that surface temperature
is asymmetric across the equator. This is due to the land distribution around
the poles, as shown in Figure \ref{fig:land_distribution}. The south pole is
covered by Antarctica, which is surrounded by sea, whilst the north pole is
covered by sea, which is then surrounded by land. The north pole has very little
permanent sea ice, as this grows and recedes throughout the year. Whilst still
moving throughout the year, Antarctica's glacial ice is multiple kilometres
thick in places and covers most of the land all year round. This reduces the
south pole's average surface temperature by 10$^\circ$C when compared to the
north pole.

In addition to capturing the surface temperature dynamics throughout a year,
the new model should also aim to represent the ice line dynamics with some
realism. The maps in Figure \ref{fig:sea_ice_march_sept} show the sea ice
concentration averaged over the months of March and September, with the colour
gradient representing the proportion of each pixel in which sea ice is present.
Due to the high heat capacity of the ocean, the minimum ice concentration lags
behind maximum temperature by approximately two months. With temperature lagging
approximately one month behind insolation, minimum ice concentration in the
northern hemisphere occurs in September, three months after the northern summer
solstice.

\begin{figure}
  \centering
  \input{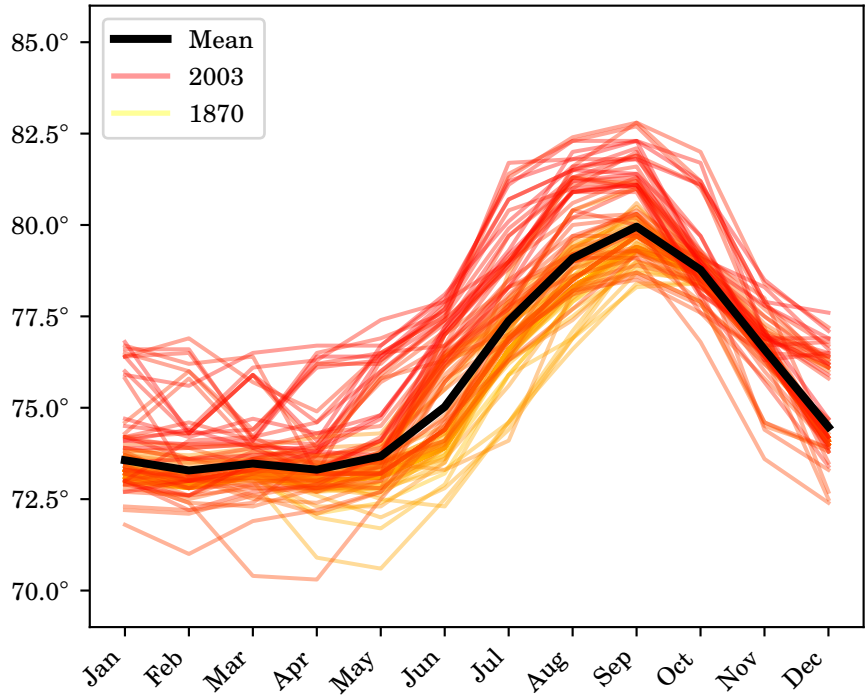}
  \caption[Yearly Sea Ice Line by Latitude]{Approximate latitude of northern sea
    ice line, corresponding to $\eta$ in the Budyko model, using Met Office data
    spanning from 1870 to 2003 \cite{sea_ice_data}. The boundary is defined as
    the latitude at which the longitudinally averaged sea ice concentration
    drops below 75\%. The monthly averages are shown for each year, starting as
    yellow and turning to red, with the average of all years shown in black.}
  \label{fig:monthly_eta}
\end{figure}

An approximation of the northern ice boundary throughout the year can be seen
in Figure \ref{fig:monthly_eta}. For this figure, the ice boundary is defined
as the highest latitude at which longitudinally averaged sea ice concentration
drops below 75\%, ignoring land mass.

\begin{figure}
  \centering
  \input{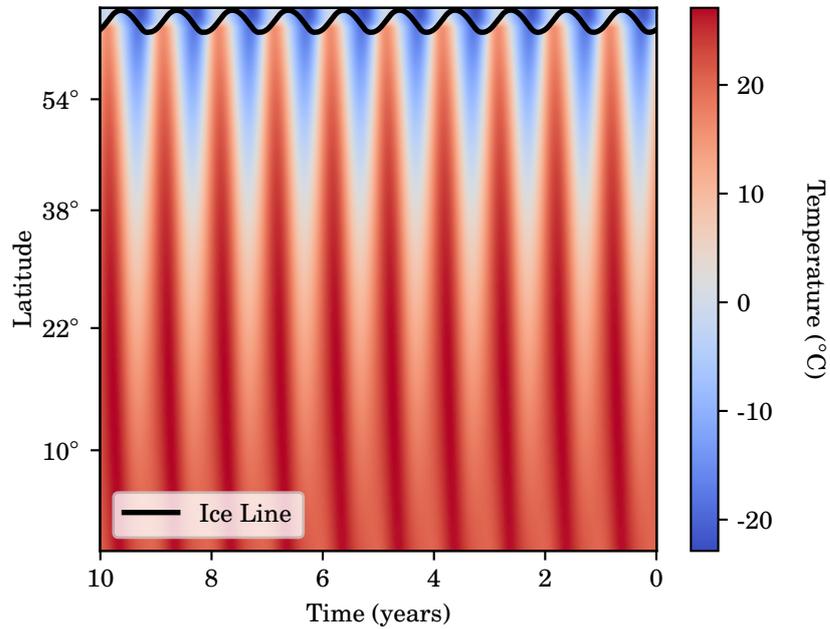}
  \caption[Budyko 10 Year Example Temperature Profile]{Temperature profile for
    example 10 year simulation using tuned parameters values as shown in Table
  \ref{tab:budyko_day_params}. Overlaid in black is the ice line position.}
  \label{fig:example_num_day_10_year_T}
\end{figure}

\begin{figure}
  \centering
  \input{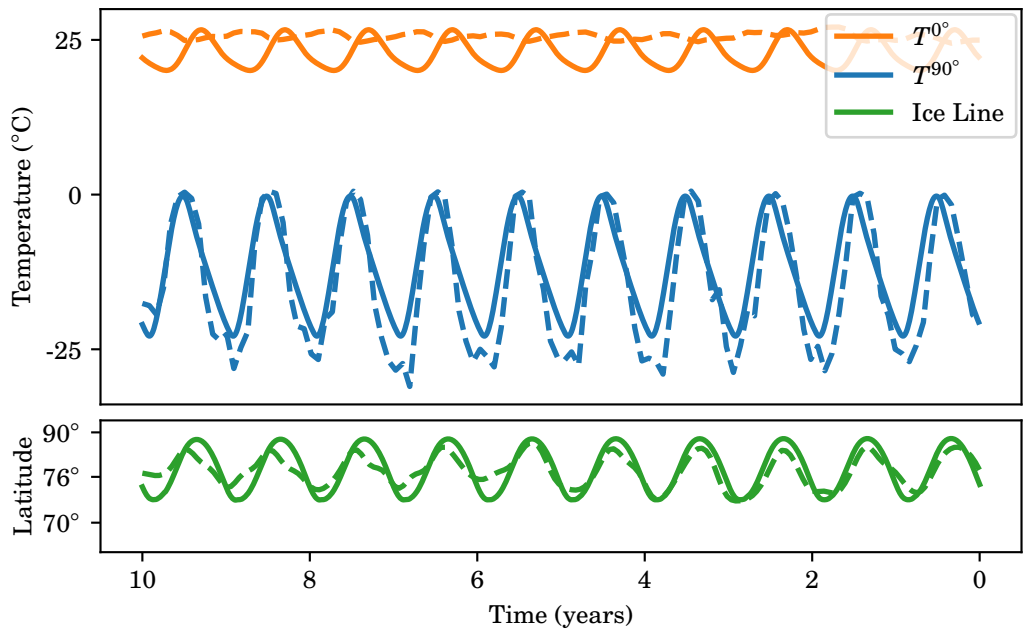}
  \caption[Budyko Temperature Profile Comparison with Data]{Key values from the
    10 year example simulation, shown in Figure
    \ref{fig:example_num_day_10_year_T}. The simulated values are shown with
    solid lines, whilst the real surface temperature data
    \cite{surface_temp_data} and ice line data \cite{sea_ice_data} are shown in
    dashed lines. The top plot shows the simulated and real temperatures at the
    north pole and at the equator. The bottom plot shows the ice line from the
    simulation along with the ice line approximated from real data.}
  \label{fig:example_num_day_10_year_compare}
\end{figure}

The approximated northern ice line follows an approximately sinusoidal path,
ranging from around 73 to 80$^\circ$. With this data as a basis, we tune the
two timescale parameters, and the free parameter $C$, which are given in Table
\ref{tab:budyko_day_params}. This allows us to simulate the seasonal changes in
temperature and ice line that occur throughout a year. We verify this model by
running a trial 10 year simulation, as shown in Figure
\ref{fig:example_num_day_10_year_T}. A comparison between three key values from
this simulation with the equivalent data is shown in Figure
\ref{fig:example_num_day_10_year_compare}.

Here we see the average daily temperature at the north pole (90$^\circ$) which
is always covered in ice, and the equator (0$^\circ$), which remains warm year
round, as well as the ice line position. The comparison shows that the model
captures seasonal variations well, however the temperature variation in the
equator has both a higher amplitude and is out of phase with the real data. A
possible explanation for this is the lack of a land-sea distribution in the
model. The non-ice albedo value $\alpha_1=0.32$ comes from the average albedo
between latitudes 0$^\circ$ and 60$^\circ$ \cite{budyko}. As we see in Figure
\ref{fig:land_distribution}, there is less land around 0$^\circ$ than the
average across 0$^\circ$ to 60$^\circ$. This means the albedo measurement for
this latitude would actually be lower than 0.32, since ocean generally has a
lower albedo than land. With a lower albedo, more of the insolation would be
absorbed at this latitude in the model and we would see a temperature curve more
closely resembling the real data. Nonetheless, this model is a significant
improvement in realism compared with the original Budyko model, which omits
seasonal variations entirely.
\begin{table}
\centering
\caption{Parameter values used for augmented Budyko models.}
\label{tab:budyko_day_params}
\begin{tabular}{c|c|c|c|c|cl}
Parameter          & $Q_\mathrm{day}$ & Two Hem & Diffusion & Land Ice & Unit
                   &  \\ \hline\hline
$C$                & 4.35             & 4.2     & 1.6      & 4.2      &
Wm$^{-2}$,$^\circ$C$^{-1}$ &\\\hline
$D$                &                  &         & 0.28     &          &
Wm$^{-2}$,$^\circ$C$^{-1}$ &\\\hline
$R$                & $4.6\times10^7$  & $4.2\times10^7$ & $4\times10^7$ &
$2.5\times10^7$ & Jm$^{-2}$,$^\circ$C$^{-1}$ &  \\\hline
$S_\mathrm{sea}$   & $2.2\times10^9$  & $10^9$ & $7.5\times10^8$ & $5\times10^8$
                   & C$^\circ$s & \\\hline
$S_\mathrm{land}$  &                  &        &          & $5\times10^{10}$ & C$^\circ$s
\end{tabular}
\end{table}
\begin{figure}
  \centering
  \input{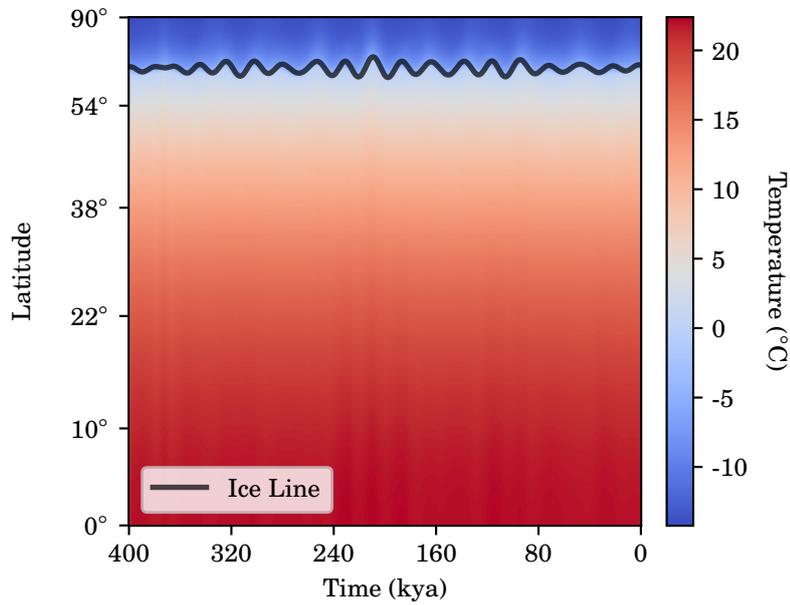}
  \caption[Budyko Daily Averaged Insolation Overlay]{Temperature profile for the
    augmented Budyko-Widiasih model that uses $Q_\mathrm{day}$. Overlaid is the
    ice line position, which is oscillating with frequencies of both 41\,kyr and
    $23$\,kyr.}
  \label{fig:budyko_num_day_insol_overlay}
\end{figure}
\begin{figure}
  \hspace{-45pt}
  \input{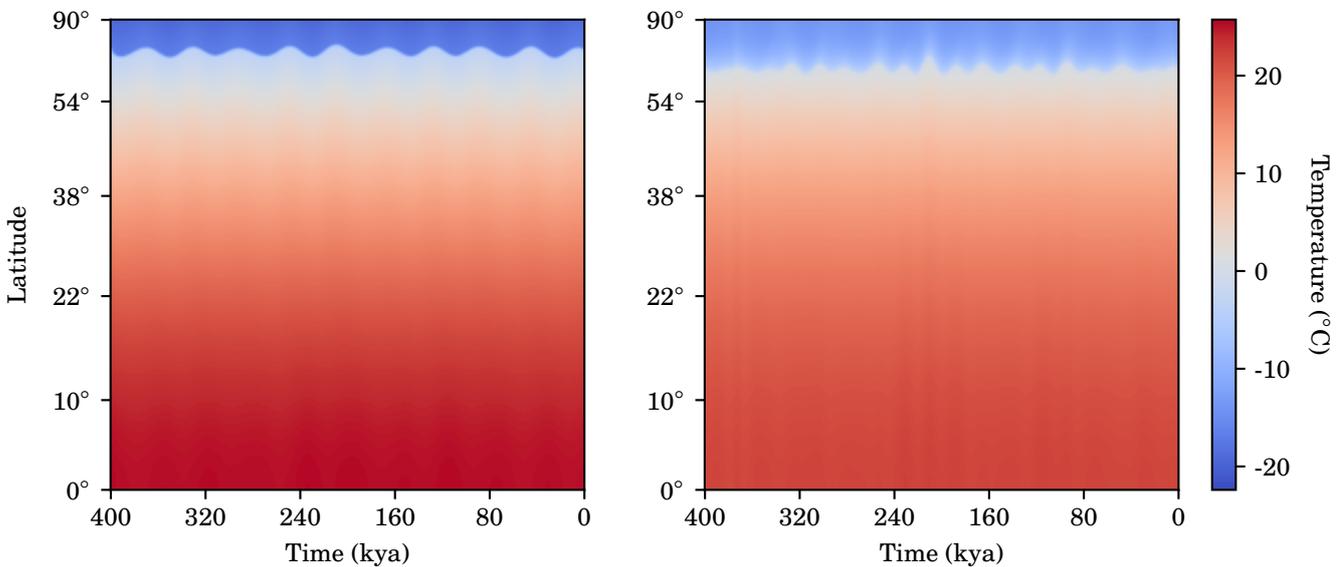}
  \caption[Budyko Augmentation Temperature Profile Comparison]{Comparison of
    the original Budyko-Widiasih model (left), which uses the annual insolation
    approximation, and the sub-year resolution model (right) which uses the
    insolation function integrated over the day $Q_\mathrm{day}$, updating the
    model throughout the year. In this plot the average yearly temperature is
    shown for comparison to the original Budyko model.}
  \label{fig:compare_budyko_400k_sim}
\end{figure}

In Figure \ref{fig:budyko_orig_400k}, we used the Budyko-Widiasih model to
simulate the past 400\,kyr of Earth's temperature profile and ice line dynamics.
This used the yearly averaged insolation approximation and showed ice line
oscillations that were dominated by the 41\,kyr period present in the input
obliquity signal, with negligible 100\,kyr oscillations from the input
eccentricity signal. In Figure \ref{fig:budyko_num_day_insol_overlay} we see
this same 400\,kyr simulation but using the sub-year resolution model, where
seasonal changes in insolation are captured. The ice line oscillations now
contain both the 41\,kyr obliquity period and the $\sim23$\,kyr precession
period. This is a significant result as we have made a model that objectively
improves the realism of the Budyko model, and in doing so, have captured the
precession signal in the ice line dynamics, better approximating the real data.
A comparison of this change is shown in Figure
\ref{fig:compare_budyko_400k_sim}. However, the model still fails to capture the
100\,kyr eccentricity signal, which is the dominant period in the ice volume
data. We therefore investigate further augmentations to the model that will
increase its realism to see if this increases the presence of eccentricity in
the ice line dynamics.

\subsection{Two Hemispheres}
The latitudinal distribution of land across Earth is not uniform, as shown in
Figure \ref{fig:land_distribution}. There is a large concentration of land
towards the top of the northern hemisphere. This is where the majority of
continental ice sheets grow during glacial periods. Although Antarctica is at
the south pole, this is largely stable over the past 800\,kyr and so the main
variation is in the northern hemisphere. This led pioneering researchers such
as Milankovitch and Budyko to focus on the role of the northern hemisphere in
explaining the glacial-interglacial cycles \cite{milankovitch,budyko}.
However, Rubincam suggests an opposing theory to Milankovitch
\cite{precession_index_opposite}, suggesting that in fact the southern
hemisphere is governing the glacial-interglacial cycles. They show that northern
ocean temperatures lag behind southern ocean temperatures and suggest that
cooled southern ocean waters flow northwards, cooling the northern hemisphere
and causing ice growth.

In order to explore this theory, and to increase the physical realism of the
Budyko model, we expand the domain to include both the northern and southern
hemispheres. This requires the addition of a second ice line, which will
be governed by a differential equation identical to the original, but with a
change of sign to account for the difference in direction. The piecewise albedo
function must also take account of this second ice line, now expressed as
\begin{equation}
  \tilde{\alpha}_\eta(y) = \begin{cases} \alpha_1, & \eta_\mathrm{S}<y<\eta_\mathrm{N}\\
                      \alpha_2, & y>\eta_\mathrm{N}\quad\text{or}\quad y<\eta_\mathrm{S} \\
                                 \frac{1}{2}(\alpha_1+\alpha_2), &
                                 y=\eta_\mathrm{N}\quad\text{or}\quad y=\eta_\mathrm{S}
                   \end{cases},
\end{equation}
where $\eta_\mathrm{N}$ and $\eta_\mathrm{S}$ are the northern and southern ice
lines respectively.

The system is now given by
\begin{equation}
  \begin{split}
    R\frac{\partial T(y,t)}{\partial t} &= Q_\mathrm{day}(y,t)(1-\tilde{\alpha}_\eta(y))
    - (A + BT(y,t)) + C(\overline{T}(t) - T(y,t)),\\
    S\frac{\mathrm{d}\eta_\mathrm{N}(t)}{\mathrm{d}t} &= T(\eta_\mathrm{N},t) - T_{\mathrm{ice}},\\
    S\frac{\mathrm{d}\eta_\mathrm{S}(t)}{\mathrm{d}t} &= T_{\mathrm{ice}} - T(\eta_\mathrm{S},t).
  \end{split}
  \label{eq:budyko_two_hem}
\end{equation}

\begin{figure}
\centering
\begin{subfigure}{.4\textwidth}
  \centering
  \includegraphics[width=\linewidth]{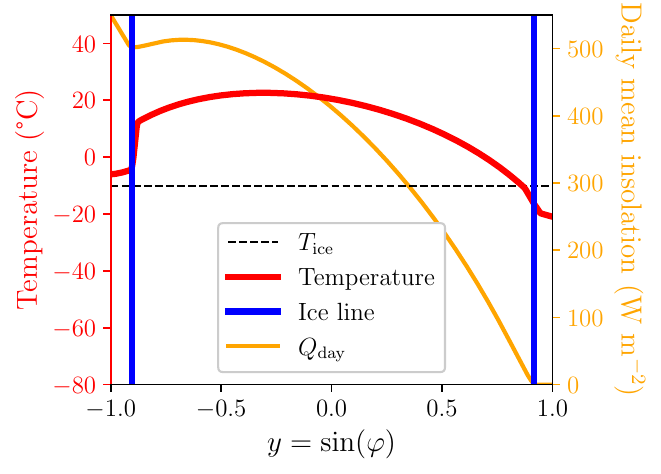}
  \vspace{-23pt}
  \caption{January}
\end{subfigure}%
\vspace{10pt}
\begin{subfigure}{.4\textwidth}
  \centering
  \includegraphics[width=\linewidth]{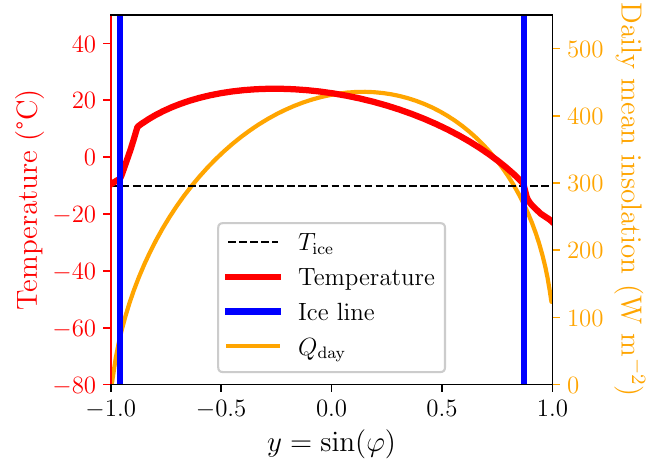}
  \vspace{-23pt}
  \caption{April}
\end{subfigure}
\vspace{10pt}
\begin{subfigure}{.4\textwidth}
  \centering
  \includegraphics[width=\linewidth]{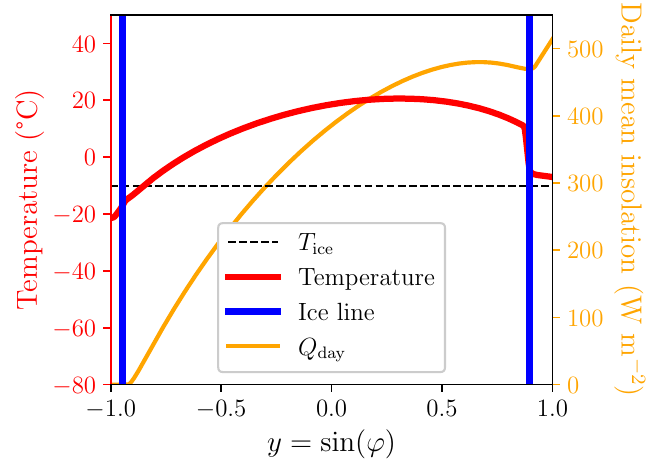}
  \vspace{-23pt}
  \caption{July}
\end{subfigure}%
\begin{subfigure}{.4\textwidth}
  \centering
  \includegraphics[width=\linewidth]{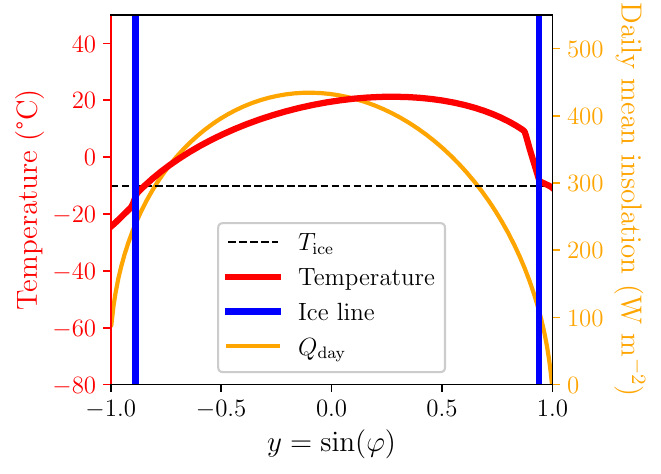}
  \vspace{-23pt}
  \caption{October}
\end{subfigure}
\caption[Two Hemisphere Budyko Simulation Snapshots]{Snapshots of our augmented
  Budyko model over a year. This model simulates both the northern and southern
  hemispheres, with two ice lines moving independently. The model parameters
  used are shown in Table \ref{tab:budyko_params}, with updated values given in
  Table \ref{tab:budyko_day_params}, which maintain physically realistic dynamics
  in the new domain. The yellow line shows how daily averaged insolation varies
  across latitudes and seasons. The red line shows the temperature profile,
  which lags behind the insolation profile by approximately one month. The ice
  lines in blue lag behind the temperature profile by approximately two months,
which is why we see the minimum ice line values around April and October. The
dashed black line shows the $-10^\circ$C threshold at which ice grows according
to the model.}
\label{fig:2_hem_budyko_demo}
\end{figure}

This change of domain once again required different values for the time
constants $R$ and $S$, as well as the free parameter $C$. The changed values are
shown in Table \ref{tab:budyko_day_params}. The results of this augmentation are
shown in Figure \ref{fig:2_hem_budyko_demo}. Here we see the temperature profile
and ice lines for both hemispheres over the course of a year. The ice lines can
be seen to behave similarly to the single hemisphere model with sub-year
resolution, only the two they are oscillating half a year out of phase with each
other. When plotting the long term dynamics of this model, we see that the ice
line oscillations are still dominated by the 41\,kyr obliquity signal and the
$23$\,kyr precession signal. However, this model does allow for a better
representation of the full Earth system, allowing us to more clearly visualise
the continuous change in seasonal insolation, temperature, and ice line position
throughout the year.

\subsection{Heat Diffusion}
When looking at Figure \ref{fig:lats_ave_surf_temp} we see a smooth temperature
profile as we move into the polar regions where ice is present. This is
contradictory to the discontinuous temperature profile we see in the previous
versions of the Budyko model, as in Figure \ref{fig:2_hem_budyko_demo}. This
results from the heat transport term failing to smooth the discontinuity in the
albedo function. Although the albedo at an ice-sea boundary would be
discontinuous, the air temperature around it would be continuous due to heat
diffusion. This suggests the need for a new augmentation that could improve the
realism of latitudinal heat transfer. This augmentation has been done
before \cite{budyko_with_diffusion,budyko_with_diffusion_2} as it is a more
realistic representation of the system. However, it has not been included
alongside the other augmentations we present here.

The diffusion term in the Budyko model works as a replacement for the
latitudinal transport term, ${C(\bar{T}(t) - T(y,t))}$, which distributes
temperature towards the global average $\bar{T}(t)$. The diffusion of heat is
modelled by
\begin{equation}
  \nabla^2T = \frac{\partial^2 T}{\partial x^2} +\frac{\partial^2 T}{\partial y^2}
  +\frac{\partial^2 T}{\partial z^2}
\end{equation}
where $\nabla^2$ is the Laplace operator and represents the second derivative
in the spatial domain.

The Laplace operator is based in Cartesian coordinates. It is therefore
necessary to transform the operator into our coordinate system, as described in
Section \ref{sec:budyko}. This coordinate system is based on spherical polar
coordinates with the origin at Earth's centre. In this coordinate system, the
Laplace operator is given by
\begin{equation}
  \nabla^2T = \frac{1}{r^{2}} \frac{\partial}{\partial r}\left(r^{2} \frac{\partial
  T}{\partial r}\right)+\frac{1}{r^{2} \sin \phi} \frac{\partial}{\partial
  \phi}\left(\sin \phi \frac{\partial T}{\partial
  \phi}\right)+\frac{1}{r^{2} \sin ^{2} \phi} \frac{\partial^{2}
  T}{\partial \gamma^{2}},
\end{equation}
where $r$ is the radial distance from Earth's centre, $\phi$ is the polar
angle measured from the $z$-axis, relating to latitude, and $\gamma$ is the
azimuthal angle measured from the $x$-axis, relating to longitude. For the
Budyko model, we treat $T$ as constant in $r$ and $\gamma$, representing a
constant Earth radius and no longitudinal variation in temperature. This
simplifies the Laplace operator to
\begin{equation}
  \nabla^2T = \frac{1}{r^{2} \sin \phi} \frac{\partial}{\partial
  \phi}\left(\sin \phi \frac{\partial T}{\partial
  \phi}\right).
\end{equation}
Since $\phi$ is measured from the $z$-axis, we can express latitude as
$\varphi = \frac{\pi}{2}-\phi$. Hence the Budyko spatial variable becomes
$y = \sin(\varphi) = \sin(\frac{\pi}{2}-\phi)=\cos\phi$.

Using the chain rule, we have that
\begin{equation}
\begin{split}
  \frac{\partial}{\partial \phi} &= \frac{\partial y}{\partial \phi}
  \frac{\partial }{\partial y}\\
  &= -\sin\phi\frac{\partial }{\partial y}.
\end{split}
\end{equation}
Substituting this into our expression for $\nabla^2T$ gives
\begin{align}
  \nabla^2T &= \frac{-\sin \phi}{r^{2} \sin \phi} \frac{\partial}{\partial
  y}\left(-\sin^2\phi \frac{\partial T}{\partial y}\right)\\
  &= \frac{1}{r^2}\frac{\partial}{\partial y}\left((1-\cos^2\phi)
  \frac{\partial T}{\partial y}\right)\\
  &= \frac{1}{r^2}\frac{\partial}{\partial y}\left((1-y^2)
  \frac{\partial T}{\partial y}\right).
  \label{eq:budyko_diffusion_term}
\end{align}

Since we are modelling both hemispheres of Earth, we require that
$\frac{\partial T}{\partial y}=0$ at the poles. This ensures that the system is
physically realistic, without a cusp in the temperature profile around the
poles. As is shown in \eqref{eq:budyko_diffusion_term}, we multiply by $1-y^2$.
This has a problematic effect when solving the model numerically as the equation
becomes stiff around the poles. To avoid this, we change the domain from $y$ to
$\varphi$, putting us in the latitude domain.

To express diffusion in terms of $\varphi$ instead of $y$, we use the
substitution $y = \sin(\varphi)$, where \mbox{$\varphi \in \left[-\frac{\pi}{2},
\frac{\pi}{2}\right]$} corresponds to latitude. We then transform the
differential operator with respect to $y$ into one with respect to $\varphi$.

Since $y = \sin(\varphi)$, we have
\begin{equation}
  \frac{\partial}{\partial y} = \frac{1}{\cos(\varphi)} \frac{\partial}{\partial
  \varphi},
\end{equation}
and \(1 - y^2 = \cos^2(\varphi)\). Substituting these into the diffusion term
\begin{equation}
  \nabla^2 T = \frac{1}{r^2} \frac{\partial}{\partial y} \left( (1 - y^2)
  \frac{\partial T}{\partial y} \right)
\end{equation}
yields
\begin{equation}
  \nabla^2 T = \frac{1}{r^2 \cos(\varphi)} \frac{\partial}{\partial \varphi}
  \left( \cos(\varphi) \frac{\partial T}{\partial \varphi} \right).
\end{equation}

This form expresses the diffusion operator in terms of \(\varphi\), allowing it
to capture latitudinal heat transport in the spherical coordinate system.

\begin{figure}
  \centering
  \includegraphics[width=0.6\linewidth]{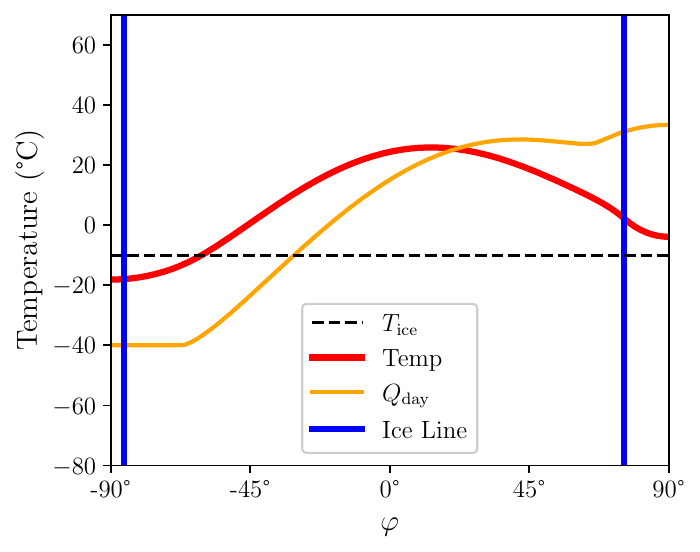}
  \caption[Budyko Model with Diffusion Term]{Snapshot from our augmented
    Budyko-Widiasih model that includes a diffusion term. The model parameters
    used are shown in Table \ref{tab:budyko_params} with updated values given in
    Table \ref{tab:budyko_day_params}. The diffusion term smooths the
    temperature profile, removing the discontinuity at the ice line present in
    the original model.}
  \label{fig:budyko_diffusion_demo}
\end{figure}

The Budyko model is now given by
\begin{align}
  R\frac{\partial T(\varphi,t)}{\partial t} &=
  Q_\mathrm{day}(\varphi,t)(1-\tilde{\alpha}_\eta(\varphi))
  - (A + BT(\varphi,t)) - D\nabla^2 T(\varphi,t),\\
  S\frac{\mathrm{d}\eta_\mathrm{N}(t)}{\mathrm{d}t} &= T(\eta_\mathrm{N},t) -
  T_{\mathrm{ice}},\\
  S\frac{\mathrm{d}\eta_\mathrm{S}(t)}{\mathrm{d}t} &= T_{\mathrm{ice}} -
  T(\eta_\mathrm{S},t),
\end{align}
where $D$ is the diffusion parameter. The result of this augmentation is shown
in Figure \ref{fig:budyko_diffusion_demo}, which uses the same parameters as
the original model given in Table \ref{tab:budyko_params}, with the changed
values given in Table \ref{tab:budyko_day_params}. In this demonstration, we 
see that we are now in the latitude domain, with the poles at $\pm\frac{\pi}{2}$
(written in terms of degrees for convenience). The temperature profile is now
smooth at the ice line, and has zero gradient at the poles, bringing the model
closer to the physical data. Solving this model over 400\,kyr, we see that
the ice line dynamics are largely similar to previous augmentations, with the
41\,kyr obliquity signal and the $23$\,kyr precession signal dominating the
oscillations. Although we have still failed to capture the 100\,kyr eccentricity
signal, we have made a model that is more physically realistic, and still
captures the precession signal in the ice line dynamics, improving upon the
original Budyko-Widiasih model.

\subsection{Albedo}
The Budyko model uses an albedo value of 0.32 for the entire ice-free portion of
Earth. This is a significantly oversimplified representation of the Earth's
albedo and so we will attempt to improve this aspect of the model. The albedo
value of 0.32 used by Budyko was determined from observational data,
incorporating the effect of ocean, land, and cloud cover, treating Earth's
surface as a homogeneous land-water mix with average cloud cover. The first
improvement to this system is bringing the albedo up to date with current
satellite data. The current estimate for global albedo is 0.29
\cite{earth_albedo}. Using the current annual insolation reaching Earth's
atmosphere, 340.375\,Wm$^{-2}$, we can expect to absorb
$(1-0.29)\times340.375=241.67$\,Wm$^{-2}$ annually. If we assume the glacial ice
to have an average albedo of 0.6 \cite{ice_albedo_approx}, we can use an albedo
value for the unfrozen portion of Earth that results in a net absorption of
241.67\,Wm$^{-2}$ for current orbital parameters. This yields $\alpha_1=0.279$,
which is notably lower than the 0.32 value used so far. Since we have reduced
both of the albedo values to match more closely with the data available, we can
expect to see a warmer temperature profile. In order to keep the model
reflecting the real world, we will change the free variable $C$, which has so
far been $C = 1.6B = 3.04$\,Wm$^{-2}$, in order to maintain an equilibrium
northern ice line of $\eta_\mathrm{N}=68^\circ$ as before. Although these
changes will cancel out for the equilibria, it will have an effect on the
dynamic behaviour of the model and brings a number of the parameters closer to
their true physical values.

Using the equilibrium solution from Section \ref{sec:budyko}, we have a cubic
polynomial for $\eta_\mathrm{N}$, and we wish to fix the most positive root at
$\eta_\mathrm{N}=68^\circ$ and find the value of $C$ that allows for this, given
the new values for $\alpha_1$ and $\alpha_2$. From rearranging for $C$ and
substituting in these values, we get $C=2.108$, which is approximately $1.1B$.
This follows a trend of decreasing $C$ as data becomes more accurate. The
initial value for $C$ started at $2.4B$ from Budyko in 1969. In 1974, Held and
Suarez suggested a lower value of $2.1B$ after introducing a heat diffusion term
\cite{budyko_with_diffusion_2}. In 2007, Tung suggested $1.6B$ to account for
the more recent value for the solar constant, taken from satellite data
\cite{tung2007topics}. Reducing $C$ has the effect of slowing the rate at which
heat energy is spread latitudinally across Earth, hence we can expect to see a
larger difference between the temperature at the equator and pole. In the case
of $\eta_\mathrm{N}=68^\circ$, the original temperatures at the pole and
the equator was $-19.1$\,C$^\circ$ and 25.2\,C$^\circ$. Once updating the albedo
in the model, it becomes $-21.9$\,C$^\circ$ and 36.1\,C$^\circ$. This new range
is further from the measured annual values of approximately $-19$\,C$^\circ$ and
27\,C$^\circ$ \cite{mean_annual_temp}. This shows that improving the realism of
the model does not necessarily mean that its solution will better represent the
real ice line dynamics, at least without additional changes.

The next level of realism to add to the model's albedo is the heterogeneity of
land and ocean. Since the current model treats Earth as a homogeneous mix of
land and ocean with respect to albedo, we are missing a potentially significant
effect of temperature distribution. Without clouds, the albedo of land ranges
from 0.05 to 0.4 \cite{land_albedo}, whilst the ocean is between 0.03 and 0.06
\cite{sea_albedo}. However, the presence of clouds is shown to increase these
values, which will be discussed later.

\begin{figure}
  \centering
  \includegraphics[width=0.85\linewidth]{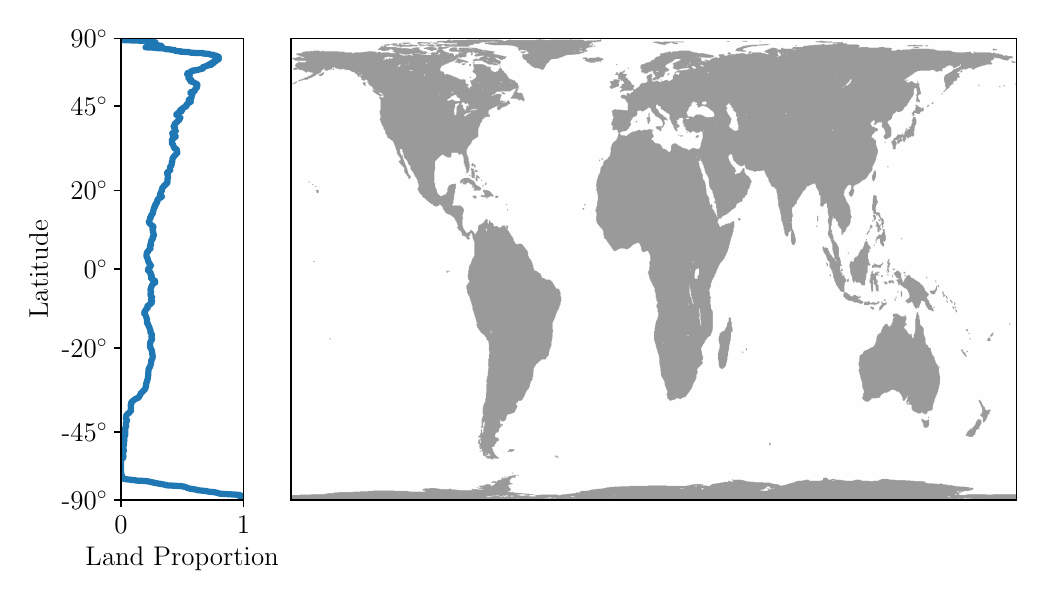}
  \caption[Latitudinal Land Density Distribution]{Distribution of land density
    across latitude. The reference map shown on the right uses the Gall-Peters
    projection \cite{gall_peters_image}. This preserves area, as well as
    maintaining the latitudinal spacing used in the Budyko model $y =
  \sin{\varphi}$.}
  \label{fig:land_distribution}
\end{figure}

By incorporating the land distribution curve shown in Figure
\ref{fig:land_distribution}, we can estimate albedo as a function of latitude.
The figure demonstrates how there is an uneven distribution of land across
Earth's latitudes. The central half of latitudes is more than 50\% ocean,
meaning it will absorb more insolation than the higher latitude regions with
more land. Although this is once again an increase in the level of physical
realism, it does not have a significant impact on the model's behaviour as heat
diffusion causes the uneven heating to smooth out over a relatively short
timescale.

Another factor that affects the distribution of albedo is the Solar Zenith Angle
(SZA). This is the angle $\mu$ between the incident sunshine and the Earth's surface
normal at a given point. The cosine of this angle is proportional to how
much irradiance reaches a given point on Earth and was derived in Section
\ref{sec:insolation}, given by
\begin{equation}
  \cos\mu = ((sin \gamma)\sin(\rho-\theta) - \cos \beta \cos \gamma
  \cos(\rho-\theta))\cos \varphi - \sin \beta \sin \varphi \cos(\rho-\theta),
  \label{eq:sza}
\end{equation}
where $\gamma$ is longitude, $\rho$ is precession, $\theta$ is Earth's angular
position in its orbit, $\beta$ is obliquity, and $\varphi$ is latitude.

\begin{figure}
  \centering
  \input{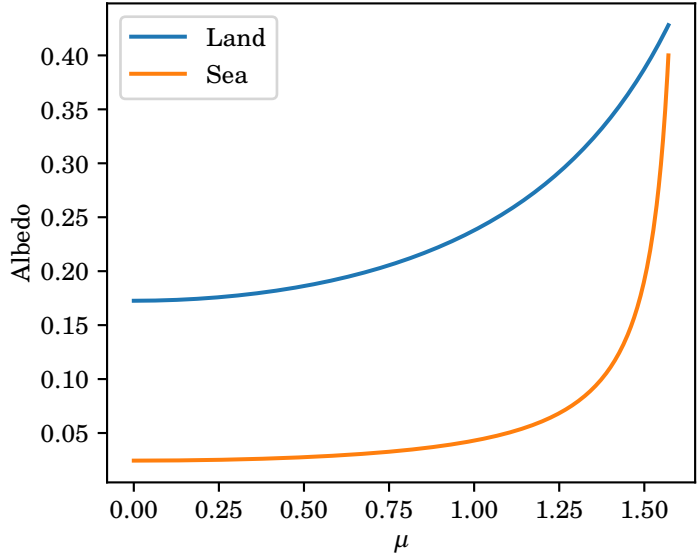}
  \caption[Albedo as a Function of Solar Zenith Angle]{Estimated albedo as a
  function of SZA $\mu$ for land and sea.}
  \label{fig:sza_for_land_sea}
\end{figure}

For a location on Earth that is in less direct sunlight, the SZA is greater,
and the proportion of irradiance that is reflected increases non-linearly with
the cosine of the SZA. This effect is particularly strong when the irradiance is
incident on the ocean. With an albedo in direct irradiance of around 0.05, this
can increase up to 0.4 for the maximum physical SZA \cite{albedo_zenith_effect}.
This effect is also present for land, with the albedo increasing by a factor of
approximately two times when the SZA is at its maximum \cite{land_zenith_range}.
This effect is shown in Figure \ref{fig:sza_for_land_sea}, where we have
estimated the average albedo of land and sea as a function of SZA.

From measuring the albedo of a range of land surfaces with a varying SZA, Yang
estimated land albedo as a function of SZA and diffuse albedo, which relates to
the surface type but is independent of SZA \cite{land_zenith_range}. Using the
data presented by Yang, we estimate the average diffuse albedo of land to be
$\alpha_{\mathrm{diff}}=0.2$. Using these estimates for diffuse albedo, and the
coefficients for the SZA dependence from Yang, we can approximate direct albedo
on land as a function of SZA $\mu$ with
\begin{equation}
  \alpha_{\mathrm{land}} = \frac{0.428}{1+1.48\cos\mu}.
\end{equation}

For the ocean albedo, we use the same method, but base our approximation on the
coefficients from Coakley \cite{ocean_albedo_sza}. This gives the ocean albedo
as
\begin{equation}
  \alpha_{\mathrm{ocean}} = \frac{0.026}{0.065+\cos\mu},
\end{equation}
where a diffuse albedo of $\alpha_{\mathrm{diff}}=0.06$ is used.

Although the SZA will also affect the albedo of the ice sheets, due to
their already high albedo, and proximity to the poles, the effect of SZA on the
ice albedo is negligible, and so we will consider the direct albedo of the ice
to be constant. SZA is defined as a function of latitude, longitude, and the
orbital configuration. However, our model is based on latitude alone, so we
must average the SZA over longitude by integrating \eqref{eq:sza} over $\gamma$
between $-\pi/2$ and $\pi/2$. This gives us the average SZA as a function of
latitude and orbital parameters, which are a function of time. This is given by
\begin{equation}
  \mu = -\sin\beta \sin\varphi \cos(\rho - \theta).
  \label{eq:sza_lat}
\end{equation}

We now have separate albedo estimates for land and ocean which are a function of
latitude $\varphi$ and time $t$. Combining this with our land proportion
function $l(\varphi)$, shown in Figure \ref{fig:land_distribution}, we can
express our updated albedo as
\begin{equation}
\alpha'_\eta(\varphi,t) = 
  \begin{cases}
    \alpha_{\mathrm{land}}(\varphi,t)l(\varphi) +
    \alpha_{\mathrm{ocean}}(\varphi,t)(1-l(\varphi)), &\quad
    \eta_\mathrm{S}<\varphi<\eta_\mathrm{N}\\
    \alpha_2, &\quad \varphi>\eta_\mathrm{N}\quad\text{or}\quad
    \varphi<\eta_\mathrm{S}\\
    \frac{1}{2}(\alpha_{\mathrm{land}}(\varphi,t)l(\varphi) +
    \alpha_{\mathrm{ocean}}(\varphi,t)(1-l(\varphi)) + \alpha_2), &\quad
    \varphi=\eta_\mathrm{N}\quad\text{or}\quad \varphi=\eta_\mathrm{S}.
  \end{cases}
\end{equation}

To scale our albedo function such that it reflects the current real world
average of 0.29 \cite{earth_albedo}, we ensure that the total reflected
insolation over the full year is $0.29Q_\varepsilon=98.71$\,Wm$^{-2}$, where the
current value for $\varepsilon$ is given in Table
\ref{tab:current_orbital_params}. Adding this correcting coefficient to the
updated albedo function, along with the previously introduced sub-year
resolution, two hemisphere domain, and diffusion heat transport, gives the
system as
\begin{equation}
  \begin{split}
    R\frac{\partial T(\varphi,t)}{\partial t} &=
    Q_\mathrm{day}(\varphi,t)(1-k\alpha'_\eta(\varphi))
    - (A + BT(\varphi,t)) - D\nabla^2 T(\varphi,t),\\
    S\frac{\mathrm{d}\eta_\mathrm{N}(t)}{\mathrm{d}t} &= T(\eta_\mathrm{N},t) -
    T_{\mathrm{ice}},\\
    S\frac{\mathrm{d}\eta_\mathrm{S}(t)}{\mathrm{d}t} &= T_{\mathrm{ice}} -
    T(\eta_\mathrm{S},t).
  \end{split}
  \label{eq:budyko_albedo}
\end{equation}

Although we have made a number of augmentations in an attempt to improve the
albedo representation in the model, once again, we do not find a significant
change in the model's behaviour. The long term ice line dynamics still fail to
capture any 100\,kyr periodicity, resembling the results shown in Figure
\ref{fig:budyko_num_day_insol_overlay}, but over two hemispheres. Because of the
added complexity that this augmentation brings, without a significant
improvement in the model's behaviour, we will exclude it from the final
augmentation.

\subsection{Land Ice}
Although we have implemented augmentations to account for the impact of land on
the distribution and magnitude of albedo, we have not yet considered how land
ice behaves differently to sea ice, which has so far been the modelled quantity.
Although we see significant sea ice fluctuations over the course of a year, as
shown in Figure \ref{fig:sea_ice_march_sept}, the glaciers on land are more
stable, allowing for changes to build up over time. So far the Budyko model, and
our augmented versions, have treated Earth's ice as entirely sea-based, with the
ice line time constant being tuned to capture the sea ice fluctuations within a
year. This has been successful in capturing the seasonal changes in temperature
and ice line, but has failed to capture the long term changes in ice volume that
we see in the benthic data.

Land-based glaciers account for approximately 97\% of Earth's ice volume
\cite{sea_ice_vs_glacial}, so it is reasonable that we should be modelling them
directly if we wish to reproduce the global ice volume data. Since we do not
wish to impose the assumption that sea ice is not an important factor in
explaining the ice volume data, we will introduce a new augmentation that
models both land and sea ice separately. This will include a separate larger time
constant for the slow moving land ice as well as the previously implemented
sub-year resolution, two hemispheres, and diffusion term. We once again chose
the time constants and free parameter $C$ to achieve reasonable behaviour from
both the temperature profile and ice lines.

This gives the system as
\begin{equation}
  \begin{split}
    R\frac{\partial T(\varphi,t)}{\partial t} &=
    Q_\mathrm{day}(\varphi,t)\left(1-\frac{1}{2}(\alpha_\mathrm{sea}(\varphi)+\alpha_\mathrm{land}(\varphi))\right)
    - (A + BT(\varphi,t)) - D\nabla^2 T(\varphi,t),\\
    S_\mathrm{sea}\frac{\mathrm{d}\eta_\mathrm{seaN}(t)}{\mathrm{d}t} &= T(\eta_\mathrm{seaN},t) -
    T_{\mathrm{ice}},\\
    S_\mathrm{sea}\frac{\mathrm{d}\eta_\mathrm{seaS}(t)}{\mathrm{d}t} &= T_{\mathrm{ice}} -
    T(\eta_\mathrm{seaS},t),\\
    S_\mathrm{land}\frac{\mathrm{d}\eta_\mathrm{landN}(t)}{\mathrm{d}t} &=
    T(\eta_\mathrm{landN},t) -
    T_{\mathrm{ice}},\\
    S_\mathrm{land}\frac{\mathrm{d}\eta_\mathrm{landS}(t)}{\mathrm{d}t} &= T_{\mathrm{ice}} -
    T(\eta_\mathrm{landS},t),
  \end{split}
  \label{eq:budyko_land_ice}
\end{equation}
where
\begin{equation}
  \alpha_x(\varphi) =
  \begin{cases}
    \alpha_1 &\quad \eta_{x\mathrm{S}}<\varphi<\eta_{x\mathrm{N}}\\
    \alpha_2, &\quad \varphi>\eta_{x\mathrm{N}}\quad\text{or}\quad
    \varphi<\eta_{x\mathrm{S}}\\
    \frac{1}{2}(\alpha_1 + \alpha_2), &\quad
    \varphi=\eta_{x\mathrm{N}}\quad\text{or}\quad \varphi=\eta_{x\mathrm{S}}.
  \end{cases}
\end{equation}
Here we have introduced the subscript placeholder $x$ to represent either land
or sea. The new parameters used in this model are given in Table
\ref{tab:budyko_day_params}. For this model, we have reverted to the simpler
albedo function, which is now two separate functions for land and sea ice. The
impact of these albedo functions are then averaged in the temperature equation,
representing an even mix of land and sea ice surface area.

\begin{figure}
\centering
\begin{subfigure}{.45\textwidth}
  \centering
  \includegraphics[width=\linewidth]{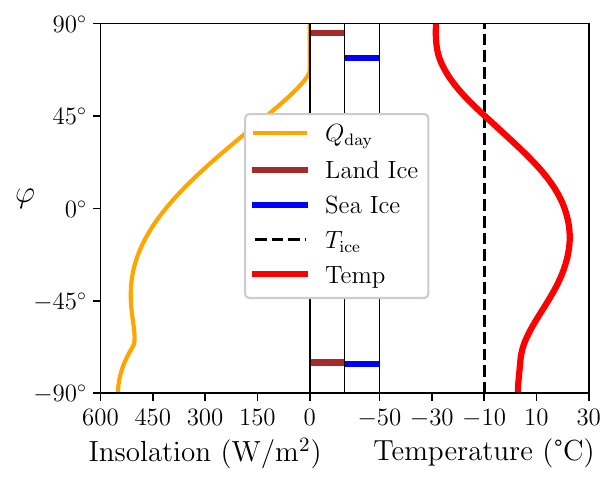}
  \vspace{-23pt}
  \caption{January}
\end{subfigure}%
\vspace{10pt}
\begin{subfigure}{.45\textwidth}
  \centering
  \includegraphics[width=\linewidth]{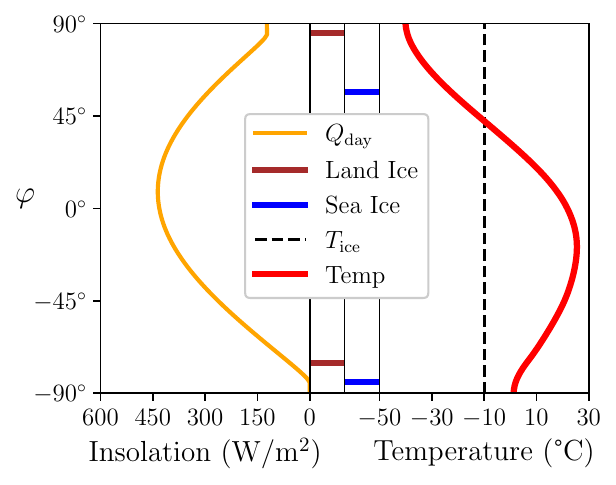}
  \vspace{-23pt}
  \caption{April}
\end{subfigure}
\vspace{10pt}
\begin{subfigure}{.45\textwidth}
  \centering
  \includegraphics[width=\linewidth]{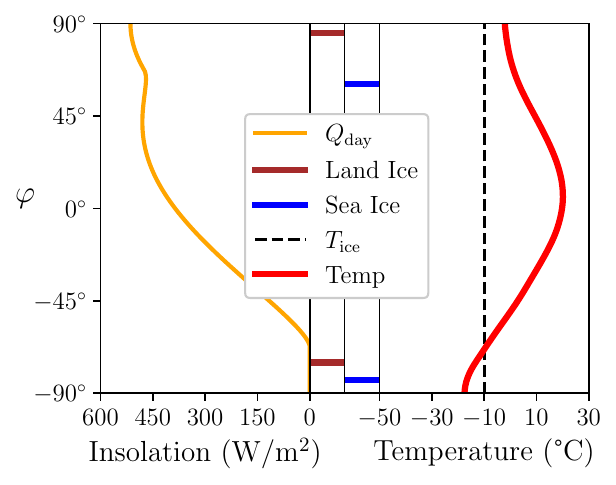}
  \vspace{-23pt}
  \caption{July}
\end{subfigure}%
\begin{subfigure}{.45\textwidth}
  \centering
  \includegraphics[width=\linewidth]{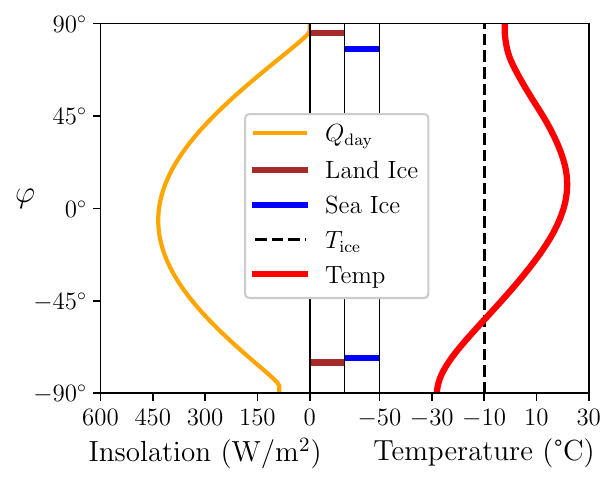}
  \vspace{-23pt}
  \caption{October}
\end{subfigure}
\caption[Two Hemisphere Land Ice Budyko Simulation Snapshots]{Snapshots of our
  augmented Budyko model over a year. This model simulates both the northern and
  southern hemispheres, uses the diffusion equation for latitudinal heat
  transport, and models separate ice lines for land and sea ice. The model
  parameters used are shown in Table \ref{tab:budyko_params}, with updated
  values given in Table \ref{tab:budyko_day_params}, which maintain physically
  realistic dynamics in the new domain. The yellow line shows how daily averaged
  insolation varies across latitudes and seasons. The red line shows the
  temperature profile, which lags behind the insolation profile by approximately
  one month. The ice lines in blue lag behind the temperature profile by
  approximately two months, which is why we see the minimum ice line values
  around April and October. The dashed black line shows the $-10^\circ$C
  threshold at which ice grows according to the model.}
\label{fig:land_ice_budyko_demo}
\end{figure}

An example of this augmented model can be seen in Figure
\ref{fig:land_ice_budyko_demo}. We can see all of the same components as in the
diffusion model, shown in Figure \ref{fig:budyko_diffusion_demo}, but with the
addition of two land-based ice lines. These lines are not visibly moving over the
single year shown, as we would expect for glacial ice, however they are able to
capture long term changes that result from the orbital parameters. We now have a
hybrid model that can show both the seasonal changes in temperature and ice line
as well as the long term changes in ice volume. However, the land ice line fails
to capture the 100\,kyr periodicity that we see in the data, once again
reproducing frequencies related to obliquity and precession, but not
eccentricity.

Although these augmentations have not been successful in capturing the 100\,kyr
periodicity in the ice volume data, they have improved the realism of the model
and have allowed us to capture the precession signal in the ice line dynamics.
This is a notable improvement upon the original Budyko model, which only
captured the impact of obliquity. We now move on to a new category of model that
is designed models ice volume as a zero dimensional system, as opposed to the
latitudinal distribution that we have been using so far.

\section{Conceptual Ice Volume Models}
\label{sec:conceptual_models}
The models we present here differ from the Budyko model in that they are all
ordinary differential equation models, meaning they produce a single output for
ice volume rather than a latitudinal distribution. This further simplifies
computation and allows for a more direct comparison to the established proxy
data.

The models we present here are intended to represent a range of complexity and
alignment with either the geochemical or astronomical theory. An approximate
representation of each model's complexity and theory alignment is shown in
Figure \ref{fig:conceptual_model_landscape}, where we briefly introduced the
models originally. All of the models rely on orbital forcing in some form. This
is either through a linear combination of the orbital parameters, or a
calculated insolation curve that depends on them, usually $Q_{65}$. In order to
standardise the comparison between models, we are using a cubic interpolation of
Laskar's \cite{laskar2004} solution for the orbital parameters in all models,
and then calculating the insolation curves from these using the derivation in
Section \ref{sec:insolation}. The simulations will all span from 800--0\,kya and
be solved using the explicit fourth order Runge-Kutta method with constant time
steps of 0.01\,kyr. In all cases, we are fitting to Bintanja's modelled ice
volume data \cite{benthic_model_contributions}, as opposed to the original
benthic $\delta^{18}$O data from Lisiecki and Raymo \cite{benthic_data} that
these models have often been compared against.

Another standardisation measure was to avoid normalising the orbital and
insolation inputs, as is often done in the original models. This helps with
reproducibility and allows the full relation between input and modelled ice
volume to be expressed in parameters. We must therefore refit each of the
models, using our prescribed timespan, ice volume data, and orbital forcing
inputs. With the intention of finding the best fit for each model, we utilise a
high performance computer cluster to optimise the models. For the models with
fewer than five parameters, we perform a full sweep of the parameters over
suitable ranges determined either through trial and error or from the source
papers. This ensures that we have found the global optimum for these models. For
the larger models, which have up to fifteen parameters, this method is not
feasible. Instead, we initialise the parameters randomly within the specified
ranges and use the BFGS minimisation algorithm from \textit{Scipy} to find the
local optimum from there. We repeat this process up to the order of 10$^5$
times until the best fits appear to converge. This large amount of compute
power gives us the opportunity to optimise the models in a way that was not
reported to have been used in any of the original models.

Since the models produce ice volume solutions in arbitrary, model-specific
units, the parameter values shown in the following tables are generally not
physically interpretable. The exception to this are the time constants, which
are in kyr, and the parameters of the VCV18 model, which is derived from
physical principles and uses degrees Celsius for its temperature variables.
The ice volume solutions are transformed according to
\begin{equation}
  \tilde{V}(t) = mV(t)+c,
\end{equation}
where $m>0$ and $c>0$ are given in the table of parameters for each model.

At the end of this section we present Figure \ref{fig:comp_models}, which shows
a flow diagram representation of each model, allowing us to compare the dynamic
structure of each. We also present Figure \ref{fig:all_model_diagrams}, which
shows the optimised ice volume solution from each model, alongside our two, that
we will introduce in the following chapters.

\subsection{Imbrie and Imbrie, 1980}
\label{sec:imbrie_imbrie_1980}
This is the simplest of the models presented in this section, developed by
Imbrie and Imbrie in 1980 \cite{imbrie}. It was one of the earliest physically
motivated approaches to this problem and serves as a good baseline for our
comparison. As with most of the models outlined here, a switching mechanism is
used to reproduce the difference in ice volume growth and decay rates. The
system is given by
\begin{equation}
  \frac{\mathrm{d}V(t)}{\mathrm{d}t} =
  \begin{cases}
    \dfrac{1+b}{T_m} (-Q_{65}(t)-V(t)), &-Q_{65}(t)\leq V(t)\\[9pt]
    \dfrac{1-b}{T_m} (-Q_{65}(t)-V(t)), &-Q_{65}(t)>V(t),
  \end{cases}
\end{equation}
where $V$ represents global ice volume, $T_m$ represents the average time
constant which is varied by the switching mechanism coefficient $0\leq b \leq
1$.

In the paper, Imbrie made the linear approximation
\begin{equation}
  Q_{65}(t) \approx \beta(t) + \alpha \varepsilon(t) \sin{(\rho(t)-\phi)},
\end{equation}
where $\alpha$ and $\phi$ are fit constants.

Since we have a more accurate calculation for $Q_{65}$ now, we can use this
instead. However, in order to test the full capacity of this model, we also
refit the parameters for the linear approximation to best fit the ice volume
data. This is no longer with the intention of approximating $Q_{65}$, but to
see the improvement that follows from the inclusion of two more parameters and
the ability to use a different input signal.

We therefore present two models; \textit{II80-Q} and \textit{II80-F}, where the
former uses our current estimate of $Q_{65}$ and latter shows the model's full
potential for explaining the ice volume data. The optimised parameters for these
models are shown in Table \ref{tab:imbrie_params}. In the original paper, this
model produces an inverted solution for ice volume, more closely resembling
global temperature which is approximately inversely proportional to ice volume.
For more direct comparison to the other models and our own, we have changed the
sign of the forcing term $Q_{65}$ and the piecewise conditions to produce a
solution for ice volume.
\begin{table}[h]
  \centering
  \caption[Parameter values for Imbrie and Imbrie, 1980]{Parameter values for
  II80-Q and II80-F.}
  \label{tab:imbrie_params}
  \begin{tabular}{c|c|c}
    Parameter & II80-Q Value & II80-F Value\\ \hline\hline
    $V(-800)$ & $-491.0$ & $-0.4111$\\ \hline
    $b$ & 0.8355 & 0.8916\\ \hline
    $T_m$ & 3.925 & 4.510\\ \hline
    $\alpha$ & & $-0.4492$\\ \hline
    $\phi$ & & 1.779\\ \hline\hline
    $m$ & $7.334\times10^5$ & $1.550\times10^6$\\\hline
    $c$ & $4.296\times10^8$ & $7.061\times10^8$
  \end{tabular}
\end{table}

Due to this model's strong reliance on insolation forcing to reproduce the ice
volume dynamics, we can describe it as aligning more closely with the
astronomical theory. However, the inclusion of a switching mechanism does imply
the presence of a non-linear geochemical response to the insolation forcing.
Imbrie and Imbrie justify the inclusion of this mechanism due to the evidence of
dynamic asymmetry in the ice volume proxy data. They suggest two possible
explanations for this effect. The first is a geometric effect, whereby a
shrinking ice sheet changes volume faster because it is more spread out,
increasing the area over which ablation can occur. The second relates to a
possible inherent instability that exists in marine-based ice sheets, leading to
a faster collapse than growth.

Due to the limitations of available data at the time, the authors manually fit
the parameters to the past 150\,kyr of data, specifically using six climate
events in this period with known timings. This produced a solution with a
greater 100\,kyr period than we find from fitting to the past 800\,kyr, however
it also results in a greater 400\,kyr period (the second period of the
eccentricity cycle), which is not present in the data. As this has a net effect
of worsening the fit to the data, we see both periods decrease in amplitude in
our optimised models.

As is shown in Figure \ref{fig:comp_models}, neither versions of this model
reproduce the past 800\,kyr of ice volume data especially well. Due to the
model's piecewise linear dynamics, and the use of $Q_{65}$, there cannot be a
significant eccentricity signal in the solution for ice volume. This led Imbrie
and Imbrie to suggest that the 100\,kyr periodicity in the data is not directly
forced by orbital variations, but emerges from non-linear responses to orbital
forcing.

\begin{figure}
  \hspace{-30pt}
\includegraphics[width=1.1\textwidth]{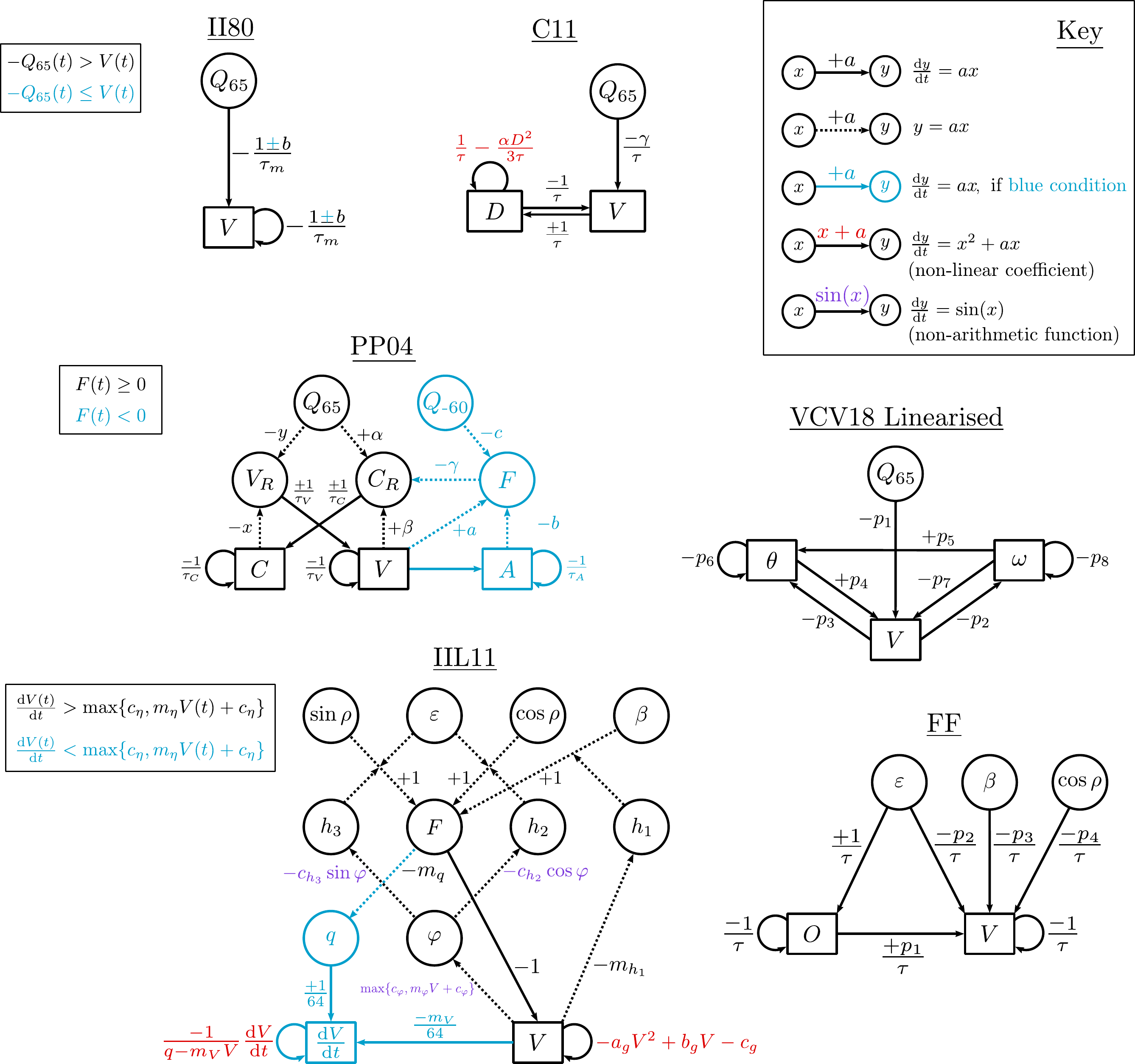}
\caption[Conceptual Model Visualisation]{Flow diagrams representing the dynamics
  of each model discussed in this section, also included is our model
  \textit{FF}, which is introduced in the following chapter. The VCV18 model
  cannot be clearly conveyed in this form so we have shown the linearised
  version, as derived in Section \ref{sec:comparison_verbitsky}. As shown in the
  key, solid arrows represent a variable impacting the rate of change of
  another, whilst the dashed arrow represents a variable instantaneously
  changing another. Anything in blue is only present in the model when the
  relevant blue condition is met. Every arrow has a scaling function associated
  with it. Functions coloured black are constant coefficients, red are
  non-linear coefficients, and purple are non-arithmetic functions of the
  variable. All parameters are positive, so the sign in front of them indicates
  whether they are positively or negatively scaling the input. For the IIL11
  model, the functions $h_{1\text{-}3}$ and $\varepsilon$ point at other arrows,
  meaning they act as the scaling functions for those arrows. We have omitted
any constant offset terms to improve clarity.}
\label{fig:all_model_diagrams}
\end{figure}

A visual representation of the model's dynamics is shown in Figure
\ref{fig:all_model_diagrams}, with the colour of the switching condition
corresponding to whether the time constant is $-\frac{1+b}{T_m}$ or
$-\frac{1-b}{T_m}$. When compared to the other models in this figure, we
see that II80 is dynamically the simplest, with only one state variable.

From these two comparisons, we conclude that this model is not sufficient to
adequately explain the ice volume data. In the following chapter, we introduce
a model we have developed, which builds on the feedforward nature of II80, but
includes a second state variable, allowing for the indirect impact of insolation
to also be captured. Another difference is that, instead of using the insolation
forcing term $Q_{65}$, we use the orbital parameters directly, allowing for
their individual impacts on the ice volume to be quantified. This model (FF) can be
seen in the bottom right of Figure \ref{fig:all_model_diagrams}. Although this
model does not use a switching mechanism, it explains more variance in the ice
volume data, with eccentricity's 400\,kyr period significantly reduced, whilst
maintaining the 100\,kyr period.

\subsection{Paillard and Parrenin, 2004}
This model is more complex than II80, comprising a switching mechanism and three
state variables. These are global ice volume $V(t)$, the extent of the Antarctic
ice sheet $A(t)$, and atmospheric CO$_2$ concentration $C(t)$ \cite{pp04}. The
system is given by
\begin{align}
  \tau_V\frac{\mathrm{d}V(t)}{\mathrm{d}t} &= V_\mathrm{R}(t) - V(t)\\
    \tau_A\frac{\mathrm{d}A(t)}{\mathrm{d}t} &= V(t) - A(t)\\
    \tau_C\frac{\mathrm{d}C(t)}{\mathrm{d}t} &= C_\mathrm{R}(t) - C(t),
\end{align}
where $\tau_V$, $\tau_A$, and $\tau_C$ are time constants and
\begin{align}
    V_\mathrm{R}(t) &= -xC(t)-yQ_{65}(t) + z\\
    C_\mathrm{R}(t) &= \alpha Q_{65}(t) + \beta V(t) - \gamma H(-F(t)) + \delta,
\end{align}
where $x$, $y$, $z$, $\alpha$, $\beta$, $\gamma$, and $\delta$ are tuned
phenomenological parameters that do not have explicit physical interpretations.

The function,
\begin{equation}
    F(t) = aV(t) - bA(t) - cQ_{\text{-}60}(t) + d,
\end{equation}
represents the ocean's bottom water state and is used to switch between glacial
and interglacial states. It can be thought of as a measure of how favourable
conditions are for forming dense, salty bottom waters around Antarctica. This is
why it includes $Q_{\text{-}60}(t)$, which is the insolation during the southern
summer solstice (February 21st) at $60^\circ$ south. It is also impacted
positively by global ice volume, which cools the bottom water, and negatively by
Antarctic ice extent.

This negative effect is due to the difference in brine rejection from sea ice
growth over shallow and deep water, which is affected by Antarctic ice extent.
When sea ice grows, it rejects salt, which sinks to the bottom. In shallow
waters, which are present around the continental shelf of Antarctica, the entire
water column can be efficiently cooled, effectively increasing the salt
concentration at the bottom. In deep water, this is less efficient, as the
entire depth must be cooled to almost its freezing point. Because a large
Antarctic ice sheet extends into the continental shelf, the sea ice growth takes
place over deeper waters, hence the negative impact of $A(t)$ on $F(t)$.
This bottom water state variable is then passed through a Heaviside function,
forming a switch mechanism within the modelled carbon cycle, where $F(t)>0$
indicates a glacial state and $F(t)<0$ indicates an interglacial state. The
switching mechanism is proposed to reflect either non-linearity in the carbon
cycle, or an interaction between bottom water stratification and
thermohaline circulation.

The authors use this novel mechanism to explain the glacial-interglacial cycles,
achieving a good fit to the data over the past 800\,kyr. They also show that the
data can be reproduced as far back as 5\,myr, which includes the mid-Pleistocene
transition. However, this requires that a small linear drift is included in the
model adding a time dependent term to the definition of $F(t)$. This term
represents the efficiency with which the dense, salty water can form at the
bottom of the southern bottom water. The authors suggest that this drift could
be related to gradual glacial erosion of the Antarctic continent, or continental
drift. We will see in the more recent model from Imbrie, Imbrie-Moore, and
Lisiecki that the mid-Pleistocene transition can be explained without the need
for this drift term. However, this is a more complex non-linear model and is
highly sensitive to parameter values.

This model uses astronomical forcing to drive the system, but is capable of
producing 100\,kyr oscillations in the absence of forcing due to the feedback
mechanisms present. The authors present their work as resolving a key gap in the
astronomical theory, explaining how the astronomical forcing can lead to large
100\,kyr oscillations in the ice volume data. Because of the feedback mechanisms
attributed to CO$_2$ and deep water stratification switching mechanism, we can
consider this model to be more geochemically aligned.

The authors tuned the parameters to qualitatively reproduce the main features of
the ice volume data, but mention that parameters $a$ and $c$ can be set to zero
without changing the timings of glaciations, only the qualitative agreement.
From fitting the parameters in this thesis, we find that the optimised model
chooses a $c$ value very close zero, effectively omitting the
$Q_{\text{-}60}(t)$ term in forcing function $F(t)$. To confirm this term is not
important to the model, we set $c=0$ and measured no notable change to the fit.
This further reduces the model's dependence on astronomical forcing and aligns
it more closely with the geochemical theory. This model will be referred to as
\textit{PP04}. The optimised parameters for this model are shown in Table
\ref{tab:pp04_params}.

As can be seen in Figure \ref{fig:all_model_diagrams}, PP04 is more
complex than II80, but shares the same piecewise linear form. The blue section
of the model is only present when the system is in a glacial termination state,
speeding up the rate at which the ice volume shrinks, similar to II80. Our
feedforward model (FF) reproduces the ice volume data almost as well as PP04,
but with a fully linear system, and using only two state variables. This is
intended to be the simplest possible model that can explain the data well,
minimising assumptions about the system and allowing for easy analysis of the
model's dynamics. We will also introduce a feedback model (FB) in Chapter
\ref{chap:feedback_model}, which simply adds a negative arrow from $V$ to $O$ in
the FF model diagram. This feedback allows for unforced oscillations to occur,
similar to how PP04 behaves. Figure \ref{fig:comp_models} shows this model to
outperform PP04 to a similar extent that it outperforms FF, suggesting that the
data does not necessitate a switching mechanism nor a third state variable.
\begin{table}[h]
\centering
\caption[Parameter values for Paillard and Parrenin, 2004]{Parameter values for PP04.}
\label{tab:pp04_params}
\begin{tabular}{c|c}
  Parameter & Value \\ \hline\hline
$V(-800)$ & $-178.9$ \\\hline
$A(-800)$ & 0.9695 \\\hline
$C(-800)$ & $-11740$ \\\hline
$\tau_V$ & 3904 \\\hline
$\tau_C$ & 611.3 \\\hline
$\tau_A$ & 20.45 \\\hline
$x$ & 19.54 \\\hline
$y$ & 820.3 \\\hline
$z$ & 107900 \\\hline
$\alpha$ & 228.5 \\\hline
$\beta$ & $496.9$ \\\hline
$\gamma$ & $130500$ \\\hline
$\delta$ & 18040 \\\hline
$a$ & 0.3828 \\\hline
$b$ & 0.701 \\\hline
$c$ & 0.0004983 \\\hline
$d$ & 16.52 \\\hline\hline
$m$ & $2.505\times10^5$\\\hline
$c$ & $1.006\times10^8$
\end{tabular}
\end{table}

\subsection{Crucifix, 2011}
\label{sec:crucifix_2011}
This model uses fewer parameters than \textit{PP04} but is not designed around
physical processes \cite{crucifix_original,crucifix_adapted}. Drawing from the
work of Saltzman and Maasch \cite{saltzman_intrinsic_first} and Tziperman et al.
\cite{tziperman2006}, Crucifix aims to produce one of the simplest possible
models that is still able to reproduce the ice volume data. Crucifix achieves
this by using the Van der Pol oscillator, which can produce free oscillations
with a period of 100\,kyr. This oscillator also incorporates $Q_{65}$ to
account for orbital forcing and is given by
\begin{align}
  \tau\frac{\mathrm{d}V(t)}{\mathrm{d}t} &= -D(t) + \beta - \gamma Q_{65}(t)\\
  \tau\frac{\mathrm{d}D(t)}{\mathrm{d}t} &= -\alpha \frac{D(t)^3}{3} + D(t) + V(t),
\end{align}
where $\alpha$, $\beta$, $\gamma$, and $\tau$ are non-physical parameters tuned
to the data. $V(t)$ is global ice volume, whilst $D(t)$ responds to ice volume
and is proposed to represent the fluctuating carbon concentration in the
atmosphere associated with ocean circulation.

The model describes a slow-fast system, where the large $\alpha$ causes $D$ to
move quickly compared to $V$. The $V$ equation controls the slow evolution of
the system whilst $D$ controls the quick transitions between two states
representing glacial and interglacial. The rate of these dynamics are set by the
time constant $\tau$, which we find to be optimally around $40$\,kyr, which is
close to the original paper's $36$\,kyr. The asymmetry of the ice volume
dynamics can be reproduced through the $\beta$ parameter. This changes the
position of the fixed point, which is a state where the system would remain at
rest in the absence of forcing. The system is attracted to this fixed point, but
can be pushed away by the $Q_{65}$ forcing. The oscillating system is governed
by a slow manifold, which is the dynamic path that the system tends to follow.
By changing the position of the fixed point along this manifold, the relative
duration of the glacial and interglacial states can be controlled, leading to
a slow growth to glacial conditions and then a more abrupt jump to interglacial
conditions, as we have seen in II80 and PP04. Crucifix emphasises the importance
of relaxation oscillations in explaining the ice volume dynamics. This was
partly the motivation to develop the feedforward model, to investigate how much
of the data can be explained by a model that is incapable of producing these
unforced oscillations.

As mentioned, Crucifix designed this model to be one of the simplest
mathematical frameworks that can reproduce the key dynamics of the ice volume
data. As we can see in Figure \ref{fig:comp_models}, the formulation of this
model is almost as simple as II80, whilst Figure \ref{fig:comp_models} shows
it to outperform most of the conceptual models compared here in explaining the
data. However, the Van der Pol oscillator is fundamentally abstract and lacks
direct physical interpretation. Although its formulation is parsimonious, the
model terms do not correspond to specific climate processes or mechanisms. For
instance, the parameter $\alpha$ controlling the fast-slow dynamics has no clear
physical basis, and the model does not offer an explanation for why the system
exhibits these relaxation oscillations in the first place. Additionally, the
$Q_{65}$ term is incorporated simply to include the effect of orbital forcing,
without consideration of how this actually influences ice sheet dynamics. Our FF
and FB models attempt to improve upon these limitations. We methodically
demonstrate the necessity of each term through fitting each subset of the
parameters, and justify the optimised parameter values by comparing them against
the potential physical data that they represent.

Although we are comparing this model with other conceptual models that aim to
explain the potential mechanisms that govern the ice volume dynamics, Crucifix
presents this model within a different context. They use their model to
demonstrate that any model with an appropriate internal periodicity around
100\,kyr could potentially reproduce ice age patterns when forced by
astronomical cycles. This raises important questions about how we can
distinguish between different mechanisms that might drive ice ages. This is
why we develop both the FF and the FB model, to first see how much of the data
can be explained without unforced oscillations, and then to see how much they
improve the fit.

Crucifix also notes that the Van der Pol model, as well as other models of this
dynamic classification, are sensitive to parameter tuning or noise in the
system, calling into question the predictive capabilities of these models. In
the following chapter, we show that our FF model is highly robust to parameter
changes. This increased reliability could suggest that it has greater predictive
power, at least in the absence of anthropogenic impacts.

We will refer to this model as \textit{C11} moving forward. The optimised
parameters for this model are shown in Table \ref{tab:vdp_params}.
\begin{table}[h]
\centering
\caption[Parameter values for Crucifix, 2011]{Parameter values for C11.}
\label{tab:vdp_params}
\begin{tabular}{c|c}
  Parameter & Value \\ \hline\hline
  $V(-800)$ & 0.5214 \\\hline
  $D(-800)$ & 1.1321 \\\hline
  $\tau$ & 40.89\\\hline
  $\alpha$ & 40.89\\\hline
  $\beta$ & 38.59\\\hline
  $\gamma$ & 0.07939 \\\hline\hline
  $m$ & $2.158\times10^7$\\\hline
  $c$ & $5.534\times10^7$
\end{tabular}
\end{table}

\subsection{Imbrie, Imbrie-Moore, and Lisiecki, 2011}
This is the most complex of the models presented in this chapter, and achieves
the best fit to the data. Unlike C11, it is developed from a physical basis and
points to the interplay between the orbital parameters as a cause of both the
100\,kyr glacial cycles in the past 800\,kyr, as well as the mid-Pleistocene
transition \cite{imbrie2011}. It is the first model to achieve a reasonable
simulation of the glacial-interglacial cycles as they transition from 41\,kyr to
100\,kyr cycles without a sliding parameter, attributing the transition instead
to a change in eccentricity amplitude. As discussed, this transition was also
reproduced by the PP04 model, but required a time-dependent term to be included
in the forcing function.

The model switches between a slower, first order \textit{Accumulation} state,
and a more rapid, second order \textit{Termination} state, relating to ice
volume growth and decay respectively. The state of the model changes according
to
\begin{equation}
  \begin{cases}
    Accumulation, & \dfrac{\mathrm{d}V(t)}{\mathrm{d}t}>\max\{-c_\eta,m_\eta
    V(t)-c_\eta\}\\[8pt]
    Termination, & \dfrac{\mathrm{d}V(t)}{\mathrm{d}t}<\max\{-c_\eta,m_\eta
    V(t)-c_\eta\}.
  \end{cases}
\end{equation}
This switching condition can be seen as a piecewise linear function separating
the phase space into two regions. The lower region of the phase space relates to
the \textit{Termination} state and the upper relates to the
\textit{Accumulation} state.

Dynamics in the \textit{Accumulation} state are governed by the first-order
equation
\begin{equation}
  \frac{\mathrm{d}V(t)}{\mathrm{d}t} = -F(V,t) + g(V),
\end{equation}
where
\begin{align}
  F(V,t) &= h_1(V) \beta(t)
        + h_2(V) \varepsilon(t)\sin\tilde{\rho}(t)
        + h_3(V) \varepsilon(t)\cos\tilde{\rho}(t),\\
        \label{eq:imbrie_forcing_fun}
  g(V) &= -a_gV(t)^3 + b_gV(t)^2 - c_gV(t) + d_g,
\end{align}
where the precession variable $\tilde{\rho}(t)$ here is defined differently to
our definition such that $\tilde{\rho}(t)=\nicefrac{3\pi}{2}-\rho(t)$. It should
be noted that, for the purpose of fitting the parameters, we have expressed the
$g(V)$ function as a cubic polynomial, whilst the authors use a piecewise linear
function that resembles a cubic polynomial, but would be able to represent other
functions. It is unlikely this simplification has impacted the fit given that
both the quadratic and cubic coefficients in the optimised model are two orders
of magnitude smaller than the linear term, suggesting that higher order terms
would not significantly improve the fit. The forcing function $F(V,t)$ comprises
a version of the three orbital parameters; eccentricity, obliquity, and
precession, but uses both the sine and cosine of precession which is then scaled
by eccentricity. The orbital variables are weighted by functions
\begin{align}
    h_1(V) &= -m_{h_1}V + c_{h_1},\\
    h_2(V) &= -c_{h_2}\cos(\varphi(V)),\\
    h_3(V) &= -c_{h_3}\sin(\varphi(V)),
  \end{align}
where
\begin{equation}
    \varphi(V) = \max\{c_\varphi,m_\varphi V+c_\varphi\}.
\end{equation}

In the \textit{Termination} state, the system is governed by the second-order
equation
\begin{equation}
    \frac{\mathrm{d}^2V(t)}{\mathrm{d}t^2} = \frac{q(F) - m_V V(t)}{64} -
    \frac{1}{q(F) - m_V V(t)}\left(\frac{\mathrm{d}V(t)}
    {\mathrm{d}t}\right)^2,
\end{equation}
where
\begin{equation}
  q(F) = -m_q F + c_q.
\end{equation}

This model only uses one state variable, ice volume. However, it is a second
order system and is highly non-linear, capable of producing a wide range
of dynamics depending on the parameter values. Another unique feature of the
model is the choice of orbital forcing terms, shown in
\eqref{eq:imbrie_forcing_fun}. Instead of the conventional $Q_{65}$ function,
the authors opt for a combination of the orbital parameters. Rather than
assuming a specific latitude and season for insolation forcing, they use
regression analysis to determine which combination of orbital parameters best
explains the rate of ice volume change. The coefficients for these terms are
then determined through fitting to the ice volume data. We use a simplified
version of this approach for our FF and FB models.

Although the model allows for both phases of the precession cycle to be
included, and with time dependent scaling factors, we find through fitting that
the variable scaling is not significant to the overall fit, and that the
$-\sin\tilde{\rho}$ term is significantly more important than the cosine term.
As mentioned, the authors definition of precession is phase shifted from the
definition used in this thesis, meaning that $-\sin\tilde{\rho}$ in the model is
equivalent to $\cos{\rho}$, using our definition. This is what we use in our FF
and FB models, removing variable scaling factors and the other phase of
precession. We also remove the multiplication with eccentricity, instead
presenting eccentricity as a separate term in the weighted sum of the three
orbital parameters. Similarly to precession, this model treats the impact of
obliquity as negatively correlating with ice volume, and the optimal parameters
ensure that obliquity always has a negative impact on ice volume. Our models
once again simplify this, using a constant negative impact of obliquity on ice
volume.

One novel component of this model comes from its explanation of the
mid-Pleistocene transition. The authors suggest that the transition did not
occur due to a change in the Earth climate system, but due to a natural decrease
in eccentricity amplitude around 900\,kya. The model demonstrates that, over the
past 1.5\,myr, stronger amplitudes in eccentricity produce a stronger impact
from precession, and so a preferential response to the 23\,kyr precession cycle
appears in the ice volume solution. This leads to a counter-intuitive amplitude
decrease in the 100\,kyr ice volume cycle as eccentricity increases.
\begin{table}[h]
  \centering
  \caption[Parameters Imbrie, Imbrie-Moore, and Lisiecki, 2011]{Parameter values for IIL11.}
\label{tab:phase_params}
\begin{tabular}{c|c}
Parameter       & Value       \\ \hline\hline
$V(-800)$       & $2.154$     \\ \hline\\[-11pt]
$\dot{V}(-800)$ & $0.06182$   \\ \hline
$m_\eta$        & 0.07904     \\ \hline
$c_\eta$        & $0.1964$   \\ \hline
$m_\varphi$     & $0.3697$    \\ \hline
$c_\varphi$     & 0.1780      \\ \hline
$m_{h_1}$       & $0.4420$   \\ \hline
$c_{h_1}$       & 4.657       \\ \hline
$c_{h_2}$       & 2.761       \\ \hline
$c_{h_3}$       & 1.702       \\ \hline
$a_g$           & $0.006153$ \\ \hline
$b_g$           & $0.004840$  \\ \hline
$c_g$           & $0.1655$   \\ \hline
$d_g$           & $1.921$     \\ \hline
$m_q$           & $12.31$    \\ \hline
$c_q$           & $23.20$     \\ \hline
$m_V$           & 4.214       \\ \hline\hline
$m$             & $1.384\times10^7$\\\hline
$c$             & $4.191\times10^7$
\end{tabular}
\end{table}

The significance of this model is that it offers a simpler explanation for the
major features of the ice volume data than many other conceptual models.
The authors suggest that the apparently complex climate behaviours might emerge
from relatively simple orbital forcing mechanisms, rather than intrinsic changes
within the Earth system, such as the carbon cycle or ocean stratification.

This model will be referred to as \textit{IIL11}. The optimised parameters for
this model are shown in Table \ref{tab:phase_params}. As with the II80 model,
the solution for this model in the original paper is inverted, approximating
global temperature instead of ice volume. We have therefore flipped sign of the
forcing term and piecewise conditions to produce a solution for ice volume.

Since this model places a strong emphasis on the orbital parameters, we can
consider it to be more aligned with the astronomical theory. However, we
acknowledge that it still requires a non-linear response to this forcing on
Earth, meaning that some geochemical processes are still at play. Our FF model
aims to more closely align with the geochemical theory, to see how much of the
ice volume data can be explained with only the simplest Earth system processes
at play. This can be seen in Figure \ref{fig:all_model_diagrams} when looking at
the IIL11 and FF models. The FF model omits the second order dynamics, switching
mechanism, and non-arithmetic and non-linear functions that are present in the
IIL11 model.

\subsection{Verbitsky, Crucifix, and Volobuev, 2018}
This model is the most recent and incorporates many of the features that we have
seen in the models so far, such as $Q_{65}$ forcing and internal feedbacks that
can be tuned to reproduce the dominant 100\,kyr period in the absence of
forcing. It derives the model from physical principles and points to the 
ratio of positive climate feedbacks and negative glaciation feedbacks as an
explanation for the glacial-interglacial dynamics. The changing of this ratio is
used as an explanation for the mid-Pleistocene transition \cite{verbitsky2018}.

The authors choose to model ice surface area, instead of volume. However, in
order to compare more directly with the other models, we will treat this as
linearly proportional to ice volume. Empirical studies often show the
relationship between ice volume and surface area to follow the power law $V
\propto S^{1.25}$ \cite{ice_surf_to_vol}. However, the approximate range of ice
volume (2.9--8.6$\times10^7$\,km$^3$) means that a linear rescaling has 
less than a 1\% error compared to the power law, which is far less than the
confidence interval we have for the ice volume data itself. The model is
therefore given by
\begin{align}
  \frac{\mathrm{d}V}{\mathrm{d}t} &= \frac{4}{5\zeta}V(t)^{3/4}(a-\varepsilon Q_{65}(t)-k\omega(t)-c\theta(t)),\\
  \frac{\mathrm{d}\theta}{\mathrm{d}t} &= \frac{1}{\zeta}V(t)^{-1/4}(a-\varepsilon
  Q_{65}(t)-k\omega(t))(\alpha\omega(t)+\beta(V(t)-V_0)-\theta(t)),\\
  \frac{\mathrm{d}\omega}{\mathrm{d}t} &= \gamma_1-\gamma_2(V(t)-V_0)-\gamma_3\omega(t).
\end{align}
The model has 3 variables, $V$, $\theta$, and $\omega$, representing glaciation
area (which we will treat as proportional to ice volume), ice basal temperature,
and climate temperature respectively.

Similar to the PP04 model, the authors are able to reproduce the mid-Pleistocene
transition through slowly varying the model parameters, affecting the ratio of
positive and negative feedbacks present in the model. This produces a period
doubling effect in the system where the authors are able to transition first
from the approximately 20\,kyr period of precession to the 40\,kyr period of
obliquity, then again to the 100\,kyr period of eccentricity. There are a number
of possible physical explanations for this gradual change in the system, such as
changes in atmospheric CO$_2$ concentration, ocean circulation, or ice sheet
dynamics. The system changes slowly in this model, as we might see in the
aforementioned physical processes, though the model displays a reasonably abrupt
transition to 100\,kyr cycles, as is seen in the ice volume data. The timing of this
onset does not align perfectly with the mid-Pleistocene transition, though this
can be attributed to the decision to only vary the model parameters linearly.

Focussing on just the past 800\,kyr, the model reproduces the data well,
matching the timings of most terminations and capturing the sawtooth nature of
the ice volume data. It is worth noting that the solution given in the paper,
and the optimised solution we present here, are not able to accurately reproduce
the Marine Isotope Stage 11 (MIS 11) interglacial, which occurs approximately
400\,kya. We will discuss this particular interglacial in the following
chapters when we introduce our models, as it is the longest and warmest
interglacial in the past 800\,kyr, and so is often a challenge for models to
reproduce, as shown by Figure \ref{fig:comp_models}. The authors note this may
be because their model puts too much emphasis on the $Q_{65}$ forcing function
and not enough on obliquity. This further justifies our decision to provide our
models with the orbital parameters as direct forcing inputs, allowing us to
determine the relative importance of each parameter in the ice volume dynamics.

Despite allowing independent weighting of the orbital parameters in our
models, Figure \ref{fig:comp_models} shows our FF model to similarly struggle
with the MIS 11 interglacial, though the FB model performs much better. This
will be examined further in Chapter \ref{chap:analysis}, where we compare our
two models, along with this model, which we show to be dynamically similar when
linearised.

This model is non-linear, which is how it is able to produce a period doubling
bifurcation by varying the parameter values, capturing the mid-Pleistocene
transition. This non-linearity comes largely from the physics of ice flow and
the relationship between ice area and thickness, which are physically
non-linear. However, we will show in Chapter \ref{chap:analysis} that this model
is approximately linear for ice volume dynamics over the past 800\,kyr, when no
variations in the model parameters are present. We demonstrate this by
linearising the model and comparing it with the original model, finding only a
4\% difference in the variance explained by the two models. Although this
reduces the physical accuracy of the model, we have shown through our
augmentations to the Budyko model that improving the physical realism does not
necessarily improve the model's ability to explain the data. The minimal
difference between the original and linearised model suggests that the
non-linearity of ice-sheet dynamics does not play a significant role in
explaining the glacial-interglacial cycles. In this thesis, we are focussed on
understanding the main drivers of the ice volume dynamics. This is why we have
chosen to produce linear models for our analysis, as the ice volume data does
not appear to necessitate non-linear dynamics.

The multiplicative nature of this original model means that changes in one
variable affect the sensitivity of the other variables, making their dynamics
all interconnected in a non-linear way. This makes it difficult to represent in
the flow diagram format that we have used in Figure
\ref{fig:all_model_diagrams}. As mentioned, the dynamics are largely unchanged
when the model is linearised, and so we have shown a linearised version of the
model instead, to allow for the relationships between the variables to be
visually represented.

Since this model relies heavily on the role of Earth-based feedbacks to both
capture the change in dynamics over the past 3\,myr and to explain the 100\,kyr
period in the past 800\,kyr of ice volume data, we can consider it to be more
geochemically aligned. We will refer to this model as \textit{VCV18}, and will
revisit it in Chapter \ref{chap:analysis}, putting it into the same form as our
feedback model and analysing the differences. The optimal values for the 11
model parameters are shown in Table~\ref{tab:verbitsky_params}.
\begin{table}
  \caption[Parameter values for Verbitsky, 2018]{Optimised parameter values for VCV18.}
  \label{tab:verbitsky_params}
  \centering
  \begin{tabular}{c|c}
    Parameter & Value \\ \hline\hline
    $S(-800)$ & 33.72 \\ \hline
    $\theta(-800)$ & 11.31 \\ \hline
    $\omega(-800)$ & 50.34 \\ \hline
    $\zeta$ & 1.472 \\ \hline
    $a$ & 1.225 \\ \hline
    $k$ & 0.008611 \\ \hline
    $c$ & 0.03058 \\ \hline
    $\alpha$ & 1.329 \\ \hline
    $\beta$ & 1.985 \\ \hline
    $\gamma_1$ & 0.03832 \\ \hline
    $\gamma_2$ & 0.1316 \\ \hline
    $\gamma_3$ & 0.2681 \\ \hline
    $S_0$ & 24.18 \\ \hline
    $\varepsilon$ & 0.002360\\ \hline \hline
    $m$ & $3.006\times10^6$ \\ \hline
    $c$ & $-1.896\times10^7$
\end{tabular}
\end{table}

\section{Conclusion}
We began this section by looking at the Budyko model, an early attempt to
study the stability of global ice volume in the latitudinal domain. This was
developed further by Widiasih, who introduced a separate equation for the ice
line and allowed us to study the system dynamics over both short term (years)
and long term (100\,kyr) timescales. We found that over the long term, the model
was able to capture the influence of obliquity on the ice line, but failed to
capture the 23\,kyr precession signal and the 100\,kyr eccentricity signal that
we see in the data. We then introduced a number of augmentations to the model to
investigate if a more realistic model would better reproduce the ice volume
data. The first of these augmentations was to increase the time resolution of
the model from years to days. This meant we could model intra-annual changes in
temperature and ice line, which allowed us to capture the seasonal impact of
precession on the ice line, an important feature of the real Earth system. The
other augmentations helped to increase the realism of the model, but did not
impact the ice line dynamics significantly. This raises the question of whether
increased realism should be the goal of these models, or whether it is more
beneficial to focus on the key mechanisms that drive the ice volume changes.

We then examined a distinct category of models focused on reproducing global ice
volume rather than its latitudinal distribution. These models all use ordinary
differential equations and are designed to capture the fundamental mechanisms
driving ice volume changes. They use different approaches and interpretations of
the underlying dynamics which aligns the models with either the geochemical or
astronomical theory. Models following the geochemical theory generate the
100\,kyr cycle through unforced oscillations, depicting an Earth system that
naturally oscillates due to internal climate feedbacks. In contrast, those
aligned with the astronomical theory emphasise orbital forcing as the primary
driver of ice volume dynamics, depicting a system that is more directly governed
by insolation changes.

Many of these models incorporate switching mechanisms to modulate ice growth and
decay rates, capturing the apparent difference in rates seen in the ice
volume data. However, the parameter values that control this switch can
significantly vary glaciation timing, such as for the PP04 model, which uses a
function of deep-water stratification to trigger the switch of regimes
\cite{pp04}. If we change one of the parameters in this function by just 1\%,
the ice volume solution entirely misses an interglacial that is present in the
data. Models, such as those from Saltzman and Maasch
\cite{saltzman_intrinsic_first}, Tziperman et al. \cite{tziperman2006}, and
Crucifix \cite{crucifix_original}, employ relaxation oscillators with a 100\,kyr
period that is also forced by an insolation function. This does a good job of
reproducing the data without a switching mechanism, but also leads to glaciation
being highly sensitive to the choice of parameters. As Crucifix notes, such
sensitivity poses challenges for predictive modelling \cite{crucifix_anti_ecc}.
However, Imbrie suggests that non-linearity and high sensitivity to physical
parameters is characteristic of the Earth system, and so should be present when
modelling it \cite{imbrie2011}. It should also be noted that the primary purpose
of these models often lies not in predicting the glacial-interglacial cycle
accurately, but in demonstrating the potential mechanisms behind them. Indeed,
Tziperman notes that successfully fitting a model to reproduce the ice volume
data does not necessarily prove the identification of physical mechanisms
\cite{tziperman2006}.

\begin{figure}
  \centering
  \input{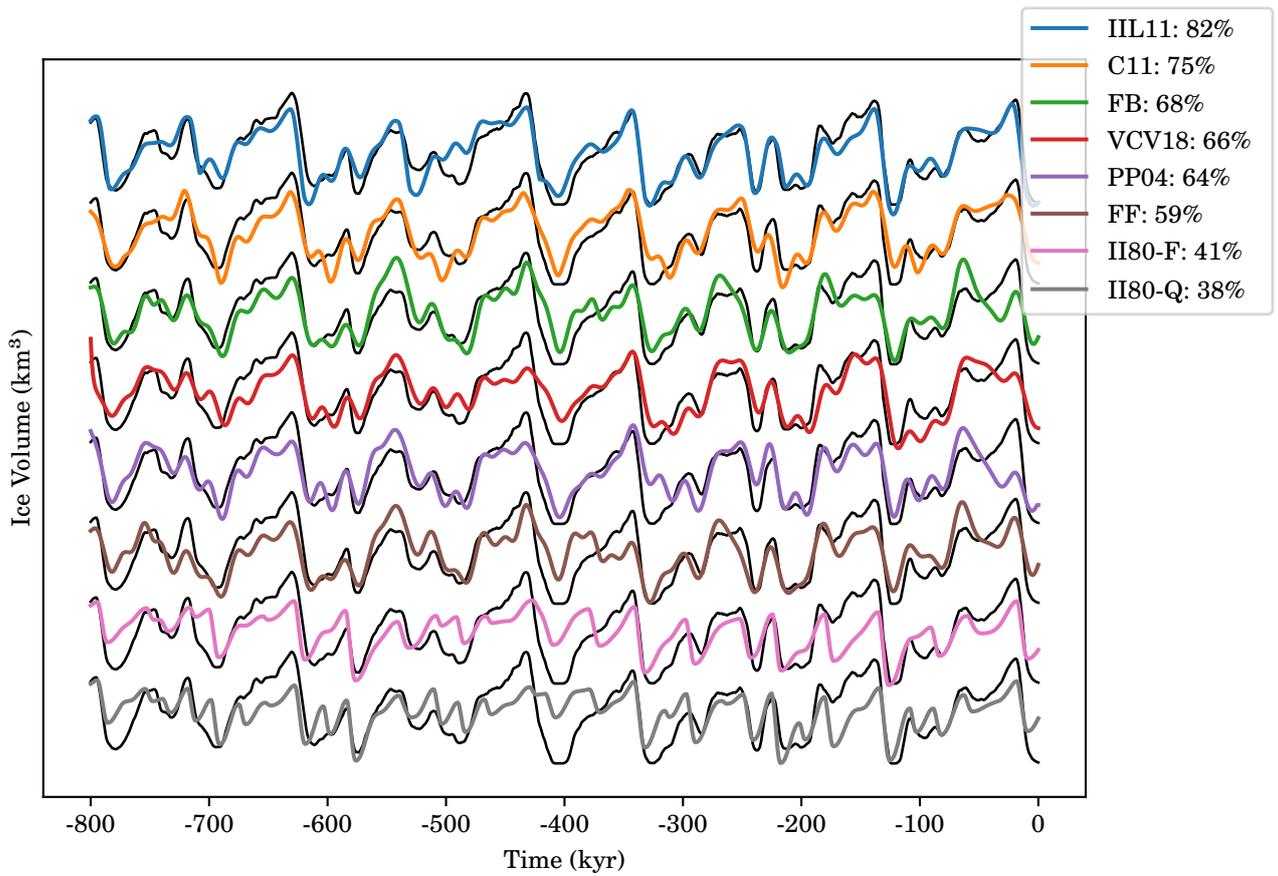}
  \caption[Conceptual Model Comparison]{Visual comparison of the 6 models
    discussed in this section and ours. The black plots are Bintanja's ice
    volume, to which each model's parameters were fit to. The variance of this
    data that is explained by each model is shown in the legend.}
  \label{fig:comp_models}
\end{figure}

In some cases, the original parameters for the models were derived from physical
principles or measurements, rather than being chosen to maximise the fit to the
data. For consistency, we used standardised inputs for all models and refit them
to the same ice volume data. Although this results in different parameter values
to those in the original papers, the qualitative behaviour of the model remains
the same in each case, except from for VCV18. This model uses physically derived
parameter values and is less focussed on maximising the fit. We explore this
model further in Section \ref{sec:comparison_verbitsky}. A comparison of all of
these models when fit to the same ice volume data and using the same insolation
and orbital inputs is shown in Figure \ref{fig:comp_models}. Here we have also
included our two models (FF and FB), which are introduced in the following two
chapters. One key takeaway from this figure is that our simple linear models are
comparable to the other models in terms of how well they can explain the data.

The two models we will introduce are designed to distil the key features of
these models into the simplest possible form, reducing the number of assumptions
made about the physical system. The first (FF) aligns with the astronomical theory,
and is incapable of producing the 100\,kyr eccentricity period without orbital
forcing. The second (FB) aligns with the geochemical theory and is able to
oscillate at 100\,kyr without the need for forcing. In Chapter
\ref{chap:analysis}, we compare the two models and revisit the VCV18 model,
showing that its linearised version performs similarly well to the original and
is dynamically similar to our FB model. We show that the FB model only makes
significant improvements on the FF model around MIS 11, therefore calling into
question how much is added by first the introduction of unforced oscillations in
the FB model, and then non-linear dynamics in the VCV18 model. This will help us
to understand what can be confidently determined from the ice volume data and
what should be considered more speculative.

All models in this category are non-linear but vary in complexity and their
ability to reproduce ice volume data. The simplest example, Imbrie and Imbrie
1980, employs a single state variable representing global ice volume but fails
to adequately reproduce the data. The remaining models use multiple state
variables, which, as we demonstrate in the following chapter, is a minimum
requirement to meaningfully reproduce the data.
\clearpage
\thispagestyle{empty}

\clearemptydoublepage
\chapter{Feedforward Model}
\label{chap:feedforward_model}

\initial{H}ere we introduce the first of our two linear models which is intended
as the simplest possible case that can still reproduce the ice volume data. We
refer to this model as the feedforward model, as it cannot produce feedback,
meaning it is incapable of oscillating without external forcing. We perform a
leave-one-out fitting to measure the necessity of each term in the model,
finding all terms meaningfully improve the fit, with precession having the
smallest impact on the solution. This analysis suggests that we have produced
the simplest linear model that is capable of reproducing the ice volume data,
and that two state variables are the minimum requirement for a system of this
nature.

We then propose a physical interpretation of this phenomenological model and
verify that the parameters are physically valid. We propose that two physical
mechanisms contribute to the ice volume dynamics; surface air temperature, and
bulk ocean temperature. These were chosen as we observed the need for a
slow-fast pair of variables that respond to eccentricity. As we saw in the
Budyko model, air temperature is relatively quick to adjust to changes in
insolation, whilst we know that the significantly higher heat capacity of the
ocean means that it can take thousands of years to equilibrate. We investigate
the validity of these proposed variables and demonstrate that eccentricity on
its own can explain the approximate range in bulk ocean temperature that we
observe in proxy data. We also show the modelled solutions for these variables
alongside the equivalent proxy data, observing good data alignment from ice
volume and surface temperature solutions, but a weaker correlation from the
ocean temperature solution.

\section{Phenomenological Model}
\label{sec:phenom_model}
We now present a simple phenomenological model that is intended to reproduce the
global ice volume data using only a linear function of the orbital parameters.
We then fit the coefficients of this model to the data and evaluate its
predictive power. Although we do not expect this model to outperform the more
complex, non-linear ice volume models from Chapter \ref{chap:models}, it will
demonstrate the extent to which a linear astronomical model can explain the
ice volume dynamics.

Before presenting the model equations, we outline the key design decisions and
their justifications, so that the construction logic is clear from the outset.

The primary observational constraint guiding the model's construction is the
power spectrum of the ice volume data shown in
Figure~\ref{fig:orbital_and_benthic_time_series_power_specs}. This spectrum
contains peaks at the periods of all three orbital parameters, suggesting that
ice volume depends on eccentricity, obliquity, and precession. Each of these
therefore appears as a forcing term in the ice volume equation, with a
coefficient to be determined by fitting. This would form the simplest possible
linear dependence on the orbital parameters, a weighted sum, filtered through a
time constant that prevents an instantaneous response.

However, a single equation forced by the three orbital parameters is
insufficient. The ice volume power spectrum contains a strong
$\sim100$\,kyr peak corresponding to eccentricity, but conspicuously
lacks the $\sim400$\,kyr peak that dominates eccentricity's own power
spectrum. This is the well-known ``400\,kyr problem''. Any model that depends
linearly on the instantaneous value of eccentricity alone will inevitably
reproduce this $400$\,kyr period, producing a poor fit to the data.

To address this, we introduce a second state variable $O(t)$ that tracks
eccentricity with a lag. The equation for $O(t)$ takes the simplest possible
form, it is driven towards the current value of eccentricity at a rate governed
by a time constant $\tau$, with no dependence on obliquity or precession. Only
eccentricity is included in this equation because it is the only orbital
parameter that modulates the total annual insolation reaching Earth, as derived
in Section~\ref{sec:eccentricity}. Obliquity and precession redistribute
insolation in latitude and season respectively, but do not change the annual
global total. The bulk ocean, with its large thermal inertia, integrates
insolation over a full year and so responds primarily to this annual total
rather than to seasonal or latitudinal redistribution. We confirmed this by
trialling a version of the model in which $O(t)$ depended on all three orbital
parameters, finding no significant improvement to the quality of the fit.

The slow variable $O(t)$ then appears alongside the instantaneous eccentricity
$\varepsilon(t)$ in the ice volume equation. Since $O(t)$ is a lagged copy of
$\varepsilon(t)$, the difference between the two approximates the time
derivative of eccentricity $\dot{\varepsilon}(t)$, scaled by the lag. More
precisely, if $p_1$ and $p_2$ are of similar magnitude but opposite sign, as
fitting confirms, then \[ p_1 O(t) + p_2 \varepsilon(t) \approx
p_1\bigl[\varepsilon(t-\Delta t) - \varepsilon(t)\bigr], \] which is
proportional to $-\dot{\varepsilon}(t)$ over a timescale $\Delta t \sim \tau$. A
signal proportional to the derivative of eccentricity retains the
$\sim100\,\mathrm{kyr}$ oscillation but has significantly reduced power at
$\sim400\,\mathrm{kyr}$, because differentiation amplifies higher frequencies
relative to lower ones. This mechanism is the key to resolving the 400\,kyr
problem within a purely linear framework, and is explored quantitatively in
Section~\ref{sec:ecc_change_approx} and Figure~\ref{fig:deps_approx_power_spec}.

As discussed in Section~\ref{sec:ocean_model}, one physical candidate for this
slow-responding variable is the bulk ocean temperature, which responds to the
magnitude changes in insolation caused by eccentricity on a 10\,kyr order
timescale. We therefore denote this variable by $O(t)$, though we emphasise that
at this stage the model is phenomenological, $O(t)$ is defined by its dynamical
role rather than by any specific physical interpretation. In
Section~\ref{sec:physical_model}, we explore a physical interpretation of this
model and assess how realistic it can be.

The model is given by
\begin{align}
  \tau\frac{\mathrm{d}O}{\mathrm{d}t}
  &= \varepsilon(t) - O(t),
  \label{eq:phenom_model_slow_var}\\[4pt]
  \tau\frac{\mathrm{d}I}{\mathrm{d}t} &= p_1 O(t) +  p_2\varepsilon(t) +
  p_3\beta(t) + p_4\cos(\rho(t)) - I(t) + p_5,
  \label{eq:phenom_model_ode}
\end{align}
where the $p_i$ coefficients are to be fitted, along with the time constant
$\tau$. This time constant is introduced since neither ice volume, nor the slow
variable, is expected to respond instantaneously to changes in the orbital
parameters. Two separate time constants were originally used for this model,
however fitting revealed the two constants to be approximately equal. This could
be due to some coupling between ice volume and the mechanism that $O(t)$
represents, causing them to change at the same rate.

\subsection{Analytical Solution}
Using a simple linear model has the benefit that the analytical solution is both
attainable and interpretable. This is useful for verifying that each term plays
an meaningful role in the model's performance. In the previous section, we found
that the optimised PP04 model had a parameter that did not contribute to the
ice volume solution. This can happen when assembling a model from a physical
perspective, rather than a data driven approach, with the physical interpretation
and validation afterwards. We now present the analytical solution to this model
and fit the coefficients to the ice volume data. Using the integrating factor
method, we solve for the slow variable to get
\begin{equation}
  O(t) = \zeta_\tau[\varepsilon(t)] +
  O_0e^{\nicefrac{-t}{\tau}},
\end{equation}
where $O_0$ is the initial condition and the functional
\begin{equation}
  \zeta_\tau[y(t)] = \frac{e^{\nicefrac{-t}{\tau}}}{\tau}\int_0^t y(u)
  e^{\nicefrac{u}{\tau}}\mathrm{d}u,
  \label{eq:zeta_functional}
\end{equation}
for some function $y(t)$.

The solution for $O(t)$ simply reproduces the eccentricity signal with a lag
approximately equal to the time constant $\tau$. This is because $O(t)$ changes
according to the difference between itself and eccentricity $\varepsilon$,
converging towards $\varepsilon$ at a rate determined by $\tau$. If $\tau=0$
then \eqref{eq:phenom_model_slow_var} would simplify to $O(t)=\varepsilon(t)$,
meaning there is no lag between the two signals. However, as $\tau$ increases,
$O(t)$ converges more slowly towards $\varepsilon$, and so the lag between the
two signals increases.

Substituting the solution for $O(t)$ into the differential equation for ice
volume gives
\begin{equation}
  \tau\frac{\mathrm{d}I}{\mathrm{d}t} = p_1\zeta_\tau[\varepsilon(t)] +
  p_2\varepsilon(t) + p_3\beta(t) + p_4\cos(\rho(t)) + p_5 +
  p_1 O_0e^{\nicefrac{-t}{\tau}}.
\label{eq:phenom_model_ode_subbed}
\end{equation}
We then solve this using the integrating factor method again to get
\begin{equation}
  I(t) = p_1\zeta_\tau\big[\zeta_\tau[\varepsilon(t)]\big] +
  p_2\zeta_\tau[\varepsilon(t)] + p_3\zeta_\tau[\beta(t)] +
  p_4\zeta_\tau[\cos(\rho(t))] + p_5 + \left(\frac{p_1 O_0 t}{\tau}
  + I_0 - p_5\right)e^{\nicefrac{-t}{\tau}}.
  \label{eq:phenom_model_sol_subbed}
\end{equation}

\begin{figure}
  \centering
  \input{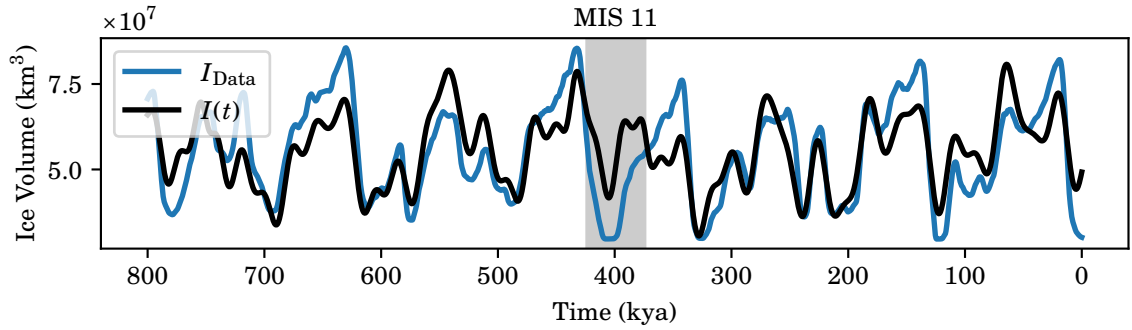}
  \caption[Feedforward Ice Volume Solution]{Our modelled ice volume $I(t)$ from
    \eqref{eq:phenom_model_sol}, alongside the ice volume data
    $I_\mathrm{Data}$. The grey region delineates Marine Isotope Stage (MIS) 11,
    around which there is a notable difference between the two curves. The model
    parameters that produce this fit are given in Table
  \ref{tab:phenom_params}.}
  \label{fig:fit_ice_vol}
\end{figure}

If we run our model for sufficiently long before our period of interest, the
$e^{\nicefrac{-t}{\tau}}$ term in $I(t)$ can be treated as zero. This gives the
asymptotic approximation as
\begin{equation}
  I(t) = p_1\zeta_\tau\big[\zeta_\tau[\varepsilon(t)]\big] +
  p_2\zeta_\tau[\varepsilon(t)] + p_3\zeta_\tau[\beta(t)] +
  p_4\zeta_\tau[\cos(\rho(t))] + p_5.
  \label{eq:phenom_model_sol}
\end{equation}
This solution is non-linear in $\tau$, so to optimise the parameters of the
model, we repeat a least squares fit for the $p_i$ coefficients whilst varying
$\tau$, guaranteeing that all of the parameters are globally optimised. These
optimal parameter values are given in Table \ref{tab:phenom_params}, with the
corresponding solution shown in Figure \ref{fig:fit_ice_vol}.
\begin{table}
  \centering
  \caption[Parameter values for the phenomenological feedforward
  model]{Parameter values for the phenomenological model given in
    \eqref{eq:phenom_model_ode}. Their roles are shown in
    Figure~\ref{fig:leave_one_out_with_diagram}B. Errors given by the 95\%
  confidence interval in the fit are shown beside the estimated value.}
  \label{tab:phenom_params}
\begin{tabular}{c|c|c|c}
  Parameter & Role & Value & Units               \\ \hline\hline
  $p_1$  & Slow Eccentricity &$1.88\pm0.05$&$\times10^9$\,km$^3$ \\ \hline
  $p_2$  & Fast Eccentricity &$-1.94\pm0.05$&$\times10^9$\,km$^3$\\ \hline
  $p_3$  & Fast Obliquity &$-1.54\pm0.06$&$\times10^9$\,km$^3$\\ \hline
  $p_4$  & Fast Precession  &$-1.9\pm0.1  $&$\times10^7$\,km$^3$\\ \hline
  $p_5$  & Constant Offset &$6.8\pm0.2  $&$\times10^8$\,km$^3$ \\ \hline
  $\tau$ & Response Rate &$14.8\pm0.4 $& kyr
\end{tabular}
\end{table}
\begin{figure}
  \centering
  \input{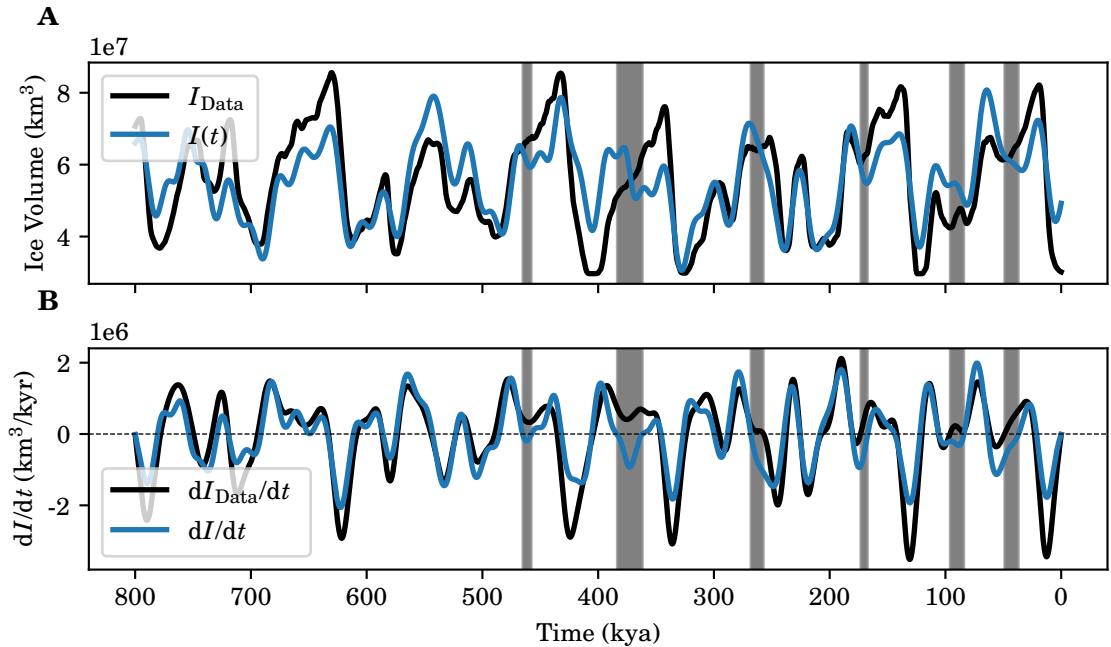}
  \caption[Ice Volume Gradient Comparison]{Comparison of the time
    derivatives of the ice volume data $I_\mathrm{Data}$ and our modelled ice
    volume $I(t)$. The gradients were calculated using a five-point stencil
    finite difference method. This the grey bars highlight where the model is
  unable to capture the direction of change in ice volume}
    \label{fig:wheen_I_gradient}
\end{figure}
\begin{figure}
  \centering
  \input{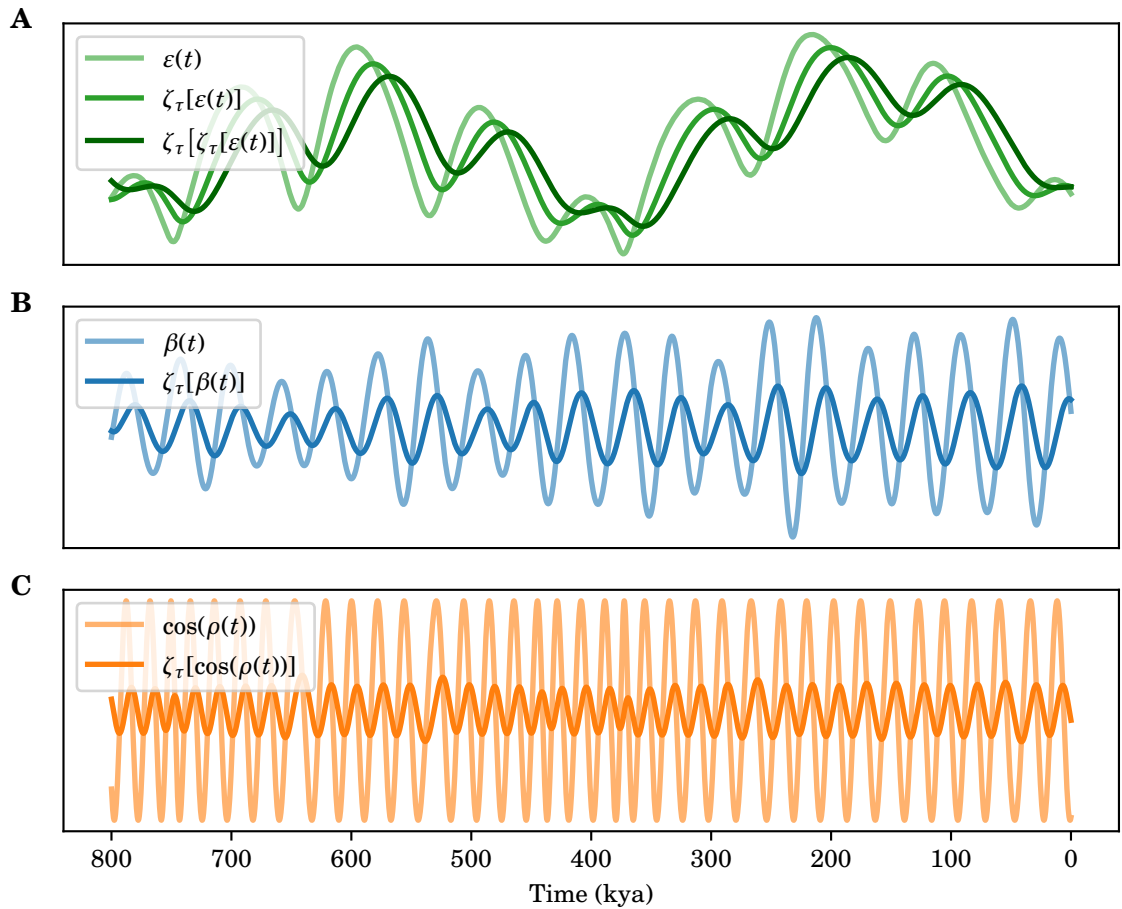}
  \caption[Effect of $\zeta_\tau$]{The qualitative effect of the functional
    $\zeta_\tau$ on eccentricity (\textbf{A}), obliquity (\textbf{B}), and the
    cosine of precession (\textbf{C}) as defined by \eqref{eq:zeta_functional},
    where $\tau=14.8$\,kyr. Each $\zeta_\tau$ term that appears in
    \eqref{eq:phenom_model_sol} is shown. Note how the higher the
    frequency of the orbital parameter, the more $\zeta_\tau$ reduces its
    amplitude.}
  \label{fig:zeta_impact}
\end{figure}

The analytical solution for ice volume is expressed in terms of the $\zeta_\tau$
functional. To demonstrate how $\zeta_\tau$ responds to its inputs,
Figure~\ref{fig:zeta_impact} shows each term in \eqref{eq:phenom_model_sol}
plotted alongside the orbital parameter it depends on. We see that the effect of
$\zeta_\tau$ is somewhat different in each case. For eccentricity, a lag
approximately equal to $\tau=14.8$\,kyr is introduced each time $\zeta_\tau$ is
applied. For obliquity and precession, the amplitude is reduced more
significantly, and the lag is shorter. This is because they oscillate at a
higher frequency than eccentricity. As a result, where the
$\zeta_\tau[\varepsilon(t)]$ curve is able to slowly follow the $\varepsilon(t)$
curve, $\zeta_\tau[\beta(t)]$ and $\zeta_\tau[\cos(\rho(t))]$ cannot reach the
extrema of their inputs before they begin to change direction. However, since
the scale factors in \eqref{eq:phenom_model_sol} will account for this effect,
the only important feature of $\zeta_\tau$ is the lag it introduces. The ice
volume solution can therefore be interpreted simply as a weighted sum of the
lagged orbital parameters, where eccentricity appears twice, with a longer lag
the second time. With this interpretation, it is clear how the feedforward model
aligns with the astronomical theory, as it is entirely dependent on orbital
forcing to produce any oscillatory behaviour.
\section{Analysis}
The ice volume solution in Figure \ref{fig:fit_ice_vol} shows good qualitative
alignment with the data. Although it does not always reach the same magnitude of
the data, Figure \ref{fig:wheen_I_gradient} shows that it almost entirely
matches the direction in which the ice volume is changing, with the notable
exception of the time interval around Marine Isotope Stage (MIS) 11. We will
discuss this anomalous time period in Section
\ref{sec:comparison_with_feedforward}. We emphasise that this model is not
intended to be a perfect fit to the data, but rather a demonstration of how much
can be explained by approximating the Earth system as having a linear dependence
on the orbital parameters. By reproducing the majority of the ice volume data
with this linear model, we propose that more complex mechanisms and non-linear
dynamics may only be needed to explain the discrepancies in magnitude around the
extrema, and possibly MIS 11.

\subsection{Parameter Necessity}
\label{sec:param_necessity}
\begin{figure}
  \hspace{-40pt}
  \input{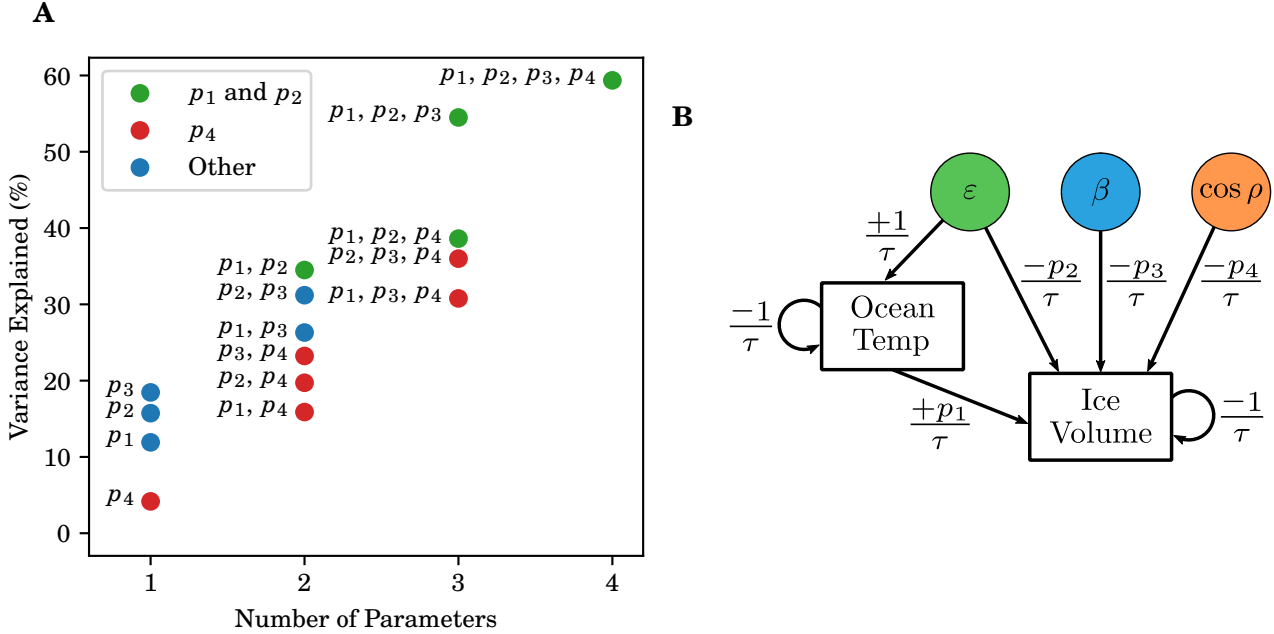}
  \caption[Feedforward Leave-One-Out]{\textbf{A}: Variance explained for all
    possible parameter combinations of the model given by
    \eqref{eq:phenom_model_ode}, with the excluded parameters set to zero. The
    constant term $p_5$ and time constant $\tau$ are always included. For each
    case, the included parameters were optimised to attain the best fit to the
    ice volume data. Cases where both $p_1$ and $p_2$ are included (green)
    produce especially good fits, whilst cases without this pair, but with $p_4$
    included (red), produce especially poor fits.\\\textbf{B}: Flow diagram
    representing the dynamics of the model and the role of each parameter.
    Whilst $p_2$, $p_3$, and $p_4$ scale the impact of the orbital parameters on
    ice volume, $p_1$ scales the impact of our ocean temperature variable, which
    is itself dependent on eccentricity. The constant offset $p_5$ is excluded
    for clarity in this diagram.}
    \label{fig:leave_one_out_with_diagram}
\end{figure}

In order to evaluate the necessity of each term in our model, we systematically
prune terms whilst evaluating the accuracy in each case. The results of this are
shown in Figure~\ref{fig:leave_one_out_with_diagram}. To measure the accuracy,
we are using the variance of the ice volume data that is explained by the model.
This calculation is given by
\begin{align}
    \text{Total Variance: } & \sigma^2_\text{Total} = \text{Var}\left[I_\text{Data}\right] \\[10pt]
    \text{Residual Variance: } & \sigma^2_\text{Residual} = \text{Var}\left[I_\text{Data} - I(t)\right]\\[10pt]
    \text{Explained Variance: } & \sigma^2_\text{Explained} = \sigma^2_\text{Total} - \sigma^2_\text{Residual} \\[10pt]
    \text{Variance Explained Ratio: } & R^2 = \frac{\sigma^2_\text{Explained}}
  {\sigma^2_\text{Total}} = \frac{\sigma^2_\text{Total} -
  \sigma^2_\text{Residual}}{\sigma^2_\text{Total}},
\end{align}
where $0\leq R^2 \leq 1$ will be expressed as a percentage variance explained.

The offset term $p_5$ and time constant $\tau$ are included in every version of
the model so that the solutions are comparable. The precession term, represented
by $p_4$, is shown to consistently contribute less to the variance explained
than the other terms. This is partly due to the nature of the ice volume curve,
in which the higher frequencies appear with smaller amplitudes. However, this
may also arise due to solid Earth forcing, such as volcanic activity, that could
affect the ice volume in a similar frequency range, introducing noise that
precession cannot account for. Although precession contributes the least, it is
still responsible for approximately 5\% of the variance in the ice volume data,
regardless of the other terms included in the model. This suggests that it is
not contributing to over-fitting. Moreover, its inclusion allows for the higher
frequencies, shown in Figure \ref{fig:benth_and_power_specs}, to be represented
in the model.

Secondly, we find that the subsets that include both $p_1$ and $p_2$,
relating to the slow and fast eccentricity terms, consistently produce the best
fit. We can see that adding both terms produces a greater improvement than the
sum of improvements that the two terms bring individually. This indicates that
the change in eccentricity represented by the pair is crucial to the model, more
so than either term on its own.
\subsection{Eccentricity Change Approximation}
\label{sec:ecc_change_approx}
\begin{figure}
  \centering
  \input{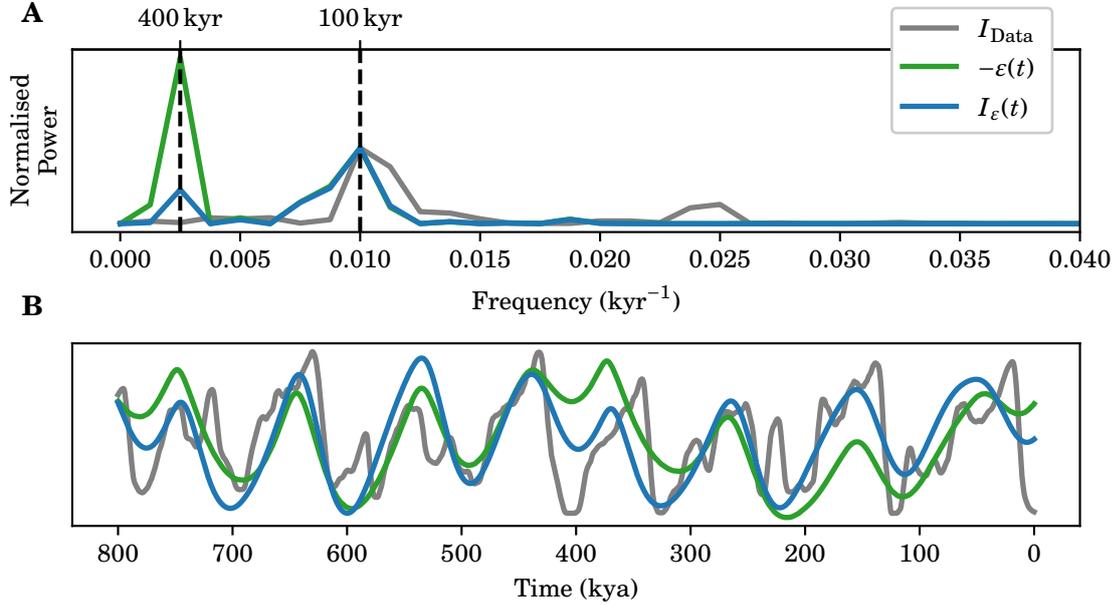}
  \caption[Eccentricity 400\,kyr Removal]{\textbf{A}: Power spectra comparison
    of the ice volume data, the negative eccentricity curve $-\varepsilon(t)$,
    and the eccentricity component of our model solution $I_\varepsilon(t)$.
    They have been normalised to equate the powers corresponding to 100\,kyr.
    Our $I_\varepsilon(t)$ power spectrum matches that of $\varepsilon(t)$ apart
    from a significant drop around the 400\,kyr period. \textbf{B}: Time series
    showing the same comparison, also normalised for qualitative comparison.
    They show how the filtered eccentricity better captures the broad behaviour
    of the ice volume curve, namely around 780, 400, 320, and 160\,kya.}
  \label{fig:deps_approx_power_spec}
\end{figure}

In order to better understand the role of the slow variable $O(t)$, we examine
how eccentricity relates to ice volume. As shown in
Figure~\ref{fig:orbital_and_benthic_time_series_power_specs}, ice volume
decreases during peaks of eccentricity. However, the degree to which ice volume
decreases appears to be independent of the magnitude of eccentricity's peaks.
This suggests that the absolute value of eccentricity may be less important than
the direction in which it is changing. By including the $O(t)$ variable
alongside $\varepsilon(t)$, we are providing the model with both current, and
lagged, values of eccentricity. The difference of these two signals therefore
approximates the change in eccentricity over time, given by the first two terms
in \eqref{eq:phenom_model_ode_subbed}.

As shown in Table \ref{tab:phenom_params}, the fit $p_1$ and $p_2$ values are
within 3\% of each other. This indicates that the optimised model only depends
on the change in eccentricity and not the instantaneous or lagged values on
their own. We could therefore simplify the model by setting $p_2=p_1$, without
losing accuracy. However, since we wish to find physical interpretations for
both the slow and fast responses to eccentricity we will keep the two terms
distinct.

By using the relatively short-term change in eccentricity as input, rather than
eccentricity itself, our model is able to produce 100\,kyr oscillations whilst
effectively removing the 400\,kyr amplitude modulation.
Figure~\ref{fig:deps_approx_power_spec}A shows the prominent 400\,kyr peak in
eccentricity's power spectrum that is significantly reduced in the eccentricity
component of our model solution $I_\varepsilon(t)$.
Figure~\ref{fig:deps_approx_power_spec}B shows this effect in the time domain,
with $I_\varepsilon(t)$ more closely matching the broad dynamics of the ice
volume data.

\subsection{Baseline Comparison}
To evaluate the performance of our feedforward model, we compare it against a
baseline model that simply uses a Fourier series to approximate the ice volume
data. This baseline model is given by
\begin{equation}
  I_{\text{FT}}(t) = a_0 + \sum_{k=1}^{3}\left[a_k\cos(2\pi f_k t) +
  b_k\sin(2\pi f_k t)\right]
\end{equation}
where $f_k$ are the three highest-magnitude frequencies in the ice volume data's
power spectrum, and $a_k$ and $b_k$ are the coefficients to be fitted. With the
constant term $a_0$, this gives a total of 7 parameters to be optimised. With 6
free parameters in our feedforward model, we can directly compare the two models
to see how well our phenomenological model performs against a standard signal
approximation technique.

\begin{figure}
  \centering
  \input{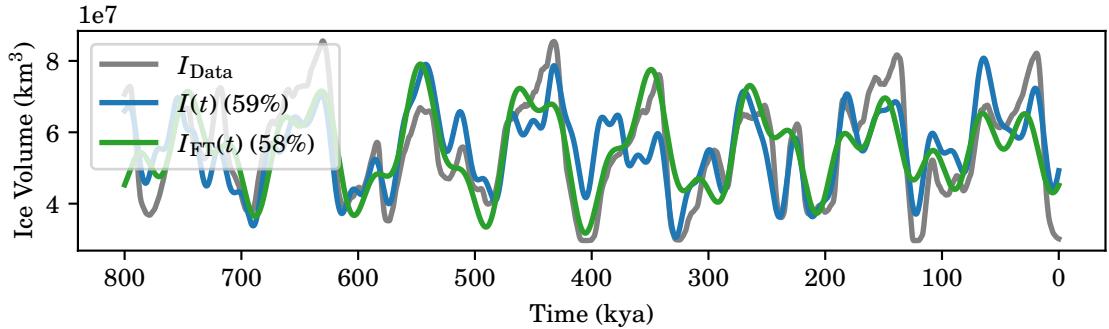}
  \caption[Feedforward vs Fourier Series]{Comparison of the ice volume data
    $I_\mathrm{Data}$ against our feedforward model $I(t)$ and a Fourier series
    baseline model $I_\mathrm{FT}(t)$ with 3 frequency components (6
    parameters), selected as the highest-magnitude frequencies in the spectrum.
    Both models achieve a similar variance explained, with the feedforward model
  achieving 59\% and the Fourier series achieving 58\%.}
  \label{fig:wheen_I_fft_comp}
\end{figure}

\begin{figure}
  \centering
  \input{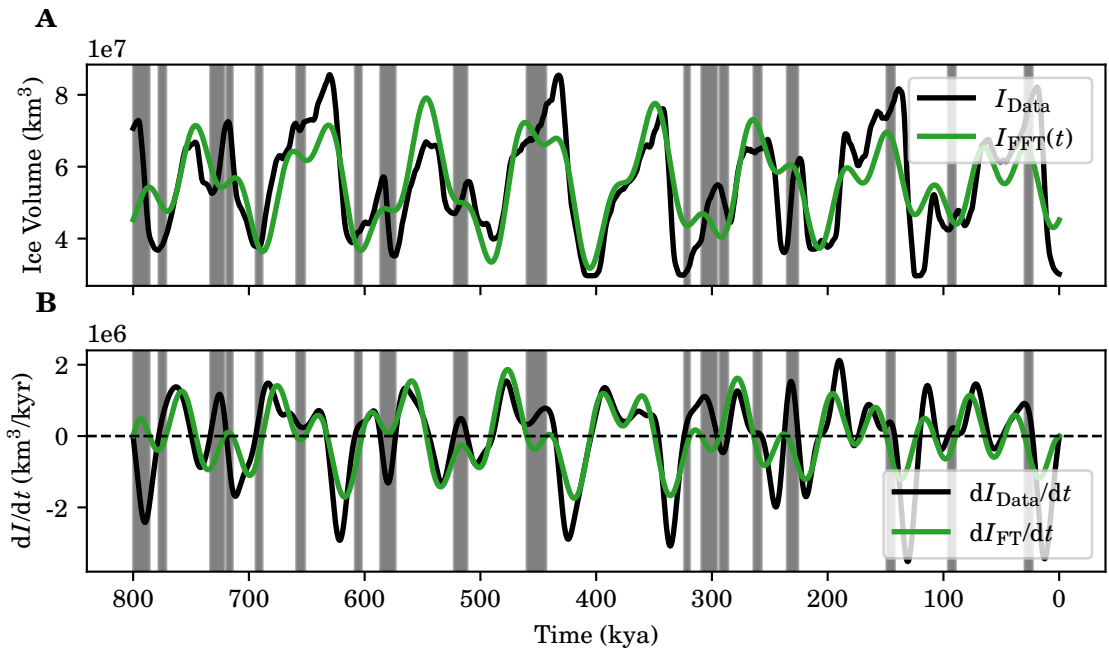}
  \caption[Fourier Series Gradient Comparison]{The Fourier series solution
    $I_\mathrm{FT}(t)$ and ice volume data $I_\mathrm{Data}$ plotted alongside
    their time derivatives. The grey bars highlight where the model is unable to
    capture the direction of change in ice volume.}
    \label{fig:fft_gradient}
\end{figure}

Figure~\ref{fig:wheen_I_fft_comp} shows both model solutions plotted against
the ice volume data. Both models achieve a similar variance explained, with our
feedforward model achieving 59\% and the Fourier series achieving 58\%. This
demonstrates that our simple phenomenological model is at least capable of
reproducing the ice volume data to the same degree as a standard signal
approximation technique. The benefit of our model is that it provides a
mechanistic explanation for the ice volume dynamics, rather than simply
approximating the signal. One way to visualise the benefit that this offers is
to once again look at the time derivative of the ice volume data and model
solutions to see if ice volume growth or ablation is being reproduced at the
correct times, as shown in Figure \ref{fig:fft_gradient}. The gradients were
computed using central finite differences after smoothing with a Gaussian filter
($\sigma = 5$\,kyr).

The disagreement in ice volume change in this figure is notably more common than
in our feedforward model, shown previously in Figure \ref{fig:wheen_I_gradient}.
This shows that our feedforward model is better able to capture the direction of
change in ice volume, an important feature when considering the physical
dynamics of the system.


\subsection{Cross-Validation}

\begin{figure}
  \centering
  \input{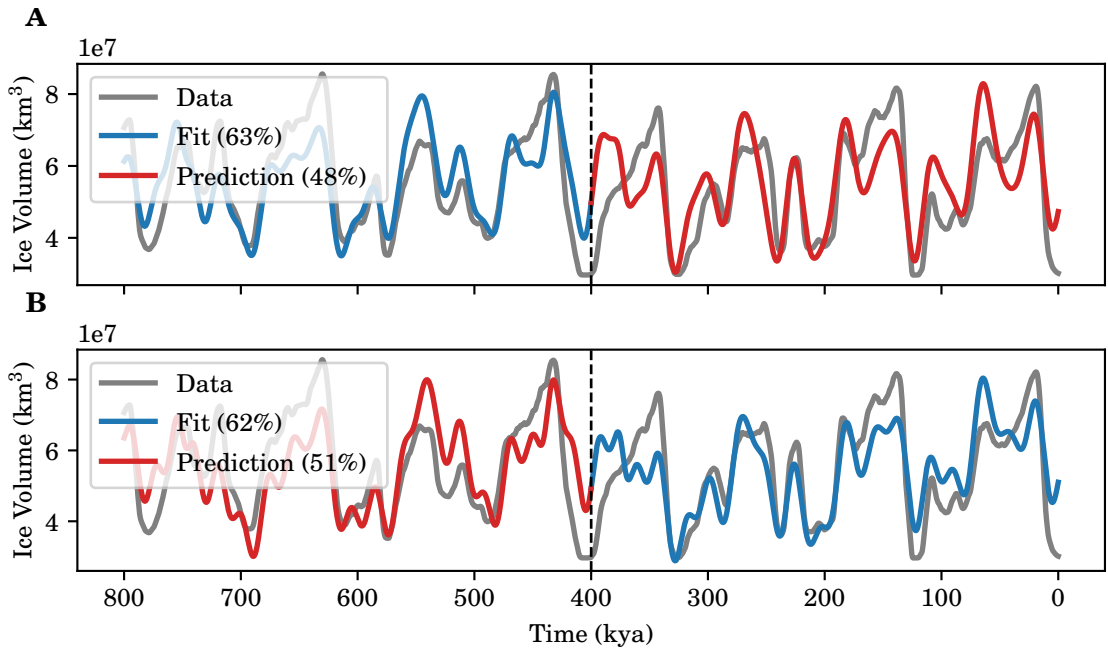}
  \caption[Cross-Validation]{Split-half cross-validation of our feedforward
    model. \textbf{A}: The model is trained on the interval 800-400 kya and used to
    predict the interval 400-0 kya. \textbf{B}: The model is trained on 400-0 kya and
    used to predict 800-400 kya. The dashed line indicates the boundary between
    training and prediction regions. Variance explained is shown for both fitted
  (blue) and predicted (red) intervals.}
  \label{fig:ff_cross_validation}
\end{figure}

To assess our model's predictive capability, we perform a split-half
cross-validation. The model is trained on the first half of the record
($800-400$\,kya) and used to predict the second half ($400-0$\,kya), and vice
versa. This tests whether the model captures the underlying dynamics of the
system rather than simply fitting noise or overfitting to the training data. The
results of this cross-validation are shown in
Figure~\ref{fig:ff_cross_validation}.

The results demonstrate that the model retains reasonable predictive power when
applied out of the training time frame. Although the variance explained is
notably lower for the predicted intervals compared to the fitted intervals, the
qualitative behaviour of the ice volume record is generally well reproduced. The
predicted curves capture the timing and broad structure of glacial cycles in
both cases. The main exception occurs in the approximately 60\,kyr following
MIS 11, where the model struggles to reproduce the observed behaviour. We
discuss this anomaly further in Section~\ref{sec:comparison_with_feedforward}.
Outside of this interval, the predicted and fitted portions of the record are
qualitatively similar, suggesting that the model has learned physically
meaningful relationships rather than artefacts of the training data.

\section{Physical Interpretation}
We now propose a physical interpretation for our feedforward model by
reformulating it to produce solutions that can be directly compared with proxy
data. This will consist of the same ice volume and ocean temperature variables,
as well as a surface temperature variable that represents the forcing terms in
the ice volume equation. The physical model is underdetermined as it requires
scaling parameters to match the proxy data. We therefore use the optimal
parameter values from our fit phenomenological model, as well as constraints
from the relevant data.

Before introducing the full physical model, we attempt to validate the
hypothesis that the slow variable $O(t)$ could represent bulk ocean temperature.
We present a simple differential equation for ocean temperature as a function of
eccentricity, using measured values to constrain the model. This is then
incorporated into the full physical model, with the unknown parameters being
estimated from the proxy ocean temperature data. We then confirm that our ocean
temperature model, that is linearly dependent on eccentricity, is capable of
explaining the range of temperatures seen in the data whilst using physically
plausible parameters. It should be noted that the primary purpose of this model
is to assess how well it can explain the ice volume data, rather than the exact
mechanisms that would produce behaviour dynamics. However, for our physical
interpretation to be worth considering, it is important to ensure that the
mechanisms we are proposing are at least physically plausible.

\subsection{Ocean Model}
\label{sec:ocean_model}
As mentioned, the slow variable $O(t)$ could represent the temperature of the
bulk ocean. Due to the high thermal inertia of Earth's ocean, it could take
thousands of years for it to equilibrate to a new temperature. This response
time is slow enough that the impact of precession and obliquity are
significantly reduced, as we see in Figure~\ref{fig:zeta_impact}. To confirm
this, we trialled a model in which $O(t)$ depended on all three orbital
parameters and found no significant improvement to the accuracy of the fit.

One of the key criticisms of the astronomical theory is that eccentricity does
not vary the magnitude of insolation enough to justify the significant 100\,kyr
period observed in the ice volume data. We propose that this is based on an
overly simplistic interpretation of eccentricity's impact on the Earth system.
By considering the ocean temperature as an intermediate mechanism, we
demonstrate that eccentricity can produce significant climatic changes without
the need for amplification.

\begin{figure}[p]
    \vspace*{-7cm}
    \hspace*{-0.7cm}
    \import{../figs/pgfs/}{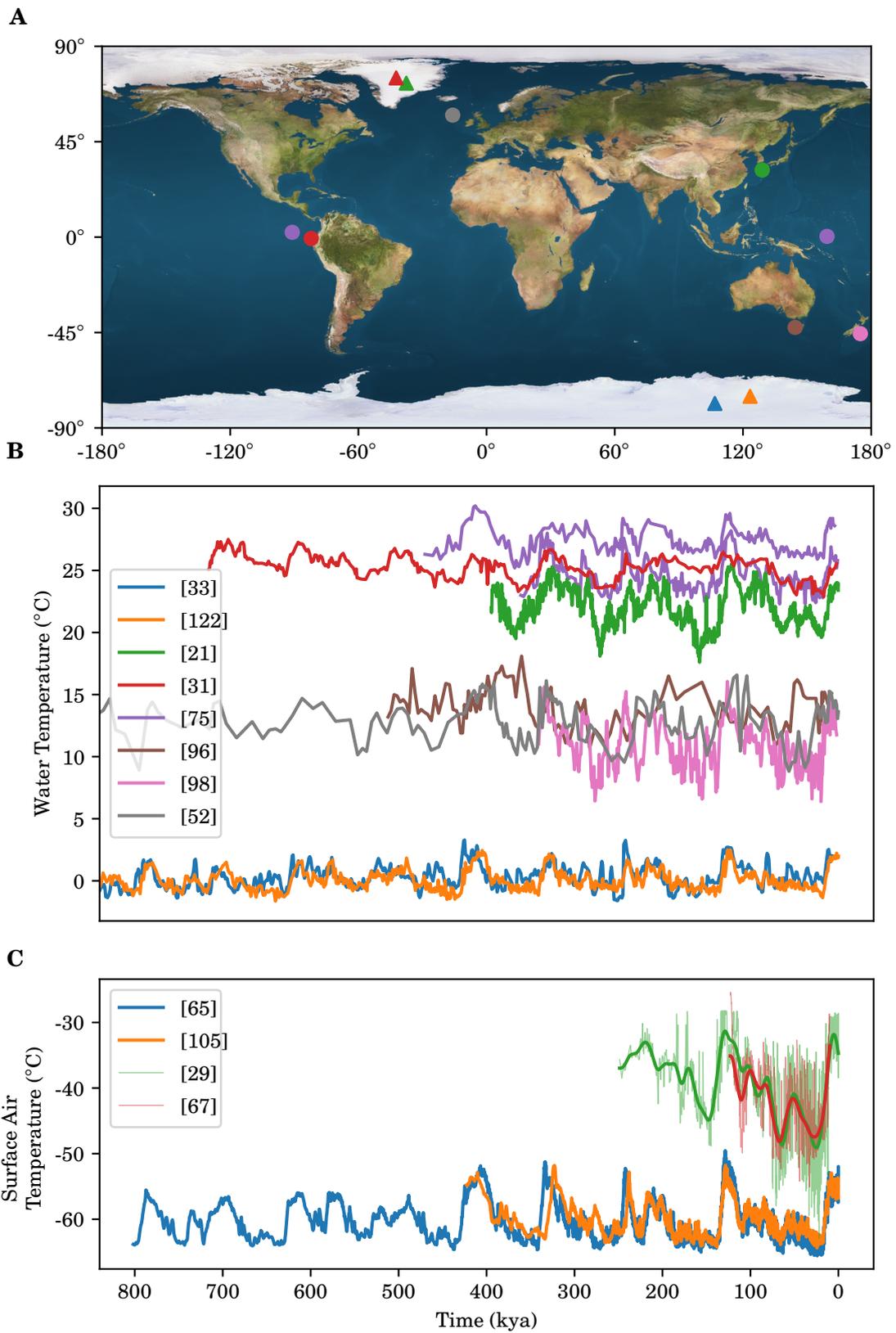}
    \vspace*{-5cm}
    \caption[Physical Model Equivalent Data]{\textbf{A}: Drilled core locations
      used in our data comparison with seafloor drill sites shown as circles and
      ice core drill sites shown as triangles~\cite{earth_map}. \textbf{B}: Proxy
      data for BWT and SWT from independent drilled cores, their colours
      correspond to the locations shown on the map. The global BWT data are
      plotted in blue and orange with the remaining plots relating to the
      regionally dependent SWT data. \textbf{C}: Regional SAT for locations in
      Antarctica and Greenland. Since the Greenland data are more noisy, a second
      order low-pass Butterworth filter was used to remove frequencies above
      1/19\,kyr$^{-1}$, as this is the highest frequency at which orbital
      parameters vary.}
    \label{fig:sea_and_air_temp_with_map}
\end{figure}

We now show how eccentricity could explain the broad dynamics of ocean
temperature without the presence of Earth system feedbacks, as shown in the
paleoclimate record in Figure \ref{fig:sea_and_air_temp_with_map}. To do this,
we will use first principles to derive a simple differential equation for ocean
temperature as a function of eccentricity. This will share the same dynamic
behaviour as the slow variable in our phenomenological model but include
parameters to rescale the solution to align with the proxy data so that it can
be used in the physically based model. Using measured values to constrain the
model sufficiently, we show that, even with conservative estimates for the
unknown parameters, the difference in energy that eccentricity provides is
enough to explain the observed range in ocean temperature without amplification.
In the following section, we use the proxy data to estimate the unknown
parameters in this model and confirm that the physical interpretation is
plausible.

In this derivation, we exclude the surface ocean since its interface with
the atmosphere causes it to behave differently than the rest of the ocean. We
will instead consider the bulk ocean temperature~$O(t)$, relating to all water
below a certain depth. We also treat the heat loss rate of the bulk ocean as
constant over time, however we will improve upon this assumption in Section
\ref{sec:fitting}.

The rate of change of the bulk ocean temperature can be expressed as
\begin{equation}
  \frac{\mathrm{d}O}{\mathrm{d}t} = \frac{\text{Total Absorbed Power}}{\text{Ocean
  Heat Capacity}} - \text{Heat Loss Rate}.
\end{equation}
The total power provided to the bulk ocean through insolation can be expressed as
\begin{equation}
  P_O(\varepsilon(t)) = \alpha \gamma A_O \overline{Q_\mathrm{E}}(\varepsilon(t)),
\end{equation}
which has units W. The $\alpha$ coefficient accounts for the absorption and
reflection that occurs whilst passing through the atmosphere. The coefficient
$0<\gamma\leq1$ accounts for the heat emitted back to the atmosphere from the
surface ocean layer before it can reach the bulk volume of the ocean, where
$\gamma=1$ implies no loss of heat to the atmosphere. The annually averaged
insolation reaching Earth $\overline{Q_\mathrm{E}}$ comes from
\eqref{eq:annual_insol}, which is then scaled by the total area of the ocean
$A_O$. For this approximate calculation we are ignoring the latitudinal
asymmetry of ocean water, treating it as uniformly distributed across Earth's
surface. All of these values, apart from the free parameter $\gamma$, are given
in Table \ref{tab:constants}.

\begingroup
\renewcommand{\arraystretch}{1.1}
\begin{table}
  \centering
\caption{Constants used to model ocean temperature as a function of eccentricity.}
\label{tab:constants}
\begin{tabular}{c|c|c|c|c}
Term                    & Symbol     & Value                 & Units     & Source\\ \hline\hline
Ocean Surface Absorption Ratio  & $\alpha$   & $0.48$                & 1         & \cite{ocean_heat_budget}  \\ \hline
Average Ocean Density   & $\rho_O$   & $1025$                & kg/m$^3$  & \cite{ocean_spec_heat}  \\ \hline
Ocean Specific Heat     & $c_O$      & $3850$                & J/kg\,$^\circ$C  & \cite{ocean_spec_heat}  \\ \hline
Ocean Volume            & $V_O$      & $1.335\times10^{18}$  & m$^3$     & \cite{ocean_vol}  \\ \hline
Ocean Surface Area      & $A_O$      & $3.619\times10^{14}$  & m$^2$     & \cite{ocean_vol}  \\ \hline
Ocean Heat Capacity     & $C_O$      & $5.268\times10^{24}$  & J/$^\circ$C&\cite{ocean_spec_heat,ocean_vol}  \\ \hline
Average Eccentricity    & $\overline{\varepsilon}$ & 0.02707 & 1         & \cite{laskar2004} \\ \hline
Average Obliquity       & $\overline{\beta}$       & 0.40739 & rad       & \cite{laskar2004} \\ \hline
Average Ocean Input Power& $P_O(\overline{\varepsilon})$     & $5.914\times10^{16}$ & J/s
&\cite{laskar2004,ocean_heat_budget,solar_constant,ocean_vol}
\end{tabular}
\end{table}
\endgroup

This power is then divided by the total heat capacity of the ocean, given as
\begin{equation}
  C_O=\rho_OV_Oc_O,
\end{equation}
which has units J/$^\circ$C. The mass of the ocean is given by $\rho_OV_O$,
where $\rho_O$ is the average density of the ocean and $V_O$ is the total
volume. Although heat capacity varies with the salinity and temperature of the
ocean, we will express the average specific heat capacity with $c_O$.

We can now substitute in our measured values to estimate the bulk ocean warming
rate. Since this will have units of $^\circ$C/s, we rescale by
$3.1536\times10^{10}$ to attain the rate in $^\circ$C/kyr.

This gives us a warming rate of
\begin{equation}
  \frac{\mathrm{d}O}{\mathrm{d}t} = \frac{P_O(\varepsilon(t))}{C_O} - l=
  \frac{353.907\gamma}{\sqrt{1-\varepsilon(t)^2}} - l
  {}\,^\circ\mathrm{C/kyr},
  \label{eq:bulk_heating_rate}
\end{equation}
where $l$ is the constant heat loss term and $\gamma$ is the absorption
constant. Since we have not formally defined the depth at which our bulk ocean
meets the ocean surface, we do not have a value for $\gamma$. Instead, we treat
this, and $l$ as free parameters to be determined through fitting. If
eccentricity is capable of producing the observed ocean temperature changes
without amplification, then we would expect a fit value of $\gamma<1$. A fit
value of $\gamma>1$ is not physically valid and would suggest that there is some
amplification mechanism that we have not accounted for.

As a simple calculation, we explore the role $\gamma$ plays in this warming rate,
using ocean temperature data as a reference. The last glacial maximum occurred
around 20\,kya, and since then our ocean has heated at an extreme rate, similar 
to other transitions into an interglacial period~\cite{lgm}. The surface water
temperature (SWT) has increased by up to 10$^\circ$C in individual
locations~\cite{sst_temp_ranges}. However, globally averaged SWT and bottom
water temperature (BWT) have both increased by around 3 to
5$^\circ$C~\cite{benthic_to_ice_vol, sst_lgm_map}. This gives us an approximate
maximum warming rate of 0.2$^\circ$C/kyr.

If we suppose that half of the incoming heat reaches the bulk ocean, meaning
that $\gamma=0.5$, then for an eccentricity range of $0\leq\varepsilon\leq0.06$,
\eqref{eq:bulk_heating_rate} would give a difference in warming rates of
0.319$^\circ$C/kyr. In Section \ref{sec:fitting}, we use the ocean temperature
proxy data to estimate $\gamma=0.59$, resulting in a similar rate as calculated
here, and also of the same scale as the 0.2$^\circ$C/kyr we see in the data.
This suggests that eccentricity could explain the observed ocean temperature
changes without the need for amplification, whereas it is commonly considered to
be orders of magnitude too weak to have a significant direct impact on the
climate, as we saw with the Budyko-Widiasih model.

It is important to note that we have used the extrema of eccentricity and ocean
temperature for this calculation. We also acknowledge that the ocean heat loss
rate will vary as a function of temperature to some degree. However, this simple
calculation shows that, given the range of eccentricity, it has the ability to
drive a change in ocean temperature on the same scale as that observed in the
data.

We are using the fit parameters from our phenomenological model, which means we
are assuming ocean temperature has a time constant of $\tau=14.8$\,kyr.
However, since the temperature of the ocean never reaches a global equilibrium,
it is difficult to determine if this is physically plausible. Van Aken states
that heat can be mixed across all depths on a time scale of hundreds to a few
thousand years~\cite{ocean_100_to_1000_timescale}. The Intergovernmental Panel
on Climate Change state that deep ocean temperatures can take thousands of years
to respond to surface temperature changes~\cite{ocean_1000_timescale}. Crucifix
estimates an even longer mixing time on the order of
10\,kyr~\cite{crucifix_original}. There is a similar level of uncertainty when
it comes to estimating the time constant for ice sheet growth and ablation.
Estimates range from 500 years on the lower end~\cite{ice_time_const_short}, but
can go up to 27\,kyr~\cite{ice_time_const_long}. Our fit time constant of
$\tau=14.8\pm0.4$\,kyr therefore seems feasible for both ocean temperature and
ice volume.

Although we have demonstrated that the observed warming rate of the ocean can be
sufficiently explained by eccentricity, we have so far treated the heat
loss rate as constant. Heat loss occurs through back radiation, convection,
conduction, and evaporation~\cite{ocean_heat_budget}. We will assume for
simplicity that these are all linearly dependent on bulk ocean temperature
$O(t)$, giving a consolidated loss term of $m O(t) + n$. Replacing the constant
loss rate in \eqref{eq:bulk_heating_rate} with this linear loss rate gives
\begin{equation}
  \frac{\mathrm{d}O}{\mathrm{d}t} = \frac{353.907\gamma}{\sqrt{1-\varepsilon(t)^2}} - mO(t) - n,
\end{equation}
which has units $^\circ$C/kyr.

As we wish to substitute $O(t)$ into our linear model, we take the first order
Taylor expansion about the average eccentricity $\overline{\varepsilon}=0.02707$
to get
\begin{equation}
  \frac{\mathrm{d}O}{\mathrm{d}t} = 9.591\gamma\varepsilon(t) + 353.8\gamma -
  mO(t) - n, 
  \label{eq:linear_approx_ocean_sol}
\end{equation}

In order to simplify this equation for use in our physical model, we write it as
\begin{equation}
  \tau \frac{\mathrm{d}O}{\mathrm{d}t} = c\varepsilon(t) - O(t) + \alpha_O,
  \label{eq:simple_ocean_model}
\end{equation}
\vspace{-5pt}
where
\begin{align}
    \tau &= \frac{1}{m},\\[2pt]
    c &= \frac{9.591\gamma}{m},\\[2pt]
    \alpha_O &= \frac{353.8\gamma - n}{m}.
    \label{eq:ocean_param_subs}
\end{align}
Once we have estimated values for $c$, and $\alpha_O$, we can estimate all of
the coefficients in \eqref{eq:linear_approx_ocean_sol}. For this equation to be
physically plausible, we would expect the resultant heat gain and heat loss
rates to be approximately equal, so as to avoid unbounded temperature change. We
also require that $0<\gamma\leq 1$ is satisfied.

Although we are treating $O(t)$ as ocean temperature, it may also incorporate
mechanisms such as long-term feedbacks in atmospheric CO$_2$. The ocean is one
of the largest carbon sinks on Earth, but holds less CO$_2$ at higher
temperatures~\cite{hot_ocean_loses_co2}. This positive correlation between ocean
temperature and atmospheric CO$_2$ concentration means it is difficult to
separate their impact on ice volume. Because of this, we will treat the $O(t)$
variable as ocean temperature for convenience when comparing with the proxy
data, though it may also be capturing the impact of other mechanisms.

\subsection{Physical Model}
\label{sec:physical_model}
Our physical model will produce the same ice volume solution $I(t)$ as the
phenomenological model, but will also produce solutions representing bulk ocean
temperature $O(t)$ and SAT $S(t)$. These three variables are then compared to
their respective proxy data to help validate the model. To remain consistent
with the phenomenological model, we require that ocean temperature positively
impacts ice volume, whilst SAT negatively impacts it. It is logical that an
increase in SAT would lead to a decrease in ice volume, however it is less
obvious how an increase in ocean temperature could lead to an increase in ice
volume.

A possible explanation for this relates to the evaporation rate from the ocean
surface. As the ocean temperature increases, the evaporation rate also
increases, leading to an increase in the moisture content of the air. This could
then lead to greater precipitation over the ice sheets, increasing their volume.
It is difficult to determine the degree to which this would impact ice volume,
especially since ocean warming could also be attributed to a decrease in ice
volume. As mentioned, this variable could also incorporate the effect of other
mechanisms, but it may also not represent global ocean temperature. One
alternative interpretation could be that it represents the ocean temperature
only at high latitudes. In this case, it is possible that increased moisture in
the atmosphere plays a more significant role.
\begin{figure}
  \centering
  \includegraphics[width=0.38\textwidth]{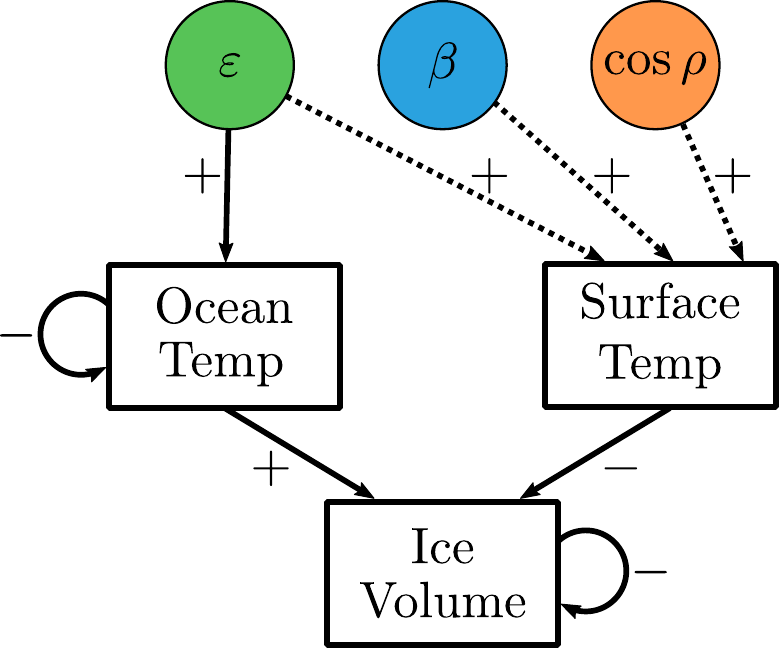}
  \caption[Physical Model Flow Diagram]{Flow diagram showing the proposed
    physical model of orbital influence through bulk ocean and surface air
    temperature. This diagram describes the same dynamical system as with the
    phenomenological model diagram shown in Figure
    \ref{fig:leave_one_out_with_diagram}B. However, it now includes surface air
    temperature as an intermediate variable. The dashed arrows between the
    orbital parameters and this variable represent that they instantaneously
    impact surface temperature, rather than its rate of change.}
  \label{fig:ode_diagram}
\end{figure}
The flow diagram in Figure~\ref{fig:ode_diagram} shows this proposed physical
model, extending the phenomenological model from Section \ref{sec:phenom_model},
shown in Figure \ref{fig:leave_one_out_with_diagram}. The governing equations
for this physical model are
\begin{align}
  \tau\frac{\mathrm{d} I}{\mathrm{d} t} &= a O(t) + b S(t) - I(t) +
  \alpha_I,
  \label{eq:ice_ode}\\
  \tau\frac{\mathrm{d} O}{\mathrm{d} t} &= c\varepsilon(t) - O(t) +
  \alpha_O,\\
  S(t) &= d\varepsilon(t) + e\beta(t) + f\cos\left({\rho(t)}\right) +
  \alpha_S,
  \label{eq:surface_steady}
\end{align}
where $a$, $b$, $c$, $d$, $e$, $f$, $\alpha_O$, $\alpha_I$, $\alpha_S$, and
$\tau$ are parameters that will be determined. In order to produce three
distinct solutions that can be compared to their relevant data, each variable
requires an offset and scaling factor. As a result, this physical model contains
10 parameters, compared to the 6 in the phenomenological model. In Section
\ref{sec:fitting}, we use the proxy data for ocean temperature and SAT to
introduce 4 extra constraints, allowing us to uniquely determine the physical
parameter values.

In order to analytically solve this system, we once again solve for the slow
variable $O(t)$ using an integrating factor. This is then substituted into the
ice volume equation, alongside the instantaneous $S(t)$. We then solve for
$I(t)$ using an integrating factor. The solution for $O(t)$ is given as
\begin{equation}
  O(t) = c\zeta_\tau[\varepsilon(t)] + \alpha_O - \alpha_O
  e^{\nicefrac{-t}{\tau}}  + O_0e^{\nicefrac{-t}{\tau}},
  \label{eq:O_sol}
\end{equation}
where $O_0$ is the initial condition for $O(t)$ and the functional
$\zeta_\tau$ is defined in \eqref{eq:zeta_functional}.

Substituting \eqref{eq:surface_steady} and \eqref{eq:O_sol} into
\eqref{eq:ice_ode} then gives
\begin{equation}
  \tau\frac{\mathrm{d} I}{\mathrm{d} t} = a c\zeta_\tau[\varepsilon(t)]
  + b d \varepsilon{(t)} + b e \beta{(t)} + b f \cos{(\rho{(t)})}- I{(t)} + (a
  \alpha_{O} +  b\alpha_{S} + \alpha_{I}) + a(O_0 - \alpha_O)e^{\nicefrac{-t}{\tau}}.
\end{equation}

The analytical solution for $I(t)$ is then
\begin{align}
  I(t) = ac\zeta_\tau\big[\zeta_\tau[\varepsilon(t)]\big] &+ bd\zeta_\tau[\varepsilon(t)] +
be\zeta_\tau[\beta(t)] + bf\zeta_\tau[\cos(\rho(t))]  \\ &+(a \alpha_{O} +  b\alpha_{S} +
\alpha_{I}) + \left(\frac{a}{\tau}\big((O_0 - \alpha_O)t - \alpha_O \tau\big) +
b\alpha_S - \alpha_I + I_0\right)e^{\nicefrac{-t}{\tau}}.
\end{align}
As before, if we solve from sufficiently long before our period of interest, we
can asymptotically approximate this as
\begin{align}
  I(t) &= ac\zeta_\tau\big[\zeta_\tau[\varepsilon(t)]\big] + bd\zeta_\tau[\varepsilon(t)] +
be\zeta_\tau[\beta(t)] + bf\zeta_\tau[\cos(\rho(t))] + (a \alpha_{O} +  b\alpha_{S} +
\alpha_{I}),\\
     &= p_1\zeta_\tau\big[\zeta_\tau[\varepsilon(t)]\big] + p_2\zeta_\tau[\varepsilon(t)] +
     p_3\zeta_\tau[\beta(t)] + p_4\zeta_\tau[\cos(\rho(t))] + p_5,
\end{align}
where we have included the parameters from the phenomenological model from
\eqref{eq:phenom_model_sol} to demonstrate how its coefficients align with those
of the physical model. As is shown by this comparison, although the physical
model has 10 parameters, the two solutions for ice volume are still equivalent.
Where we originally had single $p_i$ coefficients scaling the orbital
parameters, we now have a product of two coefficients, one scaling the
intermediate physical variable and the other scaling its contribution to ice
volume. Similarly, the $p_5$ coefficient is now represented by the weighted sum
of three offsets, one for each physical variable. In order to understand the
physical quantities that we are modelling, we review the available proxy data
for each one and compare it to our model solutions.

\subsection{Ocean Temperature}
The ocean temperature time series shown in
Figure~\ref{fig:sea_and_air_temp_with_map}B shows both BWT and SWT proxy data.
The blue and orange plots represent the global BWT and show reasonable agreement
with each other. The blue plot comes from Elderfield, who used Mg/Ca ratios to
separate the BWT and ice volume contributions to benthic foraminifera
$\delta^{18}$O data~\cite{benthic_to_ice_vol}. The orange plot comes from
Shackleton's quadratic model of BWT from $\delta^{18}$O, fit using current
samples of foraminifera for which the local BWT is known~\cite{shack_bwt_eq}.
This model was then applied to the Lisiecki and Raymo $\delta^{18}$O data, which
largely comprises the same foraminifera used to calibrate Shackleton's
model~\cite{benthic_data}.

The remaining 7 plots in Figure \ref{fig:sea_and_air_temp_with_map}B correspond
to SWT proxy data, which vary far more with location. Data sources were chosen
to incorporate drill sites from a range of latitudes, as shown in Figure
\ref{fig:sea_and_air_temp_with_map}A. The SWT data are produced through a method
similar to that used by Elderfield, with the addition of a second measure that
relies on alkenones. These are a type of ketone that some phytoplankton produce,
with more saturated alkenones being produced in higher temperatures. Since
phytoplankton live near the surface of the water to collect sunlight, the degree
of alkenone saturation found in these phytoplankton can act as a proxy for
SWT~\cite{alkenones}.

\subsection{Surface Air Temperature}
The third variable in our physical model represents globally averaged SAT. Proxy
data on the timescale we are interested in are limited, with the predominant
source being continental ice cores, as marked by triangles in Figure
\ref{fig:sea_and_air_temp_with_map}A. As snow accumulates, the top layers
slowly compress lower layers into ice, preserving the isotopic composition of
the snow at the time. The ice cores containing the longest history are located
in Antarctica and Greenland, reaching depths of approximately 3\,km.
Figure~\ref{fig:sea_and_air_temp_with_map}C depicts proxy SAT data for Greenland
and Antarctica.

The blue and orange plots are SAT at Antarctica's Dome C and Vostok
respectively. Both datasets were given as temperature anomalies from current
conditions, so we have shifted the plots to match with the present mean
temperature of -55$^\circ$C for this region of Antarctica~\cite{petit1999}. The
orange plot employs the same principle as is used to estimate global ice volume
from foraminifera shells. Since precipitation containing the lighter $^{16}$O is
more likely to evaporate, we expect to see greater proportions of $^{18}$O in
the preserved ice when temperatures were higher. Through experiment, Dansgaard
found this relationship to be approximately $T=1.45\delta + 19.7$, where
$\delta$ is the $\delta^{18}$O ratio~\cite{original_d18O_to_temp}. The blue plot
also uses this principle, but instead measures the ratio of hydrogen isotopes in
the ice. Although the isotopic ratios of hydrogen and oxygen are closely
related, they can deviate slightly due to the conditions in the region of ocean
from which the precipitation originated~\cite{ice_core_history}. Although
different isotopes were measured in the two separate cores, we see a close
agreement between the datasets.

The Greenland SAT proxies are plotted in green and red, based on cores from the
GRIP and NGRIP sites respectively. The green plot uses $\delta^{18}$O as a proxy
for SAT and has been converted to temperature using $T = -0.1925\delta^2
-11.88\delta - 211.4$~\cite{sat_from_d18O}. This model was fitted to ice core
data from GRIP so is likely to be more accurate than the generalised linear
model. This core has a $\delta^{18}$O record up to 250\,kya. However, due to ice
folding close to the bedrock, only the most recent 100\,kyr of the record is
reliable~\cite{last_10p_of_grip_corrupted}. We have shown the full record
here because we only wish to compare it to our solution qualitatively. The red
plot spans only 120\,kyr and is considered reliable for this full duration as it
is from the separate drill site, NGRIP~\cite{last_10p_of_grip_corrupted}. The
plot is also based on $\delta^{18}$O but additionally uses the nitrogen isotope
ratio from the air bubbles trapped within the ice. Whilst air can still move
around the compacting snow, the $^{15}$N isotopes are more likely to sink than
the lighter $^{14}$N. This enrichment of the lower portion of compacting snow is
inversely related to the surrounding temperature, providing additional data for
the SAT estimate~\cite{kindler2014}.

Notably, the data from Greenland provides a higher time resolution but
significantly shorter time-span. This is due to the accumulation rates around
the drill sites~\cite{ice_core_history}. With greater precipitation, there are
more data available for the same duration. However, this reduces the time span
captured by the same depth of ice.

The increased resolution in Greenland's ice core allowed for the discovery of
the high frequency climate fluctuations known as Dansgaard–Oeschger
events~\cite{DO_1,DO_2}. The cause of these fluctuations is not fully
understood, though they appear to be local to the northern hemisphere and are
believed to relate to changes in the Atlantic ocean due to freshwater
perturbations~\cite{greenland_high_amplitude}. These oscillations occur with a
period of approximately 1.5\,kyr \cite{DO_events_period}. Since our model is a
linear combination of the significantly slower orbital parameters, these
fluctuations will not appear in our solution for SAT. This is the intended
behaviour of our model since the phenomenon is intrinsic to Earth. Instead of
reproducing the SAT data exactly, we wish to show that the SAT dynamics that
result directly from the orbital variations are sufficient to explain the
majority of the ice volume data.

Since our data only represents the most extreme of environments, we are
careful not to overgeneralise to global SAT. However, it is worth noting that
$S(t)$ might better reflect polar SAT as it holds the most significant influence
on ice volume. We can therefore expect to find a reasonable qualitative agreement
between the polar SAT data and the $S(t)$ solution that best reproduces the ice
volume data.

\subsection{Fitting}
\label{sec:fitting}
To estimate each physical parameter, we use the fit $p_i$ values from our
phenomenological model along with 4 additional constraints. These relate to the
range, and average, of bulk ocean temperature and SAT. We see BWT varies between
-1$^\circ$C and 2$^\circ$C globally, whereas the mean SWT varies dependent on
location, but has an approximate range of 5$^\circ$C. Assuming these both
contribute equally to the bulk ocean temperature gives an estimated bulk ocean
temperature range of $\Delta_\mathrm{O}=4^\circ$C. The average of the SWT data
is approximately 19.4$^\circ$C whilst the averaged BWT is 0.2$^\circ$C, giving
an estimated average of $\mu_\mathrm{O}=9.8^\circ$C.

For the SAT data, we are restricted by the extreme locations of its sources,
namely Greenland and Antarctica. We assume that these reflect the qualitative
dynamics of global SAT over time but may have a differing range to the global
signal, and will certainly have a lower mean than the global average. We will
therefore employ further sources to determine the $S(t)$ equation coefficients.

Thomas estimates that SAT increased $5.8\pm1.4^\circ$C from the last glacial
maximum to pre-industrial time~\cite{surface_air_temp_range}, which was
approximately 13$^\circ$C~\cite{pre_industrial_temp}. Since these periods span
close to the minimum and maximum temperatures of the past 800\,kyr, we
use this to constrain the SAT between 7.2 and 13$^\circ$C, giving an
estimated range of $\Delta_\mathrm{S}=5.8^\circ$C and an estimated mean SAT of
$\mu_\mathrm{S}=10.1^\circ$C. We now have the necessary constraints to extract
each physical parameter from the phenomenological model coefficients, where the
equations to solve are
\begin{align}
  ac &= p_1,\\
  bd &= p_2,\\
  be &= p_3,\\
  bf &= p_4,\\
  a\alpha_O+b\alpha_S+\alpha_I &= p_5,\\
  \mathrm{Mean}[O(t)] &= \mu_\mathrm{O},\\
  \mathrm{Range}[O(t)] &= \Delta_\mathrm{O},\\
  \mathrm{Mean}[S(t)] &= \mu_\mathrm{S},\\
  \mathrm{Range}[S(t)] &= \Delta_\mathrm{S},
\end{align}
where
\begin{align}
    \mathrm{Mean}[X(t)] &= \frac{1}{800}\int_0^{800}X(t)\mathrm{d}t,\\
    \mathrm{Range}[X(t)] &= \mathrm{max}\left[X(t)\right] -
                    \mathrm{min}\left[X(t)\right],\\
    O(t) &= c\zeta_\tau[\varepsilon(t)] + \alpha_O - \alpha_O
    e^{\nicefrac{-t}{\tau}}  + O_0e^{\nicefrac{-t}{\tau}},\\
    S(t) &= d\varepsilon(t) + e\beta(t) + f\cos\left({\rho(t)}\right) +
    \alpha_S.
\end{align}
The solution to these equations produces the physical model parameters shown in
Table \ref{tab:physical_params}.
\begin{table}
  \centering
  \caption{Parameter values for the physical feedforward model.}
\label{tab:physical_params}
\begin{tabular}{c|c|c}
Parameter  & Value             & Units           \\ \hline\hline
$a$        & $2.2\times10^7$ & km$^3/^\circ$C \\ \hline
$b$        & $-2.8\times10^7$ & km$^3/^\circ$C \\ \hline
$c$        & $85$           & $^\circ$C      \\ \hline
$d$        & $70$           & $^\circ$C      \\ \hline
$e$        & $56$           & $^\circ$C      \\ \hline
$f$        & $0.69$          & $^\circ$C      \\ \hline
$\alpha_I$ & $1.4\times10^8$ & km$^3$         \\ \hline
$\alpha_O$ & $7.5$           & $^\circ$C      \\ \hline
$\alpha_S$ & $-14$          & $^\circ$C      \\ \hline
$\tau$     & $15$           & kyr
\end{tabular}
\end{table}

\begin{figure}
    \begingroup
    \thispagestyle{empty}
    \vspace*{-6cm}
  \import{../figs/pgfs/}{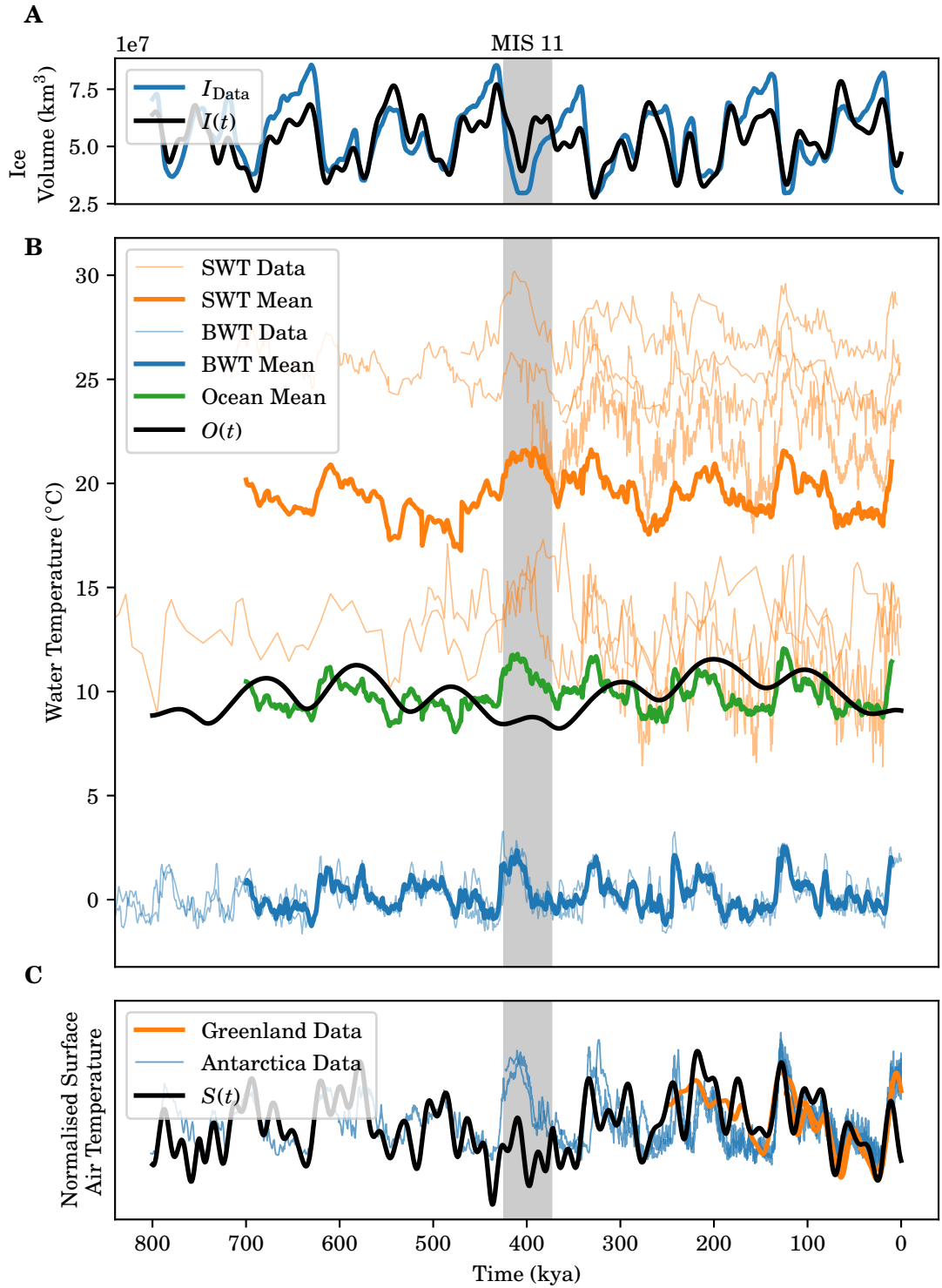}
    \vspace*{-7cm}
  \caption[Physical Model Data Comparison]{\textbf{A}: Our modelled ice volume
    $I(t)$ compared with the ice volume data $I_\mathrm{Data}$. This is the same
    solution produced by the phenomenological model. \textbf{B}: Our
    modelled bulk ocean temperature compared with the SWT and BWT data shown in
    Figure \ref{fig:sea_and_air_temp_with_map}. The mean SWT and BWT are
    averaged to give an approximate bulk ocean temperature (green) with the
    ocean solution $O(t)$ overlaid. To avoid over dependence on one data source,
    the averages are only calculated as far as 700\,kya. \textbf{C}: Our
    modelled SAT compared with data from Figure
    \ref{fig:sea_and_air_temp_with_map}. The data only reflects glacial air
    temperatures so has been normalised to align with the warmer surface solution
    $S(t)$ to show the qualitative similarity. We have also only included the
    smoothed signal for the Greenland data as it is too noisy to discern the
    large-scale dynamics. The MIS 11 interglacial period is highlighted in
  grey.}
  \label{fig:sea_air_data_sol_comp}
  \endgroup
\end{figure}

With these coefficients we can plot the proposed solutions for bulk ocean
temperature~$O(t)$ and SAT~$S(t)$, the weighted sum of which gives the same
solution for ice volume $I(t)$ shown in Figure \ref{fig:fit_ice_vol}. These
solutions are all shown together, alongside their equivalent data, in
Figure~\ref{fig:sea_air_data_sol_comp}.

First comparing the solution for $S(t)$ with with its proxy data, we see good
agreement throughout the timespan, except around MIS 11 as with the ice volume
and ocean temperature solutions. This artefact is discussed in Section
\ref{sec:comparison_with_feedforward}. The generally good accuracy of this
solution is noteworthy because we have not directly fit the $S(t)$ solution to
the SAT data. Instead, we used the ice volume data to fit these parameters, and
then linearly scaled the solution using SAT data. This suggests that surface
temperature can be well explained by the orbital parameters alone, and that it
is a significant factor in determining ice volume.

Looking now at the ocean temperature solution $O(t)$, we see that it does not
reproduce the data to the same degree, missing frequencies higher than
1/100\,kyr$^{-1}$ and failing to reproduce the steep temperature jumps seen in
the data. This is to be expected, since we have chosen to model ocean
temperature simply as a linear function of eccentricity. We found this to be the
simplest formulation of the slow variable $O(t)$ that would still allow us to
reproduce the ice volume data accurately. Although the ocean temperature
solution does not match the data exactly, it still captures the approximate
timings of the 100\,kyr interglacial periods, which is the most important aspect
when considering its impact on ice volume.

As we have discussed, the physical interpretation we present here is only one
possible explanation for the dynamics we see in the data. We have focussed on
reproducing the ice volume data with as few assumptions about the physical
mechanisms governing it. As a result, it is possible that we are only producing
the components of these mechanisms that play a role in the ice volume dynamics.
Given the simplicity of our model, it is understandable that we are unable to
accurately model multiple other physical variables, though it is also possible
that the physical variables we have proposed are incorrect. This is explored
further in Section \ref{sec:comparison_verbitsky}.

Assuming ocean temperature to be our slow variable, we can use the fit
coefficients of our physical model to convert the simplified parameters used in
\eqref{eq:simple_ocean_model} to those used in the original ocean temperature
model given by
\begin{equation}
  \frac{\mathrm{d}O(t)}{\mathrm{d}t} = 9.591\gamma\varepsilon(t) + 353.8\gamma -
    mO(t) - n,
    \label{eq:original_ocean_temp}
\end{equation}
where $\gamma$, $m$, and $n$ are unknown.

From fitting to the ice volume data, we attained values of $c = 85^\circ$C,
$\alpha_O = 7.5^\circ$C, and $\tau = 15$\,kyr. Using the substitutions given in  
\eqref{eq:ocean_param_subs}, we convert back to get
\begin{align}
    \gamma &= \frac{c}{9.591\tau} = 0.59,\\
    m &= \frac{1}{\tau} = 0.067\,\mathrm{kyr}^{-1},\\
    n &= 353.8\gamma - \frac{\alpha_O}{\tau} = 210^\circ\mathrm{C/kyr}.
\end{align}
Note that, although the constant heat loss term $n=210^\circ\mathrm{C/kyr}$
seems extreme, it is mostly cancelled out by the constant heat gain term
$353.8\gamma$, resulting in a dynamic equilibrium. With these fitted parameter
values, we can estimate the average heat loss rate of the ocean as
\begin{align}
    \mathrm{Mean}\left[mO(t)+n\right] &= m \mu_O + n,\\
    &= 0.0670\times 9.8 + 210,\\
    &= 211^\circ\mathrm{C/kyr},
\end{align}
where $\mu_O=9.8^\circ$C was estimated in Section \ref{sec:fitting}.

It is encouraging that once fitting the free parameters of the ocean model, we
attain a value of $\gamma=0.59$. This means approximately 60\% of the extra
insolation due to eccentricity is transferred as heat to the bulk ocean, which
is physically valid. If the fitting process had produced an optimal value of
$\gamma>1$, it would suggest that some form of amplification is required to
produce the observed dynamics. By attaining a value of $\gamma=0.59$, we have
therefore shown that enough energy is present in the system for eccentricity to
explain the range of ocean temperature without the need for an amplification
term. This is because, although we have not accurately reproduced the ocean
temperature data, the solution still matches the range seen in the data.

Additionally, the result of an almost constant heat loss rate lends support to
our original simple calculation in Section \ref{sec:orbital_parameters}, where
we assumed a constant heat loss rate. Aside from these assumptions, we used
measured values to estimate a temperature rise of 6.4$^\circ$C if maximum
eccentricity was sustained for 20\,kyr. This shows good agreement with the
approximate 4$^\circ$C rise expressed in the proxy data over the same period.
Although eccentricity does not remain constant, this shows it can have a direct
and significant impact on the ocean and cryosphere without the need for
amplification.

A near constant heat loss rate is physically plausible due to the omission of
the surface ocean layer when defining $O(t)$. The surface layer exchanges heat
more easily with the atmosphere due to evaporation, whilst the deeper ocean is
more insulated. It is therefore plausible that an approximately 4$^\circ$C
change in bulk ocean temperature has little impact on its heat loss rate.

\subsection{Sensitivity Analysis}
\begin{figure}
  \centering
  \input{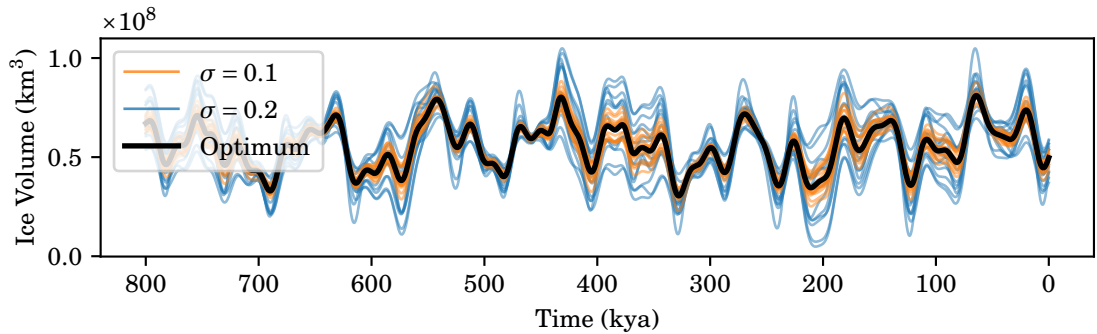}
  \caption[Feedforward Parameter Sensitivity]{Set of solutions for random
    parameter perturbations. All parameters in the physical model have been
    perturbed each iteration accept for $\alpha_I$, which was chosen as a
    function of the other parameters each time to maintain the correct offset.
    The parameters are rescaled by coefficients chosen from
    $\mathcal{N}(1,\sigma^2)$ with simulations run for $\sigma=0.1,0.2$.}
  \label{fig:param_perturb}
\end{figure}

In order to determine the overall sensitivity of the ice volume solution on the
parameters in our physical model, we perturb them randomly and show the
resultant ice volume solution. The results of this can be seen in
Figure~\ref{fig:param_perturb}. Here we have rescaled each of the parameters by
a different value drawn from the normal distribution $\mathcal{N}(1,\sigma^2)$.
The figure shows 15 iterations of this process for both $\sigma=0.1$ and
$\sigma=0.2$. In order to maintain the correct offset, the constant term
$\alpha_I$ was determined as a function of the other parameters.

In the $\sigma=0.1$ case, we see close agreement with the optimal fit in all 15
iterations. Only when the perturbations are drawn from $\mathcal{N}(1,0.2^2)$ do
we see a significant deviation from the optimal fit, though the qualitative
behaviour is still preserved. This suggests that the model is not highly
sensitive to any of the parameters, reducing the risk of overfitting.
Furthermore, it suggests that in the absence of anthropogenic forcing or
rare climatic shifts, the system is fairly predictable.

\section{Conclusion}
In this chapter, we have shown that the past 800\,kyr of global ice volume
dynamics can be approximated, with reasonable accuracy, by a linear function of
the orbital parameters. These results suggest that the data does not necessitate
a switching mechanism nor unforced oscillations, supporting the astronomical
theory.

Additionally, through estimating the annually averaged insolation reaching the
bulk ocean, we found that eccentricity is capable of producing significant
changes to the ocean and cryosphere without amplification. We propose this
important effect may have previously been overlooked due to the common use of
$Q_{65}$, which we have shown under represents the impact of eccentricity.

Another key finding from this model is the mechanism by which eccentricity's
400\,kyr period does not appear in the ice volume solution, in line with the
data. For our physical model, we propose this mechanism relates to the bulk
ocean temperature, which responds to eccentricity on a slower timescale than
SAT. By taking the difference of these two variables in our ice volume equation,
we produce a signal that resembles the change in eccentricity. This signal has
significantly less power in the 400\,kyr band which results in a significant
improvement in the fit to the ice volume data. To emphasise the significance of
this mechanism, we systematically refit the model with each subset of the terms
in the phenomenological model. We found the pair of lagged and instantaneous
eccentricity terms to be the greatest contributor to the model's accuracy,
suggesting that ice volume has more dependence on the change in eccentricity
that its absolute value.

Another interesting finding from fitting the physical model is that the
solutions for ocean temperature and surface temperature have similar ranges when
scaled by their respective coefficients in the ice volume equation. This
suggests that the roles of these two mechanisms are of similar significance in
governing the glacial-interglacial cycles.

Although we are modelling this Earth system as linear, we acknowledge that
non-linear dynamics are at play, and are more important at certain times.
Meltwater erosion and ice-albedo are two such feedbacks that could lead to the
accelerated change in ice volume expressed in the data. However, since the MPT,
there have only been eight 100\,kyr glacial-interglacial cycles. If we are to
unearth the mechanisms that govern this relatively brief period of ice volume
dynamics, there is value in imposing as few assumptions as possible. Since the
data are adequately reproduced by our model, we argue that there is
insufficient evidence to discount the astronomical theory.

\clearemptydoublepage
\chapter{Feedback Model}
\label{chap:feedback_model}

\initial{T}his chapter introduces the feedback model, which is an augmented
version of our feedforward model. This model is capable of producing unforced
oscillations, and so aligns more with the geochemical theory of climate
change. We perform an analysis of the system eigenvalues and use this to constrain
the parameters to an optimal manifold. We then perform the eigenvalue
optimisation process but with the initial conditions of the model being fit.
This allows for the phase of the unforced oscillations to be optimised. We
show that this yields a larger region of the eigenvalue space that can produce a
good fit to the data. However, the optimal eigenvalues remain the same. The model
therefore points towards a system that can resonate with the 100\,kyr period in
eccentricity, but would not produce sustained oscillations without orbital
forcing.

We then repeat the same systematic pruning process as was done for the
feedforward model, to understand the role of each parameter and the impact of
forcing on the model's dynamics. We find that the eccentricity term in the ocean
temperature equation is not crucial to the model's performance, due to the shift
from full orbital forcing to orbital resonance. This leads to how the forcing
function impacts the model's dynamics. We replace the three independently
optimised orbital parameter terms with the single $Q_{65}$ term, as is this is
commonly used in the literature. We find that the model is still able to explain
the data well. However, the dynamics of the model change from damped to
sustained oscillations.

\section{Phenomenological Model}
Here we introduce an augmentation to our feedforward phenomenological model
in which the ice volume variable appears in the equation for our slow-responding
variable. For convenience, we will refer to this as ocean temperature. This
augmentation allows for the model to produce feedback. The physical
interpretation of the feedforward model can also be applied to this feedback
model, with the only difference being that we now assume that global ice volume
has an impact on ocean temperature.

The feedback model retains the same fundamental structure as the feedforward
model, and so the design decisions described at the start of
Chapter~\ref{chap:feedforward_model} carry over. In particular, the ice volume
equation is still forced by all three orbital parameters, whilst the slow
variable equation is still driven by eccentricity alone. This is because
eccentricity is the only orbital parameter that modulates the total annual
insolation, and the high thermal inertia of the ocean filters out the seasonal
and latitudinal effects of obliquity and precession.

The key structural addition here is a single feedback term, $p_6 I(t)$, in the
ocean temperature equation, representing the influence of global ice volume on
ocean temperature. This is the minimal modification needed to allow the system
to produce unforced oscillations, moving the model from the purely astronomical
regime of the feedforward model towards a hybrid that can also accommodate
internal dynamics characteristic of geochemical models.

The inclusion of eccentricity forcing in the ocean temperature equation deserves
particular attention in the feedback model. In the feedforward model, this
forcing was essential, the pair of instantaneous and lagged eccentricity terms
was the largest contributor to the model's accuracy, as shown by the parameter
necessity analysis in Section~\ref{sec:param_necessity}. However, the feedback
model can now generate a $\sim100\,\mathrm{kyr}$ oscillation internally
through the $I$--$O$ feedback loop, and can resonate with the eccentricity
signal that enters through the ice volume equation. This reduces the model's
reliance on the lagged eccentricity mechanism to suppress the
$400\,\mathrm{kyr}$ period.

As we show in the analysis that follows, the eccentricity term in the ocean
temperature equation turns out to have a negligible impact on the optimised
feedback model. We retain it in the model formulation for generality and to
maintain structural consistency with the feedforward model, but our parameter
pruning analysis in Section~\ref{sec:leave_one_out} confirms that it can be
removed without significant loss of accuracy.

A second structural change from the feedforward model is the use of separate
time constants $\tau_I$ and $\tau_O$ for the two state variables. In the
feedforward model, fitting revealed these to be approximately equal, so a single
shared $\tau$ was used. We relax this constraint here to allow the feedback
model its full generality, since the introduction of the feedback term $p_6
I(t)$ changes the system dynamics in a way that could, in principle, require
different response rates.

As shown in Section~\ref{sec:feedback_optimisation}, fitting again yields
similar values for $\tau_I$ and $\tau_O$, and imposing $\tau_I = \tau_O$ does
not reduce the model's explanatory power. This result, together with the
underdetermined nature of the intrinsic parameters discussed later in this
chapter, indicates that it is the eigenvalues of the coupled system, rather than
the individual time constants, that are constrained by the data.

\begin{figure}
  \centering
  \includegraphics[width=0.38\textwidth]{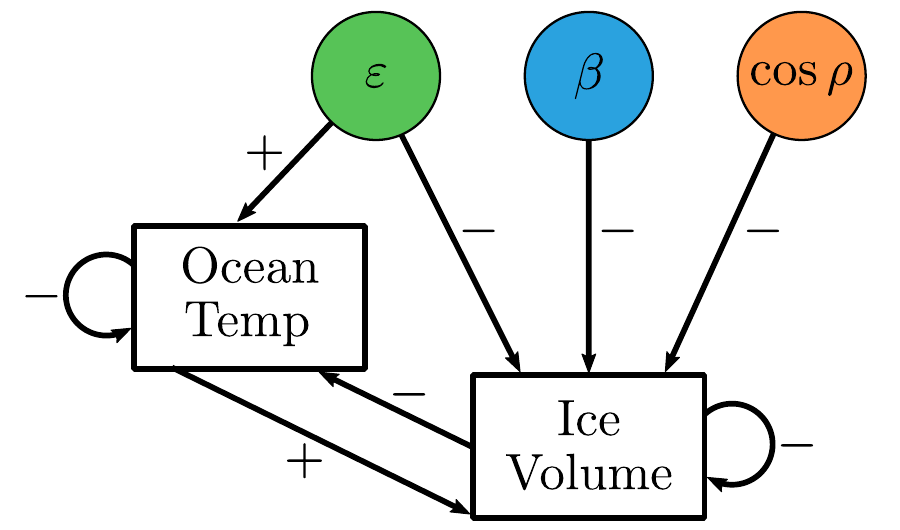}
  \caption[Feedback Model Flow Diagram]{Flow diagram showing the proposed
    dynamic behaviour of the feedback model. This system is almost identical to
    the phenomenological feedforward model shown in Figure
    \ref{fig:leave_one_out_with_diagram}B. However, it now includes a negative
    feedback mechanism from the ice volume variable to the ocean temperature
    variable.}
  \label{fig:fb_ode_diagram}
\end{figure}

The model is given by
\begin{align}
  \tau_I\frac{\mathrm{d}I}{\mathrm{d}t} &= p_1 O(t) +  p_2\varepsilon(t) +
  p_3\beta(t) + p_4\cos(\rho(t)) - I(t) + p_5, \label{eq:intrinsic_model_I}\\[4pt]
  \tau_O\frac{\mathrm{d} O}{\mathrm{d}t}
  &= \varepsilon(t) - O(t) + p_6 I(t),\label{eq:intrinsic_model_O}
\end{align}
where $\tau_I$ and $\tau_O$ are the time constants for ice volume and ocean
temperature respectively, along with parameters $p_1$\,--\,$p_6$, all of which
will be determined through fitting. This model is homologous to the feedforward
model aside from the separate time constants and the feedback term $p_6 I(t)$,
which places this model on the opposite side of the model landscape,
aligning more with the geochemical theory than the astronomical theory. This is
due to its capacity to produce unforced oscillations, which is characteristic
of geochemical models. A flow diagram of the model is shown in Figure
\ref{fig:fb_ode_diagram}, where we see it shares very similar structure to the
phenomenological feedforward model shown in Figure
\ref{fig:leave_one_out_with_diagram}B.

We will first analyse the eigenvalues of the system to understand how
they affect the underlying dynamics of the model. The eigenvalues also reveal
which parameters are interdependent, which will prove useful when optimising the
model. In the following section, we show that a globally optimum solution can be
found by sweeping over the real and imaginary parts of the model eigenvalues.
This allows us to find the optimal oscillation period and damping rate for the
model, though does not uniquely determine all of the parameters. We discuss this
issue of underdetermined parameters in the context of making a
physical interpretation of the model in Chapter \ref{chap:analysis}.

\subsection{Optimisation}
\label{sec:feedback_optimisation}
Unlike the feedforward case, the analytical solution for this model is not
straightforward. Although the system itself is linear, the solution is
non-linear in its intrinsic parameters: $p_1$, $p_6$, $\tau_I$, and $\tau_O$,
compared to the feedforward model solution, which is non-linear in just the
shared time constant $\tau$. As a result, we must solve this system numerically,
and our dynamical analysis will revolve around the eigenvalues of the system.

The eigenvalues are
\begin{equation}
  \lambda_{1,2} = - \frac{\tau_I + \tau_O}{2 \tau_I \tau_O} \pm
  \frac{\sqrt{4 p_1 p_6 \tau_I \tau_O + \tau_I^2 - 2 \tau_I
  \tau_O + \tau_O^2}}{2 \tau_I \tau_O},
  \label{eq:eigenvalues}
\end{equation}
consisting of the four intrinsic parameters. Since we are attempting to produce
an unforced oscillation, we wish to find the range of parameters that produce a
negative radicand. This leads to the condition
\begin{equation}
  p_1p_6 < -\frac{(\tau_I-\tau_O)^2}{4\tau_I\tau_O}.
  \label{eq:p1p6_opposite_signs}
\end{equation}

A second constraint is that the time constants $\tau_I$ and $\tau_O$ must be
positive. This tells us that $p_1$ and $p_6$ must be of opposite signs, and also
that the eigenvalues have a negative real part, which assures that the system is
stable.

In order to better understand the role of the unforced oscillations this model
is able to produce, we first perform a sweep over the real (Re) and imaginary
(Im) parts of $\lambda_2$, fixing them at each point and then fitting the
remaining parameters to find the best fit to the data such that
\eqref{eq:eigenvalues} is satisfied. For this sweep, we run the simulation for
an extra 800\,kyr before reaching the time range of interest. This is intended
to remove the impact of initial conditions on the solution, as was used in
fitting the feedforward model. For the feedforward model, we saw in
\eqref{eq:phenom_model_sol_subbed} that the impact of the initial conditions
decays exponentially over time. This is not necessarily the case for our
feedback model, since unforced oscillations can become sustained, the phase of
which will be determined by the initial conditions. To account for the
possibility that the model performs optimally when sustained oscillations occur,
we perform a second sweep in which initial conditions are fit each time.

The reasons for first running the sweep with a run up period instead of fitting
the initial conditions are twofold. Firstly, it is difficult to verify that the
initial conditions are accurate to the real system, or even physically
plausible. Secondly, there is a risk of overfitting when the initial conditions
are optimised. When using this approach for the feedforward model, we attained a
3\% increase in the variance explained, but the improvement was local to the
first $\sim 50$\,kyr of the solution. A drastically larger initial value for
ocean temperature allowed the ice volume solution to better match the steep drop
into an interglacial that occurs at 800\,kya. The impact of the initial
conditions exponentially decays over time, meaning it can be optimised to just
the start of the data with the rest of the solution largely unchanged. This is
the motivation to first perform a sweep with an 800\,kyr run up period, as it
allows the system to settle into its natural dynamics. However, we acknowledge
that this approach limits the capacity of the feedback model to explain the data
with unforced oscillations. We therefore perform sweeps for both cases, and
compare the results.

\begin{figure}
  \centering
  \import{../figs/pgfs/}{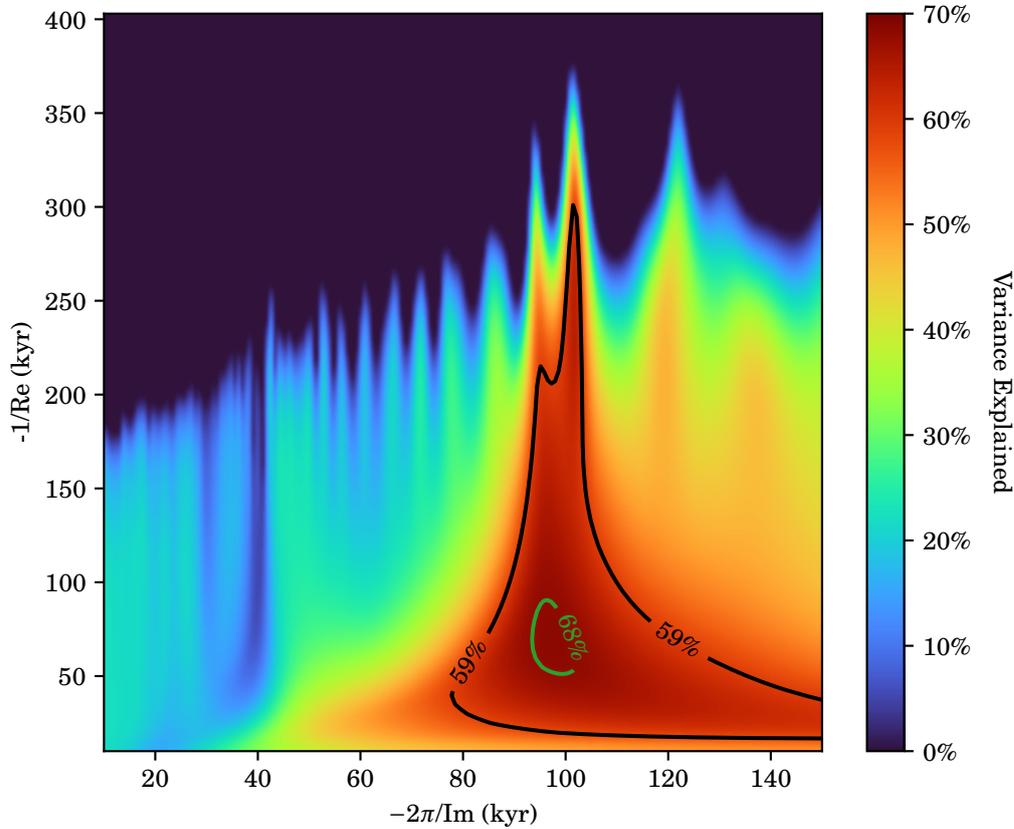}
  \caption[Real and Imaginary Eigenvalue Sweep]{Sweeping over the real and
    imaginary parts of eigenvalue $\lambda_2$ from \eqref{eq:eigenvalues}. They
    have been converted so as to be in units of kyr. The 59\% contour represents
    the maximum variance explained by the feedforward model with which unforced
  oscillations cannot occur. The 68\% contour represents the maximum variance
  explained by the feedback model.}
  \label{fig:eigenvalue_sweep}
\end{figure}

To find the local optimal solution for each eigenvalue case (which is
represented by a pixel in Figure \ref{fig:eigenvalue_sweep}), a two-dimensional
sweep of $p_1$ and $\tau_O$ is performed, using the ranges given in Table
\ref{tab:eigen_sweep_ranges}. With these two parameters determined, as well as
Re and Im for a given case, we can attain $p_6$ and $\tau_I$ by rearranging the
eigenvalue equation to get
\begin{align}
  \tau_I &= -\frac{\tau_O}{2\tau_O\mathrm{Re}+1}\label{eq:tauI_from_tauO}\\[5pt]
  p_6 &= -\frac{\tau_I\tau_O\mathrm{Im}^2}{p_1} - \frac{\tau_I}{4p_1\tau_O} +
  \frac{1}{2p_1} - \frac{\tau_O}{4p_1\tau_I}.\label{eq:p6_from_p1tauOtauI}
\end{align}
Once we have formed the intrinsic parameter constraint manifold for a given
eigenvalue case, we fit the remaining parameters. The solved system is linear in
these parameters, which relate to the external forcing terms, and are therefore
determined using a far more efficient least squares fitting method. We then have the
locally optimum parameters for each eigenvalue case, which gives a solution that
we can compare to the data and calculate the variance explained.

The linear parameters are determined using a single Gauss-Newton step. Given an
arbitrary initial guess $\mathbf{x}_0$ for the linear parameters, we compute the
Jacobian $\mathbf{J}$ of the residual $\mathbf{r} = I_{\text{model}} -
I_{\text{data}}$ using central finite difference
\begin{equation}
    J_{ij} = \frac{\partial r_i}{\partial x_j} \approx \frac{r_i(x_j + \epsilon) - r_i(x_j - \epsilon)}{2\epsilon}.
\end{equation}
The optimal linear parameters are then obtained by solving the normal equations
to get
\begin{equation}
    \mathbf{x} = \mathbf{x}_0 + (\mathbf{J}^\top \mathbf{J})^{-1} \mathbf{J}^\top \mathbf{r}.
    \label{eq:optimal_lin_params}
\end{equation}
Since the model is linear in these parameters, a single iteration is sufficient
to find the optimum for a given set of intrinsic parameters.

The eigenvalue sweep uses a resolution of 1000 points for both the real and
imaginary parts, giving 1 million cases. This ensures the optimal eigenvalues
are captured and provides a clear visualisation of the model dynamics across the
eigenvalue space. Each pixel in this sweep entails a further two dimensional
sweep of $p_1$ and $\tau_O$, using a resolution of 100 points for each
parameter. In each of these sub-cases, we solve the model a total of ten times.
Eight solutions are needed to calculate the forward and backward perturbations
of the four linear parameter initial guesses, allowing us to estimate the
Jacobian. The unperturbed parameter solution is also calculated to give the
residual. With the optimal parameters found using \eqref{eq:optimal_lin_params},
we solve the model a final time to calculate the variance explained. This gives
the total number of times we must solve the model for this sweep as $1000 \times
1000 \times 100 \times 100 \times 10 = 10^{11}$, or one hundred billion times. A
single solve of the model using the out-of-the-box \texttt{scipy} ODE solver
\texttt{solve\_ivp} takes around 0.1 seconds, meaning that solving the
model 100 billion times naively would take around 300 years. We therefore
implement several optimisations to allow for this sweep to be completed in a
reasonable time frame.

Firstly, we create a custom numerical solving method that uses a fixed time step
fourth-order Runge-Kutta method. This allows us to introduce the \texttt{numba}
library to compile the solver, which brings the solve time down from around 100
milliseconds to 20. We find that, due to the system's slow dynamics and smooth
forcing, we are able to use a time step of 1\,kyr with a root mean square error
of around $8\times 10^{-5}\%$, normalising by the range of the data. This brings
the solve time down from 20 milliseconds to 8. The next set of speed
optimisations uses the vectorisation capabilities of \texttt{numpy} to evaluate
the full set of intrinsic parameter combinations in batch. This bypasses
Python's per-element interpreter overhead by passing array operations to
pre-compiled C routines. In order for our solver and optimiser to fully benefit
from these libraries, we also create a custom vectorised least squares
fitting method that can be applied across the full set of intrinsic parameter
combinations. This estimates a Jacobian for each case in the 2-dimensional sweep
over $p_1$ and $\tau_O$. The output for a single pixel in the sweep is a
4-dimensional tensor with axes corresponding to $p_1$, $\tau_O$, state variable,
and time. Each ice volume time series from this tensor is compared to the data
to give the variance explained for each case in the $p_1$-$\tau_O$ sweep, the
best of which is then assigned to this eigenvalue case (pixel) in the final
sweep.

The total simulation time for this sweep is reduced from 300 years to around 5
years with all of these optimisations. We then parallelise the sweep across 250
nodes on a high performance computing cluster, bringing the real-life time down
to around a week.

Having completed the sweep, we now examine how the model performance varies
across the eigenvalue space. Figure \ref{fig:eigenvalue_sweep} shows the maximum
variance explained for each combination of real and imaginary parts of the
system eigenvalues. The horizontal and vertical axes correspond to the imaginary
and real parts of the eigenvalues respectively, which can be converted to
natural oscillation period and decay timescale. The steep red spike around
100\,kyr corresponds to a system with a natural period matching the dominant
glacial cycle. Good fits are achieved across decay timescales ranging from
around 10\,kyr to 300\,kyr.

The black 59\% contour represents the maximum variance explained by the
feedforward model with which unforced oscillations cannot occur, as shown in
Figure \ref{fig:unforced_response_compare}. The region within this contour is
where feedback can improve on the feedforward model, reaching a maximum of 68\%
variance explained. The global maximum for variance explained by the model
occurs at $-1/\mathrm{Re}=67.98$\,kyr and $-2\pi/\mathrm{Im}=97.68$\,kyr, which
lies within the green 68\% contour.
There is a lot to learn about the model dynamics from this sweep. The most
prominent feature is the optimal region centred around 100\,kyr on the
horizontal axis. This is an expected result as the model is designed to produce
unforced oscillations, and the most prominent period in the data is 100\,kyr. In
this region of the horizontal axis towards the bottom, we see a wide range of
oscillation periods that produce a reasonable fit to the data. This is because
the bottom of the vertical axis corresponds to heavily damped unforced
oscillations, behaving in a way dynamically similar to the feedforward model,
with the orbital parameters producing most of the oscillations. The exact period
of the heavily damped unforced oscillations are therefore less critical to the
model's performance.

As we move up the vertical axis, oscillations are less damped, meaning that
even without any external forcing, the system could oscillate for multiple
glacial-interglacial cycles. This increase in oscillation duration is
accompanied by the optimal period of these oscillations narrowing until
eventually tapering down to zero, centred around a period of 100\,kyr.
As these unforced oscillations are sustained for longer, the initial conditions
play a greater role in determining the phase, even after a run up period. As
mentioned, we will explore how the results of this sweep change when the initial
conditions are fit along with the other parameters, allowing the phase of the
unforced oscillations to be determined. As shown in Figure
\ref{fig:eigenvalue_sweep_fit_x0}, this allows a larger region in the eigenvalue
space to produce a good fit to the data. However, it is interesting to note that
the optimal eigenvalues remain the same.

\begin{figure}
  \centering
  \input{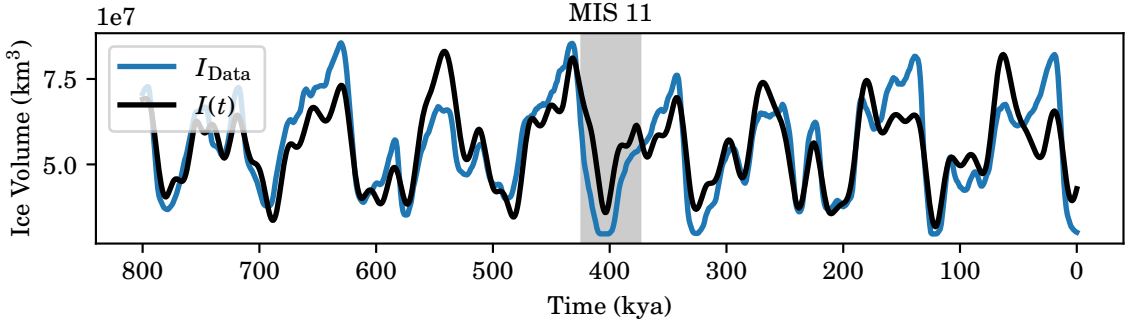}
    \caption[Feedback Ice Volume Solution]{Our modelled ice volume $I(t)$
      from \eqref{eq:intrinsic_model_I} and \eqref{eq:intrinsic_model_O},
      alongside the ice volume data $I_\mathrm{Data}$. The grey region
      delineates MIS 11, around which there is less difference between the
      two curves than the feedforward model. The model parameters that produce
    this fit are given in Table \ref{tab:optimal_params}.}
  \label{fig:fit_ice_vol_feedback}
\end{figure}

Since the optimal eigenvalues are the same in either case, we can confidently
say that the feedback model performs best when it produces damped unforced
oscillations that resonate with eccentricity at 100\,kyr, supporting a hybrid
between the geochemical and astronomical theories. We determine the optimal
parameters for the feedback model by the same sweeping approach used for Figure
\ref{fig:eigenvalue_sweep}, specifically for the case where
Re\,$=-1/67.98$\,kyr$^{-1}$ and Im\,$=-2\pi/97.68$\,kyr$^{-1}$. The optimal
parameters are given in Table \ref{tab:optimal_params}. The solution to the
model is shown in Figure \ref{fig:fit_ice_vol_feedback}. This solution explains
68\% of the variance in the data, which is an improvement on the feedforward
model, most notably around MIS 11. A full comparison of the feedback and
feedforward model is in the following chapter.
\begin{table}
\caption{Parameter values used for the optimal fit shown in Figure \ref{fig:linear_intrinsic_ice_compare}.}
\label{tab:optimal_params}
\centering
\begin{tabular}{c|c}
Parameter & Value \\ \hline \hline
$p_1$              & 0.505          \\ \hline
$p_2$              & $-25.1$        \\ \hline
$p_3$              & $-45.9$        \\ \hline
$p_4$              & $-0.718$       \\ \hline
$p_5$              & 30.1           \\ \hline
$p_6$              & $-36.3$        \\ \hline
$\tau_O$           & 71.4           \\ \hline
$\tau_I$           & 60.7          
\end{tabular}
\end{table}
\section{Analysis}

The optimal parameter values given in Table \ref{tab:optimal_params} show that
the eccentricity signal that affects ocean temperature in
\eqref{eq:intrinsic_model_O} (which has a maximum value of 0.06) is divided by
$\tau_O=71.4$\,kyr. This means that the eccentricity term in the ocean
temperature equation has a negligible impact on the system. This is confirmed by
our systematic pruning of the feedback model parameters that we perform in
Section \ref{sec:leave_one_out}. The omission of this eccentricity term is found
to have no significant impact on the model's ability to explain the data. From
this, we can conclude that the optimised feedback model is in a distinctly
different regime to the feedforward model. The parameter pruning analysis in Figure
\ref{fig:leave_one_out_with_diagram} showed that the feedforward model cannot
explain the variance in the data beyond 31\% without the eccentricity term in
the ocean temperature equation. This is in contrast to the feedback model, which
can explain 67\% of the variance without this term. This is because the feedback
model is partially using unforced oscillations to produce the 100\,kyr period in
the data, resonating with the eccentricity signal in the ice volume equation.
This therefore reduces the need for the lagged eccentricity term from the ocean
solution that reduces the 400\,kyr period in the feedforward solution.

A final observation about the optimal feedback model parameters is that the
$\tau$ values are relatively close to each other. As with the feedforward model,
we trialled refitting the model with the constraint $\tau_I=\tau_O$ and found no
change in the model's ability to explain the data. This led us to question to
what degree the intrinsic parameters are fixed by the data, or whether it is
just the eigenvalues of the system that are important. This is of interest if we
are to apply physical meaning to the feedback model parameters.

We now examine the relationship between the intrinsic parameters of the model.
To achieve this, we first perform a two-dimensional sweep over $p_1$ and $p_6$,
choosing the optimal values for the remaining $\tau_O$ and $\tau_I$ parameters,
and then switch this process to attain a sweep of $\tau_O$ and $\tau_I$. 

To attain the optimal $\tau$ values for a given $p_1$ and $p_6$, it is possible
to use our previously found optimal eigenvalues along with
\eqref{eq:tauI_from_tauO} and \eqref{eq:p6_from_p1tauOtauI} to give
\begin{equation}
  \tau_O, \tau_I = \frac{\mathrm{Re}(p_1p_6 - 1) \pm
  \sqrt{\left(\mathrm{Im}^2 + p_1 p_6 \mathrm{Re}^2\right) \left(p_1 p_6 -
  1\right)}}{\mathrm{Im}^2 + \mathrm{Re}^2},
  \label{eq:taus_from_ps}
\end{equation}
where $\tau_O$ and $\tau_I$ can be switched without affecting the model's
solution. The only constraint on $p_1$ and $p_6$ from this equation is that
\begin{equation}
  p_1p_6 < -\frac{\mathrm{Im}^2}{\mathrm{Re}^2},
  \label{eq:p1p6_condition}
\end{equation}
which guarantees that $\tau_O$ and $\tau_I$ are real.

The issue with this approach is that, as $|p_1p_6|$ increases, the optimal
$\tau$ values from \eqref{eq:taus_from_ps} diverge with one becoming very large
and the other tending to $-1/(2\mathrm{Re}+1)$. Additionally, by fixing the real
and imaginary parts of the eigenvalues, we lose all information in the region of
the sweep where \eqref{eq:p1p6_condition} is not satisfied. We therefore
constrain the four intrinsic parameters to lie within the more reasonable ranges
given in Table~\ref{tab:eigen_sweep_ranges}.

\begin{table}
  \centering
  \caption{Ranges used for the non-linear parameter sweeps shown in Figure
  \ref{fig:eigenvalue_sweep}.} \label{tab:eigen_sweep_ranges}
  \begin{tabular}{c|c}
  Parameter & Range \\ \hline\hline
  $p_1$              & 0\,--\,50           \\ \hline
  $p_6$              & $-50$\,--\,0          \\ \hline
  $\tau_O$           & 10\,--\,200         \\ \hline
  $\tau_I$           & 10\,--\,200        
  \end{tabular}
\end{table}

We know from \eqref{eq:p1p6_opposite_signs} that $p_1$ and $p_6$ must have
opposite signs. Here we have chosen to remain consistent with the signs that
emerged from the eigenvalue sweep, as well as the feedforward model, in which
the ocean temperature is modelled to positively impact ice volume, meaning $p_6$
must be negative. However, as we will show in Section
\ref{sec:both_model_physical_interpretation}, these signs can be flipped without
affecting the model's ability to explain the ice volume data, the only
difference between these options is the physical interpretation of the model and
resultant behaviour of the ocean temperature.

\begin{figure}
  \hspace{-40pt}
  \import{../figs/pgfs/}{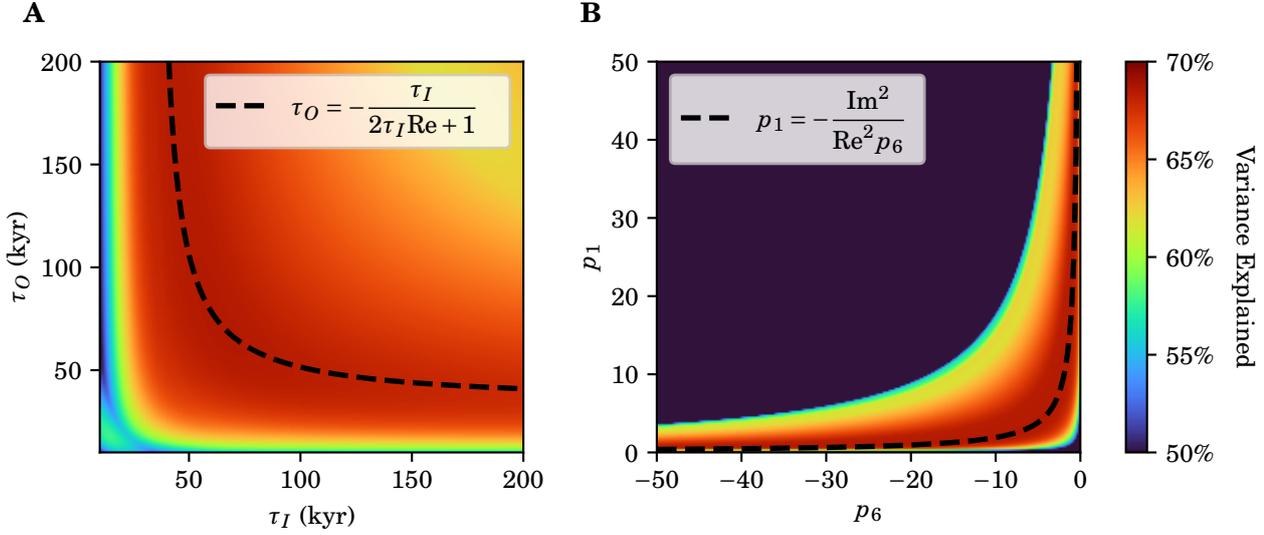}
  \caption[Intrinsic Parameter Pair Sweep]{Sweeping over pairs of intrinsic
    parameters whilst choosing the optimal values for the remaining two from a
    reasonable range. The optimal manifold in each case is shown by the dashed
    line and is calculated using the eigenvalues. The optimal relationship for
    \textbf{A} comes from \eqref{eq:tauI_from_tauO} whilst for \textbf{B} the
    optimal relationship comes from \eqref{eq:p1p6_condition}.}
  \label{fig:nonlin_sweeps}
\end{figure}

Using the parameter ranges, instead of the analytical approach given by
\eqref{eq:taus_from_ps}, we find a given value of the $p_1$--$p_6$ sweep by first
fixing $p_1$ and $p_6$ and then sweeping over the range of $\tau$ values to find
the constrained optimum. We then find the coefficients of the external forcing
terms using least squares fitting. The opposite process is then repeated for the
$\tau_O$-$\tau_I$ sweep. These two sweeps are shown in Figure
\ref{fig:nonlin_sweeps}.

This approach means that, when \eqref{eq:p1p6_condition} is satisfied, we are
not showing the best possible fit that the model can achieve, rather the
best fit that the model can achieve when the intrinsic parameters are within
reasonable bounds. This allows us to understand how the model performance
changes with the intrinsic parameters when they are constrained to a physically
plausible range.

The relationship between $\tau_I$ and $\tau_O$ is shown to follow the
curve given by \eqref{eq:tauI_from_tauO}, whilst the relationship between $p_1$
and $p_6$ follows from \eqref{eq:p1p6_condition}. The optimal relationship for
each pair occurs when we use the optimal eigenvalues Re\,$=-1/67.98$\,kyr$^{-1}$.
and Im\,$=-2\pi/97.68$\,kyr$^{-1}$. Importantly, the maximum variance explained
by the model is the same anywhere along these dashed lines, meaning that the
intrinsic parameters cannot be uniquely determined from the data, only the
eigenvalues. This also justifies the notion that the two $\tau$ values can be
set to be equal without affecting the model's performance.

The underdetermined nature of the feedback model's optimal parameters is
interesting as it is in contrast to the feedforward model, where all of the
parameters apart from $\tau$ can be globally optimised with least squares
fitting. The optimal $\tau$ value is then found using a one-dimensional sweep.
In Chapter \ref{chap:analysis}, we will explore the implications of this
difference in the context of making a physical interpretation of the model.

\subsection{Initial Conditions}

\begin{figure}
  \centering
  \import{../figs/pgfs/}{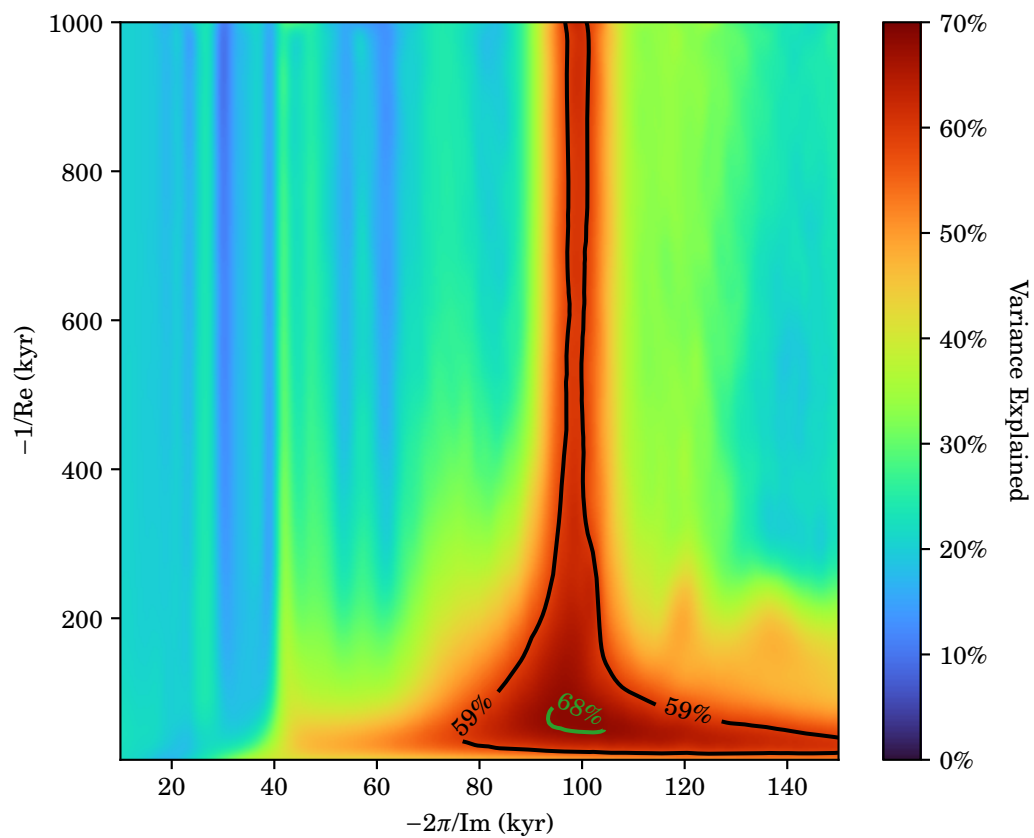}
  \caption[Real and Imaginary Eigenvalue Sweep - Fit Initial
  Conditions]{Sweeping over the real and imaginary parts of eigenvalue
    $\lambda_2$ from \eqref{eq:eigenvalues} for the model when the initial
    conditions have been fit, allowing for the phase of the unforced oscillation
    to be determined. They have been converted so as to be in units of kyr. The
    59\% contour represents the maximum variance explained by the feedforward
    model with which unforced oscillations cannot occur. The 68\% contour represents
    the maximum variance explained by the this version of the feedback model,
    the same as the original sweep.}
  \label{fig:eigenvalue_sweep_fit_x0}
\end{figure}

As mentioned, the original sweep over the eigenvalue real and imaginary parts
used a run up period to remove the impact of the initial conditions on the
solution. This was done to allow the system to settle into its natural dynamics
and to avoid overfitting. However, this approach limits the capacity of the
feedback model to explain the data with unforced oscillations. We therefore
perform a second sweep in which the initial conditions are fit along with the
other parameters, allowing the phase of potentially sustained unforced
oscillations to be optimised. The results of this sweep are shown in Figure
\ref{fig:eigenvalue_sweep_fit_x0}.

The first observation we can make from this figure is that the eigenvalue space
that can produce a good fit to the data is far larger. We have extended the range
over which the real part is swept in order to show this more clearly. In the
original sweep, where a run up period was used, the optimal region centred
around the part of the imaginary axis corresponding to a 100\,kyr period
tapered down to zero as the real part increased to around 350\,kyr. This was
because the unforced oscillations became more sustained, and so the phase of the
oscillations was determined by the arbitrarily chosen initial conditions at the
start of the run up period. In this sweep, the initial conditions are optimised
and the model is solved from 800\,kya with no run up. With the phase being
correctly set, we see similar behaviour for low real parts, but the model
remains able to explain the data up to the maximum real part. At this point, the
model is essentially producing fully sustained unforced oscillations and
therefore relies far less on the eccentricity signal in the ice volume equation to
produce the 100\,kyr period in the data.

Secondly, although it is not obvious from looking at the two sweeps in
isolation, there is also a small improvement around the rest of the imaginary
axis for low real parts. As mentioned when discussing the limitations of
fitting the initial conditions, it allows for the start of the solution to be
potentially over fit to the data. This results in a small gain in the variance
explained by the model, but only in the first $\sim 50$\,kyr of the solution.

An important result from this fit is that, although major improvements are made
for higher real parts, as well as around the rest of the imaginary axis for
lower real parts, this effect reduces to almost zero as we approach the optimal
region for the original sweep. At this point in the original sweep, we are
already in phase with the 100\,kyr period in the data since the real part is
small enough to allow the run up period to remove the impact of initial
conditions. The only improvement that fitting the initial conditions at this
point can bring is the small gains in the first $\sim 50$\,kyr of the
solution due to overfitting. However, this has less impact the more accurate the
solution already is, resulting in an improvement of just 0.3\% in the variance
explained by the model.

\subsection{Parameter Pruning}
\label{sec:leave_one_out}

In order to better understand the roles of each term in our feedback model we
perform a systematic pruning of the parameters. This is done by setting each
subset of the parameters to zero and then fitting the non-zero parameters.
This optimisation requires a choice to be made about the initial conditions of
the solution. As was done with the eigenvalue sweeps shown in Figures
\ref{fig:eigenvalue_sweep} and \ref{fig:eigenvalue_sweep_fit_x0}; we can either
use a run-up period, intended to remove the impact of the initial conditions, or
we can fit the initial conditions and avoid a run-up. Both approaches have
limitations; the run up period approach limits the capacity of the feedback
model to explain the data with sustained unforced oscillations, whilst fitting
the initial conditions can produce unphysical solutions that are not
representative of the real system.

Through experimentation, we found that optimising the initial conditions along
with the other parameters impeded the likelihood of finding the global optimum
for some subsets of parameters. This was not an issue previously, when sweeping
the eigenvalues, because the optimal parameters varied smoothly across the
eigenvalue space. Here we are changing the structure of the model each iteration
and so the optimal parameters vary significantly, complicating the optimisation.

We therefore perform this analysis using a run up period, as was done for the
first eigenvalue sweep, shown in Figure \ref{fig:eigenvalue_sweep}. This means
that the variance explained by some subsets of parameters are underestimated,
specifically those that would optimally produce fully sustained oscillations.
However, these differences are mostly small and occur for subsets with only a
few parameters. One subset performed notably better when the initial conditions
were fit, and will be discussed at the end of this section.

We have already discussed two observations about the parameters from the optimal
values given in Table \ref{tab:optimal_params}. The first was that the
eccentricity term in the ocean temperature equation is not crucial to the
model's performance, which we will more thoroughly demonstrate in this analysis.
The second was that the $\tau$ values were relatively close to each other. We
showed in Figure \ref{fig:nonlin_sweeps} that we can explain the same amount of
variance in the data with $\tau_I=\tau_O$.

We wish to investigate both the isolated role of every term in our feedback
model and their roles in the context of other terms. As we see in
\eqref{eq:intrinsic_model_I} and \eqref{eq:intrinsic_model_O}, each term is
divided by the time constants $\tau_I$ and $\tau_O$. In order to better
represent the terms that are omitted from the model, we redefine the feedback
model parameters to be divided by the time constants. This reformulated model is
given by

\begin{align}
  \frac{\mathrm{d}I}{\mathrm{d}t} &= p_1 \varepsilon(t) + p_2 \beta(t)
  + p_3 \cos(\rho(t)) + p_4 + p_5 I(t) + p_6 O(t),
  \label{eq:feedback_model_I_reformulated}\\[5pt]
  \frac{\mathrm{d}O}{\mathrm{d}t} &= p_7 I(t) + p_8 O(t) + p_9 \varepsilon(t),
  \label{eq:feedback_model_O_reformulated}
\end{align}
where $p_1$\,-\,$p_9$ are the new parameters that we will fit.

In order to further simplify the analysis of this model, we omit cases
where $p_4$ are set to zero. When the constant term $p_4$ is set to zero, the
ice volume solution cannot match the offset of the data. Cases that exclude this
term perform poorly but are not informative about the model's dynamics. Another
simplification involves omitting redundant subsets of parameters for when
$p_6=0$. When $p_6=0$, the parameters $p_7$, $p_8$, and $p_9$ have no impact on
the ice volume solution, so we only consider $p_6=0$ cases where these three
parameters are also set to zero. These two simplifications reduce the number of
cases we need to consider from 512 to 144, which is more practical for
optimisation and visualisation.
\begin{figure}
  \vspace*{-3cm}
  \hspace*{-50pt}
  \input{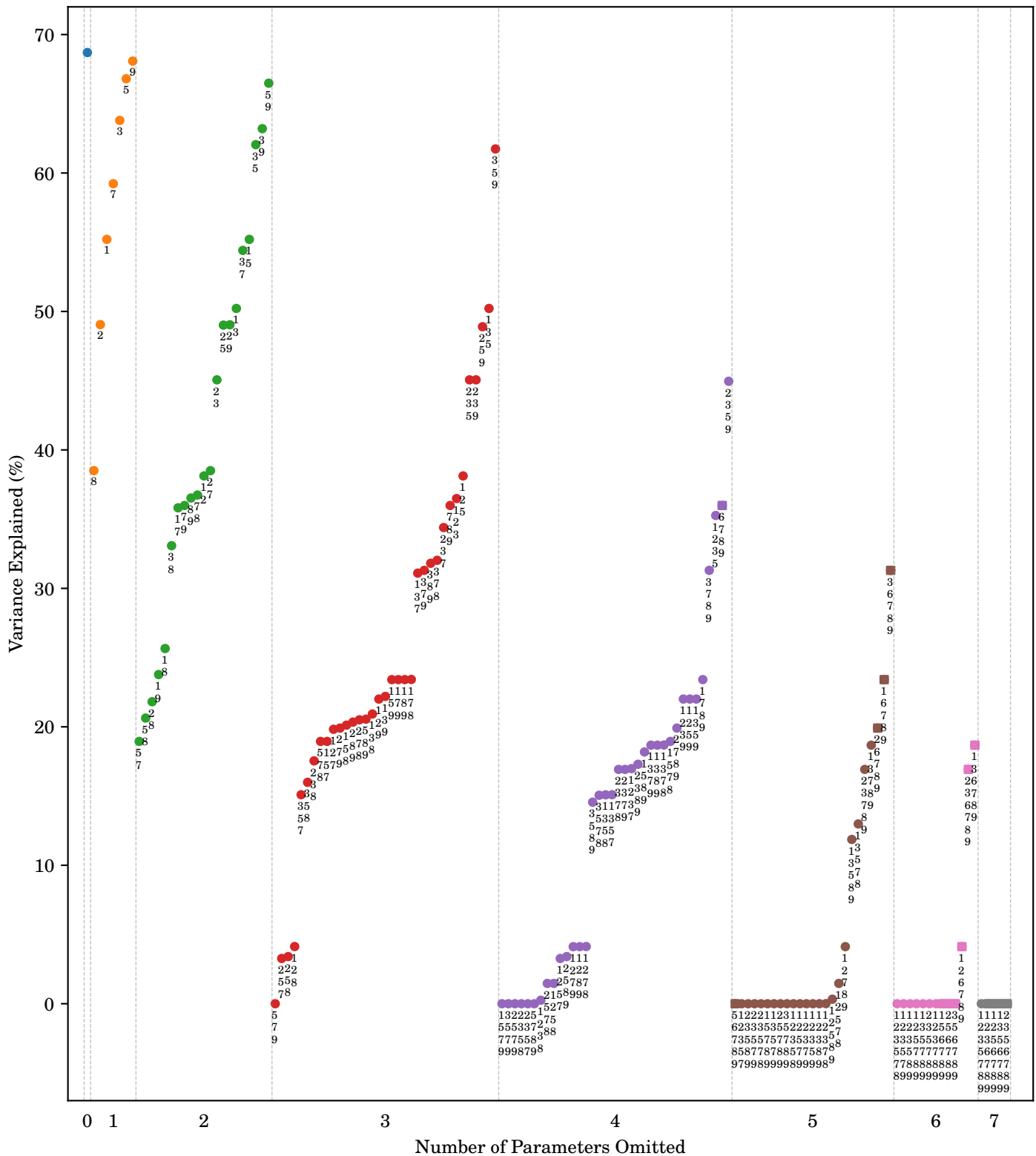}
  \caption[Feedback Leave-One-Out]{The percentage of variance explained by the
    feedback model when a given subset of parameters are set to zero. The
    numbers beneath each point represent parameters that are set to zero for
    that case. Any cases that set $p_4=0$ are omitted as they are not
    informative. Cases where $p_6=0$ are only considered when $p_7$, $p_8$, and
    $p_9$ are also set to zero, these are shown with square markers. The results
    are binned by number of parameters set to zero and ordered within that bin
    by the maximum variance the model can explain without those parameters. The
    colour of each point is to help identify which bin the case belongs to.}
    \label{fig:intrinsic_leave_one_out}
\end{figure}
\begin{figure}
  \hspace{-50pt}
  \input{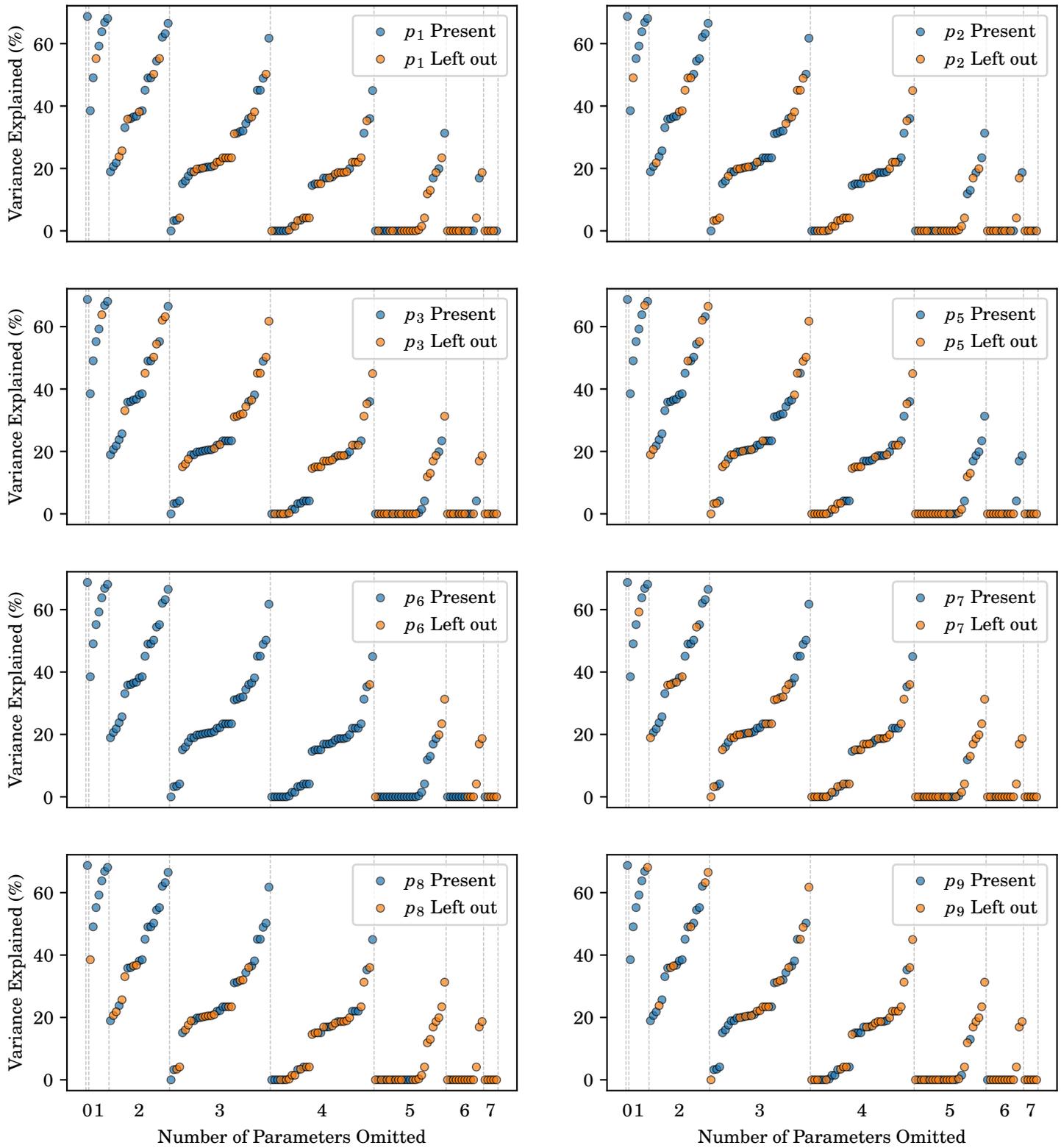}
  \caption[Feedback Leave-One-Out Individual Parameters]{Another representation
    of the data in Figure \ref{fig:intrinsic_leave_one_out}. Each sub-figure
    highlights whether the specified parameter was, or was not, present in the
    model. This allows us to isolate the effect of individual parameters across
    model complexities.}
  \label{fig:intrinsic_leave_one_out_individuals}
\end{figure}
\begin{figure}
  \centering
  \includegraphics[width=0.9\textwidth]{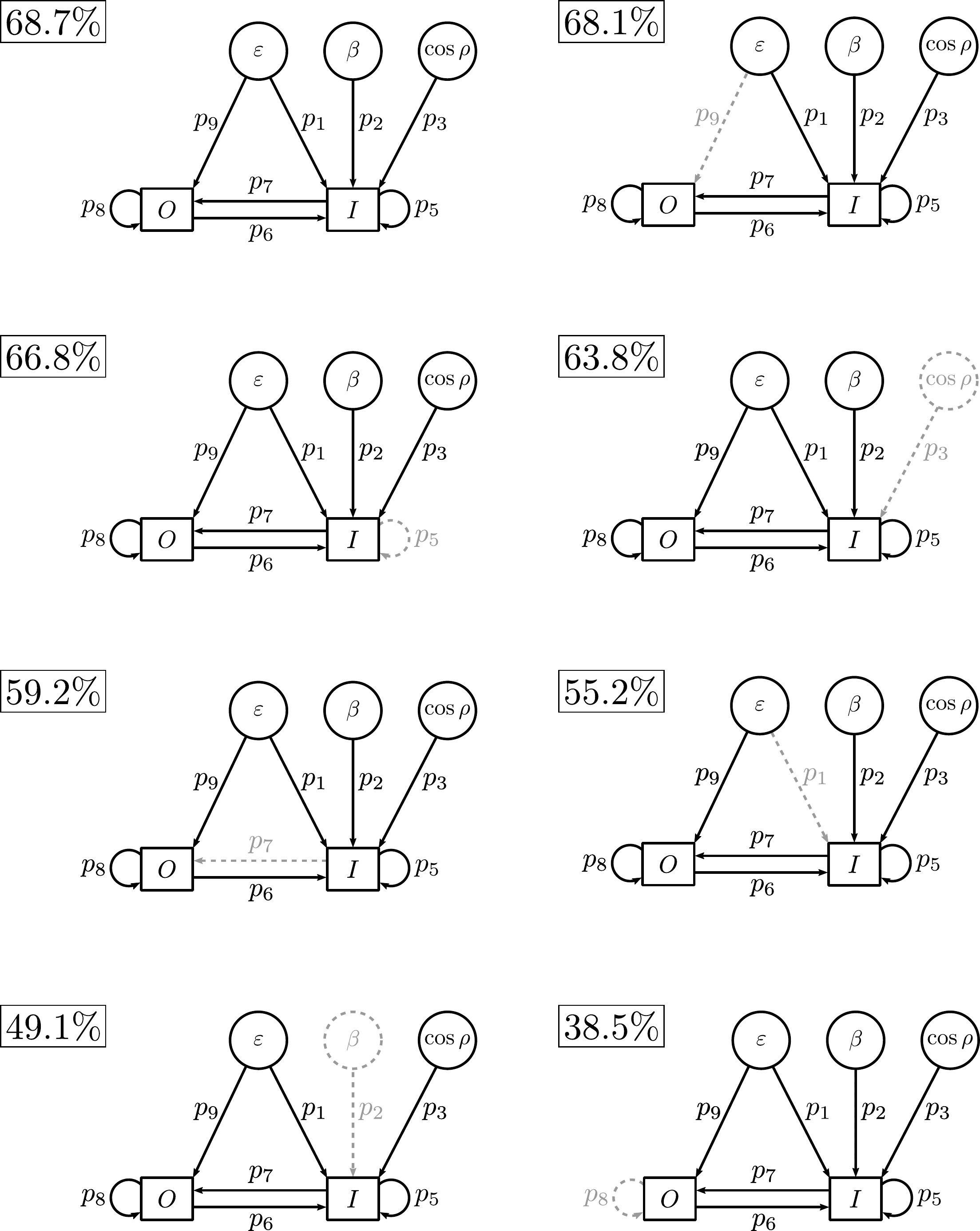}
  \caption[Feedback Leave-One-Out One Removed]{Model flow diagrams to represent
    each case of one parameter being set to zero, corresponding to the orange
    points in Figure \ref{fig:intrinsic_leave_one_out}. The removed parameter in
    each case is shown by a grey dashed line. Also included is the full model for
    reference, shown in the top left. The percentage of variance explained by
  each model is shown next to each diagram.}
  \label{fig:feedback_diagram_1_removed}
\end{figure}
\begin{figure}
  \centering
  \includegraphics[width=0.9\textwidth]{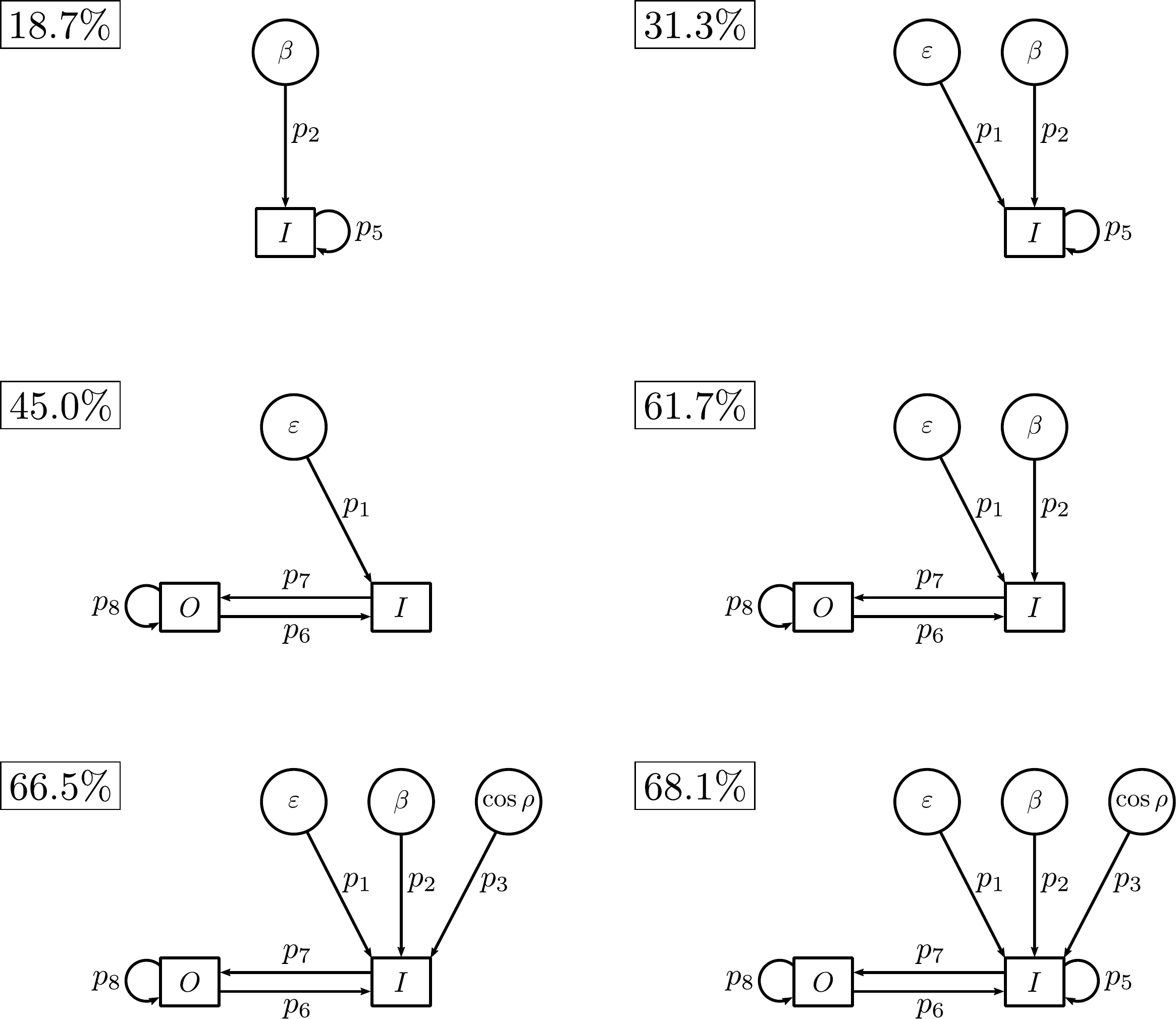}
  \caption[Feedback Leave-One-Out $n$ Removed]{Model flow diagrams to represent
    the best performing model for each bin in Figure
    \ref{fig:intrinsic_leave_one_out}. For conciseness, we have omitted the full
    model, which is shown in Figure \ref{fig:feedback_diagram_1_removed}, and
    the model with only one term as all versions explain 0\% of the variance in
    the data. The percentage of variance explained by each model is shown next
  to each diagram.}
  \label{fig:feedback_diagram_n_removed}
\end{figure}

Figure \ref{fig:intrinsic_leave_one_out} shows the percentage of variance
explained by each version of the feedback model. The annotated numbers indicate
which parameters have been set to zero in the full model given by
\eqref{eq:feedback_model_I_reformulated} and
\eqref{eq:feedback_model_O_reformulated}. The results are binned by the number
of parameters set to zero and ordered within that bin by the maximum variance
the model can explain without those parameters. The leftmost result shows the
model with no parameter set to zero, explaining the maximum variance of 68\%.
As mentioned, we have only considered cases where $p_4$ is non-zero, and cases
where $p_6$ is zero are only considered when $p_7$, $p_8$, and $p_9$ are also
set to zero, which is shown with square markers. Due to this constraint, the
cases where $p_6=0$ only emerge from the 4-parameter bin and onwards.

Figure \ref{fig:feedback_diagram_1_removed} shows the model flow diagrams for
each case where one parameter is set to zero, as well as the full model for
reference. The percentage of variance explained by each model is shown next to
each diagram. The best performing sub-model omits the eccentricity term in the
ocean temperature equation. This is to be expected as we showed from the optimal
parameter values given in Table \ref{tab:optimal_params} that this term is
significantly smaller than the other terms in the model. The next best
performing sub-model omits ice volume's self-feedback term. This is because the
coupling with ocean temperature allows for the feedback to effectively
replicated by the ocean temperature variable. The third best performing
sub-model omits the precession term in the ice volume equation. We saw this in
the feedforward model parameter pruning shown in Figure
\ref{fig:leave_one_out_with_diagram}, where the precession term was the least
important forcing term in the model. Since this frequency appears with the
lowest amplitude in the data, it is not surprising that it is the least
impactful term in the feedback model. The next best performing sub-model omits
the feedback from ice volume into ocean temperature. This is identical to the
feedforward model and so achieves the same variance explained. The remaining
sub-models do not reproduce the data well, with the worst performing sub-model
omitting the damping term in the ocean temperature equation. This effectively
reduces the model to a single variable system, achieving similar performance to
the single variable II80 model shown in Figure \ref{fig:comp_models}.

Figure \ref{fig:feedback_diagram_n_removed} shows the best performing model for
each bin in Figure \ref{fig:intrinsic_leave_one_out}. The full model is omitted
for conciseness, as well as the model with only one variable as this is not
informative. The best performing model with just two terms is shown to use just
obliquity in forcing. This is interesting as the dominant period in the data is
100\,kyr, which would suggest that eccentricity is the most important forcing
term. However, eccentricity also contains a 400\,kyr period which is not present
in the data. This is filtered out in our full models, but in the case of this
heavily pruned model, ice volume responds directly to the forcing function. This
means it is more beneficial to use obliquity, as it contains the second most
prominent period in the data and does not contain any periods that are not
present in the data. The next sub-model includes eccentricity in addition to
obliquity, which improves the fit by around 12\%, meaning that even in a direct forcing
scenario, eccentricity still plays an important role in the model. The four-term
model shows a change in regime, where obliquity is now substituted for the ocean
temperature variable, allowing for an unforced oscillation to be produced through
feedback with ice volume. This mechanism underpins the full feedback model,
where the 400\,kyr period in eccentricity is effectively filtered out because
the 100\,kyr period emerges from the feedback between the two variables
resonating with the 100\,kyr period in eccentricity. The five-term model now
reintroduces obliquity, allowing for the 41\,kyr period in the data to be
captured and so increasing the variance explained significantly. However, it
is interesting that the model benefits more from the resonance between the two
variables and the 100\,kyr period in eccentricity than it does from the 41\,kyr
period in obliquity. This shows that eccentricity on its own is not the most
important feature of the model (as we see this to be obliquity), however, its
interaction with the two physical variables is more important than obliquity's
interaction with them. The remaining two sub-models have been covered already
and provide less significant improvements to the model. This suggests that an
even more parsimonious representation of the physical system could be used,
however, the model would lose the ability to reproduce the higher frequency
components in the data.

In order to more clearly present the role of each term on the model, Figure
\ref{fig:intrinsic_leave_one_out_individuals} focusses on single parameters in
Figure \ref{fig:intrinsic_leave_one_out}. Each subplot isolates a single
parameter, showing if it is either present or absent in the model. We can see
that cases where $p_9$ is left out explains the highest amount of variance
regardless of how many terms are left out. This supports the notion that the
eccentricity term in the ocean equation is not crucial to the performance of the
feedback model. We can also see that the cases in which $p_8$ is set to zero
generally explain a low amount of the variance, showing that the damping term
in the ocean temperature equation is crucial to the model's performance for all
complexity. In contrast, the damping term in the ice volume equation ($p_5$)
appears less crucial across model complexity. As discussed, this is because the
ocean variable is able to replicate the effect of the ice volume damping term.
We can confirm this with Figure \ref{fig:intrinsic_leave_one_out}, looking at
the bin with 2 parameters omitted. Despite $p_5$ having one of the smallest
impacts on performance when removed on its own, the two worst pairs of
parameters to omit are $(p_5,p_7)$ and $(p_5,p_8)$. This is because $p_7$ allows
ice volume to feed into ocean temperature, and $p_8$ is the damping term in the
ocean temperature equation. Without both of these, ocean temperature cannot
produce a damping effect on ice volume.

Finally, we introduce an interesting case in which fitting the initial
conditions for a subset allows for a significant improvement in the model's
performance compared with the run-up approach. The subset in question defines a
model given by
\begin{align}
  \frac{\mathrm{d}I}{\mathrm{d}t} &= p_2 \beta(t) + p_4 + p_6 O(t),\\
  \frac{\mathrm{d}O}{\mathrm{d}t} &= p_7 I(t).
  \label{eq:reduced_feedback}
\end{align}
\begin{figure}
  \centering
  \includegraphics[width=0.25\textwidth]{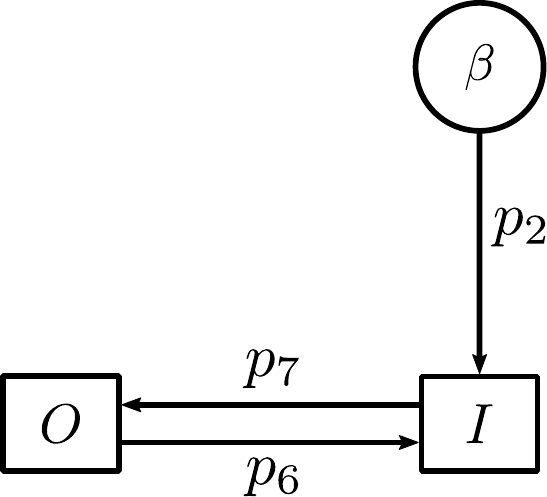}
  \caption[Minimal Feedback Model Case]{Model flow diagram for the model given
    by \eqref{eq:reduced_feedback}. This model performs abnormally well when the
    initial conditions are fit, compared to the run-up approach. The percentage
  of variance explained by this model is 56\%.}
  \label{fig:feedback_diagram_interesting_case}
\end{figure}
\begin{figure}
  \centering
  \input{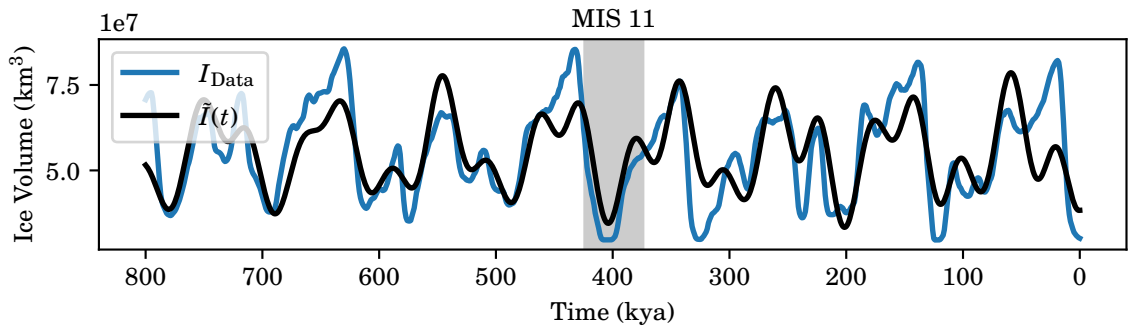}
  \caption[Pruned Feedback Ice Volume Solution]{The ice volume solution for the
    feedback model when the parameters and initial conditions are optimised
    using the inverse of the variance explained as the loss function. The model
    explains 56\% of the variance in the data but is centred around zero. Here
  we have shown the translated solution to better compare it to the data.}
  \label{fig:pruned_ice_vol_feedback}
\end{figure}
It is interesting to note that both of these equations lack a damping term,
relying on the feedback from the other variable to maintain stable oscillations.
We also see that eccentricity is not included in this subset, as the 100\,kyr
period is produced entirely through unforced oscillations. This model explains
56\% of the variance in the data, which is a significant improvement on the
run-up approach which explains only 12\%. This solution is shown in Figure
\ref{fig:pruned_ice_vol_feedback}, where we see a reasonably good fit to the
data. Although at a first glance this could provide strong support for the idea
of Earth's glacial-interglacial cycles being driven by sustained unforced
oscillations intrinsic to Earth, there are some caveats to this result. 

The first issue is that the solution is incapable of producing the correct
offset to match the data, for Figure \ref{fig:pruned_ice_vol_feedback} we have
translated the solution to better compare it to the data. This means an extra
offset in the ocean temperature equation would be needed in order to
produce a correctly offset solution, though it still lacks the higher frequency
oscillations that precession brings to the two original models. If we also added
precession to this reduced model we would achieve 59\% variance explained,
matching that of the feedforward model, though at that point we would also
be requiring the same number of parameters. The differences between these two
models is twofold. We are replacing the eccentricity term in the feedforward ice
volume equation with the feedback term $I(t)$ in the reduced feedback ocean
temperature equation. We are also replacing the eccentricity term in the
feedforward ocean temperature equation with a constant offset in the reduced
feedback ocean temperature equation. 
\begin{figure}
  \centering
  \input{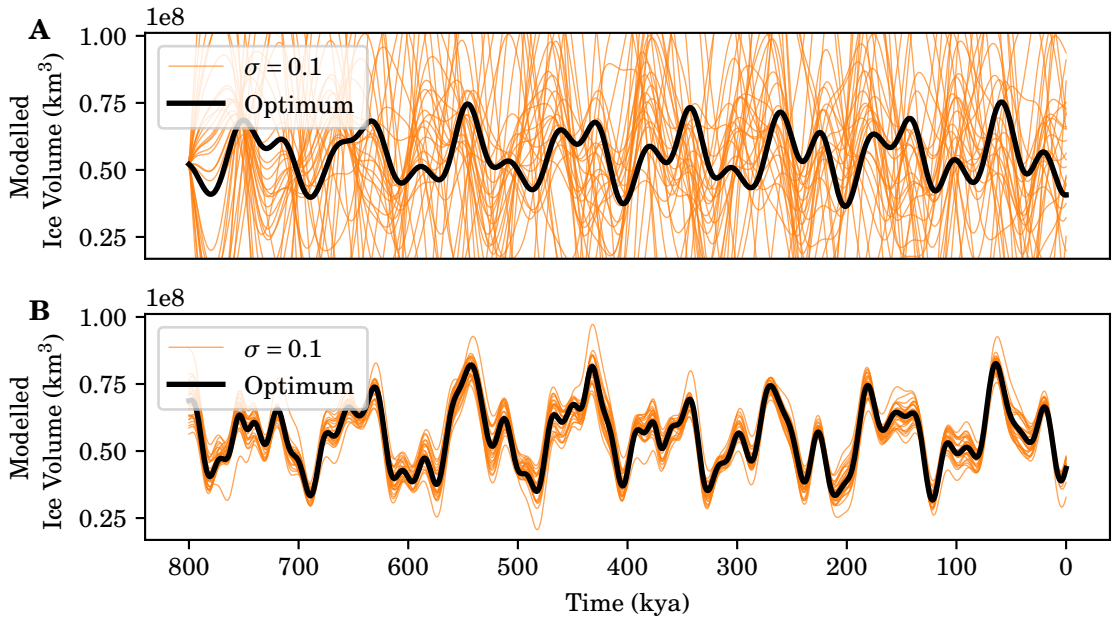}
  \caption[Feedback Parameter Perturbation Comparison]{Comparing the set of
    solutions for random parameter perturbations for \textbf{A}: the reduced
    feedback model (with fit initial conditions) given by
    \eqref{eq:reduced_feedback} and \textbf{B}: the full feedback model. All
    parameters in each model have been perturbed each iteration. The parameters
    are rescaled by coefficients chosen from $\mathcal{N}(1,\sigma^2)$ where
  $\sigma = 0.1$.}
  \label{fig:feedback_param_perturb_comparison}
\end{figure}

With the aforementioned changes to the reduced feedback model, it now has the
same number of parameters, and the same variance explained, suggesting equal
support for the two opposing theories. However, the second issue with this
reduced model, which relies heavily on unforced oscillations, is that it
requires finely tuned parameter values and initial conditions to produce the
optimal solution we see in Figure \ref{fig:pruned_ice_vol_feedback}. In the
previous chapter, Figure \ref{fig:param_perturb} showed the feedforward
model to be highly robust to parameter perturbations. This means that the
physical interpretation of the parameters has a reasonable amount of flexibility
to match measured values in the real system. For example, the optimal value of
15\,kyr for the ocean temperature time constant is greater than most estimates
of the deep ocean mixing time, with upper estimates of around 10\,kyr
\citep{crucifix_original}. However, we find that the feedforward model is able
to perform similarly well (58\% variance explained) with an ocean time constant
of 10\,kyr. This is not the case for the reduced feedback model, as is shown in
Figure \ref{fig:feedback_param_perturb_comparison}.

In this figure we show the optimal solution for the reduced feedback model and
the full feedback model. Overlaid is 30 solutions that have used parameters
that were randomly scaled by a factor of $\mathcal{N}(1,\sigma^2)$, where
$\sigma=0.1$, meaning each parameter is individually perturbed by around
$\pm10\%$ on average. We can see that the full feedback model, like the feedforward
model is able to maintain a high level of performance despite the perturbations.
This is due to its hybrid dependence on both external orbital forcing, and
internal unforced oscillations, performing optimally when it models an Earth
system that resonates with the orbital forcing. The reduced feedback model,
however, represents an Earth system that is heavily driven by unforced
oscillations, and as a result is highly sensitive to the parameter values
describing this system. Although this does not necessarily mean that the reduced
feedback model is incorrect, it does suggest that we have instead mimicked the
experiment done by Crucifix, outlined in Section \ref{sec:crucifix_2011}, where
an oscillator that produces sustained unforced oscillations is carefully tuned
to match the large amplitude oscillations in the data, with orbital forcing from
obliquity and precession (in the form of $Q_{65}$) being used to produce the
higher frequency oscillations. This was used as a criticism of such models,
arguing that they may lack a physical basis.

\subsection{Impact of Forcing}
From the analysis so far, it is clear that the choice of orbital parameters
makes a significant difference to the performance of the feedback model, so we
now investigate the impact of using the $Q_{65}$ orbital forcing function
instead of the independently optimised orbital parameters.

Up until this point, all of the parameter fitting for our models have allowed
for independently optimised orbital terms, leading to a forcing signal that best
fits the ice volume data. This is possibly not representative of the real system
as it allows for the orbital parameters to be disproportionately weighted in the
model. We therefore investigate the impact of using the $Q_{65}$ signal as our
forcing function instead of the sum of optimised orbital parameter signals.

Although this is a commonly used forcing function in the literature, it is
important to note that there is no one perfect representation of the orbital
forcing. As discussed in Section \ref{sec:Q65}, the $Q_{65}$ signal describes
the insolation at 65$^\circ$N, averaged over the summer solstice. Signals of
this nature, though sometimes with different latitudes or seasons, are used in
many models of the climate system. However, by focussing on just a single day of
the year and a single latitude, we are missing out on the potential effect of
the full insolation profile over all latitudes and times of year. This has been
our motivation for using the sum of optimised orbital parameters as our forcing
function so far, allowing the data to determine the relative importance of each
orbital parameter.

We will start once again with an eigenvalue analysis of the model when $Q_{65}$
is used as the forcing function. This will allow us to understand the landscape
of optimality for this model, and compare it to the equivalent eigenvalue sweeps
when the orbital parameters are independently optimised. Since the eigenvalues
of the feedback model only depend on the intrinsic parameters, and not the
external forcing, the same eigenvalue real and imaginary parts are used in this
sweep. However, the optimal parameters will differ to account for this new forcing
term. In order to be sure we are not missing any potential optima, we will
fit the initial conditions for this sweep, allowing the phase of the unforced
oscillations to be optimised.

\begin{figure}
  \centering
  \import{../figs/pgfs/}{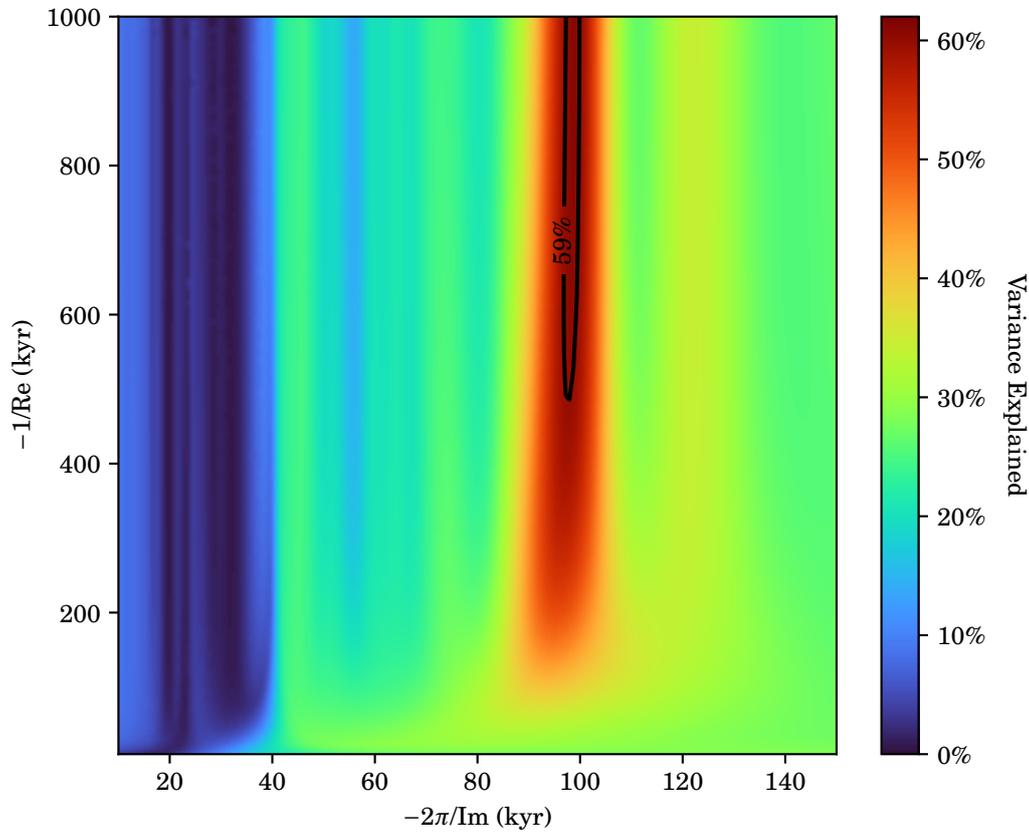}
  \caption[Real and Imaginary $Q_{65}$ Eigenvalue Sweep]{Sweeping over the real
    and imaginary parts of eigenvalue $\lambda_2$ from \eqref{eq:eigenvalues}
    for the model when $Q_{65}$ is used as the orbital forcing function.
    They have been converted so as to be in units of kyr. The 59\% contour
    represents the maximum variance explained by the feedforward model with which
  unforced oscillations cannot occur.}
  \label{fig:eigenvalue_sweep_q65}
\end{figure}
The results of this sweep are shown in Figure \ref{fig:eigenvalue_sweep_q65}.
We can see that the variance explained by the model is lower for this case, with
a maximum of 61\% compared to 68\% when the orbital parameters are independently
optimised. This is expected as the model is now forced by a single signal and so
we have lost two degrees of freedom in the model. The optimal eigenvalues are
Re\,$=-1/985.1$\,kyr$^{-1}$ and Im\,$=-2\pi/98.64$\,kyr$^{-1}$. The imaginary
part of the eigenvalue is very close to the optimal value when the orbital
parameters are independently optimised. This makes sense given the dominant
100\,kyr period in the data. However, the real part is much larger, meaning that
the model is optimal when it produces unforced oscillations that are almost
fully sustained. This is in contrast to the optimised forcing function model, in
which the model is optimal when it produces damped unforced oscillations that
resonate with the eccentricity signal in the ice volume equation.

\begin{figure}
  \centering
  \input{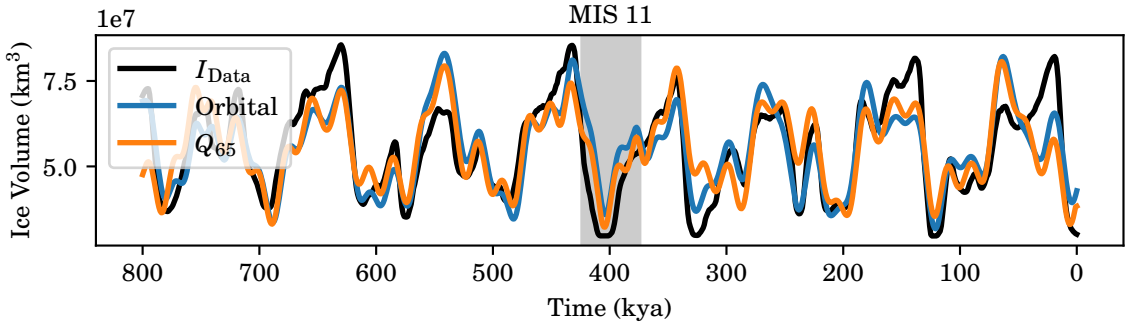}
  \caption[Comparison of Optimal Orbital Forcing and $Q_{65}$]{Comparison of the
    optimal ice volume solution when the optimal orbital parameters are used as
    the forcing function and when $Q_{65}$ is used. The variance explained by
    the models are 68\% when the forcing function comprises independently
    optimised orbital parameters and 61\% when $Q_{65}$ is used as the forcing
  function.}
  \label{fig:comp_q65_orb_sol}
\end{figure}

Despite this significant difference in underlying dynamics, the solution to the
model is very similar when $Q_{65}$ is used as the forcing function. This is
shown in Figure \ref{fig:comp_q65_orb_sol} with the ice volume solutions
appearing qualitatively similar. This is an interesting result as our optimised
model now lies in the third distinct regime for how it describes the data,
whilst still describing the data to a similar level of accuracy. The feedforward
model produces no unforced oscillations, the feedback model with optimised
orbital parameters produces damped unforced oscillations, and the feedback model
with $Q_{65}$ forcing produces sustained unforced oscillations.

\begin{figure}
  \centering
  \input{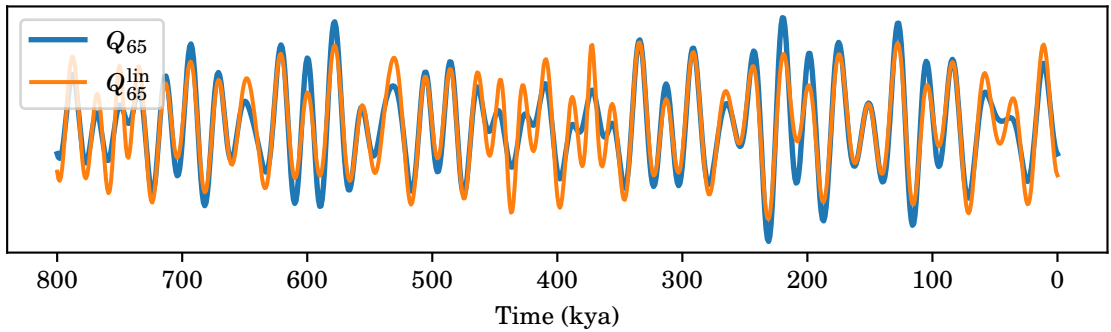}
  \caption[Linear Approximation of $Q_{65}$]{Qualitative comparison of the
    original $Q_{65}$ forcing function and the linear approximation of $Q_{65}$
    using the orbital parameters. Since these two signals have different means,
    which is accounted for by the constant term in the ice volume equation, we
    have subtracted the mean from each signal.}
  \label{fig:comp_q65_linear_approx}
\end{figure}

We will now visualise how the model's parameters and performance change as we
transition from one regime to the other. To understand how the relative
weighting of the orbital parameters changes as we transition, we will represent
the $Q_{65}$ signal with a linear approximation using the orbital parameters.
This linear approximation is compared with the original $Q_{65}$ signal in
Figure \ref{fig:comp_q65_linear_approx}. Although these signals are not
identical, their differences are relatively small compared to the forcing
function produced when the orbital parameters are independently optimised,
allowing us to treat them as equivalent for this analysis.

\begin{figure}
  \centering
  \input{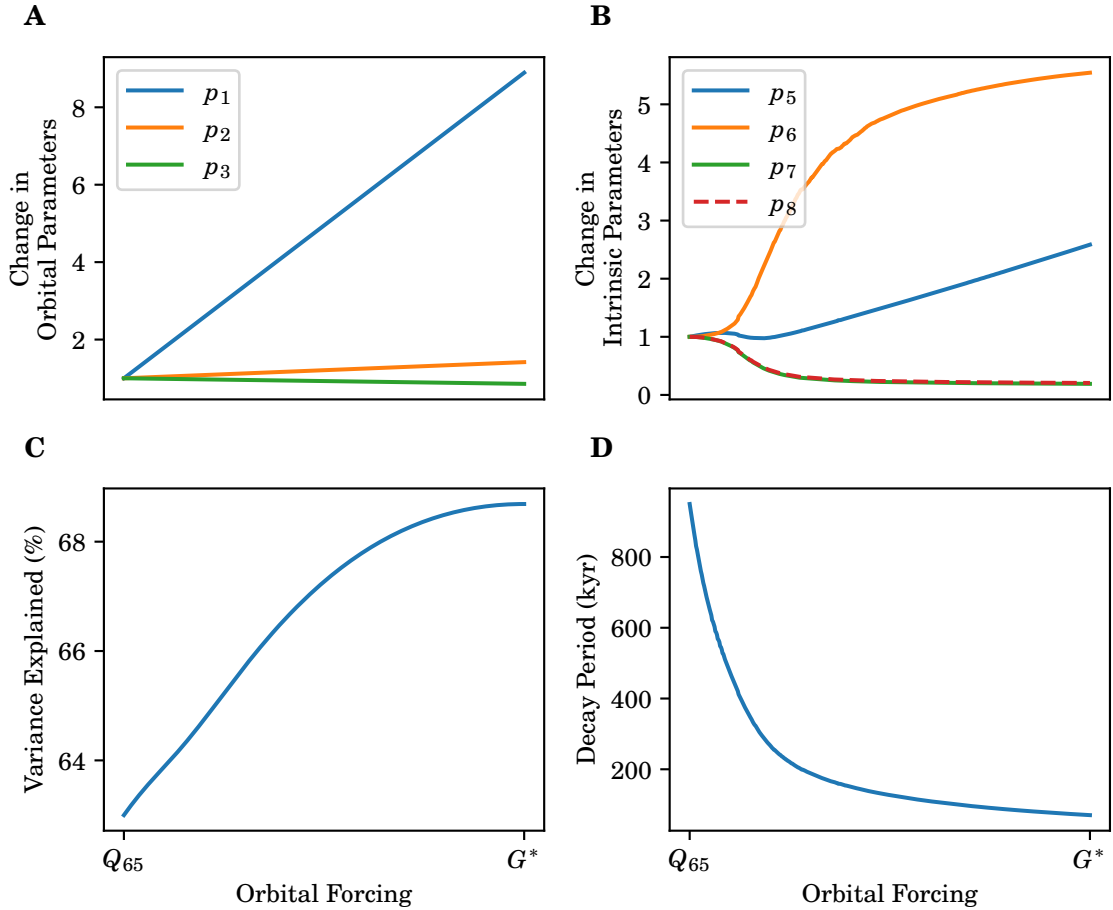}
  \caption[Sweep $Q_{65}$ to Orbital Forcing]{Sweeping over the orbital forcing,
    starting with a linear approximation of $Q_{65}$ and then optimising the
    orbital parameters to get the optimal forcing function $G^*$. The
    multiplicative change in orbital and intrinsic parameters, relative to their
    values in the $Q_{65}$ case are shown in (\textbf{A}) and (\textbf{B})
    respectively. The variance explained by the model throughout the sweep is
    shown in (\textbf{C}). The decay period of the unforced oscillation, which
    maintains an approximately constant period around 98\,kyr, is shown in
  (\textbf{D}).}
  \label{fig:sweep_q65_to_optim_orb}
\end{figure}

We can now linearly interpolate between the weights used in the linear approximation
of $Q_{65}$ and the optimal orbital parameters, optimising the remaining
parameters and initial conditions at each step. The model we are using for this
sweep is given by
\begin{align}
  \frac{\mathrm{d}I}{\mathrm{d}t} &= G(t) + p_4 + p_5 I(t) + p_6 O(t),\\[5pt]
  \frac{\mathrm{d}O}{\mathrm{d}t} &= p_7 I(t) + p_8 O(t) + p_9 \varepsilon(t),
\end{align}
where $G(t) = p_1 \varepsilon(t) + p_2 \beta(t) + p_3 \cos(\rho(t))$ changes by
linearly varying parameters $p_1, p_2$ and $p_3$ from the linear approximation
of $Q_{65}$ to the optimal orbital parameters ($G^*$). The results of this sweep
are shown in Figure \ref{fig:sweep_q65_to_optim_orb}.

There are a number of conclusions to take from this sweep. Firstly, we can see in
\textbf{A} that the weighting of obliquity ($p_2$) and precession ($p_3$) are
shown to remain relatively constant throughout the sweep. This suggests that the
impact these parameters have on the system is well captured by the $Q_{65}$
signal. On the other hand, the weighting of eccentricity ($p_1$) in the optimal
orbital forcing signal is eight times greater than in the linear approximation
of $Q_{65}$. This could suggest that the optimal orbital parameter model is
overstating the impact of eccentricity on the system, given that most models of
this nature use $Q_{65}$ as the forcing function. However, we have demonstrated
in Section \ref{sec:ocean_model} that eccentricity may have a greater impact on
the Earth system than $Q_{65}$ would suggest.

In \textbf{B}, we can see that the intrinsic parameters of the model change
significantly to accommodate the change in forcing. These changes produce the
change in regime from damped to sustained unforced oscillations. We see in
\textbf{C} that this transition results in an improvement in the variance
explained by the model. However, this does not qualitatively present as a
significant change to the solution, as shown in Figure
\ref{fig:comp_q65_orb_sol}. It is worth noting that the linear approximation of
$Q_{65}$ achieves around 63\% variance explained as opposed to the 61\% when
the original $Q_{65}$ signal is used, though once again, these differences are
not significant.

\begin{figure}
  \centering
  \input{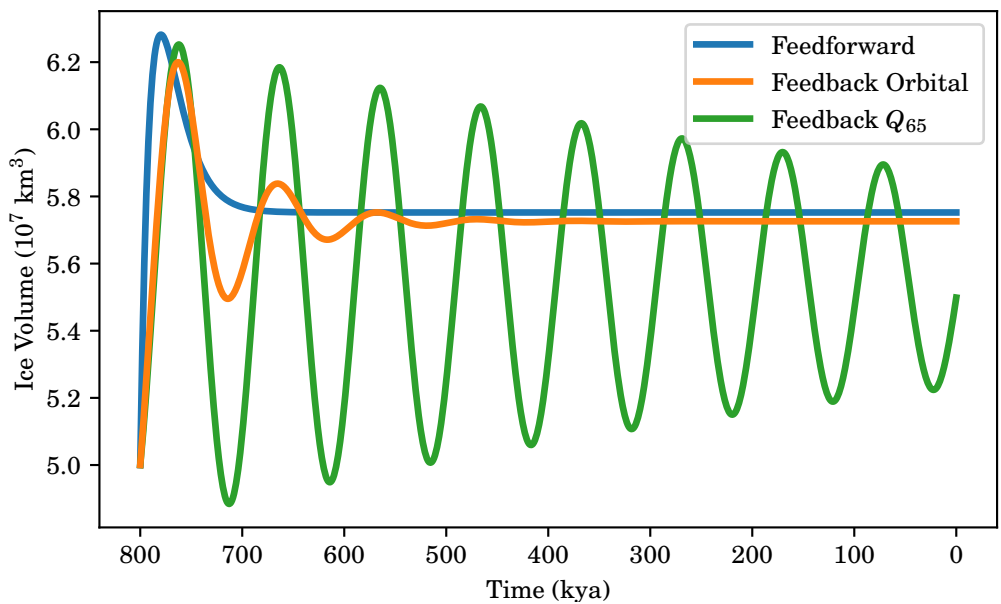}
  \caption[Unforced Response Comparison]{Comparison of the unforced responses of
    the feedforward and feedback models. The feedforward model cannot oscillate
    without external forcing, whereas the feedback model produces an unforced
    oscillation with a period of 100\,kyr. The behaviour of the feedback model
    is shown for when optimised orbital parameters are used as forcing (orange)
    or $Q_{65}$ (green).}
  \label{fig:unforced_response_compare}
\end{figure}

In \textbf{D}, we see a clear demonstration of the optimal decay period changing
as the forcing function changes. The decay period of the unforced oscillation
starts at around 950\,kyr when $Q_{65}$ is used, producing almost fully
sustained oscillations over the 800\,kyr solution period. This drops to
approximately 70\,kyr when the optimal orbital parameters are used. The time
series representation of this decay period is shown in Figure
\ref{fig:unforced_response_compare}. Here we see the unforced responses of the
feedforward and feedback models, with the feedback model using either $Q_{65}$
or optimised orbital parameters. By design, the feedforward model does not
oscillate, whilst the feedback model oscillates with a period of 100\,kyr,
decaying or sustaining depending on the forcing function used.

It is interesting to note that at the midpoint between the two forcing
functions, the variance explained and decay period are both close to their
values at the optimal orbital parameters end of the sweep. We have shown
eccentricity to have a greater impact on the Earth system than $Q_{65}$ would
suggest, however, it may not be eight times greater. This figure shows us that,
as eccentricity becomes even a few times greater than $Q_{65}$ suggests, the
Earth system rapidly appears to be better explained by damped unforced
oscillations.

This is not to say that $Q_{65}$ is an inappropriate measure of the Earth's
insolation forcing, but we see that its use can lead to markedly different
behaviour in this model. This is an important finding as $Q_{65}$ is used for
many models of this nature, as shown in Section \ref{sec:conceptual_models}. Of
the six external models that we have analysed, four use $Q_{65}$ to force the
dynamics, of those four, three are aligned with the geochemical theory. The use
of $Q_{65}$ when modelling ice volume dynamics prescribes a smaller impact of
eccentricity on the Earth system, which often entails a rejection of the
astronomical theory.

\section{Conclusion}
In this chapter we have introduced the feedback model, along with an augmented
version that uses $Q_{65}$ as the forcing function, and analysed the model
dynamics. Through eigenvalue analysis, we concluded that the globally optimised
feedback model produces damped unforced oscillations that resonate with the
eccentricity signal in the ice volume equation. We first showed this to be true
when a long run up period was used, removing the impact of initial conditions on
the solution. We then showed that this remains true when initial conditions are
fit, allowing the phase of the unforced oscillations to be optimised. While
fitting initial conditions allowed for a larger region of eigenvalue space to
produce good fits to the data, the improvement at the optimal point was minimal
(0.3\%). This ensures that the globally optimal model does not rely on over fit
initial conditions.

The feedback model contains the feedforward model as a subset of its terms. It
is therefore unsurprising that the feedback model can explain more variance in
the data (68\%) than the feedforward model (59\%). However, as we show in the
following chapter, the accuracy of the two model solutions are qualitatively
similar. It is important to note that while the variance explained measure is
effective for ranking models by performance, it does not necessarily provide a
meaningful indication of the model's absolute accuracy.

Through optimising the feedback model, as well as performing a systematic
pruning of the parameters, we made several key observations about the role each
parameter plays. We showed that the eccentricity term in the ocean temperature
equation is not crucial to the model's performance. This is in contrast to the
feedforward model, which requires eccentricity to appear in both equations to
produce a 100\,kyr period whilst reducing the 400\,kyr period. Our pruning
analysis revealed that in heavily simplified versions of the model, obliquity
alone produced better results than eccentricity alone, due to the absence of the
problematic 400\,kyr period. We also observed a significant regime change
in the four-term model, where obliquity was substituted for ocean temperature to
enable unforced oscillations, highlighting the model's preference for resonance
mechanisms when sufficient parameters are available.

We also learnt that, as with the feedforward model, the feedback model can
perform equally well with $\tau_I=\tau_O$. Expanding on this, we showed that the
four intrinsic parameters of the model are underdetermined by the data, with the
eigenvalues of the system being the only uniquely determined quantities. In the
following chapter, we explore how having underdetermined parameters impacts
making a physical interpretation of the model.

Through our analysis of reduced versions of the feedback model, we identified an
interesting subset that can explain the data without damping terms or
eccentricity forcing, relying instead on unforced oscillations and fitted
initial conditions. The simplest of this subset uses only obliquity forcing and
feedback between the variables but explained 56\% of the variance in the data.
However, it proved highly sensitive to parameter perturbations, requiring
precise fine-tuning to match the data. This is in contrast to the full feedback
model, which maintains robust performance when its parameters are perturbed, due
to its hybrid dependence on both orbital forcing and internal oscillations. The
significant difference in parameter sensitivity between these versions suggests
that, while unforced oscillations alone can reproduce much of the
glacial-interglacial pattern, a more robust explanation emerges when combining
orbital resonance with internal feedback mechanisms.

Finally, we showed that the choice of orbital forcing can have a significant
impact on the optimised model's dynamics. When $Q_{65}$ is used as the forcing
function, we see a regime change from damped to sustained unforced oscillations.
This comes with the added constraint that the initial conditions of the solution
must be fit with the rest of the parameters. This is an important finding as
$Q_{65}$ is used in many models that describe the Earth system as producing
sustained unforced oscillations. We showed that the main difference between the
two forcing terms is the weight put on the eccentricity signal, which is around
eight times greater when the orbital parameters are independently optimised. We
have shown that eccentricity may have a greater impact on the Earth system than
$Q_{65}$ would suggest. However, it may not be eight times greater. We do,
however, show that when eccentricity is even a few times greater than $Q_{65}$
suggests, the Earth system appears to be better explained by damped unforced
oscillations that resonate with orbital forcing, rather than sustained internal
oscillations. This suggests that models using $Q_{65}$ may be underestimating
the role of astronomical forcing in the Earth system, potentially leading to an
overemphasis on purely geochemical mechanisms in explaining glacial-interglacial
cycles.

\clearemptydoublepage
\chapter{Analysis of Our Models}
\label{chap:analysis}

\initial{H}ere we establish a more comprehensive understanding of the
differences and similarities between the feedback and feedforward models. We
first examine the performance of the two models and propose two augmentations to
the feedforward model that allow it to explain the data as well as the feedback
model. We then explore the implications of the underdetermined nature of the
feedback model when applying a physical interpretation to the two models. The
ocean solution from both models is a poor representation of the data, which
leads us to investigate different physical interpretations of the model
variables. For this, we re-examine the model from Verbitsky et al., which shares
similar dynamics to our feedback model, but differing variable interpretations.
We show that our feedback model resembles Verbitsky's model to the first
order, with the exception of $Q_{65}$ as the forcing function. Since all of
these models reproduce the data to similar extents, we question the degree to
which the data necessitates a non-linear model or even a model that produces
unforced oscillations.

\section{Comparison with Feedforward Model}
\label{sec:comparison_with_feedforward}
\begin{figure}
  \hspace{10pt}
  \input{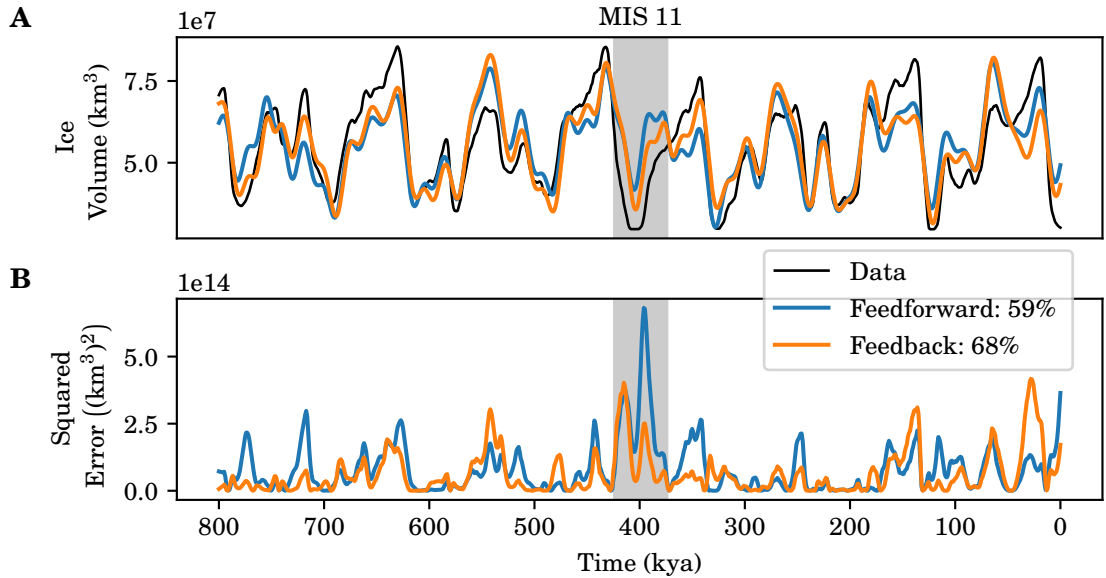}
  \caption[Feedback vs Feedforward Errors]{\textbf{A}: Comparison of the
    original linear model (blue) and the new model capable of producing
    unforced oscillations (orange). The percentage of variance explained by
    each model is shown in the legend. \textbf{B}: The squared error between the
    ice volume data and each model. The new model performs noticeably better
  around the MIS 11 timespan.}
  \label{fig:linear_intrinsic_ice_compare}
\end{figure}

The optimal solutions in Figure \ref{fig:linear_intrinsic_ice_compare}A show
that the feedback model is able to explain 68\% of the variance in the data,
whilst the feedforward model explains 59\%. The squared error between the data
and each model is shown in Figure \ref{fig:linear_intrinsic_ice_compare}B, which
shows the feedback model to produce a somewhat lower squared error curve than
the feedforward model. This improvement could indicate that one or more unforced
Earth-system oscillations play an important role in ice volume variation.
However, it is interesting to note that the improvement in the model's
performance is mostly concentrated around the MIS 11 period.

This raises the question of whether the feedback model is actually capturing
unforced oscillations in the Earth system, or if the additional parameters
just allow for a better fit to be achieved around MIS 11. It is reasonable to
assume that if one model is a better representation of the system dynamics, then
it would perform consistently better throughout the 800\,kyr period.

As we show in Figure \ref{fig:comp_models}, the extreme interglacial that occurs
during MIS 11 is often poorly reproduced by conceptual models. This region in
time coincides with a minimum in eccentricity's 400\,kyr cycle, which would be a
suitable explanation for the model failing to reproduce an interglacial of
normal magnitude at this time. However, this interglacial is both the longest
and warmest of the past 800\,kyr, leading us to posit that a physical anomaly
on Earth may have caused a departure from orbitally driven dynamics during MIS 11.
This could be from increased volcanic activity \cite{MIS11_volcano} or from ice sheet
instability \cite{mis11_ice_collapse,greenland_ice_free_mis11,glacial_instability}.

If this is the case, then we must consider that a model that improves upon the
feedforward model mostly around MIS 11 is not necessarily a better
representation of the Earth system dynamics. Since all of the conceptual models
we have discussed so far do not include specifically timed Earth-based events,
such as increased volcanic activity, they will be attempting to reproduce MIS 11 with
the same system dynamics that apply throughout the entire 800\,kyr timespan. If
MIS 11 is indeed an outlier, owing to a specific Earth-based event, then we
should not require an orbitally driven model to reproduce it accurately.

\begin{figure}
  \centering
  \input{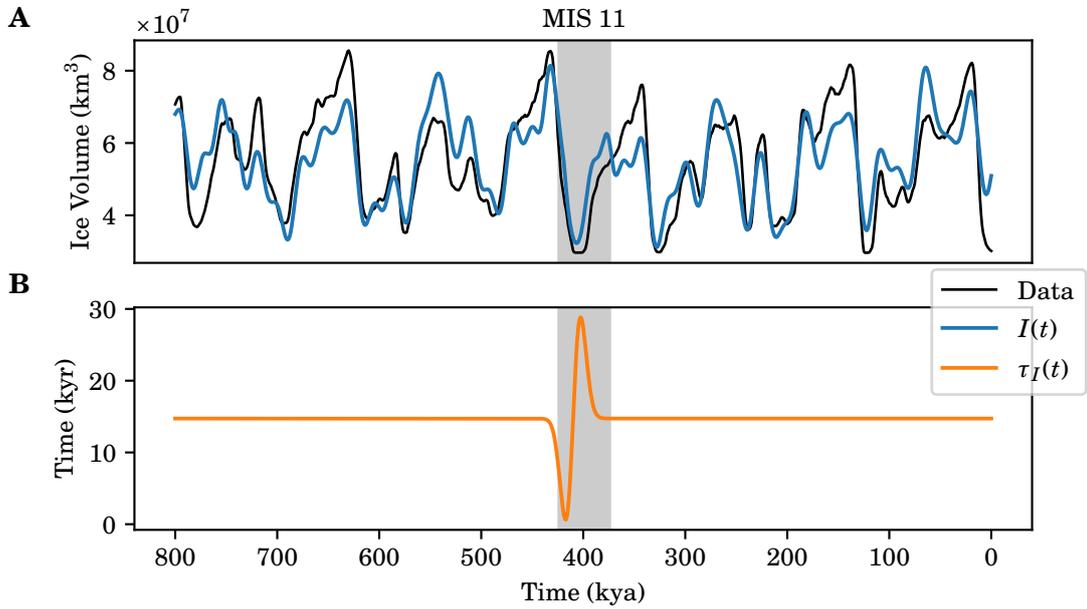}
  \caption[Possible $\tau_I$ Function]{A possible function for $\tau_I$ that
    could be used to account for MIS 11 to produce a fit of similar accuracy
  (67\%) to the feedback model.}
  \label{fig:wheen_I_tauI_fun}
\end{figure}

We propose two separate augmentations to the feedforward model to investigate
how well it could perform if we assume MIS 11 to have been impacted by a
non-recurring process on Earth, independent of orbital forcing. The first allows
the ice volume time constant $\tau_I$ to vary with time. This model assumes that
the drastic interglacial occurred due to runaway ice melting, in which we could
imagine the melting rate of the ice sheets increased. In fitting a variety of
functions for $\tau_I(t)$, we found that the best fit was achieved when
$\tau_I(t)$ first dropped, simulating an increase in ice melt rate, and then
increased again as the ice volume began to increase. This subsequent increase in
the time constant could be explained by the effect of the depleted ice sheets on
Earth's albedo. As less ice is present globally, the Earth absorbs more solar
radiation, which would slow the rate of ice growth. We propose this augmented
system to be of the form
\begin{align}
  \tau_I(t)\frac{\mathrm{d}I}{\mathrm{d}t} &= p_1 O(t) +  p_2\varepsilon(t) +
  p_3\beta(t) + p_4\cos(\rho(t)) - I(t) + p_5, 
  \label{eq:aug_ice_vol}\\[4pt]
  \tau_O\frac{\mathrm{d} O}{\mathrm{d}t} &= \varepsilon(t) - O(t),\\
  \tau_I(t) &= a + b(t+d)e^{c(t+d)^2},
  \label{eq:model_tauI_fun}
\end{align}
where $a=14.7$\,kyr, $b=3.16$, $c=-0.00919$\,kyr$^{-2}$ and $d=410$\,kyr.
Unsurprisingly, the optimal value for $a$, which sets the constant offset of
$\tau_I(t)$, is approximately equal to the value of $\tau_I$ in the original
linear model. The resultant ice volume solution is shown in Figure
\ref{fig:wheen_I_tauI_fun}A alongside the forcing function in Figure
\ref{fig:wheen_I_tauI_fun}B given by \eqref{eq:model_tauI_fun}. This solution
explains 67\% of the variance in the data, which is comparable to the feedback
model's 68\%.

\begin{figure}
  \centering
  \input{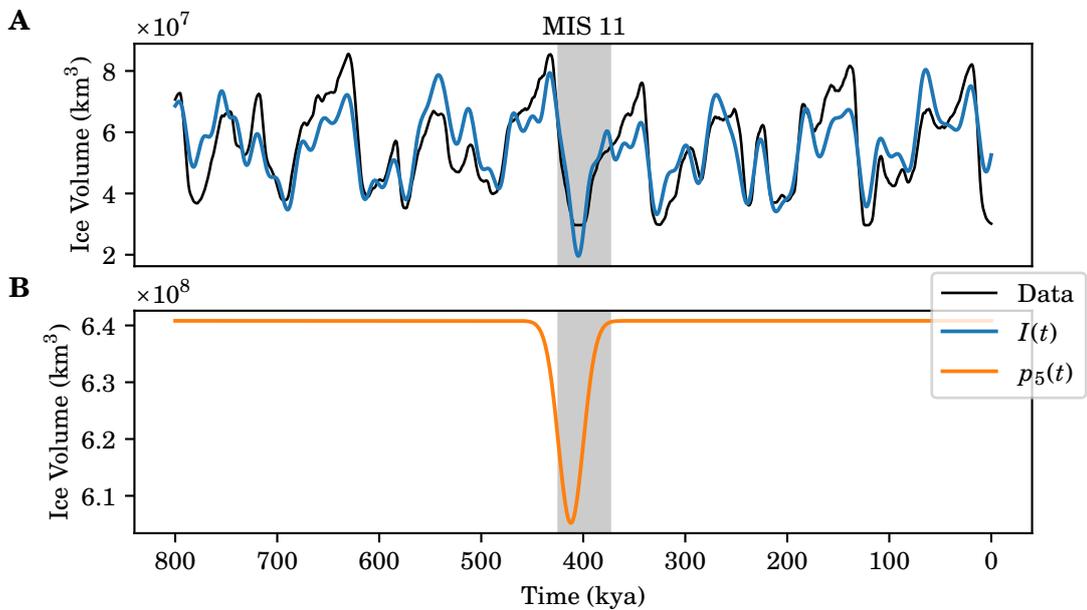}
  \caption[Possible Offset Function]{A possible function for the offset that
    could be used to account for MIS 11 to produce a fit of the same accuracy
  (68\%) to the feedback model.}
  \label{fig:wheen_I_offset_fun}
\end{figure}

The second augmentation treats the constant offset $p_5$ as a function of time.
This model assumes an alternative explanation of MIS 11 in which the dynamic
properties of the ice sheets do not change (constant $\tau_I$), but are affected
by an Earth-based force, such as volcanic eruptions. In this case, the climate
experiences a brief cooling force in addition to orbital forcing. This would
result in a model similar to \eqref{eq:aug_ice_vol}, but with $p_5$ replaced by
\begin{equation}
  p_5(t) = a + be^{c(t+d)^2},
  \label{eq:model_p5_fun}
\end{equation}
where $a=6.67\times 10^8$\,km$^3$, $b=2.44\times10^7$\,km$^3$,
$c=-0.00237$\,kyr$^{-2}$ and $d=400$\,kyr. Once again, the value for $a$ in this
case is approximately equal to the originally fit $p_5$ value in the feedforward
model, ensuring that we have not changed the model dynamics outside of MIS 11.
The resultant ice volume solution is shown in Figure
\ref{fig:wheen_I_offset_fun}A alongside the forcing function in Figure
\ref{fig:wheen_I_offset_fun}B given by \eqref{eq:model_p5_fun}. This solution
explains 68\% of the variance, matching that of the feedback model.

\begin{figure}
  \hspace{-60pt}
  \input{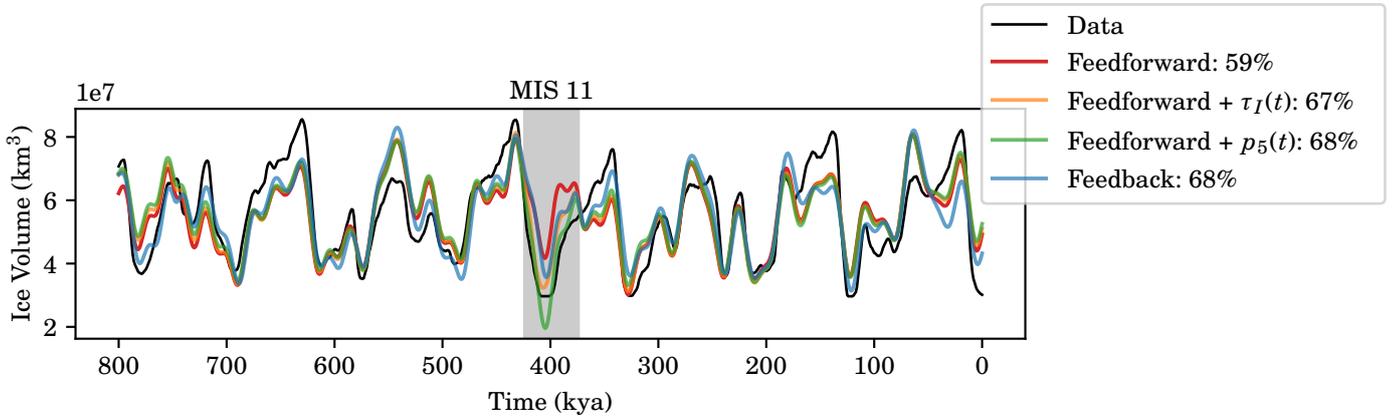}
  \caption[Comparison of Wheen Augmentations]{Comparing all 4 models to the ice
  volume data.}
  \label{fig:comp_wheen_I_augs}
\end{figure}

The comparison in Figure \ref{fig:comp_wheen_I_augs} shows each version of
the feedforward model along with the feedback model. We see all of the
solutions remain similar to each other throughout the 800\,kyr duration, only
differing around MIS 11. This is to confirm that MIS 11 is the only period that
has been impacted by the feedforward model augmentations which are able to
explain the data as well as the feedback model.

The feedback model is intended to capture a hypothesised unforced oscillation
throughout the entire duration, but more closely reflects the same effect as the
augmentations to the feedforward model. This suggests that the feedback model
may not truly capture an Earth-based cycle, but instead allows added flexibility
in the fit. It is therefore difficult to determine, from the data alone, whether
the glacial-interglacial cycles are dependent on an Earth based feedback
mechanism, or if they are a predominantly linear response to the orbital
parameters, with a non-orbitally forced event around MIS 11.

\subsection{Physical Interpretation}
\label{sec:both_model_physical_interpretation}

Since both of our models, which align with opposing theories of climate change,
are able to explain the data with similar accuracy (accounting for MIS 11), we
examine how they might be interpreted in a physical context. This could help to
determine which model is more likely to represent the Earth system dynamics.
\begin{figure}
  \hspace{-25pt}
  \input{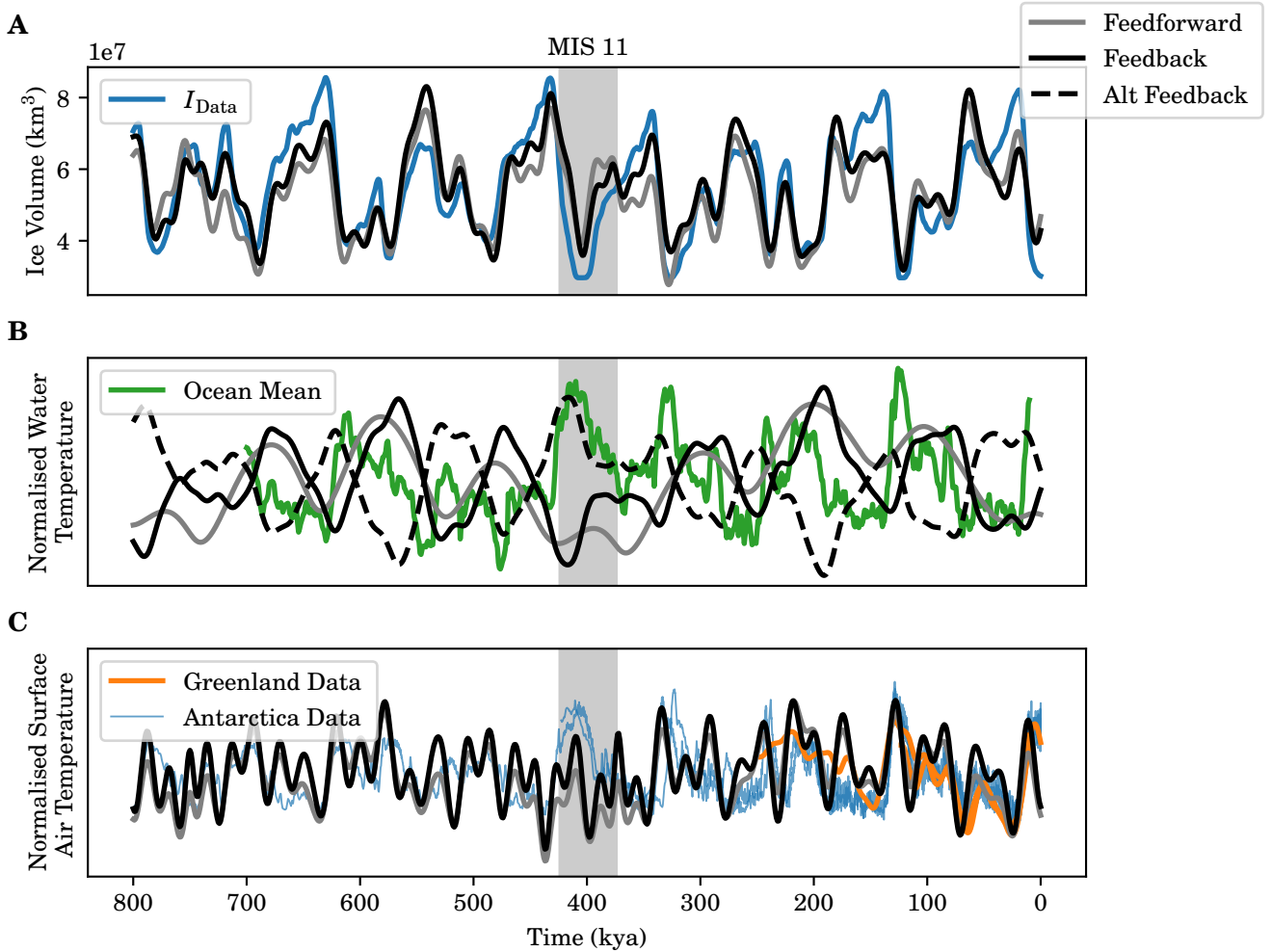}
  \caption[Physical Comparison]{Comparison of the physical interpretations of
    the feedback and feedforward models. The feedforward solutions are the same
    as those shown in Figure \ref{fig:sea_air_data_sol_comp}. Also shown are two
    solutions for the feedback model, both achieving the same optimal
    eigenvalues but with different intrinsic parameters. The feedback solutions
    align with the ice volume and surface temperature data to a similar degree
    as the feedforward model, however they differ from the feedforward ocean
    temperature solution. The two solutions for ocean temperature from the
    feedback model are inverses of each other, with the solid black line
    aligning better with the feedforward solution, which matches the data around
    either end of the data. The alternative solution, shown by the dashed line,
    aligns better in the middle, around MIS 11.}
  \label{fig:wheen_physical_comp}
\end{figure}

Because the parameters of the feedback model are underdetermined, it is
difficult to assign them a physical interpretation without including further
constraints. The choice of intrinsic parameters can fundamentally change the way
we model the interaction between ocean temperature and ice volume. One
configuration manifests as ocean temperature positively impacting ice volume and
ice volume negatively impacting ocean temperature. This is consistent with the
ocean impact suggested by the feedforward model but is physically dubious as a
warming ocean would ostensibly contribute to a decrease in ice volume. The
alternative configuration flips this relationship, meaning ocean temperature
negatively impacts ice volume and ice volume positively impacts ocean
temperature. The negative impact of ocean temperature on ice volume is more
logical and aligns better with the negative correlation between the two datasets
shown in Figure \ref{fig:sea_and_air_temp_with_map}, however the ice volume
positively impacting ocean temperature is less intuitive. One possible
interpretation of this case is that; as the ice sheets recede, the meltwater
flows into the ocean, reducing the overall temperature of the ocean.

In Figure \ref{fig:wheen_physical_comp}, we show these two opposing physical
interpretations of the feedback model, also included is the feedforward model
and the proxy data from Figure \ref{fig:sea_and_air_temp_with_map}. As mentioned,
the ice volume and surface temperature solutions from the feedback model are the
same for either interpretation, but the ocean temperature solution is inverted.
Since neither of these solutions align very well with the ocean data, it is
difficult to discern which of the configurations is more plausible. The first
configuration we mentioned is shown by the solid line and aligns reasonably well
with the solution from the feedforward model, which is to be expected as they
are both modelling a negative impact of ocean temperature on ice volume. These
appear to match the trend of the data at certain points but significantly
diverge around MIS 11. The alternative configuration produces an inverted
solution to the first configuration, producing better alignment with the data
around MIS 11, however it appears to significantly diverge from the data in
other regions.

There are two possible conclusions to draw from this physical comparison. The
first is that the ocean temperature data is not adequately captured by any of
the solutions because our interpretation of the variable is, at the very least,
incomplete. Therefore, in order to correctly model the relevant mechanisms in
the Earth system, either additional intermediate variables are needed to
describe the interaction between ocean temperature and ice volume, or the
existing variables need redefining. The second possible conclusion is that
the poor fit to the ocean data is due to either noise in the data, or an
independent Earth-based event. It is therefore reasonable to reject the second
feedback model interpretation in which ocean temperature negatively impacts ice
volume. This is because the first configuration aligns with the data better
outside of the MIS 11 region and we are positing MIS 11 to be an outlier. This
has the added benefit of being consistent with the feedforward model, which is
uniquely determined.

\subsection{Sweep Between Models}
In order to better understand the differences between these two models, we
perform a sweep from our original feedforward model to the feedback model. As a
reminder, the feedback model is given by
\begin{align}
  \tau_I\frac{\mathrm{d}I}{\mathrm{d}t} &= p_1 O(t) +  p_2\varepsilon(t) +
  p_3\beta(t) + p_4\cos(\rho(t)) - I(t) + p_5,\\
  \tau_O\frac{\mathrm{d} O}{\mathrm{d}t} &= \varepsilon(t) - O(t) + p_6 I(t),
\end{align}
whilst the feedforward model is identical, but with $p_6=0$. We therefore
transition from feedforward to feedback by changing the $p_6$ coefficient from
zero to an increasingly negative value, optimising the remaining parameters at
each step.

As we have discussed, the feedback model is underdetermined, meaning there is
no one optimum $p_6$ value. We have chosen to vary $p_6$ down to $-3.2$ as this
allows us to see both the quick transition of regimes at the start, as well as
the converging behaviour of the parameters as $p_6$ becomes more negative. For
the reasons discussed in Section \ref{sec:both_model_physical_interpretation},
$p_6$ is chosen to be negative, as this produces a feedback model that is
consistent with the feedforward model, which is uniquely determined.

Looking first at the variance explained, we see a fast increase from the
feedback model's 59\%, converging to 68\% as $p_6$ reaches $-3.2$. This is
consistent with our previous analysis, which showed that the feedback model
can better fit to the data, particularly around MIS 11. The oscillation period
is undefined for the feedforward model as it cannot oscillate without forcing,
but as we introduce the feedback mechanism, the period rapidly converges to
100\,kyr. This is because no other period is beneficial for fitting to the data.
Since the oscillation period is given by
\begin{equation}
  T_\mathrm{Osc} = \frac{4\pi\tau_I\tau_O}{\sqrt{4p_1p_6\tau_I\tau_O + (\tau_I - \tau_O)^2}},
\end{equation}
it is evident that the three other parameters can adjust to account for $p6$ to
maintain this period.

The decay time is interesting as it is shown to grow more slowly, converging
towards 55\,kyr. This final value is in agreement with the optimal decay time
range shown in Figure \ref{fig:eigenvalue_sweep}. The explanation for the slower
convergence comes from the priorities of the model as $p_6$ decreases. The
oscillation period is the most important characteristic as it affects the timing
of extrema, whilst the decay time varies the significance of unforced
oscillations on the solution. With a short decay time (around 15\,kyr), the
feedback model behaves similarly to the feedforward model. The time constant
$\tau_O$ therefore grows inversely proportional to $p_6$ in order to maintain a
constant 100\,kyr period. Since the decay time of the system is given by
\begin{equation}
  T_\mathrm{Dec} = \frac{2\tau_I\tau_O}{\tau_I + \tau_O},
\end{equation}
we can expect the slower growth of $\tau_O$ to impact the decay time.
\begin{figure}
  \hspace{-35pt}
  \input{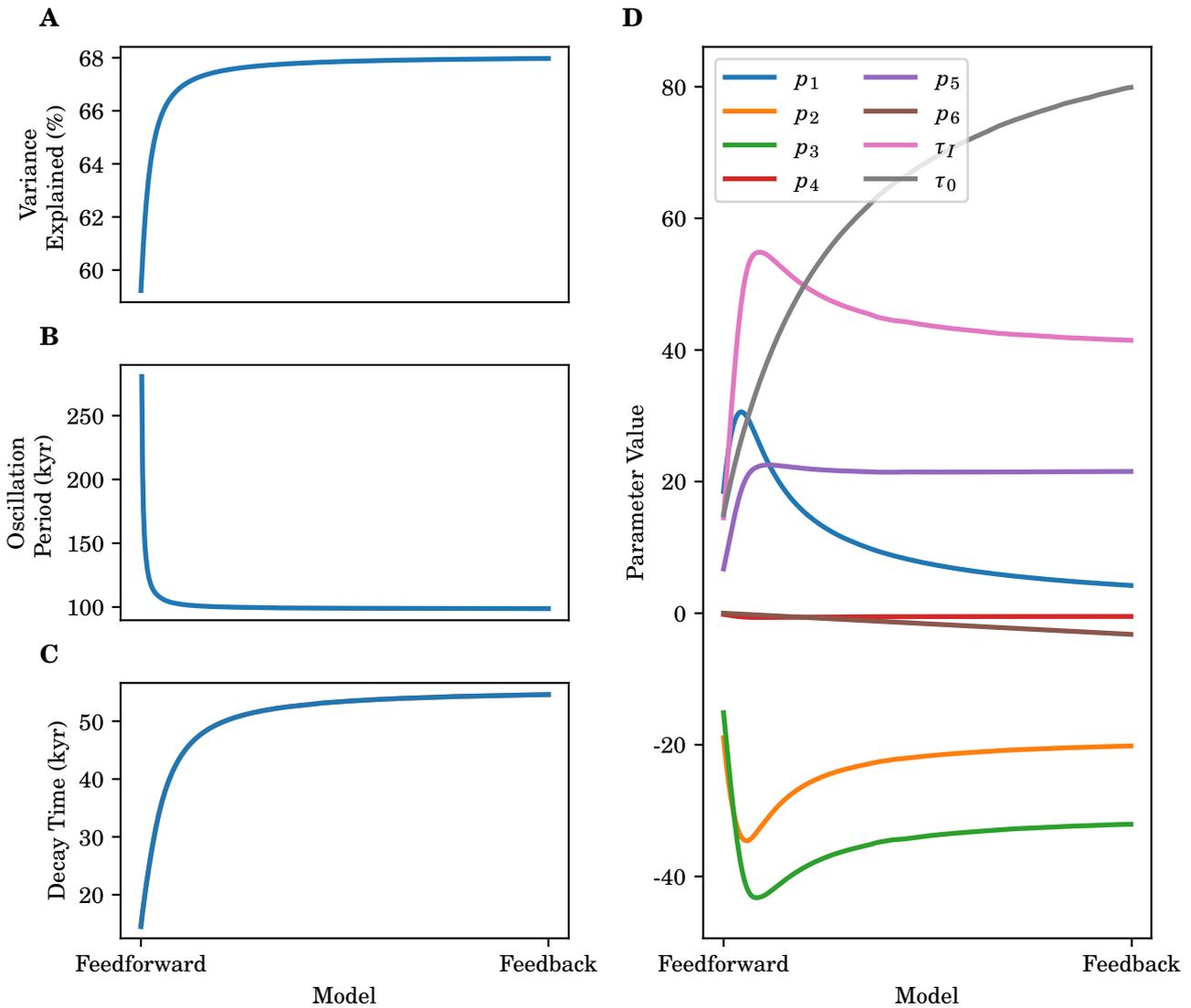}
  \caption[Sweep Feedback to Feedforward Model]{Transitioning from the
    feedforward model to the feedback model by varying the coefficient of the
    $I$ term in the ocean temperature equation ($p_6$) from zero to $-3$. The
    remaining parameters were optimised for each value of $p_6$. The variance
    explained by the model throughout the sweep is shown in \textbf{A}. The
    oscillation period and decay time are derived from the system's eigenvalues
    and are shown in \textbf{B} and \textbf{C} respectively. The optimal
  parameters as $p_6$ is varied are shown in \textbf{D}.}
  \label{fig:sweep_intrin_to_feedforward}
\end{figure}

\section{Comparison with Verbitsky et al.}
\label{sec:comparison_verbitsky}
As we showed in Section \ref{sec:both_model_physical_interpretation}, the
proposed physical interpretation of the ocean temperature variable in our
feedback model is not well supported by the data. We therefore look to the
Verbitsky model, which has a similar structure to our feedback model, but with
different variable interpretations.

For convenience, the Verbitsky et al. model (VCV18) \cite{verbitsky2018} is
given by
\begin{align}
  \frac{\mathrm{d}S}{\mathrm{d}t} &= \frac{4}{5\zeta}S(t)^{3/4}(a-\varepsilon F(t)-k\omega(t)-c\theta(t)),\\
  \label{eq:verbitsky_model_theta}
  \frac{\mathrm{d}\theta}{\mathrm{d}t} &= \frac{1}{\zeta}S(t)^{-1/4}(a-\varepsilon
  F(t)-k\omega(t))(\alpha\omega(t)+\beta(S(t)-S_0)-\theta(t)),\\
  \frac{\mathrm{d}\omega}{\mathrm{d}t} &= \gamma_1-\gamma_2(S(t)-S_0)-\gamma_3\omega(t).
  \label{eq:verbitsky_model_omega}
\end{align}
The model has 3 variables, $S$, $\theta$, and $\omega$, representing glaciation
area, basal temperature, and climate temperature respectively. For this
analysis, glacial area $S$ will be be assumed to relate linearly to ice volume.
The model is forced by $F(t)$, which is the normalised $Q_{65}$ signal.
\begin{table}
  \caption[Parameter values for Verbitsky models]{Set of parameter values
    used for the Verbitsky et al. model. The default values relate to the
    solution given in the original paper, whilst the optimised values are those
    that best reproduce the data. The linear parameters relate to the linearised
    model and are optimised to reproduce the non-linear model counterpart.}
  \label{tab:verbitsky_params_set}
  \centering
  \begin{tabular}{c|c|c|c|c|c}
    Parameter & Default & Linear Default & Optimised & Linear Optimised & Units \\ \hline\hline
    $\zeta$      & 1.0    & 0.925  & 1.10   & 0.809  & km$^{-1/2}$ \\\hline
    $a$          & 0.065  & 0.0141 & 0.0584 & 0.0311 & -- \\\hline
    $k$          & 0.005  & 0.00806& 0.00631& 0.0153 & °C$^{-1}$ \\\hline
    $c$          & 0.042  & 0.0521 & 0.0288 & 0.0262 & °C$^{-1}$ \\\hline
    $\alpha$     & 2.0    & 0.867  & 1.27   & 1.49   & -- \\\hline
    $\beta$      & 2.0    & 2.40   & 1.72   & 2.0    & °C\,km$^{-2}$ \\\hline
    $\gamma_1$   & 0.0    & 0.0    & 0.00181& 0.0882 & °C\,kyr$^{-1}$ \\\hline
    $\gamma_2$   & 0.21   & 0.120  & 0.185  & 0.0129 & °C\,km$^{-2}$\,kyr$^{-1}$ \\\hline
    $\gamma_3$   & 0.3    & 0.359  & 0.274  & 0.101  & kyr$^{-1}$ \\\hline
    $S_0$        & 12.0   & 10.5   & 25.2   & 23.1   & km$^2$ \\\hline
    $\varepsilon$& 0.11   & 0.0777 & 0.0566 & 0.118  & -- \\\hline
    $S(0)$       & 10     & 13.8   & 25.7   & 17.2   & km$^2$ \\\hline
    $\theta(0)$  & 0      & 0.0116 & 4.13   & 1.27   & °C \\\hline
    $\omega(0)$  & 2      & 11.1   & -11.4  & 0.197  & °C \\
  \end{tabular}
\end{table}

As we are comparing this model to our linear models, we will linearise around
the system's equilibrium. We can then see how much accuracy is gained from the
non-linearity of Verbitsky's model, and also how the state variables compare to
our own.

There are two equilibria for the system given by
(\ref{eq:verbitsky_model_theta}--\ref{eq:verbitsky_model_omega}). However, when
substituting in the parameter values from Table \ref{tab:verbitsky_params}, we
find that only one equilibrium is physically valid. This is given by
\begin{equation}
  \mathbf{x}^* = \begin{pmatrix}S^*\\\theta^*\\\omega^*\end{pmatrix} =
  \frac{1}{\alpha c\gamma_2 - \beta c\gamma_3 + \gamma_2 k}
  \begin{pmatrix}S_0\alpha c\gamma_2 - S_0\beta c\gamma_3 + S_0\gamma_2 k
  - a\gamma_3 + \alpha c\gamma_1 + \gamma_1 k\\a\alpha\gamma_2 - a\beta\gamma_3
+ \beta\gamma_1 k\\a\gamma_2 - \beta c\gamma_1\end{pmatrix}.
\label{eq:verbitsky_equilibria}
\end{equation}

In order to linearise around this equilibrium, we find the Jacobian, given by
\begin{equation}
  \mathbf{J} = \begin{pmatrix}\frac{3 (a - c \theta - k \omega)}{5 \zeta
    S^{\nicefrac{1}{4}}} & - \frac{4 c S^{\nicefrac{3}{4}}}{5 \zeta} & - \frac{4 k
  S^{\nicefrac{3}{4}}}{5 \zeta}\\[5pt]\frac{(a - k \omega) (- \alpha \omega +
      \beta (S_{0} - S) + 4 \beta S + \theta)}{4 \zeta S^{\nicefrac{5}{4}}} &
      \frac{- a + k \omega}{\zeta S^{\nicefrac{1}{4}}} & \frac{\alpha (a - k \omega) + k
      (- \alpha \omega + \beta (S_{0} - S) + \theta)}{\zeta
      S^{\nicefrac{1}{4}}}\\[5pt]- \gamma_{2} & 0 & - \gamma_{3}\end{pmatrix},
\end{equation}
where we have set the forcing function $F(t)$ to be 0.

To reintroduce the forcing function, we differentiate the system with respect to
$F(t)$ to get the its coefficients. This gives
\begin{equation}
  \mathbf{g} = \begin{pmatrix}- \frac{4 \varepsilon S^{\nicefrac{3}{4}}}{5
  \zeta}\\[5pt] \frac{-\varepsilon (\alpha \omega - \beta (S_{0} - S) -
\theta)}{S^{\nicefrac{1}{4}} \zeta}\\[5pt]0\end{pmatrix}.
\end{equation}

We can now construct the linearised system about the equilibrium, giving
\begin{equation}
  \frac{\mathrm{d}\left(\mathbf{x}-\mathbf{x}^*\right)}{\mathrm{d}t} =
  \frac{\mathrm{d}\mathbf{x}}{\mathrm{d}t} =
  \mathbf{J}^*\left(\mathbf{x}(t)-\mathbf{x}^*\right) + \mathbf{g}^*F(t),
\end{equation}
where the $^*$ in $\mathbf{J}^*$ and $\mathbf{g}^*$ indicates that we have
substituted in the equilibrium values for the parameters.

This produces a linear system of the form
\begin{align}
  \frac{\mathrm{d}S}{\mathrm{d}t} &= p_1F(t) + p_2\omega(t) + p_3\theta(t) + p_4,\\
  \label{eq:linearised_verbitsky_S}
  \frac{\mathrm{d}\theta}{\mathrm{d}t} &= p_5S(t) + p_6\omega(t) +
  p_7\theta(t) + p_8,\\
  \frac{\mathrm{d}\omega}{\mathrm{d}t} &= p_9S(t) + p_{10}\omega(t) + p_{11},
  \label{eq:linearised_verbitsky_omega}
\end{align}
where substituting in the equilibrium values has cancelled out the $F(t)$ term
in the second equation.

\begin{figure}
  \hspace{-20pt}
  \input{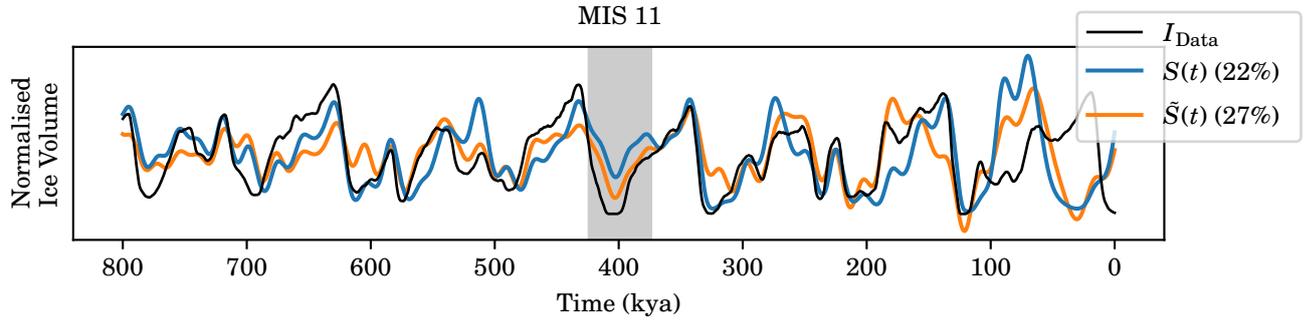}
  \caption[Verbitsky Original vs Linearised]{Ice sheet surface area solutions
    from Verbitsky's original model ($S(t)$) and the linearised version
    ($\tilde{S}(t)$). The original model uses the default parameters from Table
    \ref{tab:verbitsky_params_set}, whilst the linearised version has been refit
    to closer match the original model's solution, referred to as Linear Default
    in the same table. The degree to which these two models explain the variance
    in the ice volume data is shown in the legend.}
  \label{fig:paper_verbitsky_comp}
\end{figure}

\begin{figure}
  \centering
  \input{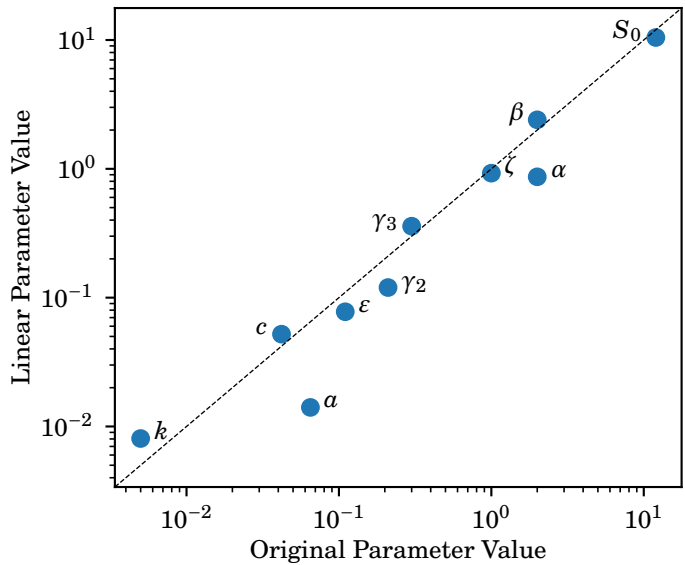}
  \caption[Verbitsky Original vs Linearised Parameters]{Parameter comparison between the default values from Verbitsky's
    model and those used for the linearised version as shown in Figure
  \ref{fig:paper_verbitsky_comp}.}
  \label{fig:verbitsky_param_compare}
\end{figure}

We show the original model ice surface area solution from Verbitsky alongside
the solution from the linearised model that has been optimised to reproduce the
same solution in Figure \ref{fig:paper_verbitsky_comp}. We can see that the
linearised model is able to reproduce the original model's solution reasonably
well, incidentally explaining 5\% more of the variance in the ice volume data.
It is important to note that the default parameters for this model are not
intended to maximise the fit to the data, but rather to represent physically
plausible values. It is therefore interesting that the linear model that
produces the same solution, achieves a slightly better fit to the data and does
so with parameters that are similar to Verbitsky's physically based values, with
the exception of $a$. A comparison of the parameters used for these solutions is
shown in Figure \ref{fig:verbitsky_param_compare}.
\begin{figure}
  \hspace{-20pt}
  \input{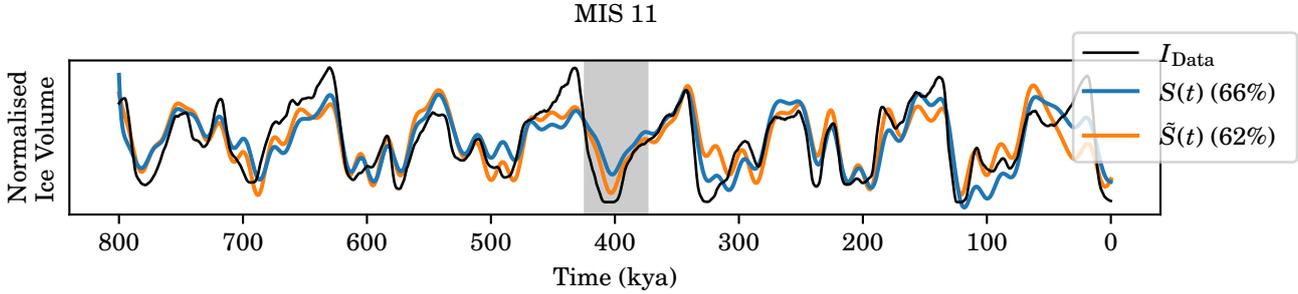}
  \caption[Verbitsky Original vs Linearised Optimised]{Ice sheet surface area
    solutions from Verbitsky's original model $S(t)$ and the linearised
    version $\tilde{S}(t)$. These use parameters that have been separately
    optimised to fit the ice volume data $I_\mathrm{Data}$ and are shown in
  Table \ref{tab:verbitsky_params}.}
  \label{fig:optim_verbitsky_comp}
\end{figure}

The linearised model's ability to reproduce the original model's solution
without needing significant parameter changes suggests that the non-linearity in
the Verbitsky model is not crucial. There is, however, a change in the
underlying dynamics of the system brought about by the parameter $a$, which
relates to the damping rate of the solution. The reduced $a$ in the linearised
case means that the model now produces sustained unforced oscillations whereas
the original model produces a damped response.

The non-linear Verbitsky model can generate 100\,kyr oscillations through a
period doubling response to the 41\,kyr obliquity signal. The authors
demonstrate this by forcing their model with just a 41\,kyr sinusoid and showing
that it produces oscillations with a period close to 100\,kyr. This period
doubling behaviour emerges from the model's non-linear dynamics, not from any
100\,kyr component in the forcing itself. The linearised model, however, is
unable to produce this period doubling behaviour, instead producing sustained
oscillations with a period of 100\,kyr. This change in system dynamics does not
affect the validity of the model, but rather aligns it more with the geochemical
theory than the non-linear model which produces damped oscillations that
interact with the orbital forcing. This change of system dynamics is similar to
what we observe when we use $Q_{65}$ as the orbital forcing function in our
feedback model. In that case, the 100\,kyr period being input to the system was
significantly reduced, which led to the model producing this period through
sustained unforced oscillations.

In Figure \ref{fig:optim_verbitsky_comp}, we show the solutions of the original
and linearised Verbitsky models that have been optimised to fit the ice volume
data. We see that they are both able to reproduce the data to a similar degree,
despite a difference in the underlying dynamics. This further supports the idea
that the non-linearity in the Verbitsky model is not crucial for explaining the
data.

\begin{figure}
  \hspace{-20pt}
  \input{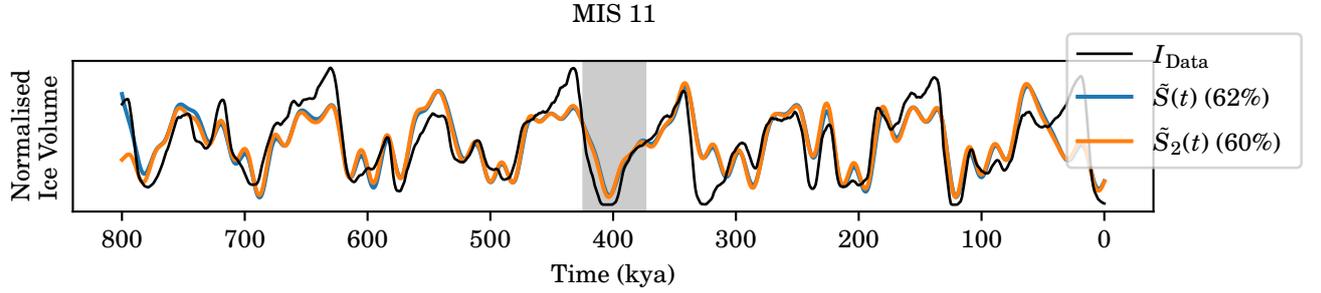}
  \caption[Verbitsky Linearised vs Two Variables]{Ice sheet surface area
    solutions from the linearised Verbitsky's model $\tilde{S}(t)$ and the
    linearised model with the quasi-steady $\omega$ assumption
    $\tilde{S}_2(t)$. These use parameters that have been separately optimised
  to fit the ice volume data $I_\mathrm{Data}$.}
  \label{fig:verbitsky_lin_simp_comp}
\end{figure}

Although the original Verbitsky model and our feedback model are not immediately
comparable, we can show that the linearised Verbitsky model is essentially the
same as our feedback model when we use $Q_{65}$ as the forcing function.
Returning to the linearised Verbitsky model given by
\eqref{eq:linearised_verbitsky_S}--\eqref{eq:linearised_verbitsky_omega}, we see
there are 3 variables, as opposed to our feedback model which uses only two.
However, it is possible to simplify this to a two dimensional system by using a
quasi-steady assumption for $\omega$. From the parameters in the Verbitsky paper,
$\omega$ has a characteristic timescale of around $1/\gamma_3\approx3$\,kyr,
$\theta$ has a characteristic timescale of $H/a\approx 13$\,kyr, and $S$ is less
clearly defined but relates to glacial growth and so is on the order of tens of
thousands of years. This makes $\omega$ a comparatively fast variable.
Therefore, if we treat $\omega$ as quasi-steady, we get
\begin{equation}
  \omega(t) = -\frac{p_9 S(t) + p_{11}}{p_{10}}.
\end{equation}
Substituting this means the linear system becomes
\begin{align}
  \frac{\mathrm{d}S}{\mathrm{d}t} &= p_1F(t) - \frac{p_2 p_9}{p_{10}}S(t) +
  p_3\theta(t) + p_4 - \frac{p_2 p_{11}}{p_{10}},\\
  \frac{\mathrm{d}\theta}{\mathrm{d}t} &= \left(p_5-\frac{p_6
  p_9}{p_{10}}\right)S(t) + p_7\theta(t) + p_8 - \frac{p_6 p_{11}}{p_{10}}.
\end{align}
The solutions of this model before and after the dimension reduction are shown
in Figure \ref{fig:verbitsky_lin_simp_comp}. We see that the reduction in
dimensionality affects only the start of the solution, leaving the majority of
the solution unchanged.

This assumption of quasi-steady $\omega$ is similar to the assumption of
quasi-steady surface temperature in the physical interpretation of our
feedforward model. Although the physical mechanisms are different, they are both
fast variables that relate to Earth's climate outside of its glacier.

Since we wish to compare the ice sheet surface area $S(t)$ from this model
to the ice variable in our feedback model, we will set the constant term in the
second equation to 0. To account for this, we add a constant term to the first
equation that is proportional to the constant term in the second equation.
This is given by
\begin{equation}
  -\frac{p_3}{p_7}\left(p_8 - \frac{p_6 p_{11}}{p_{10}}\right),
\end{equation}
making the system
\begin{align}
  \frac{\mathrm{d}S}{\mathrm{d}t} &= p_1F(t) - \frac{p_2 p_9}{p_{10}}S(t) +
  p_3\theta(t) + p_4 - \frac{p_2 p_{11}}{p_{10}} - \frac{p_3}{p_7}\left(p_8 -
  \frac{p_6 p_{11}}{p_{10}}\right),\\
  \frac{\mathrm{d}\theta}{\mathrm{d}t} &= \left(p_5-\frac{p_6
  p_9}{p_{10}}\right)S(t) + p_7\theta(t),
\end{align}
which we will rewrite with simplified parameters as
\begin{align}
  \frac{\mathrm{d}S}{\mathrm{d}t} &= p_1F(t) + p_2S(t) + p_3\theta(t) +
  p_4,\\[4pt]
  \frac{\mathrm{d}\theta}{\mathrm{d}t} &= p_5S(t) + p_6\theta(t).
\end{align}

As a reminder, our feedback model is given by
\begin{align}
  \frac{\mathrm{d}I}{\mathrm{d}t} &= \frac{p_1}{\tau_I}\varepsilon(t) +
  \frac{p_2}{\tau_I}\beta(t) + \frac{p_3}{\tau_I}\cos(\rho(t)) - \frac{1}{\tau_I}I(t) +
  \frac{p_4}{\tau_I} O(t) + \frac{p_5}{\tau_I}, \\[4pt]
  \frac{\mathrm{d} O}{\mathrm{d}t} &= 
  \frac{p_6}{\tau_O} I(t) - \frac{1}{\tau_O}O(t) +
  \frac{1}{\tau_O}\varepsilon(t),
\end{align}
which we can rewrite with simplified parameters as
\begin{align}
\frac{\mathrm{d}I}{\mathrm{d}t} &= p_1\varepsilon(t) + p_2\beta(t) +
p_3\cos(\rho(t)) + p_4 I(t) + p_5 O(t) + p_6,\\[4pt]
\frac{\mathrm{d} O}{\mathrm{d}t} &= p_7 I(t) + p_8 O(t) +
p_9\varepsilon(t).\label{eq:simple_intrinsic_model_O}
\end{align}

\begin{figure}
  \hspace{-20pt}
  \input{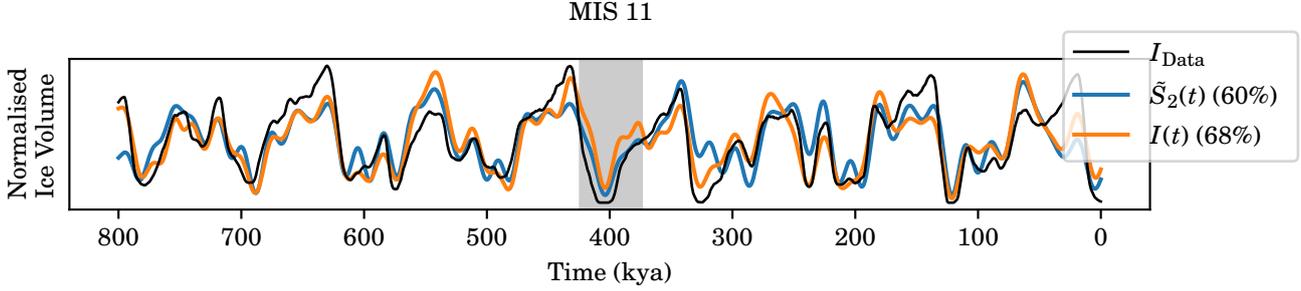}
  \caption[Verbitsky Two Variables vs Feedback]{The ice surface area solution
    from the linearised Verbitsky model with the quasi-steady $\omega$
    assumption $\tilde{S}_2(t)$ and our feedback model ice volume solution
    $I(t)$. These use parameters that have been separately optimised to fit the
    ice volume data $I_\mathrm{Data}$.}
  \label{fig:verbitsky_simp_fb_comp}
\end{figure}
\begin{figure}
  \input{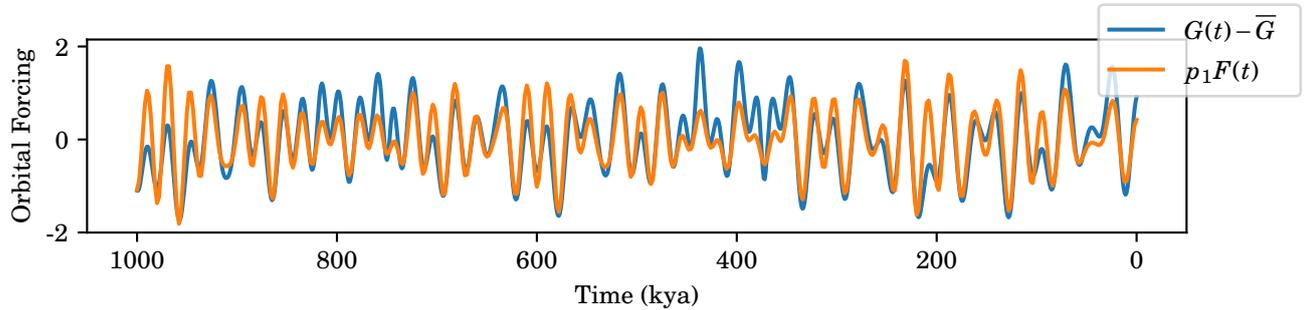}
  \caption[Forcing Function Comparison]{The orbital forcing functions used in the Verbitsky
    model ($p_1F(t)$, where $p_1=-0.671$) and our own ($G(t) = p_2\varepsilon(t) + p_3\beta(t) +
    p_4\cos(\rho(t))$, where $p_2$, $p_3$, and $p_4$ are given in Table
    \ref{tab:optimal_params}). Unlike $F(t)$, which is normalised in the
    Verbitsky model, we are summing the scaled orbital parameters, which have
    non-zero mean values. To account for this, we have subtracted the average of
  our our forcing function $\bar{G}$ to make the two functions comparable.}
  \label{fig:verbitsky_F_vs_orbitals}
\end{figure}

A comparison of this simplified linear Verbitsky model and our feedback model is
shown in Figure \ref{fig:verbitsky_simp_fb_comp}, once again showing minimal
difference between the two solutions. Comparing these two models mathematically,
we see they are very similar in structure, aside from two differences. Firstly,
the presence of the $\varepsilon(t)$ term in
\eqref{eq:simple_intrinsic_model_O}. However, as we discussed in Chapter
\ref{chap:feedback_model}, this term has a negligible impact on the solution and
is there for consistency with the feedforward model. Secondly, as mentioned, the
orbital forcing in Verbitsky's model is $F(t) = Q_{65}$, whilst our original
feedback model uses a linear combination of the three orbital parameters. These
two forcing functions can be seen in Figure \ref{fig:verbitsky_F_vs_orbitals}.

We see that both of these functions have a similar shape and amplitude, and
indeed, as Figure \ref{fig:comp_q65_orb_sol} shows, refitting the feedback model
with $Q_{65}$ as the forcing function does not significantly affect the optimal
solution. Despite these differences between the models, we can see that
Verbitsky's glaciation area $S(t)$ and basal temperature $\theta(t)$ are
analogous to our feedback model's ice volume $I(t)$ and ocean temperature $O(t)$
respectively.

This leads us to question if perhaps the ocean temperature variable in our
models might be better interpreted as basal temperature, as in the Verbitsky
model. The Verbitsky model is derived from physical principles and therefore may
be more physically plausible than our approach of producing a phenomenological
model first and then proposing a physical interpretation afterwards. It is
interesting to see that, despite these differing approaches, the models are
able to reproduce the data to a similar degree.
\begin{figure}
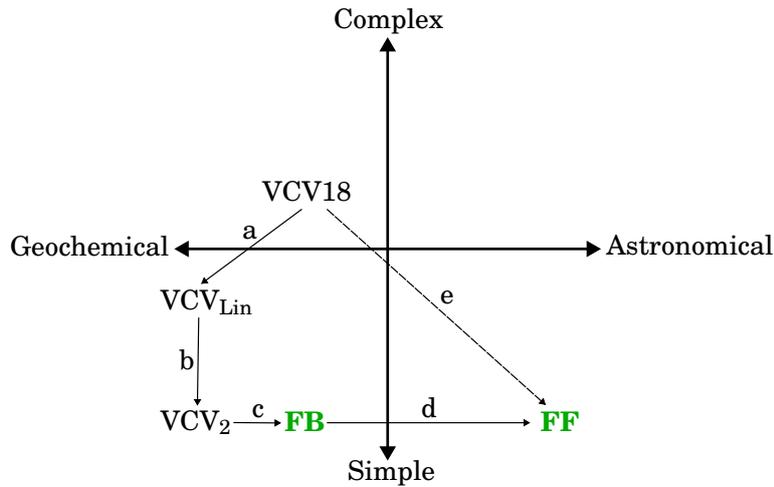

  \centering
  \def\svgwidth{0.6\textwidth}
  
  \caption[Verbitsky Augmentation Map]{A simplified representation of the model
    landscape as shown in Figure \ref{fig:benth_and_power_specs}. This time
    showing the Verbitsky model (VCV18) along with the linearised version
    (VCV$_\mathrm{Lin}$), the linear version with just two variables
    (VCV$_2$), and our two models; Feedback (FB) and Feedforward (FF). The solid
    arrows indicate a single augmentation to the previous model and the dashed
    arrow shows the comparison we are making between the start and end of these
    augmentations. The two models relating to each augmentation are shown in
    \textbf{a}: Figure \ref{fig:optim_verbitsky_comp}, \textbf{b}: Figure
    \ref{fig:verbitsky_lin_simp_comp}, \textbf{c}: Figure
    \ref{fig:verbitsky_simp_fb_comp}, \textbf{d}: Figure 
    \ref{fig:linear_intrinsic_ice_compare}, and \textbf{e}: Figure
  \ref{fig:verbitsky_orig_ff_comp}.}
  \label{fig:verbitsky_aug_map}
\end{figure}

With the similarities between the linearised Verbitsky model and our feedback
model established, we now expand the comparison to include the original
Verbitsky model and our feedforward model. Figure \ref{fig:verbitsky_aug_map}
presents the models discussed in this section, arranged according to their
alignment with the geochemical and astronomical theories and their relative
complexity. Each arrow, starting from the original Verbitsky model through to
the feedforward model, has a corresponding figure to show that no significant
changes occur between the solutions. The augmentations up to our feedback model
have allowed us to reduce the complexity of the model, whilst maintaining the
same alignment with the geochemical theory, and the same fit to the data. The
final arrow from the feedback model to the feedforward model changes the regime
to a system that cannot oscillate without forcing, aligning with the
astronomical theory. As discussed, this change leads to a loss of accuracy in
explaining the data, however the difference is largely concentrated around MIS
11 and we have explored ways in which this discrepancy could be explained.

\begin{figure}
  \hspace{-20pt}
  \input{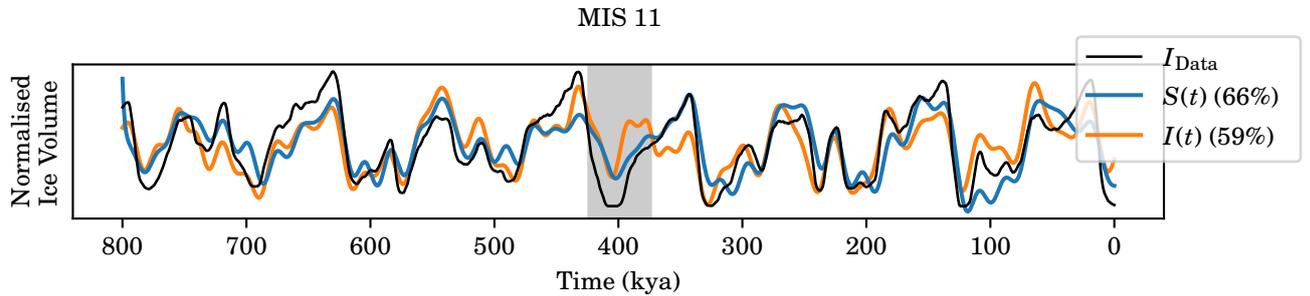}
  \caption[Verbitsky vs Feedforward]{The ice surface area solution from the
    original Verbitsky model $S(t)$ and our feedforward model ice volume
    solution $I(t)$. These use parameters that have been separately optimised to
    fit the ice volume data $I_\mathrm{Data}$.}
  \label{fig:verbitsky_orig_ff_comp}
\end{figure}

Now comparing the original Verbitsky model to our feedforward model, we are
jumping diagonally across the classification space from a non-linear, three
dimensional, geochemically aligned model to a linear, two dimensional, and
astronomically aligned model. The two solutions are shown in Figure
\ref{fig:verbitsky_orig_ff_comp}. As is to be expected, there are notable
differences between these two models, added to by the MIS 11 discrepancy in the
feedforward model. However, aside from the period around MIS 11, the only other
large discrepancy appears around 100\,kya, when both solution diverge from the
ice volume data.
\subsection{Linearised Crucifix}
For this section we have focussed on the Verbitsky model because it is of
intermediate complexity whilst not relying on a switching mechanism to represent
the Earth system. This allows it to be linearised and compared to our models.
However, the Crucifix model (C11) \cite{crucifix_original} also omits a switching mechanism
and so could be used for this comparison. To avoid repetition, we have not done
the full process of augmenting the model to show how it compares to our models.
Instead, we will just show that it can also be linearised without significantly
changing the degree to which it explains the data, further supporting the notion
that the data does not necessitate the non-linearity in these models.

For convenience, the Crucifix model is given by
\begin{align}
  \tau\frac{\mathrm{d}V(t)}{\mathrm{d}t} &= -D(t) + \beta + \gamma Q_{65}(t)\\
  \tau\frac{\mathrm{d}D(t)}{\mathrm{d}t} &= -\alpha \frac{D(t)^3}{3} + D(t) + V(t),
\end{align}
where $V(t)$ is the ice volume variable.

By performing the linearisation process as we did for the Verbitsky model, we
find that the linearised Crucifix model is given by
\begin{align}
\tau\frac{\mathrm{d}V(t)}{\mathrm{d}t} &= \frac{\alpha\beta}{3} - \beta + \gamma
Q_{65}(t) - D(t)\\
\tau\frac{\mathrm{d}D(t)}{\mathrm{d}t} &= \frac{\alpha^2\beta}{9} -
\frac{2\alpha\beta}{3} - \frac{\alpha D(t)}{3} + D(t) + V(t).
\end{align}

\begin{figure}
  \centering
  \input{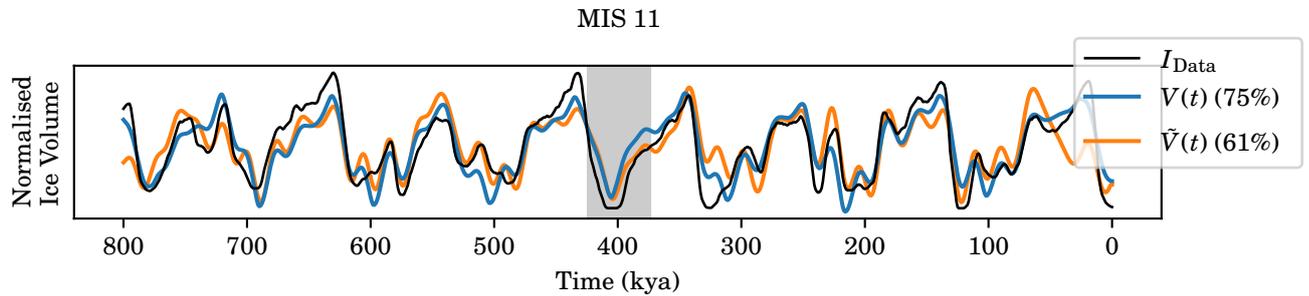}
  \caption[Crucifix Original vs Linearised]{The ice volume solution from the
    original Crucifix model $V(t)$ and the linearised version $\tilde{V}(t)$.
    These use parameters that have been separately optimised to fit the ice
  volume data $I_\mathrm{Data}$.}
  \label{fig:optim_vdp_comp}
\end{figure}

We can see these two models compared against the ice volume data in Figure
\ref{fig:optim_vdp_comp}. We see that the linearised Crucifix model explains
notably less of the variance in the data than the original model. However, the
two signals both visually align similarly with the data, apart from the most
recent interglacial, which is also poorly reproduced by the PP04 model and our
feedback model, as shown in Figure \ref{fig:comp_models}. This demonstrates how
the variance explained should be considered as more of an ordinal measurement of
performance, rather than an absolute metric.

\section{Conclusion}
In this section, we have shown that the differences between our feedforward and
feedback model solutions are mostly concentrated around MIS 11, and that an
external forcing mechanism on Earth could explain this discrepancy. This is
interesting because it prevents us from classifying either model as less valid,
despite aligning with opposing theories of climate change. Since the majority of
conceptual models presented in this paper align more with the geochemical
theory, it is significant that the data does not necessitate this alignment.

We have also shown that both the Verbitsky and Crucifix models can be linearised
without significantly impacting their ability to explain the data. Further, in
the case of the Verbitsky model, we have shown that the physical validity of the
original solution is not lost from this linearisation, with the exception of
$a$ parameter, which allowed the linear model to produce unforced oscillations.
This suggests that the non-linearity in these models is not crucial for
explaining the data.

By further simplifying the Verbitsky model, we have shown that it is
mathematically very similar to our linear, two-dimensional feedback model. This
allowed us to draw parallels between the model variables, with the ocean
temperature in our feedback model being analogous to the basal temperature in
the Verbitsky model. From this, we might infer that the ocean temperature in our
models is better interpreted as basal temperature, though this would need to be
further explored.

This mathematical similarity between the Verbitsky model and our feedback model,
when considered alongside the demonstrated relationship between feedback and
feedforward models, creates an interesting connection across the model
classification space. Despite these models traditionally being aligned with
different theoretical foundations, we observe reasonable alignment between them.
This suggests that the observational data alone neither supports one theoretical
framework over another, nor necessitates the additional complexity present in
more sophisticated models. Indeed, the extra physical mechanisms and assumptions
present in more complex models may not be justifiable from the data alone, which
may be the most appropriate basis for developing models of systems this complex
and poorly understood.

\clearemptydoublepage
\chapter{Discussion}
\label{chap:discussion}
\initial{I}n this thesis, we examined the fundamental dynamics governing Earth's
glacial-interglacial cycles through the development and analysis of conceptual
models. We designed our feedforward and feedback models to be the simplest
representations of the astronomical and geochemical theories respectively,
whilst still reproducing the data well. We evaluated these models alongside a
range of existing conceptual models to understand the necessity of features such
as feedbacks, switching mechanisms, and other non-linear dynamics. The existing
models were chosen to reflect a range of complexity and alignment with either
the astronomical and geochemical theory. By introducing this classification
framework, we compared model performance in the context of their underlying
assumptions, helping to structure our understanding of Earth system dynamics.
Finally, we compared the behaviour of our two models to evaluate the relative
importance of orbital forcing versus Earth system feedbacks in driving these
cycles. We will now present the key findings from this thesis, discuss the
implications and limitations of these results, and suggest future research
directions.

\section{Search for Minimality}
This thesis takes a first principles approach to explain ice volume dynamics
over the past 800\,kyr by identifying the minimal mechanisms needed to
reproduce key features in the data. Whilst we acknowledge that Earth's climate
system contains inherent non-linearities and complex feedbacks, our linear
models demonstrate that the fundamental patterns in ice volume data can be
captured without invoking complex non-linear dynamics. This finding suggests
that non-linear mechanisms, though present in the Earth system, may not be
essential for explaining glacial-interglacial cycles at this timescale.

The feedforward model was designed to be the simplest possible representation of
the astronomical theory that could reproduce the ice volume data. Through
systematic parameter analysis, we demonstrated that each of the terms in the
model were necessary to reproduce the data, with precession being the least
important and the two eccentricity terms being the most important. We were able
to show that this linear model, which could not produce unforced oscillations,
was able to explain a significant portion of the variance in the data. This was
a key result, as it showed that a model that omits common features in existing
models, such as a switching mechanism or feedbacks, could still perform well.

The feedback model was developed by adding a feedback term to the  feedforward
model allowing for unforced oscillations to occur. The purpose of this model was
to investigate the opposing geochemical theory whilst maintaining the
formulation of the feedforward model, omitting a switching mechanism or
non-linear dynamics. The feedback model explained significantly more variance in
the data than the feedforward model, performing better than a number of existing
models. It is interesting that this notable improvement arose from just a single
additional linear term. This demonstrates the significant impact of model theory
alignment, revealing how geochemical models can achieve better data
representation through the incorporation of feedback mechanisms, even when
implemented with minimal complexity.

Through eigenvalue analysis, we showed that the feedback model performs
optimally when it takes a hybrid approach between the geochemical and
astronomical theories, producing damped unforced oscillations that resonate with
eccentricity to boost and phase-lock the 100\,kyr period. This is in contrast to
the PP04, IIL11, and C11 models, which all produce sustained unforced
oscillations. This result supports the notion that the Earth system is driven by
orbital forcing to some degree, rather than just phase locking its internal
oscillations to the orbital frequencies.

\begin{figure}
  \centering
  \includegraphics[width=0.9\textwidth]{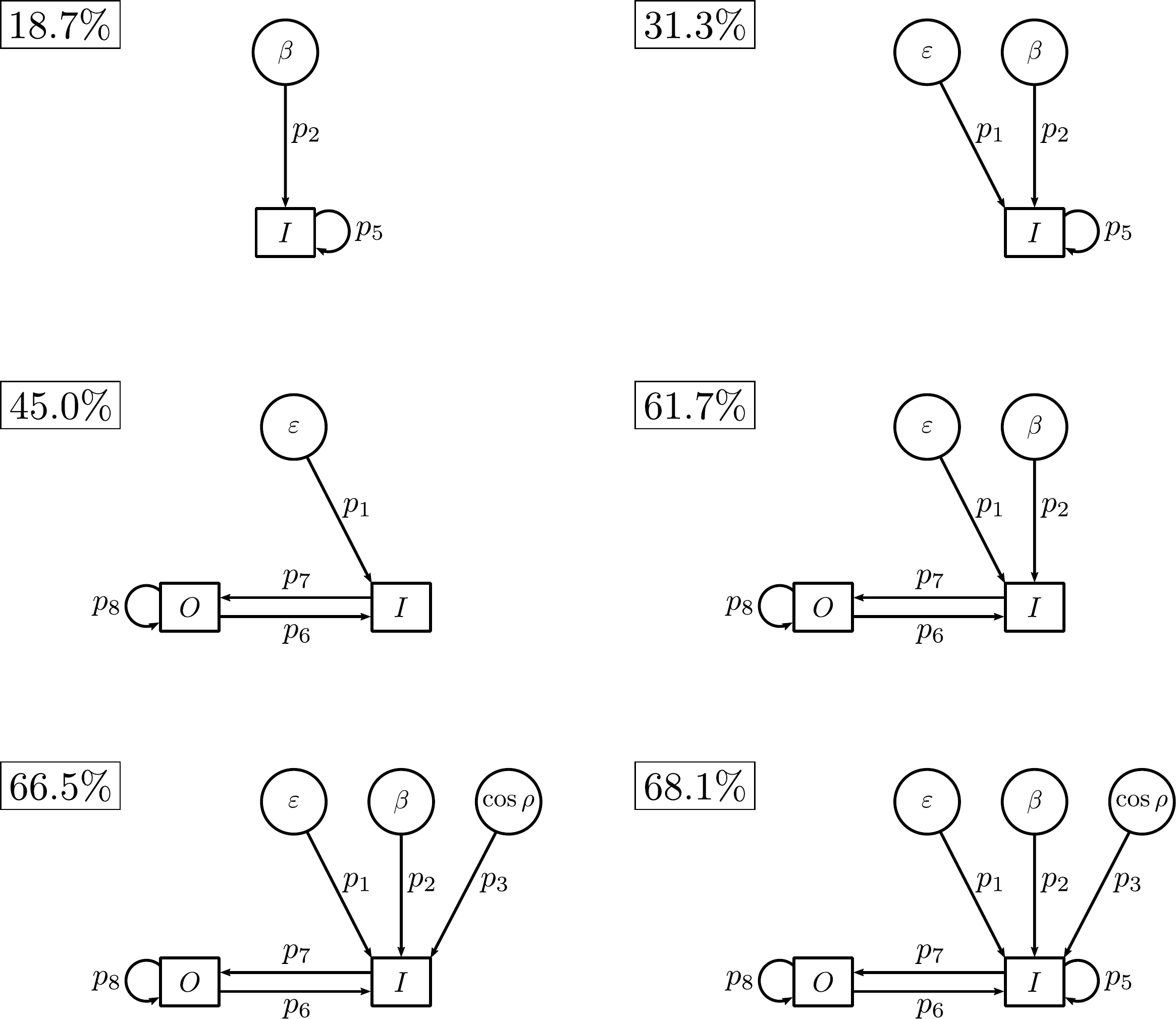}
  \caption[Feedback Leave-One-Out $n$ Removed]{Model flow diagrams to represent
    the best performing model for each bin in Figure
    \ref{fig:intrinsic_leave_one_out}. For conciseness, we have omitted the full
    model, which is shown in Figure \ref{fig:feedback_diagram_1_removed}, and
    the model with only one term as all versions explain 0\% of the variance in
    the data. The percentage of variance explained by each model is shown next
  to each diagram.}
  \label{fig:feedback_diagram_n_removed_2}
\end{figure}

As with the feedforward model, we performed a systematic parameter analysis of
the feedback model to understand the relative importance of each term. This
analysis went further and helped us to understand which mechanisms were the most
fundamental to explaining the ice volume data. Figure
\ref{fig:feedback_diagram_n_removed_2} shows the best performing model for each
number of terms allowed in the model. The first two sub-models do not reproduce
the data sufficiently. However, the third sub-model introduces the second
dynamic variable, marking a notable improvement and achieving a better fit to
the data than the II80 model. This demonstrates that a minimum of two dynamic
variables are likely needed to explain the data. An interesting finding from
this analysis is that the feedforward model, which is a subset of the feedback
model, does not appear in any of the best performing sub-models, with the
feedback mechanism appearing at the earliest stage possible. The parameter
analysis of the feedforward model demonstrated that having eccentricity feed
into both dynamic variables was the most important feature of the model, whereas
here we see the second eccentricity term being omitted from every sub-model.
This further highlights the importance of model theory alignment, with even our
highly simplified models ascribing drastically different importance to the same
mechanisms.

Although the feedback model outperformed the feedforward model, this improvement
was largely concentrated around MIS 11. Both models perform comparably
throughout the rest of the 800\,kyr record, suggesting that direct orbital
forcing may adequately explain most ice volume dynamics. If MIS 11 represents
an anomaly caused by factors outside the normal orbital mechanisms, such as
volcanic events or ice sheet instability, then the data alone cannot rule
out the simpler feedforward approach. Indeed, when we added time-dependent terms
to the feedforward model to simulate such Earth-based events during MIS 11, its
performance matched the feedback model while maintaining physical plausibility.
This finding challenges the assumption that Earth-based feedback mechanisms are
essential for explaining the ice volume data.

The improved performance around MIS 11 from the feedback model is due to the
unforced oscillations that it produces. Although they resonate with
eccentricity, the resultant signal is more consistent in amplitude than the
100\,kyr signal produced from the feedforward model, which more closely follows
the eccentricity curve. This is shown in Figure
\ref{fig:deps_approx_power_spec_2}B, where the eccentricity component of the
feedforward model $I_\varepsilon(t)$ still exhibits a decrease in amplitude
around 400\,kya which aligns with a minimum in eccentricity's 400\,kyr period.

As discussed, we developed our models in order to produce the minimal set of
mechanisms necessary to explain the ice volume data. This indirectly resulted in
producing models that were linear, a feature that is not present in any of the
models from the literature that we examined. To explore the degree to which
these models need non-linearity to explain the data, we linearised the VCV18 and
C11 models, which were chosen because they employ continuous differential
equations without regime switching mechanisms. We showed that the linearised
models were able to explain the data to a similar extent as their non-linear
counterparts, further supporting the data exhibits dynamics that do not deviate
significantly from a linear response to orbital forcing. This is a significant
result, as it suggests that the Earth system may be more predictable than
previously thought, with simple linear models being able to capture the
essential dynamics of the system.

The linearised VCV18 model maintained similar parameters to the original model,
except for the damping parameter, which changed enough to transform damped
oscillations into sustained ones. Since the model uses $Q_{65}$ as its forcing
function, which contains almost no eccentricity signal, it must produce the
100\,kyr cycle through internal feedbacks rather than resonance with
eccentricity. This raises questions about whether $Q_{65}$ adequately represents
orbital forcing, particularly given our finding that eccentricity can
significantly influence ocean temperature.

\begin{figure}
  \centering
  \input{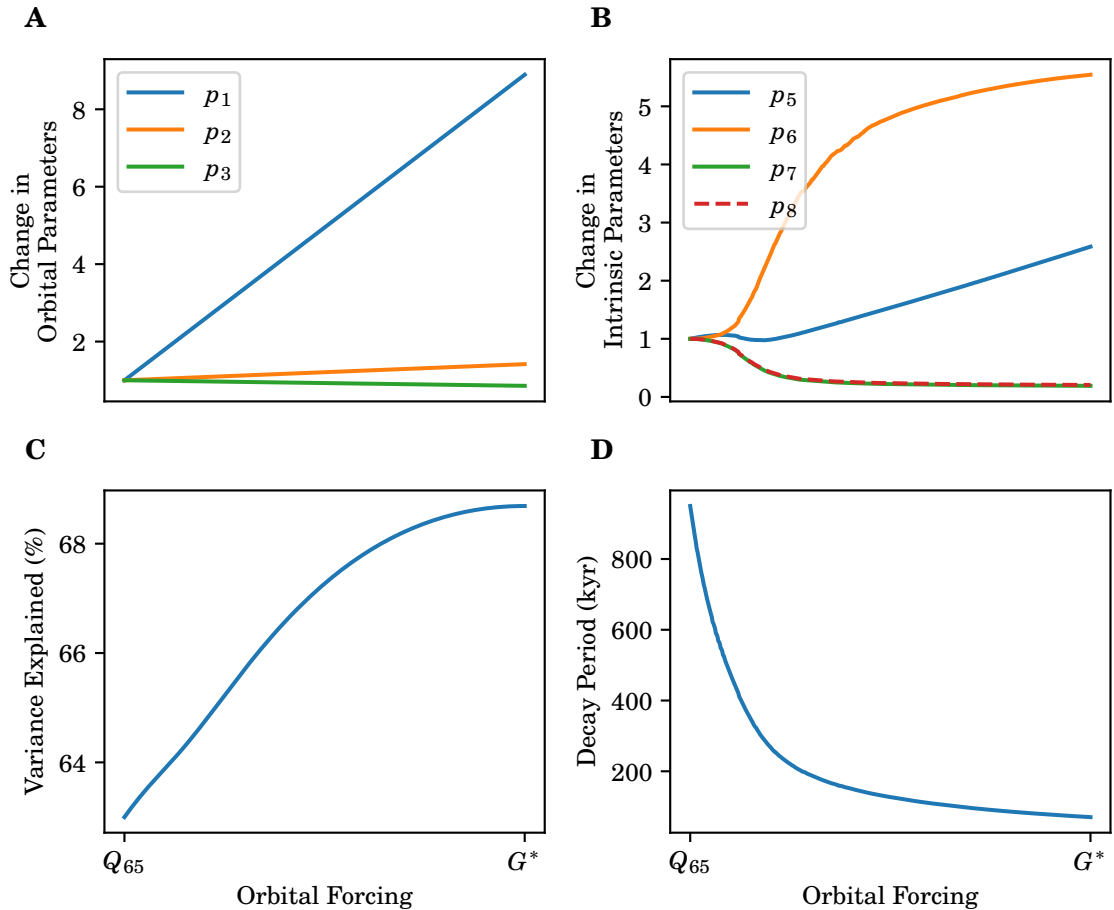}
  \caption[Sweep $Q_{65}$ to Orbital Forcing]{Sweeping over the orbital forcing,
    starting with a linear approximation of $Q_{65}$ and then optimising the
    orbital parameters to get the optimal forcing function $G^*$. The
    relative change in orbital and intrinsic parameters, relative to their
    values in the $Q_{65}$ case are shown in (\textbf{A}) and (\textbf{B})
    respectively. The variance explained by the model throughout the sweep is
    shown in (\textbf{C}). The decay rate of the unforced oscillation, which
    maintains an approximately constant period around 98\,kyr, is shown in
  (\textbf{D}).}
  \label{fig:sweep_q65_to_optim_orb_2}
\end{figure}

Figure \ref{fig:sweep_q65_to_optim_orb_2} shows how our feedback model changes
when transitioning from $Q_{65}$ to an individually weighted orbital forcing
function. With even a small increase in eccentricity's weighting, the model's
optimal damping rate shifts from producing sustained oscillations to damped
oscillations that resonate with eccentricity. This means that, although
the model could optimally produce sustained oscillations throughout the sweep,
as soon as eccentricity became more significant, the model's optimal behaviour
changed to resonance. This is consistent with our eigenvalue analysis for the
feedback model. We have therefore shown that $Q_{65}$, which is used in the
majority of models we have examined, is potentially under estimating the
significance of eccentricity. If an insolation measure that placed more emphasis
on eccentricity were used, it could significantly change the underlying dynamics
of the models that employ it.

\subsection{Limitations}
Our claim, that the data is consistent with a linear response to orbital
forcing, needs qualification. Whilst we acknowledge that there are definitely
non-linear mechanisms involved, given that the data can be reproduced by linear
models, it cannot be used to argue for any particular non-linear mechanism.

Our phenomenological modelling approach helps to identify simple governing
mechanisms, but complicates the physical interpretation of variables. This was
evident with the slow-changing variable in both of our models. Whilst ocean
temperature seemed plausible, with calculations confirming eccentricity could
drive the observed temperature changes, the optimised feedforward model
counterintuitively showed ocean temperature positively impacting ice volume. The
underdetermined feedback model offered multiple interpretations, in addition to
the feedforward model interpretation, ocean temperature could negatively impact
ice volume, but ice volume would then positively impact ocean temperature. We
suggested this could occur through cold meltwater from receding ice sheets
cooling the ocean. However, neither of these interpretations produced a good
reconstruction of the ocean temperature proxy data. The linearised VCV18 model,
which is dynamically similar to our feedback model, interpreted this slow
variable as basal temperature, offering another plausible physical
interpretation.

The VCV18 model and others from the literature were designed using physical
mechanisms from the start, combining these to explain ice volume data through
physical processes. These models were naturally non-linear since physical
quantities within the Earth system rarely change linearly. This approach made
individual variables physically interpretable and verifiable against proxy data.
Although this represents a more robust approach to model development, we have
shown that it potentially introduces more complexity than is needed to
understand the ice volume dynamics. Given that these physically driven models
still impose significant simplifications, they cannot explain the full system
dynamics, but rather the dynamics of the variables that they include. This
limitation applies to all models, with variable selection depending on the
research question. We have asked what minimal set of dynamical mechanisms drive
the glacial-interglacial cycles, rather than what physical processes are
responsible for these cycles.

\section{The Role of Eccentricity}
In order to understand how the orbital parameters influence Earth's insolation
patterns, we first derived equations for daily and yearly averaged insolation
as a function of latitude. A key finding was an apparent incongruence between the
magnitudes of orbital forcing and their representation in the ice volume data.
Specifically, eccentricity's 100\,kyr period appears far too prominent in the
data, given it can only scale global insolation by at most 0.18\%. This
observation presents two possible explanations: either insolation plays a lesser
role as a driver, with the 100\,kyr period emerging primarily from Earth-based
mechanisms, or eccentricity's impact has been misunderstood.

\begin{figure}
  \centering
  \def\svgwidth{0.6\textwidth}
  \import{../figs/}{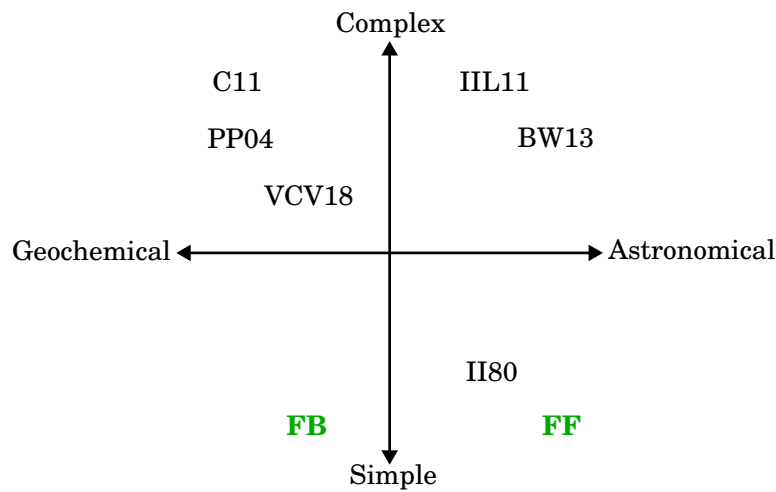}
  \caption[Conceptual Model Landscape]{A simplified representation of the
    conceptual model landscape for those discussed in this thesis. The
    horizontal axis represents the extent to which the model aligns with either
    the astronomical theory or the geochemical theory. The vertical axis
    represents the complexity of the model. The feedback (FB) and feedforward
    (FF) models that we have developed are shown in green. The models are
    \textbf{C11}: Crucifix (2011)~\cite{crucifix_original}, \textbf{IIL11}:
    Imbrie, Imbrie-Moore, and Lisiecki (2011)~\cite{imbrie2011}, \textbf{PP04}:
    Paillard and Parrenin (2004)~\cite{pp04}, \textbf{BW13}: Budyko
    (1969)~\cite{budyko} augmented by Widiasih (2013)~\cite{widiasih2013},
    \textbf{VCV18} Verbitsky, Crucifix, and Volobuev
    (2018)~\cite{verbitsky2018}, and \textbf{II80}: Imbrie and Imbrie
    (1980)~\cite{imbrie}.}
\label{fig:conceptual_model_landscape_2}
\end{figure}

In order to understand the context of this problem, we implemented six existing
models and examined them with standardised inputs and tuning conditions. We used
a classification framework to select models representing both astronomical and
geochemical theories across a range of complexities, as shown in Figure
\ref{fig:conceptual_model_landscape_2}, with our models highlighted in green.
The models were evaluated on their ability to reproduce ice volume data and
their underlying mechanisms. We first examined the updated Budyko model (BW13),
which models latitudinally varying temperature and conceptualises ice volume as
a single line representing ice sheet extent. This model demonstrates ice-albedo
feedbacks, though traditionally focuses on steady state solutions. By adding
varying orbital parameters to the insolation function, we produced an 800\,kyr
ice line solution that captured obliquity's impact but omitted eccentricity and
precession frequencies. Despite introducing several modifications to improve
realism, which helped capture precession frequencies, the model still failed to
reproduce the 100\,kyr eccentricity cycle.

The five remaining models differed from BW13 in that they were one-dimensional
and were designed with the purpose of reproducing the ice volume data, all
producing the 100\,kyr period to some extent. Three of the models, those on the
geochemical side of model landscape in Figure
\ref{fig:conceptual_model_landscape_2}, produced this period through the use of
feedback mechanisms, allowing for unforced oscillations to occur. The physical
implication of this is that the Earth system has a natural tendency to oscillate
around this frequency, and matches the phase of eccentricity through mechanisms
such as non-linear resonance. Other possible explanations for this orbital
alignment in the geochemical theory include stochastic resonance, whereby random
climate variations could help to amplify the weak eccentricity forcing, or
combination tones, whereby the higher frequency obliquity and precession signals
interact to produce a $\sim100$\,kyr signal.

The geochemical theory is underpinned by the assumption that eccentricity
is too weak to impact the climate directly. However, we used a simplified
ocean heating model to show that the ocean's high heat capacity could allow for
the eccentricity signal to have a cumulative effect over thousands of years.
This showed that the Earth system may be more sensitive to orbital forcing than
is often assumed, providing support for the astronomical theory of climate
change. Despite this supporting evidence, there is another important limitation
of the astronomical theory that this analysis did not address. Eccentricity
contains an even more prominent 400\,kyr period that is not present in the ice
volume data.

\begin{figure}
  \centering
  \input{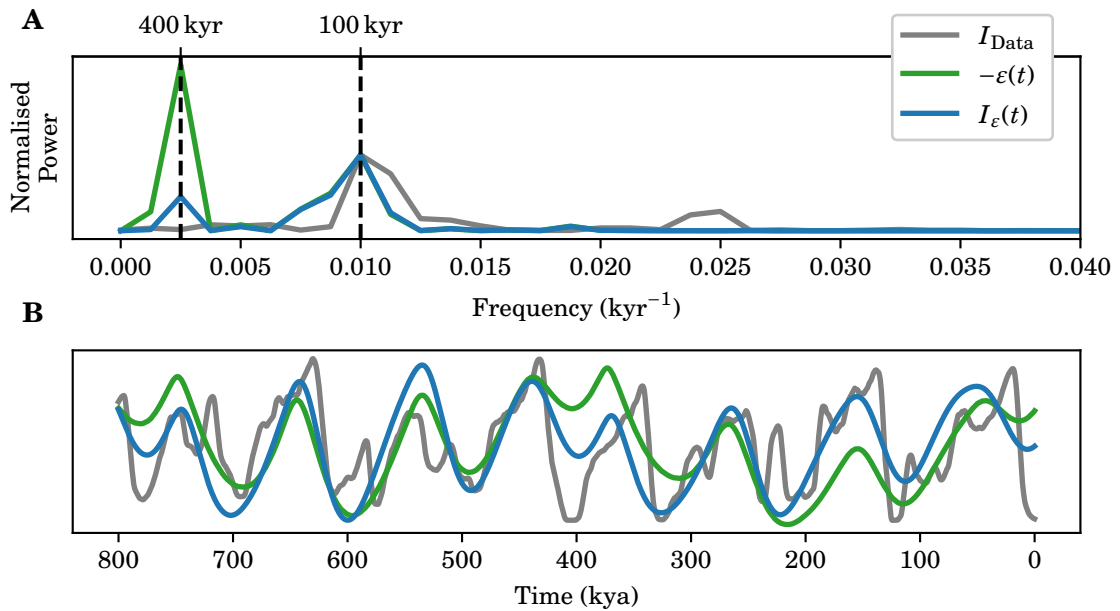}
  \caption[Eccentricity 400\,kyr Removal]{\textbf{A}: Power spectra comparison
    of the ice volume data, the negative eccentricity curve $-\varepsilon(t)$,
    and the eccentricity component of our model solution $I_\varepsilon(t)$.
    They have been normalised to equate the powers corresponding to 100\,kyr.
    Our $I_\varepsilon(t)$ power spectrum matches that of $\varepsilon(t)$ apart
    from a significant drop around the 400\,kyr period. \textbf{B}: Time series
    showing the same comparison, also normalised for qualitative comparison.
    They show how the filtered eccentricity better captures the broad behaviour
    of the ice volume curve, namely around 780, 400, 320, and 160\,kya.}
  \label{fig:deps_approx_power_spec_2}
\end{figure}

We developed the feedforward model to allow for eccentricity's 100\,kyr period
to directly impact ice volume variations, whilst minimising the effect of its
400\,kyr period. We were able to achieve this by including eccentricity twice
in the system, influencing the ice volume dynamics with differing phase lags and
opposing signs. In Figure \ref{fig:deps_approx_power_spec_2}, we show the sum of
these two terms, alongside the original eccentricity curve and the ice volume
data. The power spectrum for the eccentricity signal $\varepsilon(t)$ shows the
dominant 400\,kyr period, which does not occur in the ice volume data. On top of
this is the power spectrum for the feedforward model's eccentricity component
$I_\varepsilon(t)$, which produces the same 100\,kyr period, but significantly
reduces the 400\,kyr period, more closely matching the ice volume power
spectrum. This is further demonstrated in the time series comparison, where the
eccentricity component of the feedforward model matches the broad behaviour of
the ice volume data notably better than eccentricity itself. This addressed a
significant criticism of the astronomical theory, suggesting that the
glacial-interglacial cycles could be directly forced by the orbital parameters,
as opposed to internal feedbacks that produce unforced oscillations.

\subsection{Limitations}
This result is significant as it challenges a commonly held assumption that the
100\,kyr period in the ice volume data cannot be directly caused by
eccentricity. However, there are a number of limitations to this analysis that
should be considered. Firstly, the ocean temperature model that we used to
estimate the magnitude of eccentricity's impact was highly simplified. The model
presents a global average temperature, neglecting the effects of ocean currents
and heat transport. We treat heat loss as a single linear function of
temperature, although in reality this occurs through a combination of back
radiation, convection, conduction, and evaporation, which are non-linear
processes. We also compared warming rates for extreme eccentricity values,
whereas this effect would be more gradual in reality. However, the calculated
difference in warming rate (0.319$^\circ$C/kyr) was still safely above the
maximum rate that we see in the proxy data ($\sim0.2^\circ$C/kyr). This suggests
that the model is at least plausible, though further work is needed to improve
its realism.

Another limitation of the feedforward model relates to the two inputs of
eccentricity that require differing phase lags and signs. According to our
physical interpretation, ocean temperature is the source of the lagged
eccentricity signal feeding into ice volume, alongside the faster radiative
effects on the troposphere. The phase lag emerging from the ocean's high heat
capacity is physically reasonable, as it takes significantly longer for the
ocean to warm up from insolation changes than the air. However, the time
constant to govern this slower response, which was attained from fitting the
model to the ice volume data, was at the upper limit of estimated values from
the literature (15\,kyr). We found that a time constant of 10\,kyr can still
produce a similar fit to the data, though it is interesting that the optimal
value is this high. This finding contributed to our conclusion that the
slow-changing variable is likely not modelling ocean temperature directly, and
instead could be an abstraction of multiple physical variables, or a different
aspect of the climate system entirely. This highlights the challenge of
assigning physical meanings to abstract model variables, especially in
simplified models where variables may represent composite effects, rather than
single physical quantities.

\section{Astronomical Theory vs Geochemical Theory}

The comparison between our two models provides unique insight into the
long-standing debate between astronomical and geochemical theories. The
feedforward and feedback models represent minimal implementations of these
competing theories, allowing us to directly compare their explanatory power with
minimal confounding factors. Our analysis showed that while the feedback model
performed notably better overall, this difference was largely concentrated
around MIS 11, suggesting that for most of the 800\,kyr record, direct orbital
forcing provides an equally good explanation of ice volume dynamics as feedback
mechanisms.

The choice of orbital forcing function significantly impacts model dynamics and
theoretical interpretation. Our linearisation of the VCV18 model revealed that,
while it maintained similar performance to its non-linear counterpart, the
damping parameter shifted to produce sustained, rather than damped, oscillations.
This behaviour stems from the model's use of $Q_{65}$, which contains minimal
eccentricity signal. In contrast, our feedback model, which uses individually
weighted orbital parameters, performed optimally with damped oscillations that
resonated with eccentricity. As Figure \ref{fig:sweep_q65_to_optim_orb_2}
demonstrates, even a slight increase in eccentricity's weighting shifts the
optimal behaviour from sustained oscillations to damped, resonance-based
dynamics. This suggests that the widespread use of $Q_{65}$ may inadvertently
bias models toward geochemical explanations when a hybrid approach might be more
appropriate if eccentricity were given greater weight.

\begin{figure}
  \centering
  \def\svgwidth{0.6\textwidth}
  \import{../figs/}{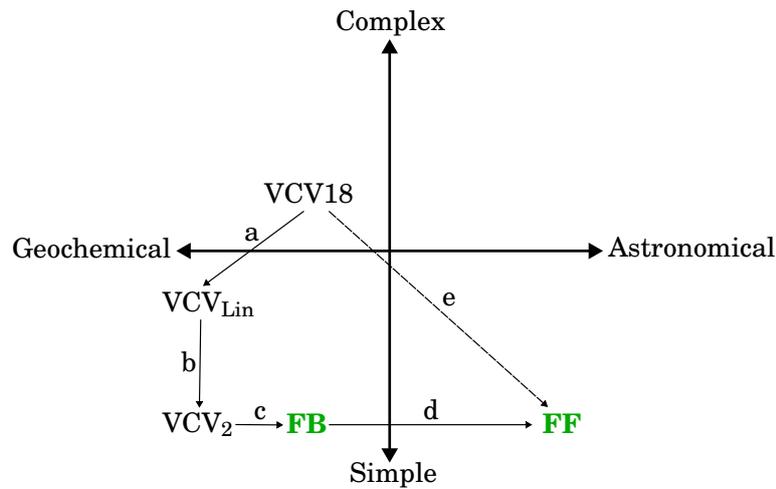}
  \caption[Verbitsky Augmentation Map]{A simplified representation of the model
    landscape as shown in Figure \ref{fig:benth_and_power_specs}. This time
    showing the Verbitsky model (VCV18) along with the linearised version
    (VCV$_\mathrm{Lin}$), the linear version with just two variables
    (VCV$_2$), and our two models; Feedback (FB) and Feedforward (FF). The solid
    arrows indicate a single augmentation to the previous model and the dashed
    arrow shows the comparison we are making between the start and end of these
    augmentations. The two models relating to each augmentation are shown in
    \textbf{a}: Figure \ref{fig:optim_verbitsky_comp}, \textbf{b}: Figure
    \ref{fig:verbitsky_lin_simp_comp}, \textbf{c}: Figure
    \ref{fig:verbitsky_simp_fb_comp}, \textbf{d}: Figure 
    \ref{fig:linear_intrinsic_ice_compare}, and \textbf{e}: Figure
  \ref{fig:verbitsky_orig_ff_comp}.}
  \label{fig:verbitsky_aug_map_2}
\end{figure}

The augmentation pathway from the VCV18 model to our feedforward model
demonstrates the connection between seemingly opposing theories. As visualized
in Figure \ref{fig:verbitsky_aug_map_2}, we systematically transformed a
non-linear, geochemically-aligned, three variable model into a linear,
astronomically-aligned, two variable model without significant loss of
explanatory power. This suggests that the fundamental divide between these
theories may be narrower than previously thought, with their differences
potentially arising from modelling choices rather than irreconcilable views of
Earth system behaviour.

Our eigenvalue analysis of the feedback model provides evidence for a hybrid
view of glacial-interglacial dynamics. The optimal eigenvalues produce damped
oscillations with a period of approximately 100\,kyr that decay over about
68\,kyr. This suggests that, while the Earth system may have an intrinsic
oscillatory behaviour, this behaviour is not self-sustaining but instead requires
orbital forcing to maintain its amplitude and phase. This hybrid mechanism,
where internal feedbacks create an oscillatory tendency that is then boosted 
and synchronised by orbital forcing, represents a middle ground between the
astronomical and geochemical theories.

The model landscape traversal in Figure \ref{fig:verbitsky_aug_map_2} reveals an
important methodological insight about theory evaluation. Our ability to
transform between models with different theoretical frameworks while maintaining
similar performance indicates that the ice volume data alone may be insufficient
to definitively resolve the astronomical versus geochemical debate. This
underscores the need for additional constraints from proxy data, physical
mechanisms, and careful consideration of model parsimony when evaluating
competing theories of climate change.

\section{Future Work}
Several promising directions for future research emerge from this work. First, a
more detailed investigation of the relationship between CO$_2$ variations and ice
volume could help clarify whether additional feedback mechanisms need to be
incorporated into our minimal models. This could involve developing a simple
carbon cycle component that maintains the overall model parsimony whilst
capturing essential CO$_2$ dynamics. Such an extension would address one of the
key limitations of our current approach while testing whether the fundamental
conclusions about linear dynamics remain valid when carbon feedbacks are
included.

The physical interpretation of model variables, particularly in the feedback
model, could be refined through more detailed comparison with proxy data. This
might involve exploring alternative interpretations, such as treating the slow
variable as basal temperature rather than ocean temperature, as suggested by the
comparison with the Verbitsky model. This would help bridge the gap between
statistical fit and physical understanding, potentially resolving the apparent
contradictions in how our variables relate to measurable Earth system
components.

Further investigation of the MIS 11 period, particularly through the lens of
volcanic activity and ice sheet instability, could help explain why this interval
deviates from typical orbital forcing responses. This might lead to insights
about potential future departures from expected glacial-interglacial patterns.
Developing methods to identify and characterise these anomalous periods could
help to improve our ability to distinguish between fundamental model limitations
and specific Earth-based events.

Our models could be extended to examine the MPT in more detail, testing whether
parameter variations can explain the shift from 41\,kyr to 100\,kyr cycles. This
would require careful consideration of which parameters should change and
whether these changes align with proposed physical interpretations. This could
help to determine whether regime shifts require fundamentally different dynamics
or simply changes in parameter values within the same framework.

Finally, the methodological approach of systematic model simplification
demonstrated in this thesis could be applied to other complex Earth system
problems. By identifying the minimal set of dynamical mechanisms that are needed
to explain observed behaviour, this approach could help to clarify which
processes are fundamental and which are secondary or incidental. One possible
Earth system problem would be the Dansgaard-Oeschger events, which are
only discussed briefly in this thesis. This approach could establish a bridge
between complex Earth system models and conceptual understanding, both of which
are essential for developing robust climate predictions.

\clearemptydoublepage

\backmatter
\bibliographystyle{siam}
\refstepcounter{chapter}
\bibliography{refs}
\clearemptydoublepage
%
%
\end{document}